\documentclass{ws-ijmpe}

\usepackage[super,compress]{cite}
\usepackage{graphicx}
\usepackage{epsfig}
\usepackage{epstopdf}
\usepackage[linktocpage,colorlinks,pdfpagemode=UseOutlines]{hyperref}

\begin{document}

\newcommand{\JP}{J/\psi}
\newcommand{\Ima}{\textrm{Im}}
\newcommand{\Rea}{\textrm{Re}}
\newcommand{\mev}{\textrm{ MeV}}
\newcommand{\gev}{\textrm{ GeV}}
\newcommand{\dtres}{d^{\hspace{0.1mm} 3}\hspace{-0.5mm}}
\newcommand{\rts}{ \sqrt s}
\newcommand{\non}{\nonumber \\[2mm]}
\newcommand{\eps}{\epsilon}
\newcommand{\half}{\frac{1}{2}}
\newcommand{\thalf}{\textstyle \frac{1}{2}}
\newcommand{\Nmass}{M_{N}} 
\newcommand{\delmass}{M_{\Delta}} 
\newcommand{\pimass}{\mu}  
\newcommand{\rhomass}{m_\rho} 
\newcommand{\piNN}{f}      
\newcommand{\rhocoup}{g_\rho} 
\newcommand{\fpi}{f_\pi} 
\newcommand{\f}{f} 
\newcommand{\nucfld}{\psi_N} 
\newcommand{\delfld}{\psi_\Delta} 
\newcommand{\fpiNN}{f_{\pi N N}} 
\newcommand{\fpiND}{f_{\pi N \Delta}} 
\newcommand{\GMquark}{G^M_{(q)}} 
\newcommand{\vecpi}{\vec \pi}
\newcommand{\vectau}{\vec \tau}
\newcommand{\vecrho}{\vec \rho}
\newcommand{\delmu}{\partial_\mu}
\newcommand{\delMu}{\partial^\mu}
\newcommand{\nn}{\nonumber}
\newcommand{\vs}{\vspace{-0.20cm}}
\newcommand{\be}{\begin{equation}}
\newcommand{\ee}{\end{equation}}
\newcommand{\ba}{\begin{eqnarray}}
\newcommand{\ea}{\end{eqnarray}}
\newcommand{\ropi}{$\rho \rightarrow \pi^{0} \pi^{0}
\gamma$ }
\newcommand{\roeta}{$\rho \rightarrow \pi^{0} \eta
\gamma$ }
\newcommand{\omepi}{$\omega \rightarrow \pi^{0} \pi^{0}
\gamma$ }
\newcommand{\omeeta}{$\omega \rightarrow \pi^{0} \eta
\gamma$ }
\newcommand{\ul}{\underline}
\newcommand{\del}{\partial}
\newcommand{\rth}{\frac{1}{\sqrt{3}}}
\newcommand{\rsix}{\frac{1}{\sqrt{6}}}
\newcommand{\sq}{\sqrt}
\newcommand{\fr}{\frac}
\newcommand{\pr}{^\prime}
\newcommand{\ov}{\overline}
\newcommand{\Gm}{\Gamma}
\newcommand{\rw}{\rightarrow}
\newcommand{\rgl}{\rangle}
\newcommand{\De}{\Delta}
\newcommand{\Dp}{\Delta^+}
\newcommand{\Dm}{\Delta^-}
\newcommand{\Dz}{\Delta^0}
\newcommand{\Dpp}{\Delta^{++}}
\newcommand{\Sg}{\Sigma^*}
\newcommand{\Sp}{\Sigma^{*+}}
\newcommand{\Sm}{\Sigma^{*-}}
\newcommand{\Sz}{\Sigma^{*0}}
\newcommand{\X}{\Xi^*}
\newcommand{\Xm}{\Xi^{*-}}
\newcommand{\Xz}{\Xi^{*0}}
\newcommand{\Om}{\Omega}
\newcommand{\Omm}{\Omega^-}
\newcommand{\kp}{K^+}
\newcommand{\kz}{K^0}
\newcommand{\pip}{\pi^+}
\newcommand{\pim}{\pi^-}
\newcommand{\piz}{\pi^0}
\newcommand{\et}{\eta}
\newcommand{\kb}{\ov K}
\newcommand{\km}{K^-}
\newcommand{\kbz}{\ov K^0}
\newcommand{\ksb}{\ov {K^*}}

\newcommand{\com}[1]{{\sf\color[rgb]{0,0,1}{#1}}}
\newcommand{\comm}[1]{{\sf\color[rgb]{1,0,0}{#1}}}
\newcommand{\ans}[1]{{\sf\color[rgb]{0,1,0}{#1}}}



\newcommand{\Psfig}[2]{\includegraphics[width=#1]{#2}}
\newcommand{\PsfigII}[2]{\includegraphics[scale=#1]{#2}}
\newcommand{\Expect}[1]{\langle #1 \rangle}
\newcommand{\SUN}[1]{\text{SU} ( #1 )}
\newcommand{\UN}[1]{\text{U} ( #1 )}

\def\tstrut{\vrule height2.5ex depth0pt width0pt} 
\def\jtstrut{\vrule height5ex depth0pt width0pt} 
\def\kev{\text{ keV}}
\def\MeV{\text{ MeV}}
\def\GeV{\text{ GeV}}
\def\fm{\text{ fm}}
\def\beq{\begin{equation}}
\def\enq{\end{equation}}
\def\beqa{\begin{eqnarray}}
\def\enqa{\end{eqnarray}}
\def\kev{\text{ keV}}
\def\mev{\text{ MeV}}
\def\gev{\text{ GeV}}
\def\fm{\text{ fm}}
\def\naive{na\"{i}ve }
\def\naively{na\"{i}vely }
\def\Kaellen{K\"{a}llen }
\def\Schr{Schr\"{o}dinger }

\def\trace{\text{tr}}

\def\phph{\phantom{-}}
\def\phzr{\phantom{0}}

\title{Weak decays of heavy hadrons into dynamically generated resonances}

\author{Eulogio Oset\footnote{The order of the authors follows the chronological order in which their contribution to the work described here was done.}} 
\address{Institute of Modern Physics, Chinese Academy of Sciences, Lanzhou 730000,China}
\address{Departamento de F\'{\i}sica Te\'orica and IFIC, Centro Mixto Universidad de Valencia-CSIC Institutos de Investigaci\'on de Paterna, Aptdo. 22085, 46071 Valencia, Spain}

\author{Wei-Hong Liang} 
\address{Department of Physics, Guangxi Normal University, Guilin 541004, China}

\author{Melahat Bayar} 
\address{Department of Physics, Kocaeli University, 41380 Izmit, Turkey}

\author{Ju-Jun Xie} 
\address{Institute of Modern Physics, Chinese Academy of Sciences, Lanzhou 730000, China}
\address{State Key Laboratory of Theoretical Physics, Institute of Theoretical Physics, Chinese Academy of Sciences, Beijing 100190, China}

\author{Lian Rong Dai}
\address{Department of Physics, Liaoning Normal University, Dalian 116029, China}

\author{Miguel Albaladejo} 
\address{IFIC, Centro Mixto Universidad de Valencia-CSIC, Institutos de Investigaci\'on de Paterna, Aptdo. 22085, 46071 Valencia, Spain}

\author{Marina Nielsen}
\address{Instituto de Fisica, Universidade de S\~ao Paulo, Caixa Postal 66318, 05389-970 S\~ao Paulo, SP, Brazil}

\author{Takayasu Sekihara}
\address{Research Center for Nuclear Physics (RCNP), Osaka University, Ibaraki, Osaka 567-0047, Japan}

\author{Fernando Navarra}
\address{Instituto de Fisica, Universidade de S\~ao Paulo, Caixa Postal 66318, 05389-970 S\~ao Paulo, SP, Brazil}

\author{Luis Roca}
\address{Departamento de F\'isica, Universidad de Murcia, E-30100 Murcia, Spain}

\author{Maxim Mai}
\address{Universit\"at Bonn, Helmholtz-Institut f\"ur Strahlen- und Kernphysik (Theorie) and Bethe Center for Theoretical Physics, D-53115 Bonn, Germany}

\author{Juan Nieves}
\address{IFIC, Centro Mixto Universidad de Valencia-CSIC, Institutos de Investigaci\'on de Paterna, Aptdo. 22085, 46071 Valencia, Spain}

\author{Jorgivan Morais Dias}
\address{Instituto de Fisica, Universidade de S\~ao Paulo, Caixa Postal 66318, 05389-970 S\~ao Paulo, SP, Brazil}

\author{Alberto Feijoo}
\address{Departament d'Estructura i Constituents de la Mat\`eria and Institut de Ci\`encies del Cosmos, Universitat de Barcelona, Mart\'i i Franqu\`es 1, 08028 Barcelona, Spain}

\author{Volodymyr K. Magas}
\address{Departament d'Estructura i Constituents de la Mat\`eria and Institut de Ci\`encies del Cosmos, Universitat de Barcelona, Mart\'i i Franqu\`es 1, 08028 Barcelona, Spain}

\author{Angels Ramos}
\address{Departament d'Estructura i Constituents de la Mat\`eria and Institut de Ci\`encies del Cosmos, Universitat de Barcelona, Mart\'i i Franqu\`es 1, 08028 Barcelona, Spain}

\author{Kenta Miyahara}
\address{Department of Physics, Graduate School of Science, Kyoto University, Kyoto 606-8502, Japan}

\author{Tetsuo Hyodo}
\address{Yukawa Institute for Theoretical Physics, Kyoto University, Kyoto 606-8502, Japan}

\author{Daisuke Jido}
\address{Department of Physics, Tokyo Metropolitan University, Hachioji 192-0397, Japan}

\author{Michael D\"oring}
\address{Department of Physics, The George Washington University, Washington, DC 20052, USA and Thomas Jefferson National Accelerator Facility, Newport News,VA, USA}

\author{Raquel Molina}
\address{Institute for Nuclear Studies and Department of Physics, The George Washington University, Washington, DC 20052, USA}

\author{Hua-Xing Chen}
\address{School of Physics and Nuclear Energy Engineering and International Research Center for Nuclei and Particles in the Cosmos, Beihang University, Beijing 100191, China}

\author{En Wang}
\address{Department of Physics, Zhengzhou University, Zhengzhou, Henan 450001, China}

\author{Lisheng Geng}
\address{State Key Laboratory of Theoretical Physics, Institute of Theoretical Physics, Chinese Academy of Sciences, Beijing 100190, China}
\address{School of Physics and Nuclear Energy Engineering and International Research Center for Nuclei and Particles in the Cosmos, Beihang University, Beijing 100191, China}

\author{Natsumi Ikeno}
\address{Department of Regional Environment, Tottori University, 680-8550, Tottori, Japan}

\author{Pedro Fern\'andez-Soler}
\address{IFIC, Centro Mixto Universidad de Valencia-CSIC, Institutos de Investigaci\'on de Paterna, Aptdo. 22085, 46071 Valencia, Spain}

\author{Zhi Feng Sun}
\address{IFIC, Centro Mixto Universidad de Valencia-CSIC, Institutos de Investigaci\'on de Paterna, Aptdo. 22085, 46071 Valencia, Spain}

\date{\today}

\maketitle

\begin{abstract}
We present a review of recent works on weak decay of heavy mesons and baryons with two mesons, or a meson and a baryon, interacting strongly in the final state. The aim is to learn about the interaction of hadrons and how some particular resonances are produced in the reactions. It is shown that these reactions have peculiar features and act as filters for some quantum numbers which allow to identify easily some resonances and learn about their nature. The combination of basic elements of the weak interaction with the framework of the chiral unitary approach allow for an interpretation of results of many reactions and add a novel information to different aspects of the hadron interaction and the properties of dynamically generated resonances. 

\end{abstract}

{\small \addtocontents{toc}{\protect\sloppy} \tableofcontents}

\newpage

\section{Introduction} \label{Intro}

In this review we give a perspective of the theoretical work
done recently on the interpretation of results from $B$, $D$,
$\Lambda_b$, $\Lambda_c$ weak decays into final states that contain
interacting hadrons, and how it is possible to obtain additional
valuable information that is increasing our understanding of hadron
interactions and the nature of many hadronic resonances. The novelty
of these processes is that one begins with a clean picture at the
quark level which allows one to select the basic mechanisms by means
of which the process proceeds. Finally, one has a final state
described in terms of quarks. To make contact with the experiments,
where mesons and baryons are observed, one must hadronize, creating pairs of $q \bar q$ and writing the new
states in terms of mesons and baryons. This concludes the primary
hadron production in these processes. After that, the interaction of
these hadrons takes place, offering a rich spectrum of resonances
and special features from where it is possible to learn much about
the interaction of these hadrons and the nature of many resonances
in terms of the components of their wave functions.

\section{The scalar sector in the meson-meson interaction}

Let us begin with  some examples  where the low-lying scalar meson
resonances are produced. This will include $B^0$ and $B^0_s$ decays
into $J/\psi$ $f_0(500)$ and $J/\psi$ $f_0(980)$ and $D^0$ decay
into $K^0$ and $f_0(500)$, $f_0(980)$ and $a_0(980)$.

The $f_0(500)$, $f_0(980)$ and $a_0(980)$ resonances have been the
subject of discussion for years with an apparently endless debate
whether they are $q \bar q $ states, tetraquarks, molecular systems,
etc.~\cite{Klempt:2007cp,Crede:2008vw}. The advent of the chiral
unitary approach in different versions has brought some light into
this issue. Our present position is the following: QCD at low
energies can be described in terms of chiral Lagrangians in which the
original quark and gluon degrees of freedom have been substituted by
the hadrons observed in experiments, mesons and baryons
\cite{Weinberg:1978kz,Gasser:1983yg,Ecker:1994gg,Bernard:1995dp}.
These Lagrangians involve pseudoscalar mesons and low-lying
baryons, while vector mesons were included in
Refs.~\refcite{Bando:1984ej,Bando:1987br,Meissner:1987ge}. The extension of these ideas to
higher energies of the order of GeV, incorporating unitarity in
coupled channels, has brought new insight into this issue and has
allowed one to provide answers to some of the questions raised
concerning the nature of many resonances. With the umbrella of the
chiral unitary approach we include works that use the coupled
channels Bethe-Salpeter equation, or the inverse amplitude method,
and by now are widely used in the baryon sector, where it was
initiated\cite{Kaiser:1995cy,Oset:1997it,Oller:2000fj,GarciaRecio:2003ks,GarciaRecio:2005hy,Hyodo:2002pk,Ikeda:2012au,Jido:2003cb,Borasoy:2005ie,
Oller:2005ig,Oller:2006jw,Borasoy:2006sr,Hyodo:2008xr,Mai:2014xna},
and the meson sector~\cite{Oller:1997ti,Oller:1998hw,Kaiser:1998fi,Locher:1997gr,Nieves:1999bx,Pelaez:2006nj,Albaladejo:2008qa,Albaladejo:2015aca}. A recent thorough review on chiral dynamics and the nature of the low lying scalar mesons, in particular the $f_0(500)$, can be seen in Ref.~\refcite{Pelaez:2015qba}.

The Bethe Salpeter (BS) equation for meson meson interaction in coupled
channels reads as:
\begin{equation}\label{eq:BSeq}
t = [1-VG]^{-1} V,
\end{equation}
where $V$ is the transition matrix potential, usually taken as the
lowest order amplitude of chiral perturbation theory (the inverse amplitude method includes explicitly terms of next order, but in the scalar sector the largest ones are generated by rescattering in the BS equation). These matrix
elements for $\pi^+ \pi^-$,  $\pi^0 \pi^0$, $K^+ K^-$,   $K^0 \bar
K^0$ can be taken for instance from Ref.~\refcite{Oller:1997ti} and can be complemented with the
matrix elements of the $\eta \eta$ channels from
Ref.~\refcite{Gamermann:2006nm}. Then the $t$ matrix provides the transition $t$
matrix from one channel to another. The diagonal $G$-matrix is constructed out of the loop
function of two meson propagators:
\begin{equation}
G_{ii} (s) = i \int\frac{d^{4}q}{(2\pi)^{4}}\frac{1}{(P-q)^{2}-m^2_1+i\varepsilon}\,\frac{1}{q^{2}-m^2_2+i\varepsilon},
\label{eq:G}
\end{equation}
where $m_{1,2}$ are the masses of the two meson in channel $i$, and where $P^2 \equiv s$ is the center of mass energy squared. This loop function can be regularized using a cutoff method or dimensional
regularization. The interesting thing about these equations in the
pseudoscalar sectors, with a suitable cut off of the order of 1 GeV
to regularize the loops, is that one obtains an excellent description of all
the observables in pseudoscalar-pseudoscalar meson interaction up to
about 1 GeV. In particular one can also look for poles in the
scattering matrix which lead to the resonances in the system. In
this sense one obtains the $f_0(500)$, the $f_0(980)$ in $\pi \pi$ ,
the $a_0(980)$ in $\pi \eta$  and the $\kappa(800)$ in $K \pi$ in
the s-wave matrix elements. Note that one neither puts the
resonances by hand in the amplitudes, nor uses a potential that
contains a seed of a pole via a CDD\cite{Castillejo:1955ed} pole term in the potential (of
the type of $a/(s-s_0)$). In this sense, these resonances appear in
the same natural way as the deuteron appears in the solution of the
Schr\"odinger equation for $NN$ scattering and qualify as dynamically generated states,
kind of molecular meson--meson states. It is also interesting to
evaluate the residues at the poles for each channel, for this tells
us the strength of each channel in the wave function of the
resonance. In this sense the $f_0(500)$ couples essentially to $\pi
\pi$. The  $f_0(980)$ couples most strongly to $K \bar K$, although
this is a closed channel, pointing to the $K \bar K$ nature of this
resonance, and it couples weakly to $\pi \pi$, the only open decay channel. The
$a_0(980)$ couples strongly to $K \bar K$ and $\pi \eta$  and the $\kappa(800)$ to
$K \pi$.

It is worth mentioning that in works where one starts with a $q \bar
q$ seed to represent the scalars and then unitarizes the models to
account for the inevitable coupling of these quarks to the meson
meson components, it turns out that the meson meson components ``eat
up'' the seed and they remain as the only relevant components of the wave
function\cite{vanBeveren:1986ea,Tornqvist:1995ay,Fariborz:2009cq,Fariborz:2009wf}.

\section{The scalar meson sector in $B$ and $D$ decays}\label{sec:scalarsectorBDdecays}

Let us begin with an example of application of the former ideas to interpret recent results from LHCb and other facilities.

The LHCb Collaboration measured the  $B^0_s$ decays into $J/\psi$ and $\pi^+ \pi^-$ and observed a pronounced peak for the $f_0(980)$\cite{Aaij:2011fx}. At the same time the signal for the $f_0(500)$ was found very small or non-existent. The Belle Collaboration corroborated these results in Ref.~\refcite{Li:2011pg}, providing absolute rates for the $f_0(980)$ production with a branching ratio of the order of $10^{-4}$.
The CDF Collaboration confirmed these latter results in Ref.~\refcite{Aaltonen:2011nk}. Further confirmation was provided by the D0 Collaboration in Ref.~\refcite{Abazov:2011hv}. Furthermore, the LHCb Collaboration has continued working in the topic and in Ref.~\refcite{LHCb:2012ae} results are provided for the $\bar B^0_s$ decay into $J/\psi$ $f_0(980)$ followed by the $\pi^+ \pi^-$ decays of the  $f_0(980)$. Here, again the $f_0(980)$ production is seen clearly while no evident signal is seen for the $f_0(500)$. Interestingly, in the analogous decay of $\bar B^0$ into $J/\psi$ and $\pi^+ \pi^-$ \cite{Aaij:2013zpt} a signal is seen for the $f_0(500)$ production and only a very small fraction is observed for the $f_0(980)$ production, with a relative rate of about (1-10)\% with respect to that of the $f_0(500)$ (essentially an upper limit is given). Further research has followed by the same collaboration and in Ref.~\refcite{Aaij:2014emv} the  $\bar B^0_s$ into $J/\psi$ and $\pi^+ \pi^-$ is investigated. A clear peak is observed once again for  $f_0(980)$ production, while the $f_0(500)$ production is not observed. The $\bar B^0$ into $J/\psi$ and $\pi^+ \pi^-$ is further investigated in Ref.~\refcite{Aaij:2014siy} with a clear contribution from the $f_0(500)$ and no signal for the $f_0(980)$.

To interpret these results we take the dominant mechanism for the
weak decay of the $B$'s into $J/\psi$ and a primary $q \bar q$ pair,
which is $d \bar d$ for $B^0$ decay and $s \bar s$ for $B^0_s$
decay. After this, this $q \bar q$ pair is allowed to hadronize into
a pair of pseudoscalar mesons and we look at the relative weights of
the different pairs of mesons. Once the production of these meson
pairs has been achieved, they are allowed to interact, for what
chiral unitary theory in coupled channels is used, and automatically
the  $f_0(500)$, $f_0(980)$ resonances are produced. We are then
able to evaluate ratios of these production rates in the different decays
studied\cite{Liang:2014tia} and we find indeed a striking dominance of the  $f_0(500)$
in the $B^0$ decay and of the $f_0(980)$ in the $B^0_s$ decay, in a
very good quantitative agreement with experiment.

\subsection{Formalism}

\begin{figure}[t!]\centering
\includegraphics[scale=0.5]{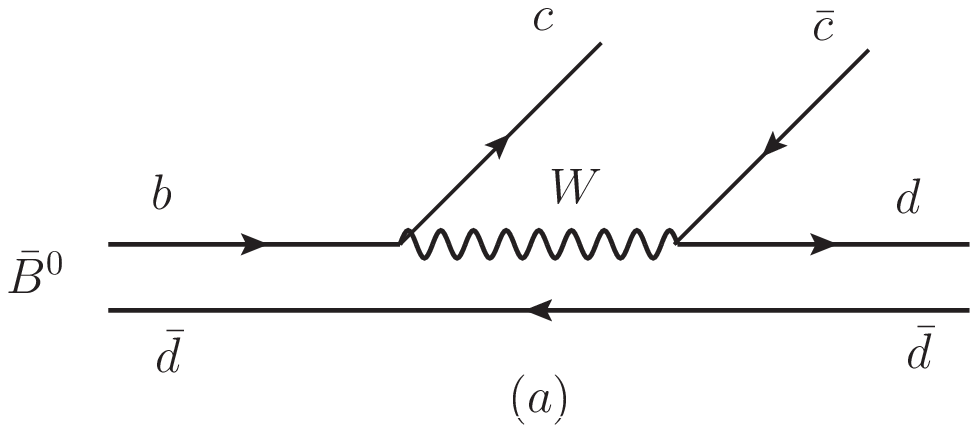}
\includegraphics[scale=0.5]{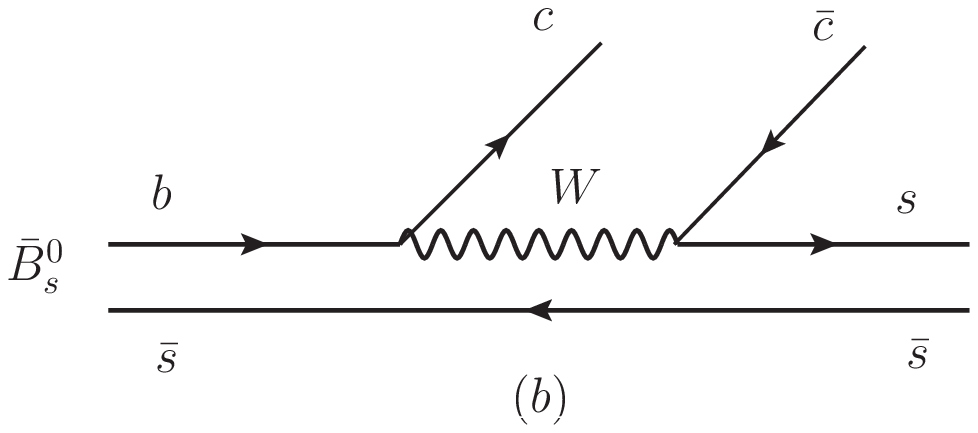}
\caption{Diagrams for the decay of $\bar B^0$ and $\bar B^0_s$ into $J/\psi$ and a primary $q\bar q$ pair, $d\bar d$ for $\bar B^0$ and $s\bar s$ for $\bar B^0_s$.\label{fig:B2Jpsipipi}}
\end{figure}

Following Ref.~\refcite{Stone:2013eaa} we take the dominant weak mechanism
for $\bar{B}^0$ and $\bar{B}^0_s$ decays (it is the same for
$B^0$ and $B^0_s$ decays) which we depict in
Fig.~\ref{fig:B2Jpsipipi}.

 In order to understand the process some very basic elements of the weak interaction are needed.
The $W^{\pm}$ connects two quarks and the strength is given by the Cabibbo-Kobayashi-Maskawa (CKM) matrix\cite{Kobayashi:1973fv,Wolfenstein:1983yz}.  The operator resulting for the exchange of the $W$ in Fig.~\ref{fig:B2Jpsipipi}(a) is given by:\cite{Buchalla:1995vs,ElBennich:2009da,Leitner:2010ai}
\begin{eqnarray}
H_{\rm W} &=& \frac{G_F}{\sqrt{2}} V_{bc} V_{cd} \bar{c}\gamma_{\mu}
(1   -   \gamma_5)b \bar{d}\gamma^{\mu} (1  - \gamma_5)c  +   h.c.
\label{eq:hweak}
\end{eqnarray}

To get a feeling of the strength of the CKM matrix elements, recall that the quarks are classified in weak doublets
\begin{equation*}
  \left(
  \begin{array}{c}
  u\\
  d
  \end{array}
  \right)
   \left(
  \begin{array}{c}
 c\\
  s
  \end{array}
  \right)
  \left(
  \begin{array}{c}
 t\\
  b
  \end{array}
  \right)
\end{equation*}

The transitions between quarks in the same doublet are Cabibbo allowed, they go roughly like the cosinus of the Cabibbo angle while from the first doublet to the second it goes like the sinus, concretely

\begin{align}\label{eq:Cabbibo}
&V_{cd}=-\sin\theta_c =-0.22534,\nonumber \\
&V_{cs}=\cos\theta_c =0.97427.
\end{align}

\begin{figure}[h!]\centering
\includegraphics[scale=0.6]{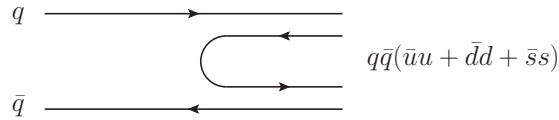}
\caption{Schematic representation of the hadronization $q\bar q \to q\bar q (u\bar u +d\bar d +s\bar s)$.\label{fig:qqbarHadronization}}
\end{figure}

The differences between the two processes in Fig.~\ref{fig:B2Jpsipipi} are: (i) $V_{cd}$ appears in the $Wcd$ vertex in $\bar B^0$ decay while $V_{cs}$ appears for the case of the $\bar B^0_s$ decay; (ii) one has a $d\bar d$ primary final hadron state in $\bar B^0$ decay and $s\bar s$ in $\bar B^0_s$ decay. Yet, one wishes to have $\pi^+ \pi^-$ in the final state as in the experiments. For this we need the hadronization. This is easily accomplished: schematically this process is as shown in Fig.~\ref{fig:qqbarHadronization}, where an extra $\bar q q$ pair with the quantum numbers of the vacuum, $\bar u u +\bar d d +\bar s s$, is added. Next step corresponds to writing the $q\bar q (\bar u u +\bar d d +\bar s s)$ combination in terms of pairs of mesons. For this purpose we define the $q\bar q$ matrix $M$,
\begin{equation}\label{eq:1}
M=\left(
           \begin{array}{ccc}
             u\bar u & u \bar d & u\bar s \\
             d\bar u & d\bar d & d\bar s \\
             s\bar u & s\bar d & s\bar s \\
           \end{array}
         \right)
\end{equation}
We can rewrite this in a different way and we see a nice property of this matrix
\begin{align}
M & = v \bar{v} =
\left(\begin{array}{c}
u \\ d \\ s
\end{array}\right)
\left(\begin{array}{ccc}
\bar{u} & \bar{d} & \bar{s}
\end{array}\right)  = \left( \begin{array}{cccc}
u\bar{u} & u\bar{d} & u\bar{s}  \\
d\bar{u} & d\bar{d} & d\bar{s} \\
s\bar{u} & s\bar{d} & s\bar{s}  \\
\end{array}\right)~, \label{eq:qqbarmatrix}
\end{align}
which fulfils:
\begin{align}\label{eq:MMqqM}
M^2 & = (v \bar{v})(v \bar{v}) = v (\bar{v} v) \bar{v} = \left(
\bar{u}u + \bar{d}d + \bar{s}s \right) M~.
\end{align}
Now, in terms of mesons, the matrix $M$ corresponds to
\begin{equation}\label{eq:phimatrix}
\phi = \left(
           \begin{array}{ccc}
             \frac{1}{\sqrt{2}}\pi^0 + \frac{1}{\sqrt{3}}\eta + \frac{1}{\sqrt{6}}\eta' & \pi^+ & K^+ \\
             \pi^- & -\frac{1}{\sqrt{2}}\pi^0 + \frac{1}{\sqrt{3}}\eta + \frac{1}{\sqrt{6}}\eta' & K^0 \\
            K^- & \bar{K}^0 & -\frac{1}{\sqrt{3}}\eta + \sqrt{\frac{2}{3}}\eta' \\
           \end{array}
         \right)~.
\end{equation}
This matrix corresponds to the ordinary one used in chiral
perturbation theory~\cite{Gasser:1983yg} with the addition of
$\frac{1}{\sqrt{3}}diag(\eta_1,\eta_1,\eta_1)$ where $\eta_1$ is a
singlet of $SU(3)$, taking into account the standard mixing between
$\eta$ and $\eta'$ \cite{Bramon:1992kr,Roca:2003uk,Gamermann:2008jh}. The $\eta'$ is omitted in the chiral Lagrangians because due to the $U_A(1)$ anomaly it is not a Goldstone Boson. 
Note also that the term
$\frac{1}{\sqrt{3}}\text{diag}(\eta_1,\eta_1,\eta_1)$ is inoperative  in the $[\phi,\partial_{\mu}\phi]$ structure. In terms of two pseudoscalars we have the correspondence:
\begin{align}
d\bar d (u\bar u +d\bar d +s\bar s) & \equiv  \left( \phi \cdot \phi \right)_{22} = \pi^- \pi^+ +\frac{1}{2}\pi^0 \pi^0
-\frac{2}{\sqrt{6}}\pi^0 \eta +K^0 \bar K^0 + \frac{1}{3}\eta \eta~,\nonumber \\
s\bar s (u\bar u +d\bar d +s\bar s) & \equiv  \left( \phi \cdot \phi \right)_{33}=K^- K^+ + K^0 \bar K^0 + \frac{1}{3}\eta \eta~,\label{eq:Mmatrix22-33}
\end{align}
where we have omitted the $\eta'$ because of its large mass. We can see that $\pi^+ \pi^-$ is only obtained in the first step in
the $\bar B^0$ decay and not in $\bar B^0_s$ decay. However, upon
rescattering of $K\bar K$ we also can get $\pi^+ \pi^-$ in the final
state, as we shall see. Yet, knowing that the $f_0(980)$ couples
strongly to $K\bar K$ and the $f_0(500)$ to $\pi \pi$, the
meson-meson decomposition of Eqs.~(\ref{eq:Mmatrix22-33}) already
tells us that the $\bar B^0$ decay will be dominated by $f_0(500)$
production and $\bar B^0_s$ decay by $f_0(980)$ production. Let us
see how the interaction proceeds.

\begin{figure}[t!]\centering
\includegraphics[scale=0.4]{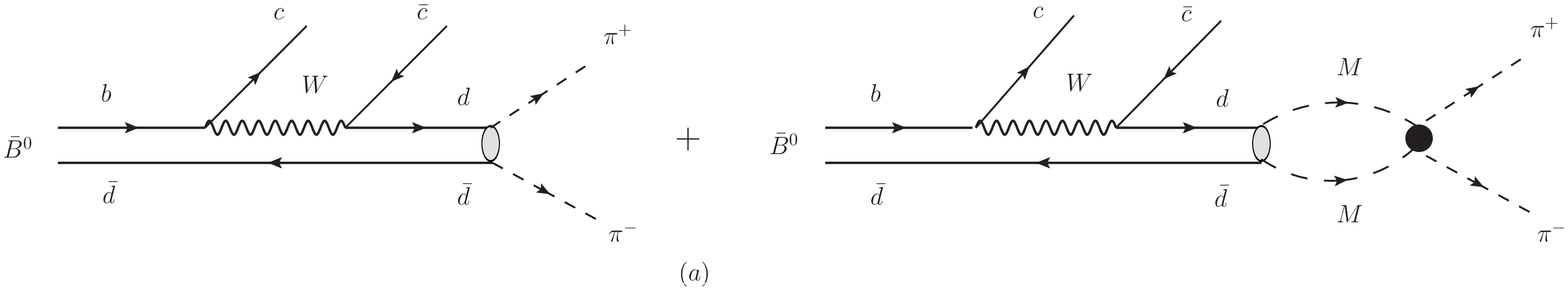}
\includegraphics[scale=0.55]{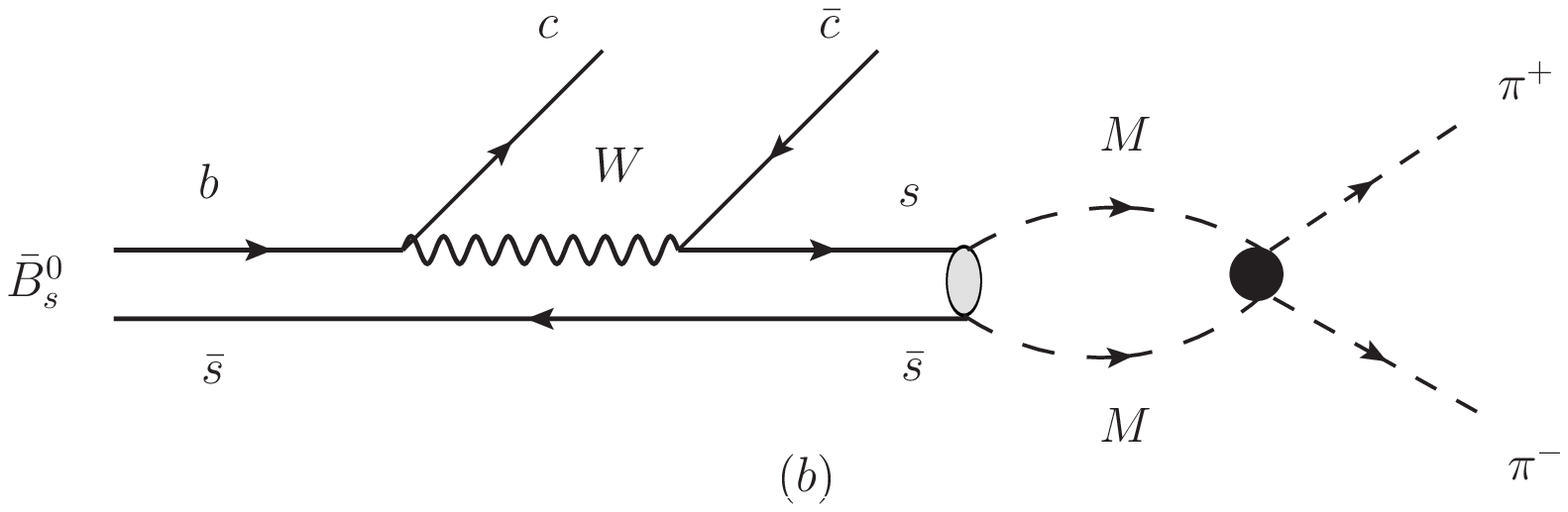}
\caption{Diagrammatic representation of $\pi^+ \pi^-$, via direct
plus rescattering mechanisms in $\bar B^0$ decay (a), and via
rescattering for $\bar B ^0_s$ decay (b).\label{fig:fig3}}
\end{figure}

Let us call $V_P$ the production vertex which contains all dynamical
factors common to both reactions. The $\pi^+ \pi^-$ production will
proceed via primary production or final state interaction as
depicted in Fig.~\ref{fig:fig3}.

The amplitudes for $\pi^+ \pi^-$ production are given by
\begin{align}\label{eq:amplitude-B0pipi}
t(\bar B^0 \to J/\psi \pi^+ \pi^-) & = V_P V_{cd} (1+G_{\pi^+ \pi^-}
t_{\pi^+ \pi^- \to \pi^+ \pi^-} +2 \frac{1}{2} \frac{1}{2} G_{\pi^0
\pi^0} t_{\pi^0 \pi^0 \to \pi^+ \pi^-} + \nonumber \\
&~~ G_{K^0 \bar K^0} t_{K^0 \bar K^0 \to \pi^+ \pi^-}
+2 \frac{1}{3} \frac{1}{2} G_{\eta \eta} t_{\eta \eta \to \pi^+ \pi^-}) ,\nonumber \\
t(\bar B^0_s \to J/\psi \pi^+ \pi^-) & = V_P V_{cs} ( G_{K^+ K^-}
t_{K^+ K^- \to \pi^+ \pi^-} +  G_{K^0 \bar K^0} t_{K^0 \bar K^0 \to
\pi^+ \pi^-}  + \nonumber \\
&~~ 2 \frac{1}{3} \frac{1}{2} G_{\eta \eta} t_{\eta \eta \to \pi^+
\pi^-} )~,
\end{align}
where $G_i$ are the loop functions of two meson propagators defined
above in Eq. (\ref{eq:G}). In Ref.~\refcite{Liang:2014tia} a cut off $\Lambda
=600$ MeV is taken, as needed in the enlarged space with respect to
Ref.~\refcite{Oller:1997ti}, including the $\eta \eta$ channel.

Note also that with respect to the weights of the meson-meson
components in Eqs.~(\ref{eq:Mmatrix22-33}) we have added a factor
$1/2$ for the propagation of the $\pi^0 \pi^0$ and $\eta \eta$
states which involve identical particles, and a factor of two for the two possible combinations to create two identical particles in the case of $\pi^0\pi^0$ or $\eta\eta$.

 One comment is in order concerning Eq. (\ref{eq:amplitude-B0pipi}), since in principle the $t$-matrices have left hand cut contributions while the form factors accounting for final state interaction which appear in the $B$ decay amplitudes do not have it. In Ref. \cite{Oller:2000ug} the problem of the form factors and its relationship to the chiral unitary aproach is addressed. A link is stablished there between the form factors and the $t$ matrices in the on shell factorization that we employ through our calculations, Eq. (\ref{eq:BSeq}). The left hand cut contributions to the $t$ matrix are smoothly dependent on the energy for physical energies \cite{nsd} and is usually taken into account by means of a constant added to the $G$ function.
  It is also interesting to recall the Quantum Mechanical version of this issue, which can be found in Ref. \cite{YamagataSekihara:2010pj}, and is basically equivalent to our approach using the on shell factorized $t$ matrices in Eq. (\ref{eq:amplitude-B0pipi}).

One final element of information is needed to complete the formula
for $d\Gamma/dM_{inv}$, with $M_{inv}$ the $\pi^+ \pi^-$ invariant
mass, which is the fact that in a $0^- \to 1^- 0^+$ transition we
shall need an $L'=1$ for the $J/\psi$ to match angular momentum
conservation. Hence, $V_P = A~p_{J/\psi} \cos \theta$, and we assume
$A$ to be constant (equal to $1$ in the calculations). Thus,
\begin{equation}\label{eq:dGamma}
\frac{d \Gamma}{d M_{inv}}=\frac{1}{(2\pi)^3}\frac{1}{4M_{\bar
B_j}^2}\frac{1}{3}p_{J/\psi}^2 p_{J/\psi} \tilde{p}_{\pi}
{\overline{ \sum}} \sum \left| \tilde{t}_{\bar B^0_j \to J/\psi
\pi^+ \pi^-} \right|^2,
\end{equation}
where the factor $1/3$ is coming from the integral of $\cos^2
\theta$ and $\tilde{t}_{\bar B^0_j \to J/\psi \pi^+ \pi^-}$ is
$t_{\bar B^0_j \to J/\psi \pi^+ \pi^-}/(p_{J/\psi} \cos \theta)$,
which depends on the $\pi^+ \pi^-$ invariant mass. In
Eq.~(\ref{eq:dGamma}) $p_{J/\psi}$ is the $J/\psi$ momentum in the
global CM frame ($\bar B$ at rest) and $\tilde{p}_{\pi}$ is the pion
momentum in the $\pi^+ \pi^-$ rest frame,
\begin{equation}\label{eq:pJpsi}
p_{J/\psi}=\frac{\lambda^{1/2}(M_{\bar B}^2, M_{J/\psi}^2,
M_{inv}^2)}{2M_{\bar
B}},~~~~~~~\tilde{p}_{\pi}=\frac{\lambda^{1/2}(M_{inv}^2, m_{\pi}^2,
m_{\pi}^2)}{2M_{inv}},
\end{equation}
with $\lambda(a,b,c)$ the K\"{a}llen function.

\subsection{Results}
\begin{figure}[ht!]\centering
\includegraphics[scale=0.3]{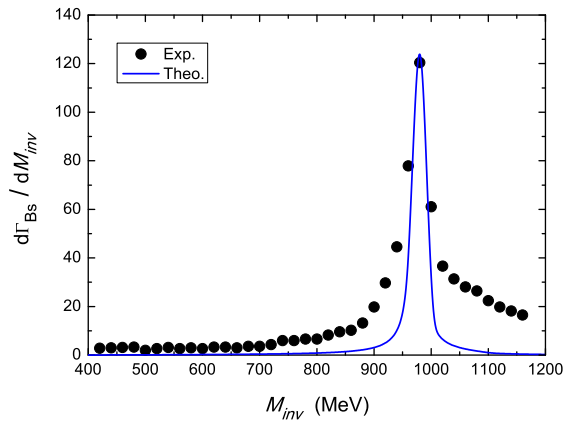} 
\caption{$\pi^+ \pi^-$ invariant mass distribution for the $\bar
B^0_s \to J/\psi \pi^+ \pi^-$ decay, with arbitrary normalization and folded with a $20\ \text{MeV}$ resolution, compared with the data\cite{Aaij:2014emv}.\label{fig:dGammaBs2Jpsipipi}}
\end{figure}

In Fig.~\ref{fig:dGammaBs2Jpsipipi} we show the $\pi^+ \pi^-$
invariant mass distribution for the case of the $\bar B^0_s \to
J/\psi \pi^+ \pi^-$ decay, comparing the results with the data of
Ref.~\refcite{Aaij:2014emv} where more statistics has been accumulated than
in the earlier run of Ref.~\refcite{Aaij:2011fx}. The data are collected in
bins of 20 MeV and the theoretical results are compared with the
results in Fig.~14 of Ref.~\refcite{Aaij:2014emv}. We can see that the
agreement, up to an arbitrary normalization, is quantitatively good.
We observe an appreciable peak for $f_0(980)$ production and
basically no trace for $f_0(500)$ production. The agreement is even
better with the dashed line in Fig.~14 of Ref.~\refcite{Aaij:2014emv} where
a small background has been subtracted. At invariant masses above
the $f_0(980)$ peak, contribution from higher energy resonances,
which we do not consider, is expected~\cite{Aaij:2014emv}.

The second of Eqs.~(\ref{eq:amplitude-B0pipi}) tells us why the
$f_0(500)$ contribution is so small. All intermediate states
involved, $K\bar K, \eta\eta$, have a mass in the $1~{\rm GeV}$
region and the $G$ functions are small at lower energies.
Furthermore, the coupling of the $f_0(500)$ to both $K\bar K$ and
$\eta \eta$ is also extremely small, such that the $t$ matrices
involved have also small magnitudes.

\begin{figure}[ht!]\centering
\includegraphics[scale=0.3]{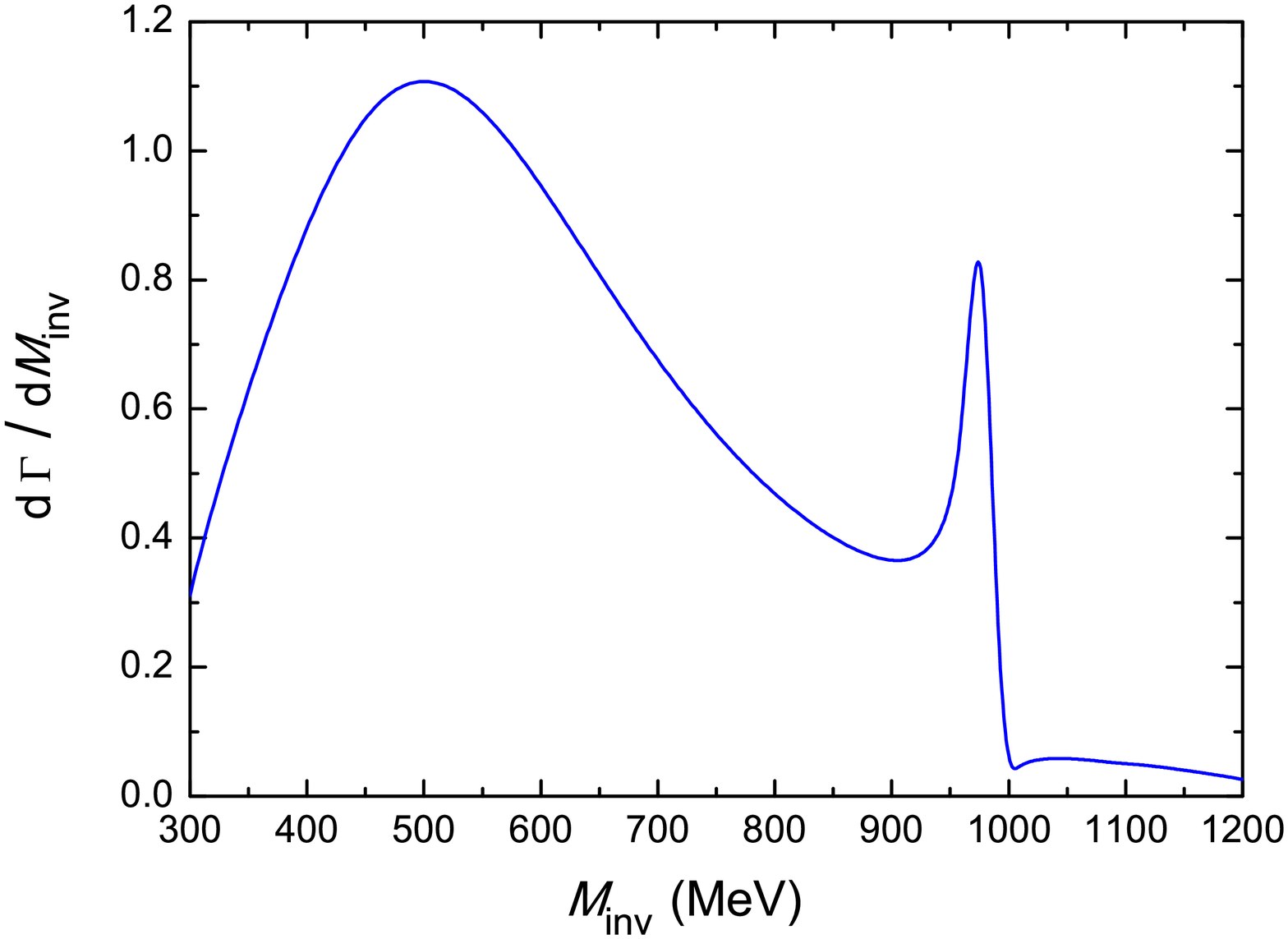}
\caption{$\pi^+ \pi^-$ invariant mass distribution for the $\bar B^0
\to J/\psi \pi^+ \pi^-$ decay, with arbitrary
normalization. In a recent work,\cite{liangrio} there are small corrections of the order of 10$\%$ with respect to this figure, from considering the singlet contribution in Eq.~\eqref{eq:phimatrix}, omitted in the work\cite{Liang:2014tia} reviewed here.\label{fig:dGammaB2Jpsipipi}}
\end{figure}

Note that in this decay we could have also $J/\psi$ and vector meson
production, but the $s\bar s$ component would give $\phi$ production
which does not decay to $\pi \pi$. The case is quite different for
the $\bar B^0 \to J/\psi \pi^+ \pi^-$ decay, because now we can also
produce $J/\psi \rho$ ($\rho \to \pi^+ \pi^-$) decay and in fact
this takes quite a large fraction of the $J/\psi \pi^+ \pi^-$ decay,
as seen in Ref.~\refcite{Aaij:2014siy}. We shall address this point in the
next section. We plot our relative $S$-wave $\pi^+ \pi^-$
production for the $\bar B^0 \to J/\psi \pi^+ \pi^-$ decay in
Fig.~\ref{fig:dGammaB2Jpsipipi}.

We can see that the $f_0(500)$ production is clearly dominant. The
$f_0(980)$ shows up as a small peak. A test can be done to compare
the results: If we integrate the strength of the two resonances over
the invariant mass distribution we find
\begin{equation}\label{eq:ratio-B2Jpsif0}
  \frac{{\mathcal{B}}[\bar B^0 \to J/\psi f_0(980), f_0(980)\to \pi^+\pi^-]}
       {{\mathcal{B}}[\bar B^0 \to J/\psi f_0(500), f_0(500)\to \pi^+\pi^-]} 
       = 0.033\pm 0.007,
\end{equation}
with an admitted 20\% uncertainty from the decomposition of the
strength in Fig.~\ref{fig:dGammaB2Jpsipipi} into the two resonances.
The most recent experimental result\cite{Aaij:2014siy} is:
\begin{equation}\label{eq:LHCb2014}
\left( 0.6^{+0.7+3.3}_{-0.4-2.6} \right) \times 10^{-2}~.
\end{equation}
The central value that we obtain is five times bigger than the
central value of the experiment in Eq.~(\ref{eq:LHCb2014}), yet, by
considering the errors in Eq.~(\ref{eq:LHCb2014}) we get a band for
the experiment of $0 \sim 0.046$ and our results are within this
band. \footnote{Alternatively, the results of Eq. (\ref{eq:LHCb2014}) can be interpreted as providing an upper limit for this ratio, in which case we can state that our results are below this upper limit.} Let us note that in the work of \cite{Daub:2015xja}, where a form factor is used, obtained using experimental phase shifts, one has a dip for the $f_0(980)$ following some enhancement in the strength of the distribution. We obtain a small, but neat peak  for the $f_0(980)$, but also followed by a dip, which is not seen  in the $B^0_s$ decay.

There is another point to consider. The
normalization of Figs.~\ref{fig:dGammaBs2Jpsipipi} and
~\ref{fig:dGammaB2Jpsipipi} is arbitrary but the relative size is
what the theory predicts. It is easy to compute
\begin{equation}
\frac{\Gamma(B^0\to J/\psi f_0(500))}{\Gamma(B^0_s \to J/\psi f_0(980))}\simeq \left( 4.5 \pm 1.0 \right)\times 10^{-2}.
\label{eq:ratio-th}
\end{equation}
This number is in agreement within errors with the band of $(2.08
\sim 4.13)\times 10^{-2}$ that one obtains from the branching
fractions of $9.60^{+3.79}_{-1.20} \times 10^{-6}$ for $\bar B^0 \to
J/\psi f_0(500)$\cite{Aaij:2013zpt} and $3.40^{+0.63}_{-0.16}
\times 10^{-4}$ for $\bar B^0_s \to J/\psi f_0(980)$\cite{LHCb:2012ae}.

Added to the results obtained for many other processes, as quoted in
the Introduction, the present reactions come to give extra support
to the idea originated from chiral unitary theory that the
$f_0(500)$ and $f_0(980)$ resonances are dynamically generated from
the interaction of pseudoscalar mesons and could be interpreted as a
kind of molecular states of meson-meson with the largest component
$\pi \pi$ for the $f_0(500)$ and $K\bar K$ for the $f_0(980)$.

 Note that, while a better quantitative agreement in the shape of 
Fig \ref{fig:dGammaBs2Jpsipipi} is obtained in \cite{Daub:2015xja} by using experimental $\pi \pi$ phase shifts in a big range of energies, the approach given here provides the basic features and allows to relate different decays processes without introducing further parameters.

So far we have assumed that $V_P$ is constant up to the $P$-wave
factor. Actually there is a form factor for the transition that
depends on the momentum transfer. Then it could be different for
$f_0(500)$ or $f_0(980)$ production. However, the work in
Refs.~\refcite{Li:2012sw,Ochs:2013gi,Kang:2013jaa} indicates that the form factors for
primary productions prior to the final state interaction, are rather
smooth. This point gives us an excuse to elaborate on this issue and place our approach in a broader context. This is done in the next subsection.

\subsection{Relationship to other approaches}\label{sec:relotherapproaches}

Referring to the diagram in Fig.~\ref{fig:B2Jpsipipi}(b), the weak decay of a $b$ quark will proceed via the exchange of a  $W^{\pm}$ which in one vertex will connect a $b$ and $c$ quark, and in the other vertex connect a $c$ and $s$ quark and the strength is given by the CKM matrix\cite{Wolfenstein:1983yz} elements.  The operator resulting for the exchange of the $W$ is given\cite{Buchalla:1995vs,ElBennich:2009da,Leitner:2010ai} by:
\begin{equation}
H_{\rm W} = \frac{G_F}{\sqrt{2}} V_{cs} V_{bc} \bar{c}\gamma_{\mu}
(1   -   \gamma_5)b \bar{s}\gamma^{\mu} (1  - \gamma_5)c +   h.c. \label{eq:hweakA}
\end{equation}
The theoretical study of these process requires the evaluation of the quark matrix elements of this operator for which many different approaches are followed.  Quark models in different versions are one of the options\cite{Albertus:2010hi,Albertus:2011xz,Albertus:2012jt,Albertus:2014xwa,Albertus:2014gba}. 
Another approach using elements of QCD under the factorization approximation is followed in weak $B$ and $D$ decays into two final mesons \cite{Furman:2005xp,ElBennich:2006yi,Cheng:2005nb,ElBennich:2008xy,Leitner:2010fq,Delepine:2005px}. $B$ decays are also addressed in Ref.~\refcite{Colangelo:2010bg} using light cone QCD sum rules under the factorization assumption. A different approach to  $B^0$ into $J/\psi$ and $\pi^+ \pi^-$ decay was followed in Ref.~\refcite{Sayahi:2013tza} using the  QCD-improved factorization approach.

Theoretical work on these issues is also done in Ref.~\refcite{Bediaga:2003zh} for the semileptonic $D$
decays using QCD sum rules. The light-front quark model is used again in Ref.~\refcite{Ke:2009ed}
to calculate form factors for $D$ decays. A Nambu-Jona-Lasinio type model is used in 
Ref.~\refcite{Achasov:2012kk} to study semileptonic $D$ decays. Estimations based on a simple model where the hadronic current is taken to be
the Noether current associated with a minimal linear sigma model are also available for semileptonic $D$ decays\cite{Fariborz:2011xb,Fariborz:2014mpa}. Research along similar lines is
done in Ref.~\refcite{Wang:2009azc}.
Light-cone sum rules are used to evaluate the form factors appearing in different weak processes\cite{Shi:2015kha,Meissner:2013hya,Wang:2015paa,Meissner:2013pba,Wang:2015uea,Kang:2013jaa,Doring:2013wka}. 

Apart from the hard processes that involve the weak transition and the hadronization, and that in QCD are considered in terms of the Wilson coefficients, one has to take into account
the meson final state interaction. In some cases this is done using the Omn\`es representation\cite{Doring:2013wka,Meissner:2013hya,Daub:2015xja}, which have the advantage of preserving all good properties of unitarity and analyticity of the amplitudes.  In
other cases Breit--Wigner or Flatt\'e structures are implemented and parametrized to account for the resonances observed in the experiment \cite{Shi:2015kha}. This latter procedure is known to have problems some times concerning these mentioned properties. Reference \refcite{Doring:2013wka} represents a hybrid approach insofar that unitarized chiral interactions are used to parameterize the $\pi K,\,\eta K$
amplitude, that is then fed into a dispersion approach to study semileptonic $B$ decays. For this, the two-channel inverse amplitude method of Ref.~\refcite{Oller:1998hw} is considered that contains next-to-leading order
contact terms, and that is supplemented with a resonance term to account for the $K_0^*(1430)$. The amplitude is fitted to $\pi K$
phase shift data. To guarantee the correct analytic structure, this amplitude serves then as input for a twice-subtracted
Muskhelishvili-Omn\`es relation in the coupled $\pi K$ and $\eta K$ channels. Additionally, the form factor is matched to the value
and slope of the one-loop ChPT result of the strangeness-changing form factors at $s=0$~\cite{Gasser:1984ux}.

In contrast to these pictures, in the present study we treat the meson--meson interaction using the chiral unitary approach.

In Fig.~\ref{fig:B2Jpsipipi}(b), after hadronization, Fig.~\ref{fig:fig3}(b), we have two mesons in the final state, in $S=0$, and we want to study their interaction. For this purpose, we encompass all the information of the hard transition part into  a constant factor and, up to an arbitrary normalization, we obtain invariant mass distributions which are  linked to the meson--meson interaction. 
The use of a constant $V_P$ factor  in our approach gets
support from the work of Ref.~\refcite{Daub:2015xja}. The evaluation of
the matrix elements in these processes is difficult and
problematic, and we have given a sketch of the many different theoretical approaches for it. There are however some cases where the
calculations can be kept under control. For the case of semileptonic decays with two pseudoscalar mesons in the final state with 
small recoil, namely when the final pseudoscalars move
slowly, it can be explored in the heavy meson chiral perturbation theory\cite{Manohar:2000dt}. Detailed calculations for the case
of semileptonic decay are done in Ref.~\refcite{Kang:2013jaa}. There one can see
that for large values of the invariant mass of the lepton
system the form factors can be calculated and the relevant ones in s wave that we need here are smooth in
the range of the invariant masses of the pairs of mesons. In the present case the lepton system would be replaced by the $J/\psi$ which is very massive and extrapolating the results of Ref.~\refcite{Kang:2013jaa} to this case one can conclude that the dependence of the s-wave matrix elements on the meson baryon invariant mass should be smooth. There
is also another limit, at large recoil, where an approach
that combines both hard-scattering and low-energy interactions has been developed and is also available\cite{Meissner:2013hya}, but
this is not the case here.

  There is also empirical information on the smoothness of these primary form factors. 
Yet, in Ref.~\refcite{Li:2012sw} this form factor is evaluated for $B$ decays and it is found that $F^{\sigma}_{B^0_s}(m^2_{J/\psi})/F^{f_0}_{B^0_s}(m^2_{J/\psi})=1$, where $\sigma, f_0$ stand for the $f_0(500), f_0(980)$. In Ref.~\refcite{Ochs:2013gi} the same results are assumed, as well as in Ref.~\refcite{Stone:2013eaa}, where by analogy $F^{\sigma}_{B^0}(m^2_{J/\psi})/F^{f_0}_{B^0}(m^2_{J/\psi})$ is also assumed to be unity. In addition, in Ref.~\refcite{Stone:2013eaa} it is also found from analysis of the experiment that $F^{f_0}_{B^0_s}(m^2_{J/\psi})/F^{\sigma}_{B^0}(m^2_{J/\psi})$ is compatible with unity. 

All that one needs to apply our formalism is that the form factors for the primary production of hadrons prior to their final state interaction are smooth compared to the changes induced by this final state interaction. This is certainly always true in the vicinity of a resonance coming from this final state interaction, but the studies quoted above tell us that one can use a relatively broad range, of a few hundred MeV, where we still can consider these primary form factors smooth compared to the changes induced by the final state interaction. 
\section{Vector meson production}\label{Sec:VectorProduction}
\subsection{Formalism for vector meson production}

At the quark level, we have
\begin{eqnarray} \label{eq:vector}
&&|\rho^0> = \frac{1}{\sqrt{2}}(u\bar{u} - d\bar{d});~~ |\omega> =
\frac{1}{\sqrt{2}} (u\bar{u} + d\bar{d}); ~~|K^{*0}> = d\bar{s}.
\end{eqnarray}

The diagrams of Fig. \ref{fig:B2Jpsipipi} without the hadronization
can serve to study the production of vector mesons, which are
largely $q \bar{q}$ states\cite{Pelaez:2003dy,Aceti:2012dd,Xiao:2012vv}. Since
we were concerned up to now only about the ratio of the scalars, the
factor $V_{P}$ was taken arbitrary. The spin of the particles
requires now $L'=0,2$, and with no rule preventing $L'=0$, we assume
that it is preferred; hence, the $p_{J/\psi} \cos\theta$ is not
present now. Then we find immediately the amplitudes associated to
Fig. \ref{fig:B2Jpsipipi},
\begin{align}
&t_{\bar{B}^{0} \rightarrow J/\psi \rho^{0}}= -\frac{1}{\sqrt{2}}\tilde{V}^{\prime}_{P}~ V_{cd},
~&&t_{\bar{B}^{0} \rightarrow J/\psi \omega}= \frac{1}{\sqrt{2}}\tilde{V}'_{P} ~V_{cd},
~&&t_{\bar{B}^{0}_{s} \rightarrow J/\psi \phi}= \tilde{V}'_{P} ~V_{cs}, \nonumber \\
&t_{\bar{B}^{0} \rightarrow J/\psi \bar{K}^{*0}}= \tilde{V}'_{P} ~V_{cs},
~&&t_{\bar{B}^{0}_{s} \rightarrow J/\psi K^{*0}}= \tilde{V}'_{P} ~V_{cd},
\label{eq:tmatM}
\end{align}
where $(-\frac{1}{\sqrt{2}})$ is the $\rho^{0}$ component in $d
\bar{d}$ and $(\frac{1}{\sqrt{2}})$ that of the $\omega$ and
$\tilde{V}'_{P}$ is the global factor for the processes, different
to $V_P$ used for the scalar sector. In order to determine
$\tilde{V}'_{P}$ versus $V_P$ in the scalar production, we use the
well-measured ratio\cite{LHCb:2012ae,Agashe:2014kda}:
\begin{equation}
\frac{\Gamma_{\bar{B}^{0}_{s}\rightarrow J/\psi f_{0} (980);f_{0}
(980)\rightarrow \pi^{+} \pi^{-}} }{\Gamma_{\bar{B}^{0}_{s}
\rightarrow J/\psi \phi}}=(13.9 \pm 0.9) \times 10^{-2}.
\label{eq:BR1}
\end{equation}

The width for $J/\psi V$ vector decay is now given by
\begin{equation}
\Gamma_{V_{i}}=\frac{1}{8 \pi}\frac{1}{m_{\bar B^{0}_i}^2} \left|
t_{\bar B^0_i \to J/\psi V_{i}}\right|^2 p_{J/ \psi}.
\label{eq:Gam1}
\end{equation}
Equations (\ref{eq:tmatM}) allow us to determine ratios of vector production with respect to the $\phi$,
\begin{eqnarray}
&& \frac{\Gamma_{\bar{B}^{0}\to J/\psi \rho^{0}}}{\Gamma_{\bar{B}^{0}_{s} \to J/\psi \phi}}= \frac{1}{2} \left| \dfrac{V_{cd}}{V_{cs}} \right|^2
\dfrac{m_{\bar{B}^{0}_{s}}^{2}}{m_{\bar{B}^{0}}^{2}} \dfrac{p_{\rho^{0}}}{p_{\phi}} =0.0263, \nonumber\\
&& \frac{\Gamma_{\bar{B}^{0}\to J/\psi \omega}}{\Gamma_{\bar{B}^{0}_{s} \to J/\psi \phi}}= \frac{1}{2} \left|
\dfrac{V_{cd}}{V_{cs}} \right|^2  \dfrac{m_{\bar{B}^{0}_{s}}^{2}}{m_{\bar{B}^{0}}^{2}} \dfrac{p_{\omega}}{p_{\phi}} =0.0263,  \\
&& \frac{\Gamma_{\bar{B}^{0}\rightarrow J/\psi \bar{K}^{*0}}}{\Gamma_{\bar{B}^{0}_{s} \rightarrow J/\psi \phi}}=
\dfrac{m_{\bar{B}^{0}_{s}}^{2}}{m_{\bar{B}^{0}}^{2}} \dfrac{p_{\bar{K}^{*0}}}{p_{\phi}} =0.957, \nonumber\\
&&\frac{\Gamma_{\bar{B}_{s}^{0}\rightarrow J/\psi
K^{*0}}}{\Gamma_{\bar{B}^{0}_{s} \rightarrow J/\psi \phi}}=  \left|
\dfrac{V_{cd}}{V_{cs}} \right|^2   \dfrac{p_{ K^{*0}}}{p_{\phi}}
=0.0551.\nonumber  \label{eq:ratio1}
\end{eqnarray}

By taking as input the branching ratio of $\bar{B}^{0}_{s}
\rightarrow J/\psi \phi$,
\begin{equation} BR(\bar{B}^{0}_{s}
\rightarrow J/\psi \phi)=(10.0^{+3.2}_{-1.8})\times 10^{-4},
\label{eq:BR2}
\end{equation}
we obtain the other four branching ratios
\begin{eqnarray}
&&BR(\bar{B}^{0} \rightarrow J/\psi \rho^{0})=(2.63^{+0.84}_{-0.47})\times 10^{-5}, \nonumber\\
&&BR(\bar{B}^{0} \rightarrow J/\psi \omega)=(2.63^{+0.84}_{-0.47})\times 10^{-5}, \nonumber\\
&&BR(\bar{B}^{0} \rightarrow J/\psi \bar{K}^{*0})=(9.57^{+3.1}_{-1.7})\times 10^{-4}, \nonumber\\
&&BR(\bar{B}_{s}^{0} \rightarrow J/\psi K^{*0})=(5.51^{+1.7}_{-1.0})\times 10^{-5}.
  \label{eq:BR3}
\end{eqnarray}
The experimental values are\cite{Agashe:2014kda}:
\begin{eqnarray}
&&BR(\bar{B}^{0} \rightarrow J/\psi \rho^{0})=(2.58 \pm 0.21)\times 10^{-5}, \nonumber\\
&&BR(\bar{B}^{0} \rightarrow J/\psi \omega)=(2.3 \pm 0.6)\times 10^{-5}, \nonumber\\
&&BR(\bar{B}^{0} \rightarrow J/\psi \bar{K}^{*0})=(1.34\pm 0.06)\times 10^{-3}, \nonumber\\
&&BR(\bar{B}_{s}^{0} \rightarrow J/\psi K^{*0})=(4.4\pm 0.9)\times 10^{-5}.
  \label{eq:BR4}
\end{eqnarray}
We can see that the agreement is good within errors, taking into
account that the only theoretical errors in Eq. (\ref{eq:BR3}) are
from the experimental branching ratio of Eq. (\ref{eq:BR2}). The
rates discussed above have also been evaluated using perturbative
QCD in the factorization approach in Ref.~\refcite{Liu:2013nea}, with good
agreement with experiment.  Our approach exploits flavor symmetries
and the dominance of the weak decay mechanisms of Fig.
\ref{fig:B2Jpsipipi} to calculate ratios of rates with good accuracy
in a very easy way.

The next step is to compare the $\rho$ production with $\rho
\rightarrow \pi^{+} \pi^{-}$ decay with $\bar{B}^{0} \rightarrow
J/\psi f_{0}; f_{0}\rightarrow \pi^{+} \pi^{-} ( f_{0} \equiv
f_{0}(500) ,f_{0}(980) $). In an experiment that looks for
$\bar{B}^{0} \rightarrow J/\psi \pi^{+} \pi^{-}$, all these
contributions will appear together, and only a partial wave analysis
will disentangle the different contributions. This is done in
Refs.~\refcite{Aaij:2013zpt,Aaij:2014siy} following the method of \cite{Zhang:2012zk}. There (see Fig. 13 of Ref.~\refcite{Aaij:2014siy}) one observes a peak of the $\rho$ and a $f_{0}(500)$
distribution, with a peak of the $\rho^{0}$ distribution about a
factor $6$ larger than that of the $f_{0}(500)$. The $f_{0}(980)$
signal is very small and only statistically significant states are shown in the figure. Since only an upper limit was determined for the $f_{0}(980)$ it is not shown.

In order to compare the theoretical results with these experimental
distributions, we convert the rates obtained in Eqs. (\ref{eq:BR3})
into $\pi^{+} \pi^{-}$ distributions for the case of the
$\bar{B}^{0} \rightarrow J/\psi \rho^{0}$ decay and $K^{-} \pi^{+}$
for the case of the $\bar{B}^{0} \rightarrow J/\psi \bar{K}^{*0}$
decay. For this purpose, we multiply the decay width of the
$\bar{B}^{0}$ by the spectral function of the vector mesons. We find:
\begin{equation}
\frac{d \Gamma_{\bar{B}^{0} \rightarrow J/\psi \rho^{0}}}{d
M_{inv}(\pi^{+} \pi^{-})}=-\frac{1}{\pi}2 M_{\rho}~{\rm Im}
\dfrac{1}{M_{inv}^{2}-M_{\rho}^{2}+i~M_{\rho} \Gamma_{\rho}(M_{inv})
} \Gamma_{\bar{B}^{0} \rightarrow J/\psi \rho^{0}} ,
 \label{eq:ratrho}
\end{equation}
where
\begin{eqnarray}
&&\Gamma_{\rho}(M_{inv})= \Gamma_{\rho} \left (\dfrac{p_{\pi}^{\rm
off}}{p_{\pi}^{\rm on}} \right )^{3}, ~~~p_{\pi}^{\rm
off}=\dfrac{\lambda^{1/2}(M_{inv}^{2},m_{\pi}^{2},m_{\pi}^{2})}{2M_{inv}}
\theta (M_{inv}-2m_{\pi}), \nonumber \\
&& p_{\pi}^{\rm
on}=\dfrac{\lambda^{1/2}(M_{\rho}^{2},m_{\pi}^{2},m_{\pi}^{2})}{2M_{\rho}}.
\label{eq:ratrhovwhere}
\end{eqnarray}
and for the case of the $\bar{B}^{0} \rightarrow J/\psi \bar{K}^{*0} ~(\bar{K}^{*0}\rightarrow  \pi^{+} K^{-})$,
we have
\begin{eqnarray}
\frac{d \Gamma_{\bar{B}^{0} \rightarrow J/\psi
\bar{K}^{*0};\bar{K}^{*0}\rightarrow  \pi^{+} K^{-}}}{d
M_{inv}(\pi^{+} K^{-})} &=& -\frac{1}{\pi}\frac{2}{3}~{\rm Im}
\dfrac{2 M_{K^{*}}}{M_{inv}^{2}-M_{K^{*}}^{2}+i~M_{K^{*}}
\Gamma_{K^*}(M_{inv}) } \nonumber \\
&& \times \Gamma_{\bar{B}^{0} \rightarrow J/\psi \bar{K}^{*0}} ,
 \label{eq:KbarStr}
\end{eqnarray}
with similar formulas for $\Gamma_{K^{*}}$, $p^{\rm off}$ and $p^{\rm on}$. In Eqs. (\ref{eq:ratrho}) and (\ref{eq:KbarStr}) we have taken into
account that $\rho^{0}$ decays only into $\pi^{+} \pi^{-}$, while $\bar{K}^{*0} $ decays into $\pi^{+} K^{-}$ and $\pi^{0} \bar{K}^{0}$
with weights $2/3$ and $1/3$, respectively. Expressions for $\bar B^0_s \to J/\psi K^{*0}; K^{*0} \to \pi^- K^+$ are readily obtained from the previous ones with the obvious changes.

\subsection{Results}

\begin{figure}[ht!]\centering
\includegraphics[scale=0.3]{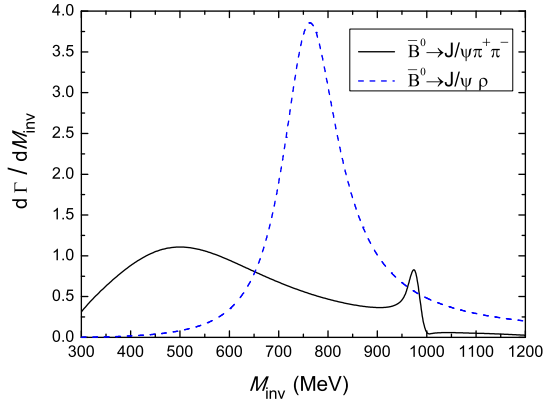}
\caption{$\pi^+ \pi^-$ invariant mass distributions for the $\bar
B^0 \to J/\psi \pi^+ \pi^-$ ($S$ wave) (solid line) and $\bar B^0
\to J/\psi \rho$, $\rho \to \pi^+ \pi^- $ ($P$ wave) decays, with
arbitrary normalization, and folded with a $20\ \text{MeV}$ resolution.} \label{fig:dGammaB2f_rho}
\end{figure}

In Fig. \ref{fig:dGammaB2f_rho} we show our predictions for
$f_{0}(500)$, $f_{0}(980)$, and $\rho^{0}$ production in  $\bar B^0
\to J/\psi \pi^+ \pi^-$, taken from Ref.~\refcite{Bayar:2014qha}.

The relative strengths and the shapes of the $f_{0}(500)$ and $\rho$
distributions are remarkably similar to those found in the partial
wave analysis of Ref.~\refcite{Aaij:2014siy}. However, our $f_{0}(500)$ has
a somewhat different shape since in the analysis of
Ref.~\refcite{Aaij:2014siy}, like in many experimental papers, a
Breit-Wigner shape for the $f_{0}(500)$ is assumed, which is
different to what the $\pi \pi$ scattering and the other production
reactions demand\cite{Ablikim:2004qna,Li:2003zi,Pelaez:2015qba}. It is interesting to remark that we have only considered the $\rho$ contribution without paying any attention to $\rho-\omega$ mixing. This is done explicitly in \cite{Daub:2015xja} and it leads to a peculiar shape, different to the one obtained in the electromagnetic form factor of the pion \cite{Oller:2000ug}. This new interesting shape is corroborated by a recent work \cite{whstudent}.
 It is also interesting to mention that, although small, we see a signal of the $f_0(980)$ in the distribution of Fig. \ref{fig:dGammaB2f_rho}, while in \refcite{Daub:2015xja} only a small bump is seen in this region. Let us mention to this respect that in the $J/\psi \to \omega \pi^+ \pi^-$ decay, similar to the $B_0$ decay here, one observes clearly the $f_0(980)$ peak \cite{Wu:2001vz,Augustin:1988ja} and there is a good agreement with the theoretical work of \refcite{Roca:2004uc}  done along similar lines as here. It would be most interesting to see what one finds in the present case when more statistics is gathered. 

\begin{figure}[ht!]\centering
\includegraphics[scale=0.3]{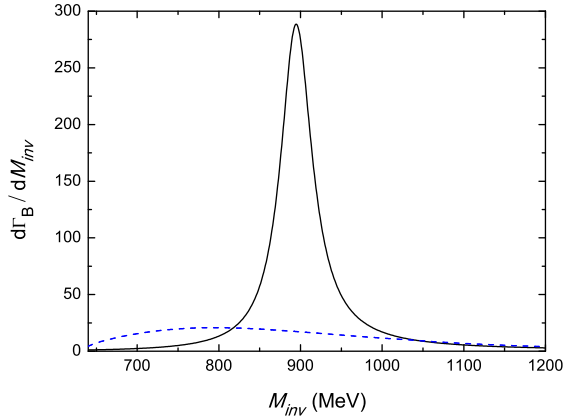}
\caption{ $\pi^+ K^-$ invariant mass distributions for the $\bar B^0
\to J/\psi \bar K^{*0}$, $\bar K^{*0} \to \pi^+ K^-$ (solid line)
and $\bar B^0 \to J/\psi \bar \kappa$, $\bar \kappa \to \pi^+ K^-$
(dashed line), with arbitrary normalization.} \label{fig:Kap1}
\end{figure}

\begin{figure}[ht!]\centering
\includegraphics[scale=0.3]{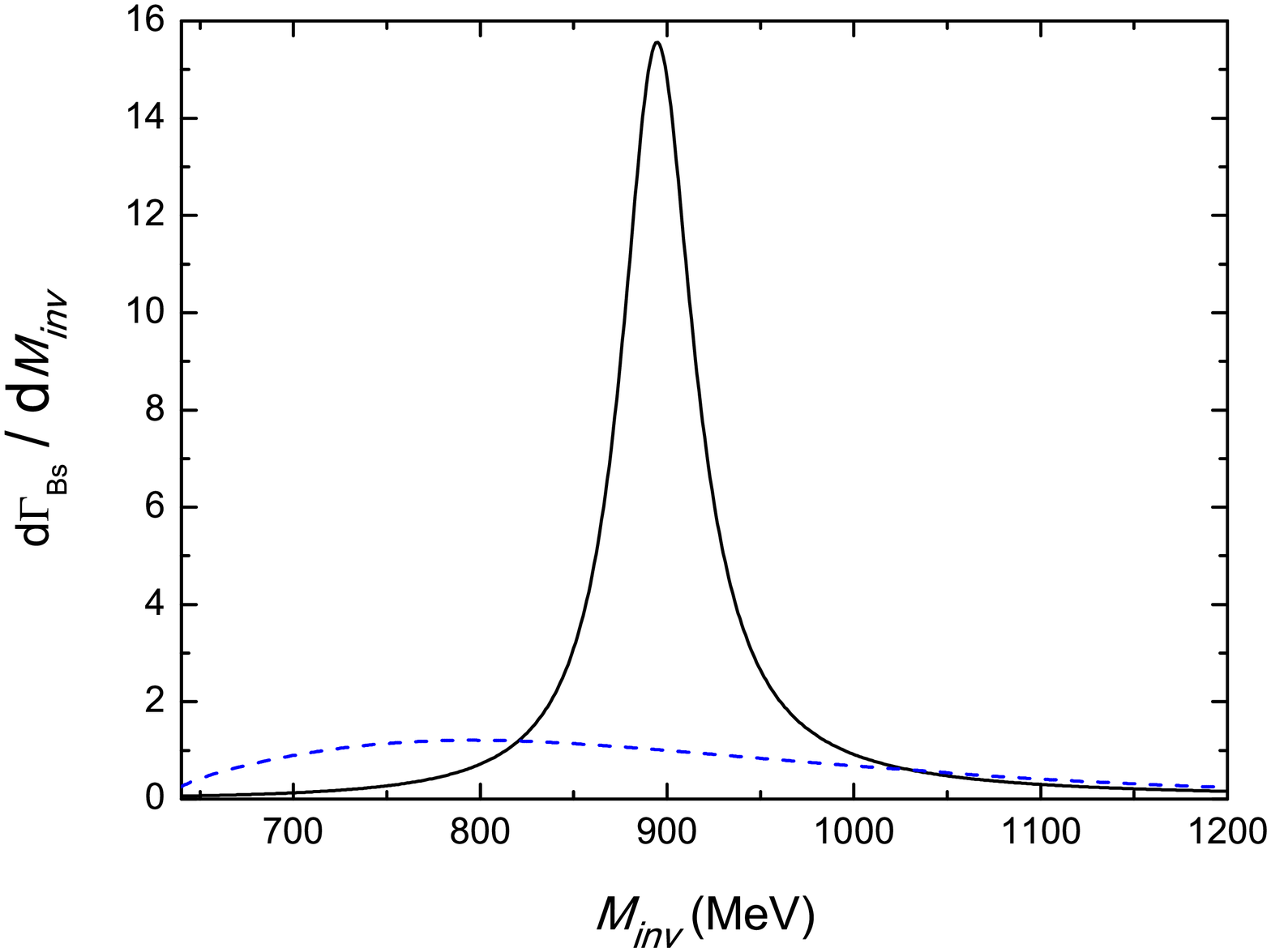}
\caption{ $\pi^- K^+$ invariant mass distributions for the $\bar
B^0_s \to J/\psi K^{*0}$, $K^{*0} \to \pi^- K^+$ (solid line) and
$\bar B^0_s \to J/\psi \kappa$, $\kappa \to \pi^- K^+$ (dashed
line), with arbitrary normalization.} \label{fig:Kap2}
\end{figure}

In Fig. \ref{fig:Kap1} we show the results for the Cabbibo allowed
$\bar B^0 \to J/\psi \pi^+ K^-$, superposing the contribution of the
$\bar \kappa$ and $\bar K^{*0}$ contributions and in Fig.
\ref{fig:Kap2}  the results for the Cabbibo suppressed $\bar B_{s}^0
\to J/\psi \pi^- K^+$, with the contributions of $\kappa$ and
$K^{*0}$. The $\kappa (800)$ scalar contribution is calculated in
Ref.~\refcite{Bayar:2014qha} in the same way as described in the former
subsection.

The narrowness of the  $K^{*}$ relative to the $\rho$, makes the
wide signal of the scalar $\kappa$ to show clearly in regions where
the $K^{*0}$ strength is already suppressed. While no explicit
mention of the  $\kappa$ resonance is done in these $\bar B$ decays,
in some analyses, a background is taken that resembles very much the
$\kappa$ contribution that we have in Fig. \ref{fig:Kap1}
\cite{Abe:2002haa}. The  $\kappa(800)$ appears naturally in chiral
unitary theory of $\pi K$ and coupled channel scattering as a broad
resonance around $800$ MeV, similar to the $f_0(500)$ but with
strangeness \cite{Oller:1998hw}. In $D$ decays, concretely in the $D^{+}
\rightarrow K^- \pi^{+} \pi^{+}$ decay, it is studied with attention
and the links to chiral dynamics are stressed
\cite{Aitala:2002kr,Pennington:2007se}. With the tools of partial
wave analysis developed in Ref.~\refcite{Aaij:2014siy}, it would be
interesting to give attention to this $S$-wave resonance in future
analysis.

\section{The low lying scalar resonances in the $D^0$ decays into $K^0_s$ and $f_0(500)$, $f_0(980)$, $a_0(980)$}\label{sec:lowscalarD0toK0S}

\subsection{Formalism}

\begin{figure}[htbp]\centering
\includegraphics[scale=0.8]{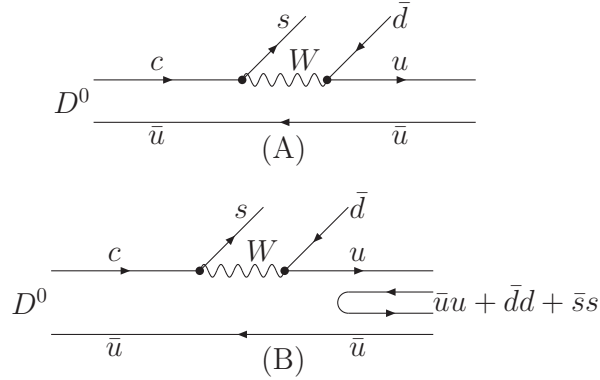}
\caption{(A): Dominant diagram for $D^0 \to \bar{K}^0 u \bar{u}$
and (B): hadronization of the $u \bar{u}$ to give two mesons.\label{Fig:feynLights}}
\end{figure}

The process for $D^0 \to K^0_s R$ proceeds at the elementary quark
level as depicted in Fig.~\ref{Fig:feynLights}(A). The process is Cabibbo
allowed, the $s\bar{d}$ pair produces the $\bar{K}^0$, which will
convert to the observed $K^0_s$ through time evolution with the weak
interaction. The remaining $u\bar{u}$ pair gets hadronized adding an
extra $\bar{q}q$ with the quantum numbers of the vacuum, $\bar{u}u +
\bar{d}d + \bar{s}s$. This topology is the same as for the
$\bar{B}_s \to J/\psi s\bar{s}$ (substituting the $s\bar{d}$ by
$c\bar{c}$)~\cite{Stone:2013eaa}, that upon hadronization of the
$s\bar{s}$ pair leads to the production of the
$f_0(980)$~\cite{Liang:2014tia}, which couples mostly to the hadronized $K
\bar{K}$ components.

The hadronization is implemented as discussed previously.
Hence upon hadronization of the $u\bar{u}$ component we shall have
\begin{equation}
u\bar{u}(\bar{u}u + \bar{d}d + \bar{s}s)  \equiv  (\phi \cdot
\phi)_{11} = \frac{1}{2}\pi^0 \pi^0 + \frac{1}{3} \eta \eta +
\frac{2}{\sqrt{6}} \pi^0 \eta  + \pi^+\pi^- + K^+ K^-~,
\label{eq:phiphi11}
\end{equation}
where we have omitted the $\eta'$ term because of its large mass.
This means that upon hadronization of the $u\bar{u}$ component we
have $D^0 \to \bar{K}^0 PP$, where $PP$ are the different meson
meson components of Eq.~(\ref{eq:phiphi11}). This is only the first
step, because now these mesons will interact among themselves
delivering the desired meson pair component at the end: $\pi^+
\pi^-$ for the case of the $f_0(500)$ and $f_0(980)$, and $\pi^0
\eta$ for the case of the $a_0(980)$.

\begin{figure*}[htbp]
\begin{center}
\includegraphics[scale=0.7]{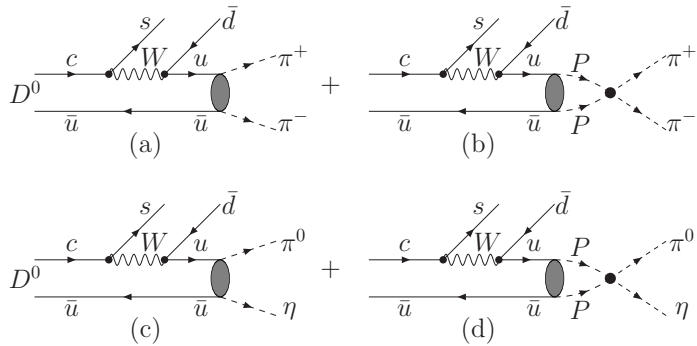}
\caption{ Diagrammatic representation of $\pi^+ \pi^-$ and $\pi^0
\eta$ production. (a) direct $\pi^+ \pi^-$ production, (b) $\pi^+
\pi^-$ production through primary production of a $PP$ pair and
rescattering, (c) primary $\pi^0 \eta$ production, (d) $\pi^0 \eta$
produced through rescattering.} \label{Fig:mesonmesonFSI}
\end{center}
\end{figure*}

The multiple scattering of the mesons is readily taken into account
as shown diagrammatically in Fig.~\ref{Fig:mesonmesonFSI}.

Analytically we shall have
\begin{eqnarray}
t(D^0 \to \bar{K}^0 \pi^+ \pi^-) &=& V_P (1 + G_{\pi^+
\pi^-}t_{\pi^+\pi^- \to \pi^+\pi^-}  + 2 \frac{1}{2} \frac{1}{2}
G_{\pi^0\pi^0} t_{\pi^0 \pi^0 \to \pi^+ \pi^-}  \nonumber \\ && + 2
\frac{1}{3} \frac{1}{2}G_{\eta \eta} t_{\eta \eta \to \pi^+ \pi^-} +
G_{K^+K^-} t_{K^+ K^- \to \pi^+ \pi^-}), \label{eq:fzero}
\end{eqnarray}
and
\begin{eqnarray}
t(D^0 \to \bar{K}^0 \pi^0 \eta) = V_P (\sqrt{\frac{2}{3}} +
\sqrt{\frac{2}{3}} G_{\pi^0 \eta} t_{\pi^0 \eta \to \pi^0 \eta}  +
G_{K^+K^-} t_{K^+ K^- \to \pi^0 \eta}), \label{eq:azero}
\end{eqnarray}
where $V_P$ is a production vertex common to all the terms, and that encodes the underlying dynamics. $G$ is the loop function of two
mesons~\cite{Oller:1997ti} and $t_{ij}$ are the transition scattering
matrices between pairs of pseudoscalars~\cite{Oller:1997ti}. The $f_0(500)$,
$f_0(980)$, and $a_0(980)$ are produced in $s$-wave where $\pi^0
\pi^0$, $\pi^+ \pi^-$ have isospin $I=0$, hence these terms do not
contribute to $\pi^0 \eta$ production ($I=1$) in
Eq.~(\ref{eq:azero}). Note that in Eq.~(\ref{eq:fzero}), as in former sections, we introduce
the factor $\frac{1}{2}$ extra for the identity of the particles for
$\pi^0 \pi^0$ and $\eta \eta$, and a factor 2 for the two possible combinations to produce the two identical particles.

The $t$ matrix is obtained as discussed before and the matrix
elements of the potential can be found in Ref.~\refcite{Xie:2014tma}.

Finally, the mass distribution for the decay is given by
Eq.~(\ref{eq:dGamma}) changing appropriately the variables. However,
since we have a transition $0^- \to 0^- 0^+$ we need $L=0$ now and
the corresponding factor to $\frac{1}{3} p^2_{J/\psi}$ of
Eq.~(\ref{eq:dGamma}) is omitted.

\subsection{Results}

\begin{figure}[htbp]
\begin{center}
\includegraphics[scale=0.44]{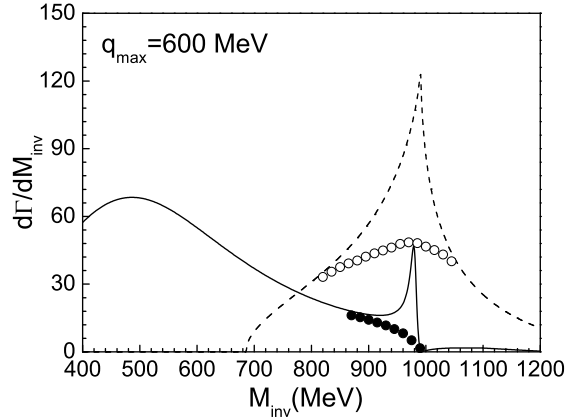}
\caption{The $\pi^+ \pi^-$ (solid line) and $\pi^0 \eta$ (dashed
line) invariant mass distributions for the $D^0 \to \bar{K}^0 \pi^+
\pi^-$ decay and $D^0 \to \bar{K}^0 \pi^0 \eta$ decay, respectively.
A smooth background is plotted below the $a_0(980)$ and $f_0(980)$
peaks.} \label{Fig:dgamr600}
\end{center}
\end{figure}

In Fig.~\ref{Fig:dgamr600}, we show the results for this process. We
have taken the cut off $q_{\rm qmax} = 600$ MeV as in
Ref.~\refcite{Liang:2014tia}. We superpose the two mass distributions
$d\Gamma /dM_{\rm inv}$ for $\pi^+ \pi^-$ (solid line) and $\pi^0
\eta$ (dashed line). The scale is arbitrary, but it is the same for
the two distributions, which allows us to compare $f_0(980)$ with
$a_0(980)$ production. As we discussed before, it is a benefit of
the weak interactions that we can see simultaneously both the $I=0$
$f_0(980)$ and $I=1$ $a_0(980)$ productions in the same $D^0 \to
\bar{K}^0 R$ decay.

When it comes to compare with the experiment we can see that the
$f_0(980)$ signal is quite narrow and it is easy to extract its
contribution to the branching ratios by assuming a smooth
background. For the case of the $\pi^0 \eta$ distribution we get a
clear peak that we associate to the $a_0(980)$ resonance, remarkably
similar in shape to the one found in the
experiment~\cite{Rubin:2004cq}. Not all the
strength seen in Fig.~\ref{Fig:dgamr600} can be attributed to the
$a_0(980)$ resonance. The chiral unitary approach provides full
amplitudes and hence also background. In order to get a ``$a_0(980)$''
contribution we subtract a smooth background fitting a phase space
contribution to the lower part of the spectrum. The remaining part
has a shape with an apparent width of $80$ MeV, in the
middle of the $50 - 100$ MeV of the PDG~\cite{Agashe:2014kda}. Integrating the
area below these structures we obtain
\begin{eqnarray}
R = \frac{\Gamma(D^0 \to \bar{K}^0 a_0(980), a_0(980) \to \pi^0
\eta)}{\Gamma(D^0 \to \bar{K}^0 f_0(980), f_0(980) \to \pi^+\pi^-)}
 = & 6.7 \pm 1.3 , \label{ratioth}
\end{eqnarray}
where we have added a $20\%$ theoretical error due to uncertainties
in the extraction of the background.

Experimentally we find from the PDG and the
Refs.~\refcite{Muramatsu:2002jp,Rubin:2004cq},
\begin{eqnarray}
\Gamma(D^0 \to \bar{K}^0 a_0(980), a_0(980) \to \pi^0 \eta)
&=& (6.5 \pm 2.0) \times 10^{-3}, \\
\Gamma(D^0 \to \bar{K}^0 f_0(980), f_0(980) \to \pi^+\pi^-) &=&
(1.22^{+0.40}_{-0.24}) \times 10^{-3} .
\end{eqnarray}

The ratio that one obtains from there is
\begin{eqnarray}
R = 5.33^{+2.4}_{-1.9} . \label{ratioex}
\end{eqnarray}

The agreement found between Eq.~(\ref{ratioth}) and
Eq.~(\ref{ratioex}) is good, within errors. This is, hence, a
prediction that we can do parameter free.

It should not go unnoticed that we also predict a sizeable fraction
of the decay width into $D^0 \to \bar{K}^0 f_0(500)$, with a
strength several times bigger than for the $f_0(980)$. The $\pi^+
\pi^-$ distribution is qualitatively similar to that obtained in
Ref.~\refcite{Liang:2014tia} for the $\bar{B}^0 \to J/\psi \pi^+\pi^-$ decay,
although the strength of the $f_0(500)$ with respect to the
$f_0(980)$ is relatively bigger in this latter decay than in the
present case (almost $50\%$ bigger). A partial wave analysis is not
available from the work of Ref.~\refcite{Muramatsu:2002jp}, where the
analysis was done assuming a resonant state and a stable meson,
including many contributions, but not the $K^0_s f_0(500)$. Yet, a
discussion is done at the end of the paper~\cite{Muramatsu:2002jp}
in which the background seen is attributed to the $f_0(500)$. With
this assumption they get a mass and width of the $f_0(500)$
compatible with other experiments. Further analyses in the line
of~Ref.~\refcite{Aaij:2014siy} would be most welcome to separate this
important contributions to the $D^0 \to K^0_s \pi^+ \pi^-$ decay.

\subsection{Further considerations}

\begin{figure}[htbp]
\begin{center}
\includegraphics[scale=0.7]{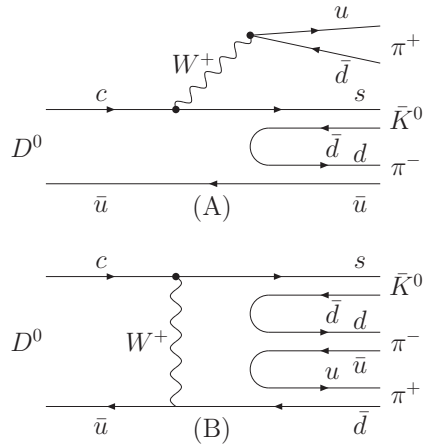}
\caption{External emission diagram [(A)] and the $W$-exchange
diagram [(B)] for $D^0 \to \bar{K}^0 \pi^+ \pi^-$ decay}
\label{wexchange}
\end{center}
\end{figure}

Our results are based on the dominance of the quark diagrams of
Fig.~\ref{Fig:feynLights}. In the weak decay of mesons the diagrams are
classified in six different
topologies~\cite{Chau:1982da,Chau:1987tk}: external emission,
internal emission, $W$-exchange, $W$-annihilation, horizontal
$W$-loop and vertical $W$-loop. As shown in
Ref.~\refcite{Cheng:2010vk}, only the internal emission graph
(Fig.~\ref{Fig:feynLights} of the present work) and
$W$-exchange~\footnote{The $W$-exchange and $W$-annihilation are
often referred together as weak annihilation diagrams.} contribute
to the $D^0 \to \bar{K}^0 f_0(980)$ and $D^0 \to \bar{K}^0a_0(980)$
decays. In Ref.~\refcite{Dedonder:2014xpa} the $D^0 \to \bar{K}^0 \pi^+\pi^-$
decay is studied. Hence, only the $D^0 \to K^0_s f_0(980)$ decay can
be addressed, which is accounted for by proper form factors and
taken into account by means of the $M_2$ ($K^0_s[\pi^+\pi^-]_s$)
amplitude of that paper, which contains the tree level internal emission, and
$W$-exchange (also called annihilation mechanism). We
draw the external emission and $W$-exchange diagrams pertinent to
the $D^0 \to \bar{K}^0\pi^+\pi^-$ decay, as shown in
Fig.~\ref{wexchange}.

A discussion of the relevance of these diagrams is done in
Ref.~\refcite{Xie:2014tma} in connection to the work of Ref.~\refcite{Dedonder:2014xpa}. The
conclusion drawn there is that because the absorption diagrams
involve two quarks of the original wave function, unlike the other
mechanisms that have one of the quarks as spectators, these diagrams
are small.

\section{$\bar B^0$ decay into $D^0$ and $f_0(500)$, $f_0(980)$, $a_0(980)$, $\rho$ and $\bar B^0_s$ decay into $D^0$ and  $\kappa(800)$, $K^{*0}$}

 In this section we report on the
decay of $\bar B^0$ into $D^0$ and $f_0(500)$, $f_0(980)$, and
$a_0(980)$. At the same time we study the decay of $\bar B^0_s$ into
$D^0$ and $\kappa(800)$. We also relate the rates of production of
vector mesons and compare $\rho$ with $f_0(500)$ production and
$K^{*0}$ with $\kappa(800)$ production. Experimentally there is
information on $\rho$ and $f_0(500)$ production in
Ref.~\refcite{Kuzmin:2006mw} for the $\bar B^0$ decay into $D^0$ and
$\pi^+ \pi^-$. There is also information on the ratio of the rates
for $B^0 \to \bar D^0 K^+ K^-$ and $B^0 \to \bar D^0 \pi^+
\pi^-$~\cite{Aaij:2012zka}. We investigate all these rates and
compare them with the experimental information, following the work of Ref.~\refcite{Liang:2014ama}.

\subsection{Formalism}

\begin{figure}[htbp]
\begin{center}
\includegraphics[scale=0.65]{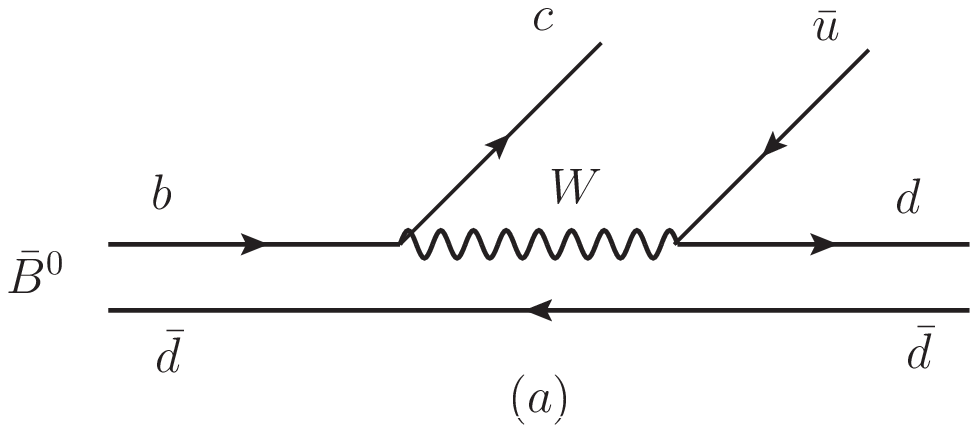}
\includegraphics[scale=0.65]{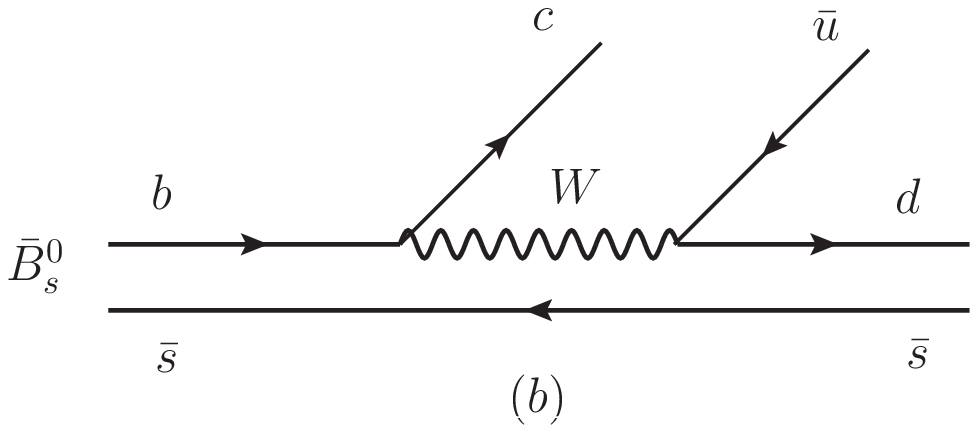}
\caption{Diagrammatic representations of $\bar{B}^0 \to D^0 d
\bar{d}$ decay (a) and $\bar{B}^0_s \to D^0 d \bar{s}$ decay (b).}
\label{Fig:btodqqbar}
\end{center}
\end{figure}

 We show in
Fig.~\ref{Fig:btodqqbar} the dominant diagrams for $\bar{B}^0$
[Fig.~\ref{Fig:btodqqbar} (a)] and $\bar{B}^0_s$
[Fig.~\ref{Fig:btodqqbar} (b)] decays at the quark level. The
mechanism has the $b \to c$ transition, needed for the decay, and
the $u \to d$ vertex that requires the Cabibbo favored $V_{ud}$ CKM matrix element ($V_{ud} =
\cos\theta_c$). Note that these two processes have the same two weak
vertices. Under the assumption that the $\bar{d}$ in
Fig.~\ref{Fig:btodqqbar} (a) and the $\bar{s}$ in
Fig.~\ref{Fig:btodqqbar} (b) act as spectators in these processes,
these amplitudes are identical.

\subsubsection{$\bar{B}^0$ and $\bar{B}^0_s$ decay into $D^0$ and a vector}

Figure~\ref{Fig:btodqqbar} (a) contains $d\bar{d}$ from where the
$\rho$ and $\omega$ mesons can be formed. Figure~\ref{Fig:btodqqbar}
(b) contains $d\bar{s}$ from where the $K^{*0}$ emerges.

Hence, by taking as reference the amplitude for $\bar{B}^0 \to D^0
K^*$ as $V'_P p_D$, we can write by using Eq. (\ref{eq:vector}) the
rest of the amplitudes as
\begin{eqnarray}
&&t_{\bar{B}^0 \to D^0 \rho^0}  =  -\frac{1}{\sqrt{2}} V'_P p_D ,
\label{bzerotodrho} \\
&&t_{\bar{B}^0 \to D^0 \omega}  =  \frac{1}{\sqrt{2}} V'_P p_D ,
\label{bzerotodomega} \\
&&t_{\bar{B}^0 \to D^0 \phi}  = 0 ,
\label{bzerotodphi} \\
&&t_{\bar{B}^0_s \to D^0 K^{*0}}  =   V'_P p_D , \label{bzerotodkstar}
\end{eqnarray}
where $V'_P$ is a common factor to all $\bar B^0 (\bar B^0_s) \to D^0 V_i$ decays, with $V_i$ being a vector meson,
and $p_D$ the momentum of the $D^0$ meson in the rest frame of
the $\bar{B}^0$ (or $\bar{B}^0_s$). The factor $p_D$ is included to account for a necessary $P$-wave
vertex to allow the transition from $0^- \to 0^- 1^-$. Although
parity is not conserved, angular momentum is, and this requires the angular momentum
$L=1$. Note that the angular momentum needed here is different than
the one in the $\bar{B}^0 \to J/\psi V_i$, where
$L=0$. Hence, a mapping from the situation there to
the present case is not possible.

The decay width is given by an expression equivalent to that of Eq.~\eqref{eq:Gam1}.

\subsubsection{$\bar{B}^0$ and $\bar{B}^0_s$ decay into $D^0$ and a pair of pseudoscalar mesons}

In order to produce a pair of mesons, the final quark-antiquark pair $d\bar{d}$ or
$d\bar{s}$ in Fig.~\ref{Fig:btodqqbar} has to hadronize into two mesons. The flavor
content, which is all we need in our study, has been discussed in former sections: we must add a $\bar{q}q$
pair with the quantum numbers of the vacuum, $\bar{u}u + \bar{d}d +
\bar{s}s$.

Following the developments in the former sections, we can write
\begin{eqnarray}
d\bar{d} (\bar{u}u + \bar{d}d + \bar{s}s) \to (\phi \cdot \phi)_{22}
&=& \pi^- \pi^+ +\frac{1}{2} \pi^0\pi^0 +\frac{1}{3}\eta\eta
 - \sqrt{\frac{2}{3}}\pi^0 \eta + K^0 \bar{K}^0, \label{phiphi22}\\
d\bar{s} (\bar{u}u + \bar{d}d + \bar{s}s) \to (\phi \cdot \phi)_{23}
&=& \! \pi^- K^+ - \! \frac{1}{\sqrt{2}} \pi^0 K^0, \label{phiphi23}
\end{eqnarray}
where we have neglected the terms including $\eta'$ that has too
large mass to be relevant in our study.

\begin{figure*}[htbp]
\begin{center}
\includegraphics[scale=0.45]{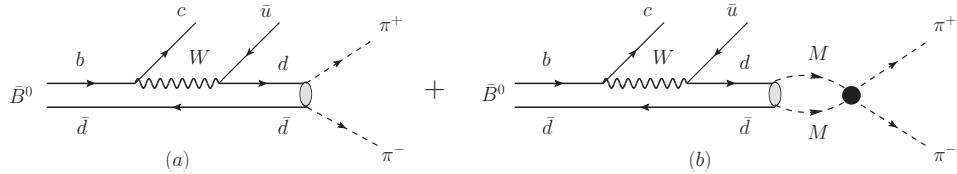}
\caption{Diagrammatic representation of the final state interaction
of the two mesons produced in a primary step. (a) Direct meson-meson production, (b) meson-meson production through rescattering.} \label{Fig:btodpipi}
\end{center}
\end{figure*}

Eqs.~\eqref{phiphi22} and \eqref{phiphi23} give us the weight for
pairs of two pseudoscalar mesons. The next step consists of letting
these mesons interact, which they inevitably will do. This is done
following the mechanism of
Fig.~\ref{Fig:btodpipi}.

The $f_0(500)$ and $f_0(980)$ will be observed in the $\bar{B}^0$
decay into $D^0$ and $\pi^-\pi^+$ final pairs, the $a_0(980)$ in
$\pi^0 \eta$ pairs and the $\kappa(800)$ in the $\bar{B}^0_s$ decay
into $D^0$ and $\pi^- K^+$ pairs. Then we have for the corresponding
production amplitudes
\begin{eqnarray}
t(\bar{B}^0 \to D^0 \pi^- \pi^+) &=& V_P  (1 + G_{\pi^- \pi^+}
t_{\pi^- \pi^+ \to \pi^- \pi^+}  + 2~\frac{1}{2}\frac{1}{2}
G_{\pi^0\pi^0} t_{\pi^0\pi^0 \to \pi^- \pi^+}  \nonumber \\
&&  +2~ \frac{1}{3}\frac{1}{2} G_{\eta \eta} t_{\eta \eta \to \pi^-
\pi^+}+ G_{K^0 \bar{K}^0} t_{K^0 \bar{K}^0 \to \pi^- \pi^+} ),
\label{bzerotodpipi}
\end{eqnarray}
where $V_P$ is a common factor of all these processes, $G_i$ is the
loop function of two meson propagators, and we have included the
factor $\frac{1}{2}$ and $2$ in the intermediate loops involving a pair of
identical mesons, as done in the previous decays. The elements of the scattering matrix $t_{i \to
j}$ are calculated in former sections following the
chiral unitary approach of Refs.~\refcite{Oller:1997ti,Guo:2005wp}. Note that the use
of a common $V_P$ factor in Eq.~(\ref{bzerotodpipi}) is related to
the intrinsic $SU(3)$ symmetric structure of the hadronization $\bar
u u + \bar d d + \bar s s$, which implicitly assumes that we add an
$SU(3)$ $\bar q q$ singlet.

Similarly, we can also produce $K^+ K^-$ pairs and we have
\begin{align}
& t(\bar{B}^0 \to D^0 K^+ K^-) = V_P  ( G_{\pi^- \pi^+} t_{\pi^-
\pi^+ \to K^+ K^-} + 2~\frac{1}{2}\frac{1}{2} G_{\pi^0\pi^0}
t_{\pi^0\pi^0 \to K^+ K^-} \nonumber \\
&  +2~ \frac{1}{3}\frac{1}{2} G_{\eta \eta} t_{\eta \eta \to K^+
K^-} - \sqrt{\frac{2}{3}}G_{\pi^0 \eta} t_{\pi^0 \eta \to K^+ K^-} + G_{K^0 \bar{K}^0} t_{K^0 \bar{K}^0 \to K^+ K^-})~.\label{bzerotodkk}
\end{align}
In the same way we can write\footnote{It is worth noting that
$\pi^+\pi^-$, $\pi^0 \pi^0$, and $\eta \eta$ are in isospin $I=0$,
while $\pi^0 \eta$ is in $I=1$.}
\begin{eqnarray}
t(\bar{B}^0 \to D^0 \pi^0 \eta) = V_P  ( -\sqrt{\frac{2}{3}} -
\sqrt{\frac{2}{3}}G_{\pi^0 \eta} t_{\pi^0 \eta \to \pi^0 \eta} +
G_{K^0 \bar{K}^0} t_{K^0 \bar{K}^0 \to \pi^0 \eta} ),
\label{bzerotodpieta}
\end{eqnarray}
and taking into account that the amplitude for $\bar{B}^0_s \to c
\bar u + d \bar s$ in Fig.~\ref{Fig:btodqqbar} (b) is the same as
for $\bar{B}^0 \to c \bar u + d \bar d$ of Fig.~\ref{Fig:btodqqbar}
(a), and using Eq.~(\ref{phiphi23}) to account for hadronization, we
obtain
\begin{align}
t(\bar{B}^0_s \to D^0 \pi^- K^+) = V_P  \big( 1 +                    
                     & G_{\pi^- K^+} t_{\pi^-K^+ \to \pi^- K^+}  \label{bstodpik} \\
- \frac{1}{\sqrt{2}} & G_{\pi^0 K^0} t_{\pi^0 K^0\to \pi^- K^+} \big) \nonumber,
\end{align}
where the amplitudes $t_{\pi^- K^+ \to \pi^- K^+}$ and $t_{\pi^0 K^0
\to \pi^- K^+}$ are taken from Ref.~\refcite{Guo:2005wp}.

As in the former section, we have the transition $0^- \to 0^- 0^+$
for $\bar{B}^0 \to D^0 f_0$, and now we need $L=0$. The differential
invariant mass width is given again by Eq.(\ref{eq:dGamma}) removing
the factor $\frac{1}{3} p_{J/\psi}^2$ and adopting the appropriate
masses.

\subsection{Numerical results}

In the first place we look for the rates of $\bar{B}^0$ and
$\bar{B}^0_s$ decay into $D^0$ and a vector. By looking at
Eqs.~(\ref{bzerotodrho}), (\ref{bzerotodomega}), and
(\ref{bzerotodkstar}), we have
\begin{eqnarray}
&&\frac{\Gamma_{\bar{B}^0 \to D^0 \rho^0}}{\Gamma_{\bar{B}^0 \to D^0
\omega}}  =  \left[\frac{p_D(\rho^0)}{p_D(\omega)} \right]^3 = 1 , \label{rhotoomega} \\
&&\frac{\Gamma_{\bar{B}^0 \to D^0 \rho^0}}{\Gamma_{\bar{B}^0_s \to D^0
K^{*0}}}  = \left(\frac{M_{\bar{B}^0_s}}{M_{\bar{B}^0}}\right)^2 \frac{1}{2} \left[\frac{p_D(\rho^0)}{p_D(K^{*0})} \right]^3 \simeq \frac{1}{2} , \label{rhotokstar} \\
&&\Gamma_{\bar{B}^0 \to D^0 \phi} = 0 .
\end{eqnarray}

Experimentally there are no data in the PDG~\cite{Agashe:2014kda} for the
branching ratio $Br({\bar{B}^0 \to D^0 \phi})$ and we find the
branching ratios for $B^0 \to \bar{D}^0
\rho^0$~\cite{Kuzmin:2006mw}, $B^0 \to \bar{D}^0 \omega$~\cite{Lees:2011gw,
Blyth:2006at}, and $B^0_s \to \bar{D}^0
\bar{K}^{*0}$~\cite{Aaij:2014baa,Aaij:2013pua,Kuzmin:2006mw}, as the following (note
the change $\bar{B}^0 \to B^0$ and $D^0 \to \bar{D}^0$, $\bar{B}^0_s
\to B^0_s$, $K^{*0} \to \bar{K}^{*0}$):
\begin{eqnarray}
Br(B^0 \to \bar{D}^0 \rho^0) &=& (3.2 \pm 0.5) \times 10^{-4}, \label{gamrrhoexp} \\
Br(B^0 \to \bar{D}^0 \omega) &=& (2.53 \pm 0.16) \times 10^{-4}, \label{gamromegaexp} \\
Br(B^0_s \to \bar{D}^0 \bar{K}^{*0}) &=& (3.5 \pm 0.6) \times 10^{-4}.
\label{gamrkstarexp}
\end{eqnarray}

The ratio $\frac{\Gamma_{\bar{B}^0 \to D^0
\rho^0}}{\Gamma_{\bar{B}^0 \to D^0 \omega}}$ is fulfilled, while the
ratio $\frac{\Gamma_{\bar{B}^0 \to D^0 \rho^0}}{\Gamma_{\bar{B}^0_s
\to D^0 K^{*0}}}$ is barely in agreement with data. The branching
ratio of Eq. (\ref{gamrkstarexp}) requires combining ratios obtained
in different experiments. A direct measure from a single experiment
is available in Ref.~\refcite{Aaij:2011tz}:
\begin{eqnarray}
\frac{\Gamma_{\bar{B}^0_s \to D^0 K^{*0}}}{\Gamma_{\bar{B}^0 \to D^0
 \rho^0}} = 1.48 \pm 0.34 \pm 0.15 \pm 0.12,\label{ratioBtoDexp}
\end{eqnarray}
which is compatible with the factor of $2$ that we get from
Eq.~(\ref{rhotokstar}). However, the result of
Eq.~(\ref{gamrkstarexp}), based on more recent measurements from
Refs.~\refcite{Aaij:2014baa,Aaij:2013pua}, improve on the result of
Eq.~(\ref{ratioBtoDexp})\cite{TimGershon}, which means that our prediction
for this ratio is a bit bigger than experiment.

\begin{figure*}[htbp]
\begin{center}
\includegraphics[scale=0.32]{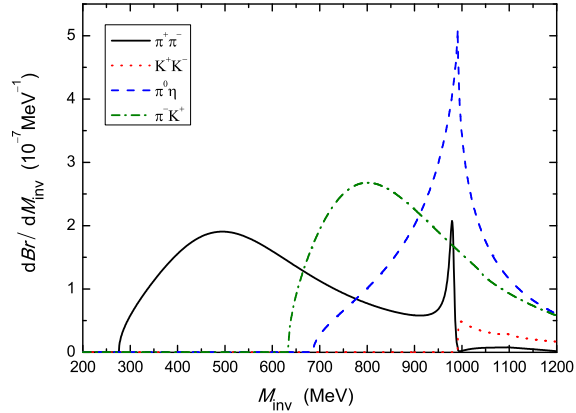}
\caption{ Invariant mass distributions for the
$\pi^+\pi^-$, $K^+ K^-$, and $\pi^0 \eta$, and $\pi^- K$ in
$\bar{B}^0 \to D^0 \pi^+ \pi^-$,~$ D^0 K^+ K^-$, $D^0 \pi^0
\eta$, and $\bar{B}^0_s \to D^0 \pi^- K^+$ decays. The normalization
is such that the integral over the $f_0(500)$ signal gives the experimental branching ratio of Eq. (\ref{sigmabran}).}
\label{Fig:dgamrdminv}
\end{center}
\end{figure*}

We turn now to the production of the scalar resonances. By using
Eqs.~(\ref{bzerotodpipi})-(\ref{bstodpik}),
we obtain the mass
distributions for $\pi^+\pi^-$, $K^+ K^-$, and $\pi^0 \eta$ in
$\bar{B}^0$ decays and $\pi^- K$ in $\bar{B}^0_s$ decay. The
numerical results are shown in Fig.~\ref{Fig:dgamrdminv}.

The normalization for all the processes is the same. The scale is
obtained demanding that the integrated $f_0(500)$ distribution has
the normalization of the experimental branching ratio of Eq.
(\ref{sigmabran}). From Fig.~\ref{Fig:dgamrdminv}, in the $\pi^+
\pi^-$ invariant mass distribution for $\bar B^0 \to D^0 \pi^+
\pi^-$ decay, we observe an appreciable strength for $f_0(500)$
excitation and a less strong, but clearly visible, for the
$f_0(980)$. In the $\pi^0 \eta$ invariant mass distribution, the
$a_0(980)$ is also excited with a strength bigger than that of the
$f_0(980)$. Finally, in the $\pi^- K^+$ invariant mass distribution,
the $\kappa(800)$ is also excited with a strength comparable to that
of the $f_0(500)$. We also plot the mass distribution for $K^+ K^-$
production. It begins at threshold and gets strength from the two
underlying $f_0(980)$ and $a_0(980)$ resonances, hence we can see an
accumulated strength close to threshold that makes the distribution
clearly different from phase space.

There is some experimental information to test some of the
predictions of our results. Indeed in Ref.~\refcite{Kuzmin:2006mw} (see
Table II of that paper) one can find the rates of production for
$f_0(500)$ [it is called $f_0(600)$ there] and $f_0(980)$.
Concretely,
\begin{eqnarray}
Br[\bar B^0 \to D^0 f_0(500)] \cdot Br[f_0(500) \to \pi^+ \pi^-] &=& (0.68 \pm 0.08) \times 10^{-4}, \label{sigmabran} \\
Br[\bar B^0 \to D^0 f_0(980)] \cdot Br[f_0(980) \to \pi^+ \pi^-] &=&
(0.08 \pm 0.04) \times 10^{-4},
\end{eqnarray}
where the errors are only statistical. This gives
\begin{eqnarray}
\left. \frac{Br[\bar B^0 \to D^0 f_0(980)] \cdot Br[f_0(980)\to
\pi^+ \pi^-]}{Br[\bar B^0 \to D^0 f_0(500)] \cdot Br[f_0(500)\to
\pi^+ \pi^-]} \right|_{\rm Exp.} &=& 0.12 \pm 0.06.
\label{eq:ratioBtof0Exp}
\end{eqnarray}

From Fig.~\ref{Fig:dgamrdminv} it is easy to estimate our
theoretical results for this ratio by integrating over the peaks of
the $f_0(500)$ and $f_0(980)$. To separate the $f_0(500)$ and
$f_0(980)$ contributions, a smooth extrapolation of the curve of
Fig. \ref{Fig:dgamrdminv} is made from 900 to 1000 MeV. We find
\begin{equation}\label{eq:ratioBtof0Theo}
\left. \frac{Br[\bar B^0 \to D^0 f_0(980)] \cdot Br[f_0(980) \to \pi^+ \pi^-]}{Br[\bar B^0 \to D^0 f_0(500)] \cdot Br[f_0(500)\to \pi^+ \pi^-]} \right|_{\rm Theo.}=0.08,
\end{equation}
with an estimated error of about 10\%. As we can see, the agreement
of the theoretical results with experiment is good within errors.

It is most instructive to show the $\pi^+ \pi^-$ production combining the $S$-wave and
$P$-wave production. In order to do that, we evaluate $V_P$ of Eq. (\ref{bzerotodpipi}) and $V'_P$ of Eq. (\ref{bzerotodrho}), normalized to obtain the branching fractions
given in Eqs.~(\ref{sigmabran}) and (\ref{gamrrhoexp}), rather than
widths. We shall call the parameters  $\tilde{V}_P$ and $\tilde{V}'_P$, suited to this normalization.

We obtain $\tilde{V}_P = (8.8 \pm 0.5) \times 10^{-2} ~{\rm MeV}^{-1/2}$ and
 $\tilde{V}'_P = (6.8 \pm 0.5) \times 10^{-3} ~{\rm MeV}^{-1/2}$.

To obtain the $\pi^+ \pi^-$  mass distribution for the $\rho$, we need to convert
the total rate for vector production into a mass distribution. For that we follow the formalism developed in Section \ref{Sec:VectorProduction}.

\begin{figure}[tbhp]
\begin{center}
\includegraphics[scale=0.35]{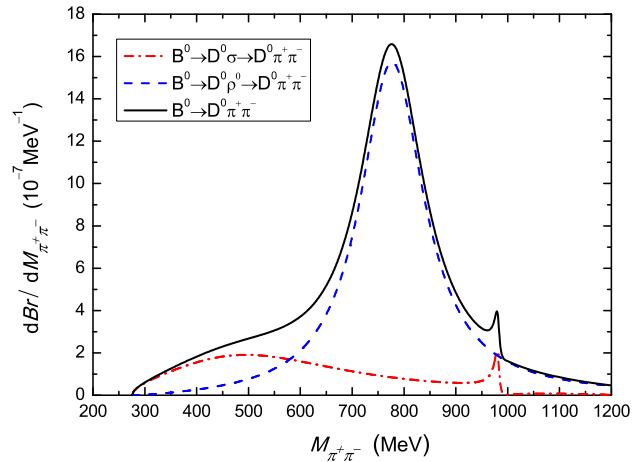}
\caption{Invariant mass distribution for $\pi^+\pi^-$
in $\bar{B}^0 \to D^0 \pi^+ \pi^-$ decay. The normalization is the same as in Fig. \ref{Fig:dgamrdminv}.} \label{Fig:dgamrdminvpipi}
\end{center}
\end{figure}

We show the results for the $\pi^+ \pi^-$ production in $\bar{B}^0
\to D^0 \pi^+ \pi^-$ in Fig.~\ref{Fig:dgamrdminvpipi}. We see a
large contribution from the $f_0(500)$ and a larger contribution
from the $\rho^0 \to \pi^+ \pi^-$ production.  We can see that the
$f_0(500)$ is clearly visible in the distribution of $\pi^+ \pi^-$
invariant mass in the region of $400 \sim 600$ MeV.

\begin{figure}[tbh]
\begin{center}
\includegraphics[scale=0.35]{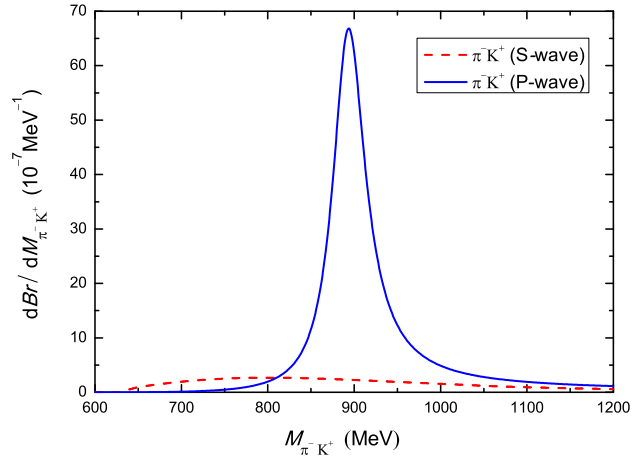}
\caption{Invariant mass distribution for $\pi^- K^+$
in $\bar{B}^0_s \to D^0 \pi^- K^+$ decay. The normalization
is the same as in Fig. \ref{Fig:dgamrdminv}.}
\label{Fig:dgamrdminvpika}
\end{center}
\end{figure}

The $V_P$ and $V'_P$ obtained by fitting the branching ratios of $f_0(500)$ and $\rho$ production
 can be used to obtain the strength of
$K^{*0}$ production versus $\kappa(800)$ production in the $\bar{B}^0_s \to D^0 \pi^- K^+$ decay.
For this we use Eqs.~(\ref{bzerotodrho})-(\ref{bzerotodkstar})
and recall that the
rate for $K^{*0} \to \pi^- K^+$ is $\frac{2}{3}$ of the total
$K^{*0}$ production. The results for $K^{*0} \to \pi^- K^+$ and
$\kappa(800) \to \pi^- K^+$ production are shown in
Fig.~\ref{Fig:dgamrdminvpika}, where we see a clear peak for
$K^{*0}$ production, with strength bigger than that for $\rho^0$ in
Fig.~\ref{Fig:dgamrdminvpipi}, due in part to the factor-of-2 bigger strength in Eq.~(\ref{rhotokstar}) and the smaller $K^{*0}$
width. The $\kappa(800)$ is clearly visible in the lower part of the
spectrum where the $K^{*0}$ has no strength.

Finally, although with more uncertainty, we can also estimate the
ratio
\begin{eqnarray}
\frac{\Gamma(B^0 \to \bar{D}^0 K^+ K^-)}{\Gamma(B^0 \to \bar{D}^0
\pi^+ \pi^-)} = 0.056 \pm 0.011 \pm 0.007 \label{kktopipi}
\end{eqnarray}
of Ref.~\refcite{Aaij:2012zka}. This requires an extrapolation of our
results to higher invariant masses where our results would not be
accurate, but, assuming that most of the strength for both reactions
comes from the region close to the $K^+ K^-$ threshold and from the
$\rho^0$ peak, respectively, we obtain a ratio of the order of $0.03
\sim 0.06$, which agrees qualitatively with the ratio of
Eq.~(\ref{kktopipi}).

\section{$\bar{B}^0$ and $\bar{B}^0_s$ decays into $J/\psi$ and $f_0(1370)$, $f_0(1710)$, $f_2(1270)$, $f'_2(1525)$, $K^*_2(1430)$}
\subsection{Vector- Vector interaction}

In this section we describe the $\bar{B}^0$ and $\bar{B}^0_s$ decays into $J/\psi$ together with $f_0(1370)$, $f_0(1710)$, $f_2(1270)$, $f'_2(1525)$, or $K^*_2(1430)$. The latter are resonances that are dynamically generated in the vector-vector interaction, which we briefly discuss now. In these interactions, an interesting surprise was found when using the local hidden gauge
Lagrangians and, with a similar treatment to the one of the scalar
mesons, new states were generated that could be associated with
known resonances. This study was first conducted in the work of
Ref.~\refcite{Molina:2008jw}, where the $f_0(1370)$ and $f_2(1270)$
resonances were shown to be generated from the $\rho \rho$
interaction provided by the local hidden gauge
Lagrangians~\cite{Bando:1984ej,Bando:1987br,Meissner:1987ge} implementing
unitarization. The work was extended to $SU(3)$ in
Ref.~\refcite{Geng:2008gx} and eleven resonances where dynamically
generated, some of which were identified with the $f_0(1370)$,
$f_0(1710)$, $f_2(1270)$, $f'_2(1525)$ and $K^*_2(1430)$. The idea
has been tested successfully in a large number of reactions and in
Ref.~\refcite{Xie:2014gla} a compilation and a discussion of these works have been
done.

\subsection{Formalism}
\begin{figure*}[htbp]
\begin{center}
\includegraphics[scale=0.6]{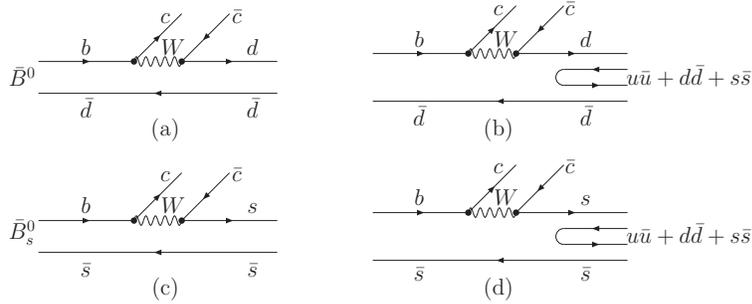}
\caption{ Basic diagrams for $\bar{B}^0$ and $\bar{B}^0_s$ decay
into $J/\psi$ and a $q\bar{q}$ pair [(a) and (c)], and hadronization
of the $q \bar{q}$ components [(b) and (d)].} \label{Fig:feynMedium}
\end{center}
\end{figure*}

As done in former sections we take the dominant mechanism for the
decay of $\bar{B}_0$ and $\bar{B}^0_s$ into a $J/\psi$ and a
$q\bar{q}$ pair. Posteriorly, this $q\bar{q}$ pair is hadronized
into vector meson-vector meson components, as depicted in
Fig.~\ref{Fig:feynMedium}, and this hadronization is implemented as has already been explained in former sections.

In this sense the hadronized $d\bar{d}$ and $s\bar{s}$ states in
Fig.~\ref{Fig:feynMedium} can be written as
\begin{eqnarray}
d\bar{d}(\bar{u}u + \bar{d}d + \bar{s}s) &=& (M\cdot M)_{22}, \\
s\bar{s}(\bar{u}u + \bar{d}d + \bar{s}s) &=& (M\cdot M)_{33} .
\end{eqnarray}
However, now it is convenient to establish the relationship of these
hadronized components with the vector meson-vector meson components
associated to them. For this purpose we write the matrix $M$ which
has been defined in Eq.~(\ref{eq:qqbarmatrix}) in terms of the nonet
of vector mesons:
\begin{equation}\label{eq:vectormatrix}
V = \!\! \left( \!\!
           \begin{array}{ccc}
             \frac{\sqrt{2}}{2}\rho^0 +  \frac{\sqrt{2}}{2} \omega  & \rho^+ & K^{*+} \\
             \rho^- & - \frac{\sqrt{2}}{2} \rho^0 +  \frac{\sqrt{2}}{2} \omega  & K^{*0} \\
            K^{*-} & \bar{K}^{*0} & \phi \\
           \end{array}
       \!\!  \right)~,
\end{equation}
and then we associate
\begin{eqnarray}
d\bar{d}(\bar{u}u + \bar{d}d + \bar{s}s) &\equiv& (V\cdot V)_{22}  = \rho^-\rho^+ +
 \frac{1}{2}\rho^0\rho^0 + \frac{1}{2} \omega \omega -\rho^0 \omega + K^{*0}\bar{K}^{*0},  \label{eq:ddbarhadronization} \\
s\bar{s}(\bar{u}u + \bar{d}d + \bar{s}s) &\equiv& (V\cdot V)_{33} =
K^{*-}K^{*+} + K^{*0}\bar{K}^{*0} + \phi \phi .
\label{eq:ssbarhadronization}
\end{eqnarray}

\begin{figure*}[htbp]
\begin{center}
\includegraphics[scale=0.7]{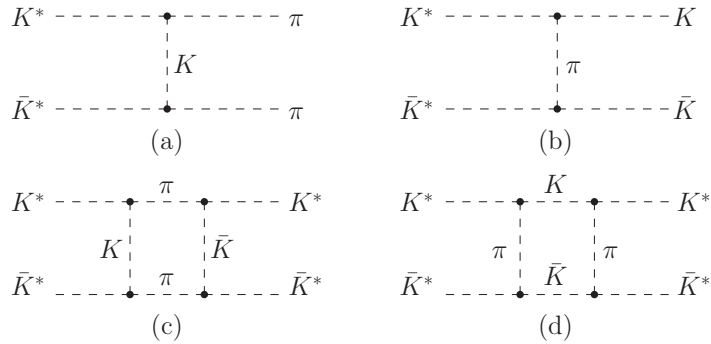}
\caption{ Decay mechanisms of $K^*\bar{K}^* + \pi\pi$, $K\bar{K}$
[(a) and (b)] and box diagrams considered\cite{Geng:2008gx} to
account for these decays [(c) and (d)].} \label{Fig:box}
\end{center}
\end{figure*}

In the study of Ref.~\refcite{Geng:2008gx} a coupled channels unitary
approach was followed with the vector meson-vector meson states as
channels. However, the approach went further since, following the
dynamics of the local hidden gauge Lagrangians, a vector
meson-vector meson state can decay into two pseudoscalars, $PP$.
This is depicted in Figs.~\ref{Fig:box} (a), (b).

In Ref.~\refcite{Geng:2008gx} these decay channels are taken into account
by evaluating the box diagrams depicted in Figs.~\ref{Fig:box} (c),
(d), which are assimilated as a part, $\delta \tilde{V}$, of the
vector vector interaction potential $\tilde{V}$. This guarantees
that the partial decay width into different channels could be taken
into account.

\begin{figure}[htbp]
\begin{center}
\includegraphics[scale=0.9]{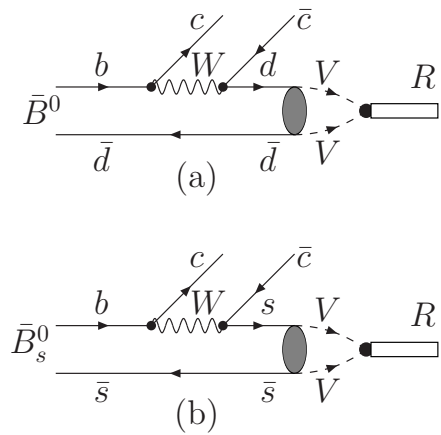}
\caption{ Mechanisms to generate the vector vector resonances
through $VV$ rescattering. The dot of the vertex $RVV$ indicates the
coupling of the resonance to the different $VV$ components.}
\label{Fig:jpsiR}
\end{center}
\end{figure}

Since we wish to have the resonance production and this is obtained
through rescattering, the mechanism for $J/\psi$ plus resonance
production is depicted in Fig.~\ref{Fig:jpsiR}.

The amplitudes for $J/\psi R$ production are then given by
\begin{eqnarray}
t(\bar{B}^0 \to J/\psi f_0) &=& \tilde{V}_P V_{cd} p_{J/\psi} {\rm
cos}\theta (G_{\rho^- \rho^+} g_{\rho^-\rho^+, f_0} + 2
\frac{1}{2}\frac{1}{2} G_{\rho^0\rho^0}g_{\rho^0\rho^0,f_0}
 \nonumber \\
 && + 2 \frac{1}{2}\frac{1}{2}G_{\omega\omega}g_{\omega\omega,f_0} + G_{K^{*0}\bar{K}^{*0}}g_{K^{*0}\bar{K}^{*0},f_0}), \label{eq:bzerojpaifzero} \\
t(\bar{B}^0_s \to J/\psi f_0) &=& \tilde{V}_P V_{cs} p_{J/\psi} {\rm
cos}\theta (G_{K^{*0}\bar{K}^{*0}}g_{K^{*0}\bar{K}^{*0},f_0}  \nonumber \\
 && +
G_{K^{*-} K^{*+}} g_{K^{*-}K^{*+}, f_0}  + 2
\frac{1}{2}G_{\phi\phi}g_{\phi\phi,f_0} ) ,
\label{eq:bszerojpaifzero}
\end{eqnarray}
where $G_{VV}$ are the loop functions of two vector mesons that we
take from Ref.~\refcite{Geng:2008gx} and $g_{VV,f_0}$ the couplings of $f_0$ to
the pair of vectors $VV$, defined from the residues of the resonance
at the poles
\begin{eqnarray}
t_{ij} \simeq \frac{g_i g_j}{s-s_R},
\end{eqnarray}
with $t_{ij}$ the transition matrix from the channel $(VV)_i$ to
$(VV)_j$. These couplings are also tabulated in Ref.~\refcite{Geng:2008gx}.
The formulas for the decay amplitudes to $J/\psi f_2$ are identical,
substituting $f_0$ by $f_2$ in the formulas and the factor
$\tilde{V}_P$ by a different one $\tilde{V}'_P$ suited for the
hadronization into a tensor. The magnitudes $\tilde{V}_P$ and
$\tilde{V}'_P$ represent the common factors to these different
amplitudes which, before hadronization and rescattering of the
mesons, are only differentiated by the CKM matrix elements
$V_{cd}~,V_{cs} $.

Note that as in former cases we include a factor $1/2$ in the $G$
functions for the $\rho^0 \rho^0$, $\omega \omega$, and $\phi \phi$
cases and a factor 2 for the two combinations to create these states, to account for the identity of the particles. The factor
$p_{J/\psi} {\rm cos}\theta$ is included there to account for a
$p$-wave in the $J/\psi$ particle to match angular momentum in the
$0^- \to 1^-0^+$ transition. The factor $p_{J/\psi}$ can however play some role
due to the difference of mass between the different resonances.

\begin{figure}[htbp]
\begin{center}
\includegraphics[scale=0.8]{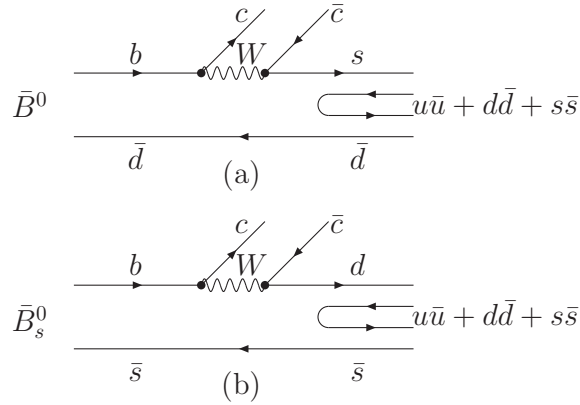}
\caption{ Mechanisms for $\bar{B}^0 \to J/\psi \bar{K}^{*}_2(1430)$
and $\bar{B}^0_s \to J/\psi K^{*}_2(1430)$.} \label{Fig:Kstar}
\end{center}
\end{figure}

The case for $\bar{B}^0 \to J/\psi \bar{K}^{*}_2(1430)$ decay is
similar. The diagrams corresponding to Figs.~\ref{Fig:feynMedium} (b), (d)
are now written in Fig.~\ref{Fig:Kstar}.

In analogy to Eqs. (\ref{eq:ddbarhadronization}),
(\ref{eq:ssbarhadronization}) we now have
\begin{eqnarray} \label{eq:kstar}
s\bar{d}(u\bar{u} + d\bar{d} + s\bar{s}) &\equiv& (V\cdot V)_{32} = K^{*-} \rho^+
+ \bar{K}^{*0} (-\frac{ \rho^0}{\sqrt{2}} + \frac{\omega}{\sqrt{2}} )  + \bar{K}^{*0} \phi , \\
d\bar{s}(u\bar{u} + d\bar{d} + s\bar{s}) &\equiv& (V\cdot V)_{23} =
\rho^- K^{*+} + (-\frac{ \rho^0}{\sqrt{2}} + \frac{\omega}{\sqrt{2}}
) K^{*0} + K^{*0} \phi ,
\end{eqnarray}
and the amplitudes for production of $J/\psi \bar{K}^*_2(1430)$ will
be given by
\begin{eqnarray}
t(\bar{B}^0 \to J/\psi \bar{K}^*_2) &=& \tilde{V}'_P p_{J/\psi} {\rm
cos}\theta V_{cs} ( G_{K^{*-} \rho^+} g_{K^{*-}\rho^+, \bar{K}^*_2}
 - \frac{1}{\sqrt{2}} G_{\bar{K}^{*0} \rho^0} g_{\bar{K}^{*0}
\rho^0, \bar{K}^*_2} \nonumber \\
&& + \frac{1}{\sqrt{2}} G_{\bar{K}^{*0} \omega}g_{\bar{K}^{*0}
\omega, \bar{K}^*_2} + G_{\bar{K}^{*0} \phi}g_{\bar{K}^{*0} \phi,
\bar{K}^*_2} ), \label{eq:bzerojpsikstar} \\
t(\bar{B}^0_s \to J/\psi K^*_2) &=& \tilde{V}'_P p_{J/\psi} {\rm
cos}\theta V_{cd} ( G_{K^{*+} \rho^-}g_{K^{*+}\rho^-, K^*_2}
 - \frac{1}{\sqrt{2}} G_{\bar{K}^{*0} \rho^0} g_{\bar{K}^{*0}
\rho^0, K^*_2} \nonumber \\
&& + \frac{1}{\sqrt{2}} G_{\bar{K}^{*0} \omega}g_{\bar{K}^{*0}
\omega, K^*_2} \!\! + \! G_{\bar{K}^{*0} \phi}g_{\bar{K}^{*0} \phi,
K^*_2} ) . \label{eq:bszerojpsikstar}
\end{eqnarray}

In Ref.~\refcite{Xie:2014gla} these amplitudes are written in terms of the
isospin amplitudes which are generated in Ref.~\refcite{Geng:2008gx}. The width
for these decays will be given by
\begin{eqnarray}
\Gamma = \frac{1}{8\pi M^2_{\bar{B}}} |t|^2 p_{J/\psi} , ~~~{\rm
with} ~~~p_{J/\psi} = \frac{\lambda^{1/2}(M^2_{\bar{B}}, M^2_{J/\psi},
M^2_R)}{2 M_{\bar{B}}}
\end{eqnarray}
with $M_R$ the resonance mass, and in $|t|^2$ we include the factor
$1/3$ for the integral of ${\rm cos}\theta$, which cancels in all
ratios that we will study.

The information on couplings and values of the $G$ functions,
together with uncertainties is given in Table V of
Ref.~\refcite{MartinezTorres:2009uk} and Table I of Ref.~\refcite{Geng:2009iw}. The
errors for the scalar mesons production are taken from
Ref.~\refcite{Geng:2009iw}.

\subsection{Results}

In the PDG we find rates for $\bar{B}^0_s \to J/\psi
f_0(1370)$~\cite{LHCb:2012ae}, $\bar{B}^0_s \to J/\psi
f_2(1270)$~\cite{LHCb:2012ae} and $\bar{B}^0_s \to J/\psi
f'_2(1525)$~\cite{Aaij:2011ac}. We can calculate $10$ independent
ratios and we have two unknown normalization constants $\tilde{V}_P$
and $\tilde{V}'_P$. Then we can provide eight independent ratios
parameter free. From the present experimental data we can only get
one ratio for the $\bar{B}^0_s \to J/\psi f_2(1270) [f'_2(1525)]$.
There is only one piece of data for the scalars, but we should also
note that the data for $\bar{B}^0_s \to J/\psi f_0(1370)$ in
Ref.~\refcite{LHCb:2012ae} and in the PDG, in a more recent
paper~\cite{Aaij:2014emv} is claimed to correspond to the
$f_0(1500)$ resonance. Similar ambiguities stem from the analysis of
Ref.~\refcite{LHCb:2012ad}.

The data for $f_2(1270)$~\cite{LHCb:2012ae} and
$f'_2(1525)$~\cite{Aaij:2011ac} of the PDG are
\begin{eqnarray}
\Gamma(\bar{B}^0_s \to J/\psi f_2(1270)) & = &(10^{+5}_{-4}) \times
10^{-7}, \\
\Gamma(\bar{B}^0_s \to J/\psi f'_2(1525)) & = & (2.6 ^{+0.9}_{-0.6})
 \times 10^{-4}.
\end{eqnarray}

However, the datum for $\Gamma(\bar{B}^0_s \to J/\psi f'_2(1525))$
of the PDG is based on the contribution of only one helicity
component $\lambda = 0$, while $\lambda = \pm 1$ contribute in
similar amounts.

This decay has been further reviewed in Ref.~\refcite{Aaij:2014emv} and
taking into account the contribution of the different helicities a
new number is now provided,~\footnote{From discussions with S. Stone
and L. Zhang.}
\begin{eqnarray}
\Gamma(\bar{B}^0_s \to J/\psi f_2(1270)) = \!\! (3.0 ^{+1.2}
_{-1.0}) \! \times  \! \! 10^{-6}, \label{br1525}
\end{eqnarray}
which is about three times larger than the one reported in the PDG
(at the date of this review).

The results are presented in Table~\ref{ratios} for the eight ratios
that we predict, defined as,
\begin{eqnarray}
R_1 &=& \frac{\Gamma[ \bar{B}^0 \to J/\psi f_0(1370)]}{\Gamma[
\bar{B}^0 \to J/\psi f_0(1710)]} , ~R_2 =  \frac{\Gamma[ \bar{B}^0
\to J/\psi f_2(1270)]}{\Gamma[ \bar{B}^0 \to J/\psi f'_2(1525)]} ,
\nonumber \\
R_3 & = & \frac{\Gamma[ \bar{B}^0 \to J/\psi f_2(1270)]}{\Gamma[
\bar{B}^0 \to J/\psi \bar{K}^*_2(1430)]} , ~ R_4 =  \frac{\Gamma[
\bar{B}^0 \to J/\psi f_0(1710)]}{\Gamma[ \bar{B}^0_s \to J/\psi
f_0(1710)]} , \nonumber \\
R_5 & = & \frac{\Gamma[ \bar{B}^0 \to J/\psi f_2(1270)]}{\Gamma[
\bar{B}^0_s \to J/\psi f_2(1270)]} , ~ R_6 = \frac{\Gamma[
\bar{B}^0_s \to
J/\psi f_0(1370)]}{\Gamma[ \bar{B}^0_s \to J/\psi f_0(1710)]}, \nonumber \\
R_7 &=&  \frac{\Gamma[ \bar{B}^0_s \to J/\psi f_2(1270)]}{\Gamma[
\bar{B}^0_s \to J/\psi f'_2(1525)]} , ~ R_8 =  \frac{\Gamma[
\bar{B}^0_s \to J/\psi f_2(1270)]}{\Gamma[ \bar{B}^0_s \to J/\psi
K^*_2(1430)]} . \nonumber
\end{eqnarray}

\begin{table}[t]
\begin{center}
\caption{Ratios of $\bar{B}^0$ and $\bar{B}^0_s$ decays.}
\begin{tabular}{|c|c|c|}
\hline \hline Ratios  & Theory & Experiment \\
\hline $R_1 $ & $6.2 \pm 1.6$ & $--$ \\ \hline $R_2 $  & $13.4 \pm
6.7$ &$--$
\\ \hline
$R_3 $  & $(3.0 \pm 1.5) \times 10^{-2}$ & $--$ \\
\hline
$R_4 $  & $(7.7 \pm 1.9) \times 10^{-3}$ & $--$ \\
\hline
$R_5$   & $(6.4 \pm 3.2) \times 10^{-1}$ & $--$ \\
\hline
$R_6 $  & $(1.1 \pm 0.3) \times 10^{-2}$ & $--$ \\
\hline
$R_7 $  & $(8.4 \pm 4.6) \times 10^{-2}$ & $(1.0 \sim 3.8) \times 10^{-2}$ \\
\hline
$R_8 $  & $(8.2 \pm 4.1) \times 10^{-1}$ & $--$ \\
\hline \hline
\end{tabular} \label{ratios}
\end{center}
\end{table}

Note that the different ratios predicted vary in a range of
$10^{-3}$, such that even a qualitative level comparison with future
experiments would be very valuable concerning the nature of the
states as vector vector molecules, on which the numbers of the
Tables are based.

The errors are evaluated in quadrature from the errors in
Refs.~\refcite{MartinezTorres:2009uk,Geng:2009iw}. In the case of $R_7$, where we
can compare with the experiment, we put the band of experimental
values for the ratio. The theoretical results and the experiment
just overlap within errors.

From our perspective it is easy to understand the small ratio of
these decay rates. The $f_2(1270)$ in Ref.~\refcite{Geng:2008gx} is
essentially a $\rho \rho$ molecule while the $f'_2(1525)$ couples
mostly to $K^* \bar{K}^*$. If one looks at
Eq.~(\ref{eq:bszerojpaifzero}) one can see that the $\bar{B}^0_s \to
J/\psi f_0(f_2)$ proceeds via the $K^* \bar{K}^*$ and $\phi \phi$
channels, hence, the $f_2(1270)$ with small couplings to
$K^*\bar{K^*}$ and $\phi \phi$ is largely suppressed, while the
$f'_2(1525)$ is largely favored.

\section{Learning about the nature of open and hidden charm mesons}

The interaction of mesons with charm has also been addressed from
the perspective of an extension of the chiral unitary approach.
Meson meson interactions have been studied in
many works\cite{Kolomeitsev:2003ac,Guo:2006fu,Gamermann:2006nm,Guo:2009ct,Cleven:2010aw,Cleven:2014oka},
and a common result is that there are many states that are
generated dynamically from the interaction which can be associated
to some known states, while there are also predictions for new
states. Since then there have been ideas on how to prove that the
nature of these states corresponds to a kind of molecular structure of
some channels. The idea here is to take advantage of the information
provided by the $B$ and $D$ decays to shed light on the nature of
these states. We are going to show how the method works with two
examples, one where the $D_{s0}^{\ast+}(2317)$  resonance is
produced and the other one where some $X,Y,Z$ states are produced.

The very narrow charmed scalar meson $D_{s0}^{\ast+}(2317)$ was
first observed in the $D_s^+\pi^0$ channel by the BABAR
Collaboration \cite{Aubert:2003fg,Aubert:2003pe} and its existence was confirmed by CLEO
\cite{Besson:2003cp}, BELLE \cite{Krokovny:2003zq} and FOCUS \cite{Vaandering:2004ix}
Collaborations. Its mass was commonly measured as $2317 \MeV$, which
is approximately $160 \MeV$ below the prediction of the very
successful quark model for the charmed mesons \cite{Godfrey:1985xj,Godfrey:1986wj}. Due to its
low mass, the structure of the meson $D_{s0}^{\ast\pm}(2317)$ has
been extensively debated. It has been interpreted as a $c\bar{s}$
state \cite{Dai:2003yg,Bali:2003jv,Dougall:2003hv,Hayashigaki:2004gq,Narison:2003td}, two-meson molecular state
\cite{Barnes:2003dj,Szczepaniak:2003vy,Kolomeitsev:2003ac,Guo:2006fu,Gamermann:2006nm,Guo:2009ct,Cleven:2010aw,Cleven:2014oka},
$K-D$- mixing \cite{vanBeveren:2003kd}, four-quark states \cite{Cheng:2003kg,Terasaki:2003qa,Maiani:2004vq,Bracco:2005kt}
or a mixture between two-meson and four-quark states \cite{Browder:2003fk}.
Additional support to the molecular interpretation  came recently
from lattice QCD simulations \cite{Mohler:2013rwa,Liu:2012zya,Lang:2014yfa,Torres:2014vna}.
In previous lattice studies of the $D_{s0}^{\ast}(2317)$, it was
treated as a conventional quark-antiquark state and  no states with
the correct mass  (below  the  $KD$ threshold) were found. In
Refs.~\refcite{Mohler:2013rwa,Lang:2014yfa}, with the introduction of $KD$  meson
correlators and using the effective range formula, a bound state is
obtained about 40 MeV below the $KD$ threshold. The fact that the
bound state appears with the $KD$ interpolator may be interpreted as
a possible $KD$ molecular structure, but more precise statements
cannot be done. In Ref.~\refcite{Liu:2012zya} lattice QCD results for
the $KD$ scattering length are extrapolated to physical pion masses
by means of unitarized chiral perturbation theory, and by means of
the Weinberg compositeness condition
\cite{Weinberg:1965zz,Baru:2003qq} the amount of $KD$ content in the
$D_{s0}^*(2317)$ is determined, resulting in a sizable fraction of
the order of 70\% within errors. Yet, one should take this result with caution since it results from using one of the Weinberg compositeness\cite{Weinberg:1965zz} conditions in an extreme case. A reanalysis of the lattice spectra
of Refs.~\refcite{Mohler:2013rwa,Lang:2014yfa} has been recently done in
Ref.~\refcite{Torres:2014vna}, going beyond the effective range approximation
and making use of the three levels of Refs.~\refcite{Mohler:2013rwa,Lang:2014yfa}. As a
consequence, one can be more quantitative about the nature of the
$D_{s0}(2317)$, which appears with a $KD$ component of about 70\%,
within errors. Further works relating LQCD results and the $D^\ast_{s0}(2317)$ resonance can be found in Refs.~\refcite{Altenbuchinger:2013vwa,Altenbuchinger:2013gaa}.

In addition to these lattice results, and more precise ones that
should be available in the future, it is very important to have some
experimental data that could be used to test the internal structure
of this exotic state.

\section[$D_{s0}^{\ast\pm}(2317)$ and $KD$ scattering from $B^0_s$ decay]{\boldmath$D_{s0}^{\ast\pm}(2317)$ and $KD$ scattering from $B^0_s$ decay}\label{sec:Ds02317_Alba}

Here we propose to use the experimental $KD$ invariant mass
distribution of the weak decay of $\bar{B}_s^0\to D_s^-(DK)^+$ to
obtain information about the internal structure of the
$D_{s0}^{\ast+}(2317)$ state~\footnote{Throughout this work, the
notation $(DK)^+$ refers to the isoscalar combination $D^0 K^+ + D^+
K^0$.}. There are not yet experimental data for the decay
$\bar{B}^0_s \to D_s^-(DK)^+$. However, based on the 1.85\% and
1.28\% branching fractions for the decays $\bar{B}^0_s \to
D_s^{*+}D_s^{*-}$ and $\bar{B}^0_s \to
D_s^{+}D_s^{*-}+D_s^{*+}D_s^{-}$, the branching fraction for the
$\bar{B}^0_s \to D_s^-D_{s0}^{\ast+}$ decay, should not be so
different from that and be seen through the channel $\bar{B}^0_s \to
D_s^-(DK)^+$. It is worth stressing that in the reactions $B^0 \to
D^- D^0 K^+$ and $B^+ \to \bar{D}^0 D^0 K^+$ studied by the BABAR
Collaboration \cite{Lees:2014abp}, an enhancement in the invariant
$DK$ mass in the range $2.35-2.50\ \text{GeV}$ is observed, which
could be associated with this $D_{s0}^{\ast+}(2317)$ state. It is
also interesting to mention that, in the reaction $B^0_s \to
\bar D^0 K^- \pi^+$, the LHCb Collaboration also finds an
enhancement close to the $KD$ threshold in the $\bar D^0 K^-$
invariant mass distribution, which is partly associated to the
$D_{s0}^*(2317)$ resonance \cite{Aaij:2014baa}.

\begin{figure*}[htbp]
\centering
\includegraphics[scale=0.5]{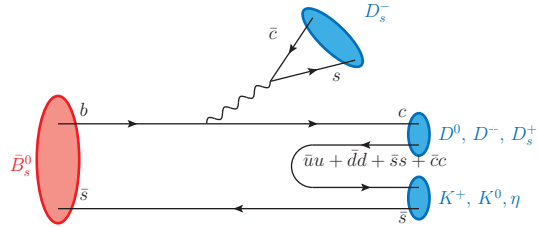}
\caption{Mechanism for the decay $\bar{B}^0_s \to D_s^-(DK)^+$.\label{fig:diag}}
\end{figure*}

In Fig.~\ref{fig:diag} we show the mechanism for the decay
$\bar{B}_s^0\to D_s^- (DK)^+$. One takes the dominant mechanism for
the weak decay of the $\bar{B}_s^0$ into $D_s^-$ plus a primary
$c\bar{s}$ pair. The hadronization of the initial $c\bar{s}$ pair is
achieved by inserting a $q\bar q$ pair with the quantum numbers of
the vacuum: $u\bar u +d\bar d +s\bar s + c \bar c$, as shown in
Fig.~\ref{fig:diag}. Therefore, the $c\bar{s}$ pair is hadronized
into a pair of pseudoscalar mesons. This pair of pseudoscalar mesons
is then allowed to interact to produce the $D_{s0}^{\ast+}(2317)$
resonance, which is considered here as mainly a $DK$ molecule
\cite{Gamermann:2006nm}. The idea is similar to the one used in former
sections for the formation of the $f_0(980)$ and $f_0(500)$ scalar
resonances in the decays of $B^0$ and $B^0_s$.

\subsection{Formalism}\label{sec:formalismDKAlbaladejo}

Here the $D_{s0}^{\ast+}(2317)$ is considered as a bound state of $D
K$ and one looks at the shape of the $DK$ distribution close to
threshold of the $\bar{B}^0_s \to D_s^-(DK)^+$ decay.

\subsubsection{Elastic $DK$ scattering amplitude}\label{subsec:Ampli}

We follow here the developments of Ref.~\refcite{Albaladejo:2015kea}. Let us start by discussing the $S$-wave amplitude for the isospin $I=0$ $DK$ elastic scattering,
which we denote $T$. It can be written as
\begin{equation}\label{eq:TEFT}
T^{-1}(s) = V^{-1}(s)-G(s)\Rightarrow T(s)=V(s)(1+G(s)T(s))~,
\end{equation}
where $G(s)$ is a loop function which in dimensional regularization can be written as

\begin{align}
16\pi^2 G(s) =& a(\mu) +  \log\frac{m_D m_K}{\mu^2} +
\frac{\Delta}{2s}\log\frac{m_D^2}{m_K^2} + \frac{\nu}{2s} \left(
\log\frac{s-\Delta+\nu}{-s+\Delta+\nu} +
\log\frac{s+\Delta+\nu}{-s-\Delta+\nu}
\right) \label{eq:GloopSubtracted}~, \\
& \Delta = m_D^2-m_K^2~, \quad \nu =
\lambda^{1/2}(s,m_D^2,m_K^2)~.\nonumber
\end{align}
In Eq.~\eqref{eq:TEFT}, $V(s)$ is the potential, typically extracted
from some effective field theory, although a different approach will
be followed here.

The amplitude $T(s)$ can also be written in terms of the phase shift $\delta(s)$ and/or
effective range expansion parameters,
\begin{equation}\label{eq:scleAlb}
T(s)= -\frac{8\pi\sqrt{s}}{p_K \text{ctg}\delta - i p_K}\simeq-\frac{8\pi \sqrt{s}}{\displaystyle
\frac{1}{a}+\frac{1}{2}r_0 p_K^2 -i p_K} ,
\end{equation}
with
\begin{equation}\label{eq:pk}
p_K(s) = \frac{\lambda^{1/2}(s,M_K^2,M_D^2)}{2\sqrt{s}}~,
\end{equation}
the momentum of the $K$ meson in the $DK$ center of mass system. In Eq.~\eqref{eq:scleAlb}, $a$ and $r_0$ are the
scattering length and the effective range, respectively.

Taking the potential of Ref.~\refcite{Gamermann:2006nm} for $DK$ scattering, we find the $D_{s0}^{\ast+}(2317)$ resonance below the $DK$ threshold, the
latter being located roughly above $2360\ \text{MeV}$. This means
that the amplitude has a pole at the squared mass of this state,
$M^2 \equiv s_0$, so that, around the pole,
\begin{equation}\label{eq:Tpole}
T(s) = \frac{g^2}{s - s_0} + \text{regular terms},
\end{equation}
where $g$ is the coupling of the state to the $DK$ channel. From
Eqs.~\eqref{eq:TEFT} and \eqref{eq:Tpole}, we see that (the
following derivatives are meant to be calculated at $s=s_0$):
\begin{equation}
\frac{1}{g^2} = \frac{\partial T^{-1}(s) }{\partial s} = \frac{\partial V^{-1}(s)}{\partial s}
- \frac{\partial G(s)}{\partial s}~.
\end{equation}
We have thus the following exact sum rule,
\begin{equation}\label{eq:sumrule}
1 = g^2\left(- \frac{\partial G(s)}{\partial s} + \frac{\partial V^{-1}(s)}{\partial s} \right)~.
\end{equation}
In Ref.~\refcite{Gamermann:2009uq} it has been shown, as a generalization of the Weinberg
compositeness condition \cite{Weinberg:1965zz} (see also Ref.~\refcite{Sekihara:2014kya} and
references therein), that the probability $P$ of finding the channel under study (in this case,
$DK$) in the wave function of the bound state is given by:
\begin{equation}\label{eq:P_def}
P = - g^2 \frac{\partial G(s)}{\partial s}~,
\end{equation}
while the rest of the r.h.s. of Eq.~\eqref{eq:sumrule} represents
the probability of other channels, and hence the probabilities add
up to 1. If one has an energy independent potential, the second term
of Eq.~\eqref{eq:sumrule} vanishes, and then $P=1$. In this case,
the bound state is purely given by the channel under consideration.
These ideas are generalized to the coupled channels case in
Ref.~\refcite{Gamermann:2009uq}.

Let us now apply these ideas to the case of $DK$ scattering. From Eq.~\eqref{eq:TEFT} it can be
seen that the existence of a pole implies
\begin{align}
V^{-1}(s) & \simeq G(s_0) + \alpha (s-s_0) + \cdots~,\\
\alpha & \equiv \left. \frac{\partial V^{-1}(s)}{\partial s} \right\rvert_{s=s_0}~,
\end{align}
in the neighborhood of the pole. Assuming that the energy dependence
in a limited range of energies around $s_0$ is linear in $s$, we can
now write the amplitude as
\begin{equation}\label{eq:expansionTG}
T^{-1}(s) = G(s_0) - G(s) + \alpha(s-s_0)~,
\end{equation}
and the sum rule in Eq.~\eqref{eq:sumrule} becomes:
\begin{equation}\label{eq:sumrule2}
P_{DK} = 1 - \alpha g^2~.
\end{equation}
In this way, the quantity $\alpha g^2$ represents the probability of finding other components
beyond $DK$ in the wave function of $D_{s0}^{\ast+}(2317)$. The following relation can also be
deduced from Eqs.~\eqref{eq:sumrule2} and \eqref{eq:P_def}:
\begin{equation}\label{eq:rel_P_alpha}
\alpha = - \frac{1-P_{DK}}{P_{DK}} \left. \frac{\partial G(s)}{\partial s} \right\rvert_{s=s_0}.
\end{equation}

We can now link this formalism with the results of
Ref.~\refcite{Torres:2014vna}, where a reanalysis is done of the energy levels
found in the lattice simulations of Ref.~\refcite{Lang:2014yfa}. In
Ref.~\refcite{Torres:2014vna}, the following values for the effective range
parameters are found:
\begin{equation}\label{eq:a0_r0_valuesSasa}
a_0 = -1.4 \pm 0.6\ \text{fm}~,\quad r_0 = -0.1 \pm 0.2\ \text{fm}~.
\end{equation}

Also, in studying the $D_{s0}^{\ast+}(2317)$ bound state, a binding
energy $B = M_D + M_K - M_{D_{s0}^{\ast+}} = 31 \pm 17\ \text{MeV}$
is found in Ref.~\refcite{Torres:2014vna}. We can start from the hypothesis that
a bound state exists in the $DK$ channel, with a mass
$M_{D_{s0}^{\ast+}} = 2317\ \text{MeV}$ (the nominal one), and with
a probability $P_{DK}=0.75$. This implies, from
Eq.~\eqref{eq:rel_P_alpha}, that one has a value $\alpha = 2.06\cdot 10^{-3}\
\text{GeV}^{-2}$. Then, for the subtraction constant in the $G$
function, Eq.~\eqref{eq:GloopSubtracted}, one takes, as in
Ref.~\refcite{Gamermann:2006nm}, the value $a(\mu) = -1.3$ for $\mu = 1.5\
\text{GeV}$, with this input we obtain the $D K$ invariant mass
distribution in next subsection. Note that $\partial G(s)/\partial
s$ does not depend on $\mu$ or $a(\mu)$, and it is a convergent function.

\subsubsection{Decay amplitude and invariant $DK$ mass distribution in the $\bar{B}^0_s \to D_s^-(DK)^+$ decay}\label{subsec:DiffWid}

Let us first show how the amplitude for the decay $\bar{B}^0_s \to
D_s^-(DK)^+$ decay is obtained, and its relation to the $DK$ elastic
scattering amplitude studied above. The basic mechanism for this
process is depicted in Fig.~\ref{fig:diag}, where, from the
$\bar{s}b$ initial pair constituting the $\bar{B}^0_s$, a $\bar{c}s$ pair
and a $\bar{s}c$ pair are created. The first pair produces the
$D_s^-$, and the $DK$ state emerges from the hadronization of the
second pair. The hadronization mechanism has been explained in former sections
but we must include the $c\bar{c}$ pair in the hadronization. To
construct a two meson final state, the $c\bar{s}$ pair has to
combine with another $\bar{q}q$ pair created from the vacuum. Extending Eq.~\eqref{eq:qqbarmatrix} to include the charm quark, we
introduce the following matrix,
\begin{align}
M  = v \bar{v} = \left(\begin{array}{c} u \\ d \\ s \\ c
\end{array}\right)
\left(\begin{array}{cccc}
\bar{u} & \bar{d} & \bar{s} & \bar{c}
\end{array}\right)  = \left( \begin{array}{cccc}
u\bar{u} & u\bar{d} & u\bar{s} & u\bar{c} \\
d\bar{u} & d\bar{d} & d\bar{s} & d\bar{c} \\
s\bar{u} & s\bar{d} & s\bar{s} & s\bar{c} \\
c\bar{u} & c\bar{d} & c\bar{s} & c\bar{c} \\
\end{array}\right)~,
\end{align}
which fulfils:
\begin{align}
M^2 = (v \bar{v})(v \bar{v}) = v (\bar{v} v) \bar{v} = \left(
\bar{u}u + \bar{d}d + \bar{s}s + \bar{c}c \right) M~,
\end{align}
which is analogous to Eq.~\eqref{eq:MMqqM}. The first factor in the last equality represents the $\bar{q}q$ creation. In analogy again with Eq.~\eqref{eq:phimatrix}, this matrix $M$ is in
correspondence with the meson matrix $\phi$:
\begin{equation}\label{eq:phimatrixAlba}
\phi = \left( \begin{array}{cccc}
\frac{\eta}{\sqrt{3}} + \frac{\pi^0}{\sqrt{2}} + \frac{\eta'}{\sqrt{6}} & \pi^+ & K^+ & \bar{D}^0 \\
\pi^- & \frac{\eta}{\sqrt{3}} - \frac{\pi^0}{\sqrt{2}} + \frac{\eta'}{\sqrt{6}} & K^0 & D^- \\
K^- & \bar{K}^0 & \frac{\sqrt{2}\eta'}{\sqrt{3}}-\frac{\eta}{\sqrt{3}} & D_s^- \\
D^0 & D^+ & D_s^+ & \eta_c
\end{array}\right)~.
\end{equation}
\begin{figure}\centering
\includegraphics[scale=0.5]{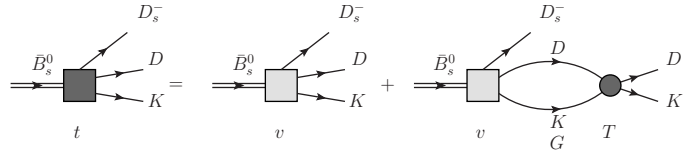}
\caption{Diagrammatical interpretation of Eq.~\eqref{eq:DecayAndFSI}, in which $DK$ final state
interaction is taken into account for the decay $\bar{B}^0_s \to D_s^-(DK)^+$. The dark square
represents the amplitude $t$ for the process, in which the final state interaction is already
taken into account. The light square represents the bare vertex for the process, denoted by $v$.
Finally, the circle represents the hadronic amplitude for the elastic $DK$
scattering.\label{fig:diag2}}
\end{figure}
The hadronization of the $c\bar{s}$ pair proceeds then through the matrix element
$\left(M^2\right)_{43}$, which in terms of mesons reads:
\begin{equation}
(\phi^2)_{43} = K^+ D^0 + K^0 D^+ + \cdots~,
\end{equation}
where only terms containing a $KD$ pair are retained, since coupled
channels are not considered here. We note that this $KD$
combination has $I=0$, as it should, since it is produced from a
$c\bar{s}$, which has $I=0$, and the strong interaction
hadronization conserves isospin.

Let $t$ be the full amplitude for the process $B^0_s \to D_s^- (DK)^+$, which already takes into
account the final state interaction of the $DK$ pair. Also, let us denote by $v$ the bare vertex
 for the same reaction. To relate $t$ and $v$, that is, to take into account the final state
interaction of the $DK$ pair, as sketched in Fig.~\ref{fig:diag2}, we write:
\begin{equation}\label{eq:DecayAndFSI}
t = v + v G(s) T(s)=v(1+G(s)T(s))~.
\end{equation}
From Eq.~\eqref{eq:TEFT}, the previous equation can also be written as:
\begin{equation}
t = v \frac{T(s)}{V(s)}~.
\end{equation}
Because of the presence of the bound state below threshold, this
amplitude will depend strongly on $s$ in the kinematical window
ranging from threshold to $100\ \text{MeV}$ above it. Hence, the
differential width for the process under consideration is given by:
\begin{equation}\label{eq:FinalDifWid}
\frac{\text{d} \Gamma}{d\sqrt{s}} = \frac{1}{32\pi^3
M_{\bar{B}^0_s}^2} p_{D_s^-} \tilde{p}_K \left\lvert
 t \right\rvert^2 = \mathcal{C} p_{D_s^-} \tilde{p}_K \left\lvert \frac{T(s)}{V(s)} \right\rvert^2~,
\end{equation}
where the bare vertex $v$ has been absorbed in $\mathcal{C}$, a
global constant, and where $p_{D^-_s}$
is the momentum of the $D_s^-$ meson in the rest frame of the
decaying $\bar{B}^0_s$ and $\tilde{p}_K$ the momentum of the kaon in
the rest frame of the $DK$ system.

\subsection{Results}\label{sec:resultsDKAlbaladejo}

\begin{table}[t]\centering
\renewcommand{\arraystretch}{1.3}
\begin{tabular}{|c|cll|} \hline
 & Central Value & $5\ \%$ & $10\ \%$ \\ \hline
$10^3\ \alpha\ (\text{GeV}^{-2})$ & $2.06$ & $^{+0.17}_{-0.40}$ & $^{+0.10}_{-1.09}$
 \\ \hline
$M_{D^\ast_{s0}}\ (\text{MeV})$ & $2317$ & $^{+14}_{-24}$ & $^{+21}_{-73}$ \\ \hline
$a(\mu)$ & $-1.30$ & $^{+0.15}_{-0.37}$ & $^{+0.27}_{-0.49}$ \\ \hline
$|g|\ (\textrm{GeV})$ & $11.0$ & $^{+1.0}_{-0.6}$ & $^{+2.2}_{-1.1}$ \\ \hline
$a_0\ (\textrm{fm})$ & $-1.0$ & $^{+0.2}_{-0.2}$ & $^{+0.4}_{-0.5}$ \\ \hline
$r_0\ (\textrm{fm})$ & $-0.14$ & $^{+0.06}_{-0.03}$ & $^{+0.16}_{-0.04}$ \\ \hline
$P_{DK}$ & $0.75$ & $^{+0.07}_{-0.06}$ & $^{+0.16}_{-0.11}$ \\ \hline
\end{tabular}
\caption{Fitted parameters ($\alpha$, $M_{D_{s0}^{\ast+}}$ and $a(\mu)$)  and
predicted quantities ($|g|$, $a_0$, $r_0$, $P_{DK}$) for $\mu=1.5\ \text{GeV}$. The second column shows the
central value of the fit, whereas the third (fourth) column presents the errors
(estimated by means of MC simulation) when the experimental error is 5\% (10\%).
\label{tab:res}}
\end{table}

We want to investigate the influence of the $D_{s0}^{\ast+}(2317)$
state in the $(DK)^+$ scattering amplitude. For this purpose, we
generate synthetic data from our theory for the differential decay
width for the process with Eqs.~\eqref{eq:FinalDifWid} and
\eqref{eq:expansionTG}. We generate 10 synthetic points in a range
of $100\ \text{MeV}$ starting from threshold, using the input
discussed above and assuming $5\%$ or $10\%$ error. The idea is to
use now these generated points as if they where experimental
data and perform the inverse analysis to obtain information on the
$D_{s0}^{\ast+}(2317)$.

\begin{figure}[t]\centering
\includegraphics[scale=0.8]{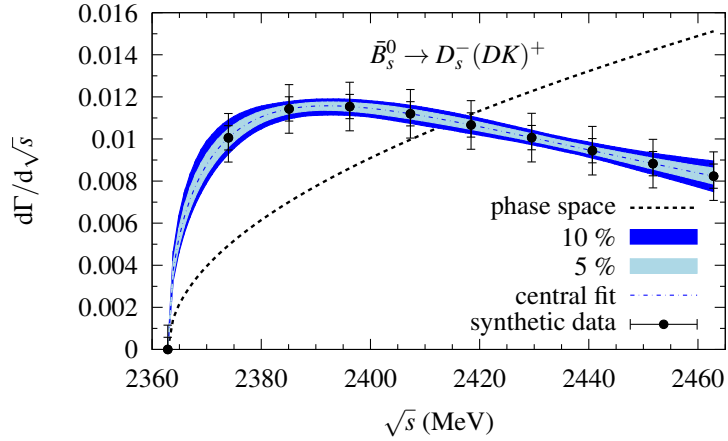}
\caption{Differential decay width for the reaction $\bar{B}^0_s \to D^-_s (DK)^+$. The synthetic
data (generated as explained in the text) are shown with black points. The smaller (larger)
error bars correspond to a 5\% (10\%) experimental error. The dash-dotted line represents the
theoretical prediction obtained with the central values of the fit. The light (dark) bands
correspond to the estimation of the error (by means of a MC simulation) when fitting the data
with 5\% (10\%) experimental error. The dashed line corresponds to a phase space distribution
normalized to the same area in the range examined.\label{fig:DifWid}}
\end{figure}

The generated synthetic data are shown in Fig.~\ref{fig:DifWid}. As
explained, we consider two different error bars, the smaller one
corresponding to 5\% experimental error and the larger one to 10\%. A phase space distribution ({\it i.e.}, a differential
decay width proportional to $p_{D_s^-}p_K$, but with no other
kinematical dependence of dynamical origin) is also shown in the
figure (dashed line). The first important information extracted from the figure is that the data are clearly incompatible
with the phase space distribution. This points to the presence of a
resonant or bound state below threshold. Two error bands are shown
in the same figure, the lighter and smaller (darker and larger) one
corresponding to 5\% (10\%) experimental error.  The fitted
parameters ($a(\mu)$, $M_{D_{s0}^{\ast+}}$, and $\alpha$) are shown
in Table~\ref{tab:res}. We also show the masses obtained and, by looking at the upper error, we observe that experimental data with a 10\% error, which is
clearly feasible with nowadays experimental facilities, can clearly
determine the presence of a bound $DK$ state, corresponding to the $D_{s0}^\ast(2317)$, from the $DK$ distribution.

We can also determine $P_{DK}$, the probability of finding the $DK$ channel in the
$D_{s0}^{\ast+}(2317)$ wave function. It is shown in the last row of Table~\ref{tab:res}. The central value $P_{DK}=0.75$ is the same as the initial one, but we are here
interested in the errors, which are small enough even in the case of a 10\% experimental error.
This means that with the analysis of such an experiment one could address with enough accuracy
the question of the molecular nature of the state ($D_{s0}^{\ast+}(2317)$, in this case).

Finally, it is also possible to determine other parameters related with $DK$ scattering, such as
the scattering length ($a_0$) and the effective range ($r_0$). They are also shown in
Table~\ref{tab:res}. They are compatible with the lattice QCD studies presented in
Refs.~\refcite{Lang:2014yfa,Torres:2014vna}. Namely, the results from Ref.~\refcite{Torres:2014vna} are shown in
Eqs.~\eqref{eq:a0_r0_valuesSasa}, and their mutual compatibility is clear.

\section{Predictions for the $\bar B^0 \to \bar K^{*0} X (YZ)$ and $\bar B^0_s \to \phi X (YZ)$ with $X(4160)$, $Y(3940)$, $Z(3930)$}\label{Sec:B2KstarXYZ}

The $XYZ$ resonances with masses in the region around 4000 MeV have
posed a challenge to the common wisdom of mesons as made from $q
\bar q$. There has been intense experimental work done at the BABAR,
BELLE, CLEO, BES and other collaborations, and many hopes are placed
in the role that the future FAIR facility with the PANDA
collaboration and J-PARC will play in this field.  There are early
experimental reviews on the topic
\cite{Barnes:2006xq,Asner:2008nq,Godfrey:2008nc,ChengPing:2009vu}
and more recent ones
\cite{Olsen:2012zz,Lange:2013sxa,Liu:2013waa,Guo:2014pya,Olsen:2014qna}.
From the theoretical point of view there has also been an intensive
activity trying to understand these states. There are quark model
pictures \cite{Ortega:2012rs,Vijande:2014cfa} and explicit
tetraquark structures \cite{Esposito:2014rxa}. Molecular
interpretations have also been given
\cite{Branz:2009yt,Branz:2010sh,Yang:2009fj,Dong:2012hc,Wang:2013cya,Cleven:2013sq,Wang:2013kra,
Aceti:2014kja,Aceti:2014uea}. The introduction of heavy quark spin
symmetry (HQSS) \cite{Nieves:2012tt,HidalgoDuque:2012pq,Albaladejo:2015dsa} has brought
new light into the issue. QCD sum rules have also made some
predictions
\cite{Nielsen:2009uh,Khemchandani:2013iwa,Nielsen:2014mva}. Strong
decays of these resonances have been studied to learn about the
nature of these states \cite{Dong:2013iqa,Ma:2014zva}, while very
often radiative decays are invoked as a tool to provide insight into
this problem
\cite{Liang:2009sp,Branz:2010gd,Aceti:2012cb,Ma:2014ofa,Dong:2014zka},
although there might be exceptions\cite{Guo:2014taa}. It has even been speculated that some states
found near thresholds of two mesons could just be cusps, or
threshold effects \cite{Swanson:2014tra}. However, this speculation
was challenged in Ref.~\refcite{Guo:2014iya} which showed that the near
threshold narrow structures cannot be simply explained by
kinematical threshold cusps in the corresponding elastic channels
but require the presence of $S$-matrix poles.
Along this latter point one should also mention a recent work on possible effects of singularities on the opposite side
of the unitary cut enhancing the cusp structure for states with mass
above a threshold \cite{Szczepaniak:2015eza}. Some theoretical reports on these issues can be found in other works\cite{Zhu:2007wz,Brambilla:2010cs,Drenska:2010kg}.

So far, in the study of these $B$ decays the production of $XYZ$
states has not yet been addressed and we show below some reactions where these states can be produced, evaluating
ratios for different decay modes and estimating the absolute rates\cite{Liang:2015twa}.
This should stimulate experimental work that can shed light on the
nature of some of these controversial states.

\subsection{Formalism}
\begin{figure}[tbh]
\begin{center}
\includegraphics[scale=0.55]{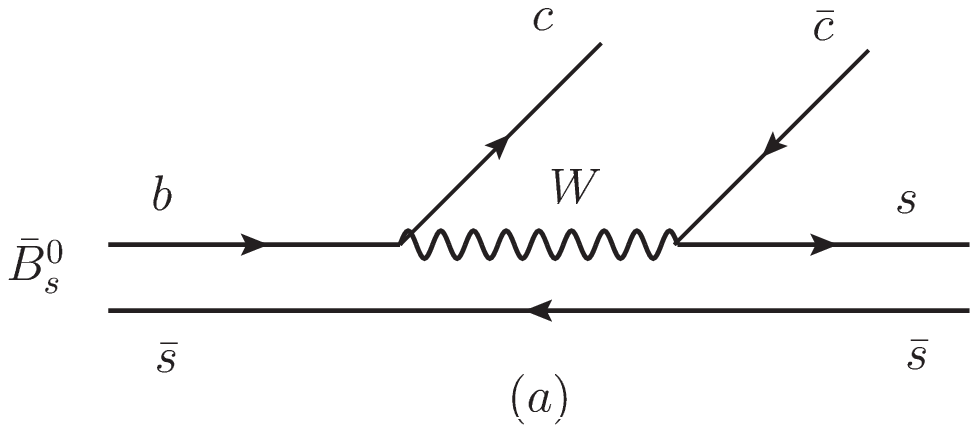}
\includegraphics[scale=0.55]{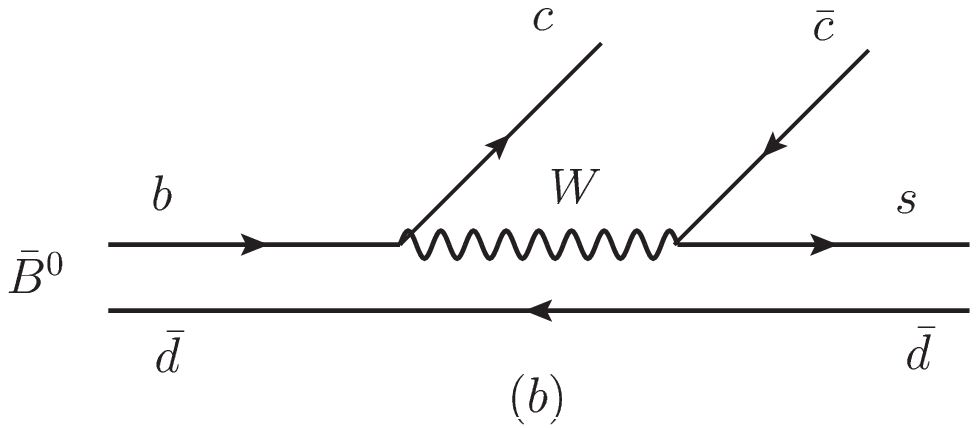}
\end{center}
\caption{Diagrams at the quark level for $\bar B^0_s$ (a) and $\bar B^0$ (b) decays into $c\bar c$ and a $q\bar q$ pair.}
\label{Fig:B2JpsiXYZ}
\end{figure}

Following the formalism developed in the former sections, we plot in
Fig. \ref{Fig:B2JpsiXYZ} the basic mechanism at the quark level for
$\bar B^0_s (\bar B^0)$ decay into a final $c \bar c$ and another
$q\bar q$ pair.
The $c\bar c$ goes into the production of a $J/\psi$ and the $s\bar
s$ or $s\bar d$ are hadronized to produce two mesons which are then
allowed to interact to produce some resonant states. Here, we shall
follow a different strategy and allow the $c\bar c$ to hadronize
into two vector mesons, while the $s\bar s$ and $s\bar d$ will make
the $\phi$ and $\bar K^{*0}$ mesons respectively. Let us observe
that, apart for the $b \to c$ transition, most favored for the
decay, we have selected an $s$ in the final state which makes the
$c\to s$ transition Cabibbo allowed. This choice magnifies the decay
rate, which should then be of the same order of magnitude as the
$\bar B_s^0 \to J/\psi f_0(980)$, which also had the same diagram at
the quark level prior to the hadronization of the $s\bar s$ to
produce two mesons.

\begin{figure*}[ht]
\begin{center}
\includegraphics[scale=0.55]{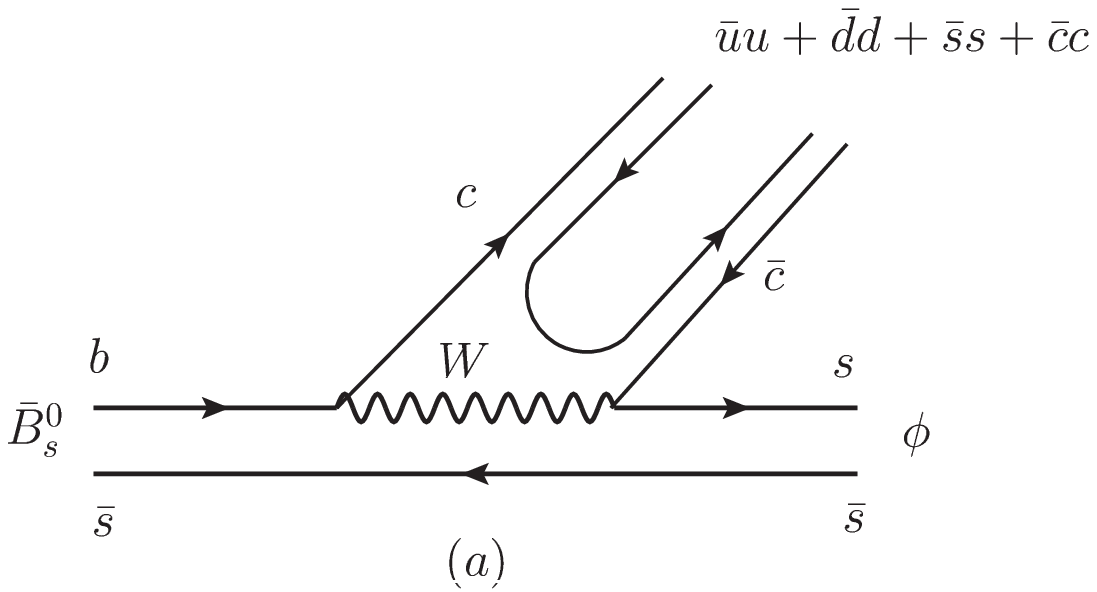}
\includegraphics[scale=0.55]{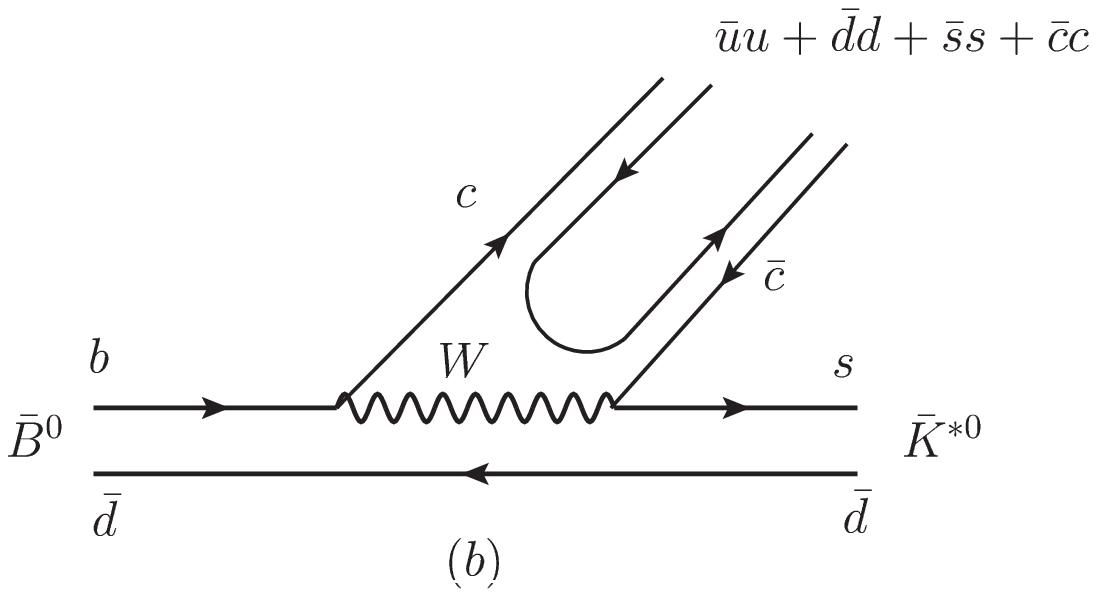}
\end{center}
\caption{Hadronization of the $c\bar c$ pair into two vector mesons
for $\bar B^0_s$ decay (a) and $\bar B^0$ decay (b).} \label{Fig:2}
\end{figure*}

In the next step, one introduces a new $q\bar q$ state with the
quantum numbers of the vacuum, $\bar uu +\bar d d +\bar s s +\bar c
c$, and see which combinations of mesons appear when added to $c\bar
c$. This is depicted in Fig. \ref{Fig:2}. For this we follow the
steps of the former section, and we have
\begin{equation}\label{eq:MM44}
c\bar c (\bar uu +\bar d d +\bar s s + \bar c c)\equiv (M \cdot M)_{44} \equiv (V \cdot V)_{44}
\end{equation}
and
\begin{equation}\label{eq:VV44-1}
(V \cdot V)_{44}= D^{*0} \bar D^{*0} + D^{*+} D^{*-} +D^{*+}_s  D^{*-}_s + J/\psi J/\psi.
\end{equation}
Note that we have produced an $I=0$ combination, as it should be
coming from $c\bar c$ and the strong interaction hadronization,
given the isospin doublets ($D^{*+}, -D^{*0}$), ($\bar
D^{*0}, \bar D^{*-}$). The $J/\psi J/\psi$ component is
energetically forbidden and hence we can write
\begin{equation}\label{eq:VV44-2}
(V \cdot V)_{44} \to \sqrt{2} ( D^* \bar D^*)^{I=0} + D^{*+}_s  D^{*-}_s.
\end{equation}

The vector mesons produced undergo interaction and we use the work of Ref.~\refcite{Molina:2009ct}, where an extension of
the local hidden gauge approach\cite{Bando:1984ej,Bando:1987br,Nagahiro:2008cv,Meissner:1987ge} is adopted, and where some $XYZ$ states are dynamically generated. Specifically, in Ref.~\refcite{Molina:2009ct} four
resonances were found, that are summarized in Table
\ref{Tab:XYZstate}, together with the channel to which the resonance
couples most strongly, and the experimental state to which they are
associated.
\begin{table}[ht]
     \renewcommand{\arraystretch}{1.2}
\centering
\begin{tabular}{cccccc}
\hline\hline
Energy [MeV]~   & $I^G [J^{PC}]$ & Strongest & Experimental\\
& & channel &state\\
\hline
$3943-i7.4$   & ~$0^+ [0^{+~+}]$~ & $D^* \bar D^*$ & $Y(3940)$ \cite{Abe:2004zs}\\
$3945-i0$  & $0^- [1^{+~-}]$ & $D^* \bar D^*$ & ?~$Y_P$ \\
$3922-i26$  & $0^+ [2^{+~+}]$ & $D^* \bar D^*$  & $Z(3930)$ \cite{Uehara:2005qd}\\
$4169-i66$  & $0^+ [2^{+~+}]$ & $D^*_s \bar D^*_s$& $X(4160)$ \cite{Abe:2007sya}\\
\hline\hline
\end{tabular}
\caption{States found in a previous work\cite{Molina:2009ct}, the channel to which
they couple most strongly, and the experimental states to which they
are associated\cite{Godfrey:2008nc,Agashe:2014kda}. $Y_P$ is a
predicted resonance. \label{Tab:XYZstate}}
\end{table}
In Ref.~\refcite{Molina:2009ct}, another state with $I=1$ was found, but
this one cannot be produced with the hadronization of $c\bar c$.
Some of these resonances have also been claimed to be of $D^* \bar
D^*$ or $D^*_s \bar D^*_s$ molecular nature\cite{Branz:2010gd,Gutsche:2010jf,Branz:2009yt} using for it the Weinberg
compositeness condition\cite{Weinberg:1965zz,Baru:2003qq,Sekihara:2014kya} and also using QCD sum rules\cite{Khemchandani:2013iwa,Wang:2014gwa,Nielsen:2009uh}, HQSS\cite{HidalgoDuque:2012pq,Albaladejo:2015dsa} and phenomenological potentials\cite{Liu:2009ei}.

\begin{figure*}[ht]
\begin{center}
\includegraphics[scale=0.5]{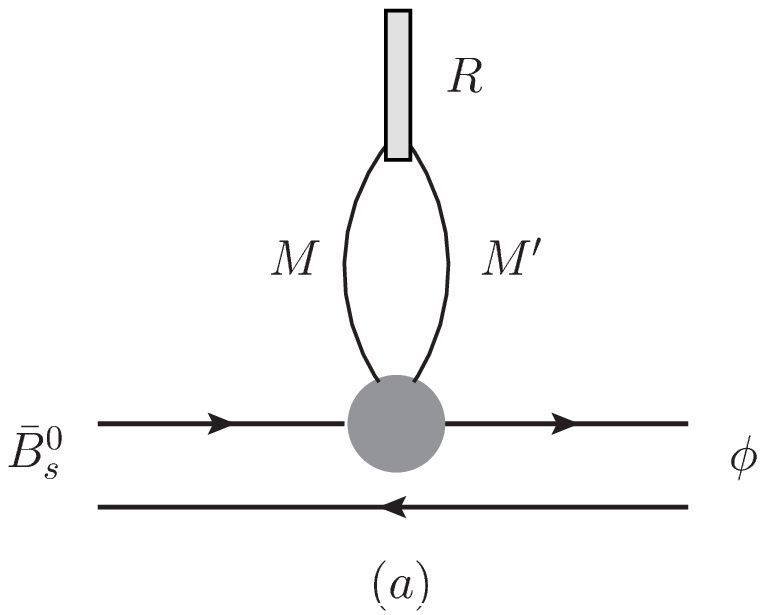}
\includegraphics[scale=0.5]{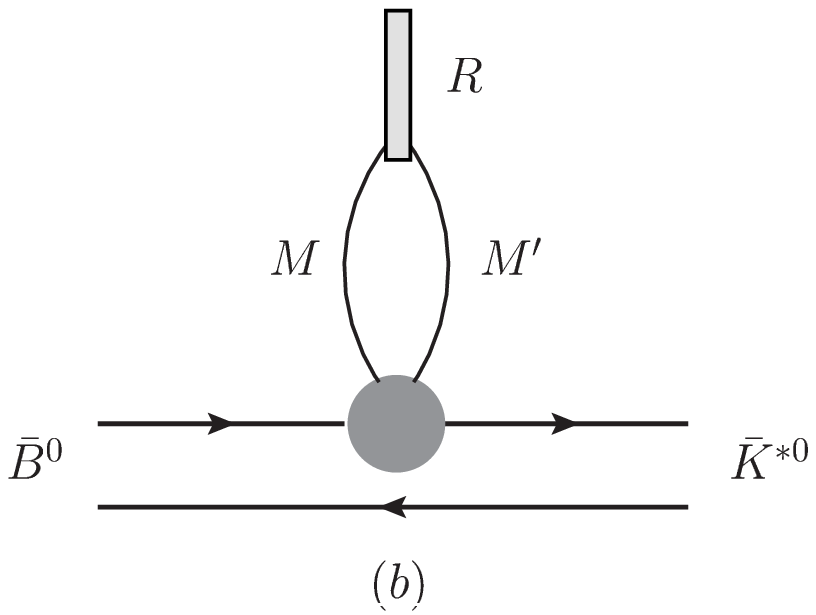}
\end{center}
\caption{Diagrammatic representation of the formation of the
resonances $R (X,Y,Z)$ through rescattering of $M M'$ ($D^* \bar
D^*$ or $D^*_s \bar D^*_s$) and coupling to the resonance.}
\label{Fig:3}
\end{figure*}

The final state interaction of the $D^* \bar D^*$ and $D^*_s \bar D^*_s$ proceeds diagrammatically as depicted in Fig. \ref{Fig:3}. Starting from Eq. (\ref{eq:VV44-2}) the analytical expression for the formation of the resonance $R$ is given by
\begin{equation}\label{eq:amplituT}
t(\bar B^0_s \to\phi R) = V_P (\sqrt{2} g_{D^* \bar D^*, R} G_{D^* \bar D^*}  + g_{D^*_s \bar D^*_s, R} G_{D^*_s \bar D^*_s}),
\end{equation}
where $G_{MM'}$ is the loop function of the two intermediate meson propagators and $g_{MM', R}$ is the coupling of the resonance to the $MM'$ meson pair.

The formalism for $\bar B^0 \to \bar K^{*0} R$ runs parallel since
the hadronization procedure is identical, coming from the $c\bar c$,
only the final state of $q\bar q$ is the $\bar K^{*0}$ rather than
the $\phi$. Hence, the matrix element is identical to the one of
$\bar B^0_s \to \phi R$, only the kinematics of different masses changes.

There is one more point to consider which is the angular momentum
conservation. For $J^P_R =0^+,2^+$, we have the transition $0^- \to
J^P~1^-$. Parity is not conserved but the angular momentum is. By
choosing the lowest orbital momentum $L$, we see that $L=0$ for
$J^P=1^+$ and $L=1$ for $J^P=0^+,2^+$. However, the dynamics will be
different for $J^P=0^+, 1^+, 2^+$. This means that we can relate
$\bar B^0_s \to Y(3940) \phi$ with $\bar B^0 \to Y(3940) \bar
K^{*0}$, $\bar B^0_s \to Z(3930) \phi$ with $\bar B^0 \to Z(3930)
\bar K^{*0}$, $\bar B^0_s \to X(4160) \phi$ with $\bar B^0 \to
X(4160) \bar K^{*0}$ and $\bar B^0_s \to Y_P \phi$ with $\bar B^0
\to Y_P \bar K^{*0}$, but in addition we can relate $\bar B^0_s \to
Z(3930) \phi$ with $\bar B^0_s \to X(4160) \phi$, and the same for
$\bar B^0 \to Z(3930) \bar K^{*0}$ with $\bar B^0 \to X(4160) \bar
K^{*0}$. Hence in this latter case we also have a $2^+$ state for
both resonances and the only difference between them is the
different coupling to $D^* \bar D^*$ and $D^*_s \bar D^*_s$, where
the $Z(3930)$ couples mostly to $D^* \bar D^*$, while the $X(4160)$
couples mostly to $D^*_s \bar D^*_s$.

The partial decay width of these transitions is given by
\begin{equation}\label{eq:width}
\Gamma_{R_i}=\frac{1}{8\pi} \frac{1}{m^2_{\bar B_i^0}} \left| t_{\bar B_i^0 \to \phi (\bar K^{*0}) R_i} \right|^2 P^{2L+1}_{\phi (\bar K^{*0})},
\end{equation}
which allows us to obtain the following ratios, where the different
unknown constants $V_P$, which summarize the production amplitude at
tree level, cancel in the ratios:
\begin{eqnarray}
&& R_1 \equiv \frac{\Gamma_{\bar B^0_s \to \phi
R^{J=0}}}{\Gamma_{\bar B^0 \to K^{*0} R^{J=0}}},~~ R_2 \equiv
\frac{\Gamma_{\bar B^0_s \to \phi R^{J=1}}}{\Gamma_{\bar B^0 \to
K^{*0} R^{J=1}}}, R_3 \equiv\frac{\Gamma_{\bar B^0_s \to \phi
R^{J=2}_1}}{\Gamma_{\bar B^0
\to K^{*0} R^{J=2}_1}}, \nonumber \\
&& R_4 \equiv\frac{\Gamma_{\bar B^0_s \to \phi
R^{J=2}_2}}{\Gamma_{\bar B^0 \to K^{*0} R^{J=2}_2}}, ~~R_5 \equiv
\frac{\Gamma_{\bar B^0_s \to \phi R^{J=2}_1}}{\Gamma_{\bar B^0_s \to
\phi R^{J=2}_2}}, \nonumber
\end{eqnarray}
where $R^{J=0}$, $R^{J=1}$, $R^{J=2}_1$ and $R^{J=2}_2$ are the $Y(3940)$, $Y_P$, $Z(3930)$ and $X(4160)$, respectively.

\subsection{Results}

The couplings $g_{MM', R}$ and the loop functions $G_{MM'}$ in Eq.
(\ref{eq:amplituT}) are taken from Ref.~\refcite{Molina:2009ct}, where the
dimensional regularization was used to deal with the divergence of
$G_{MM'}$, fixing the regularization scale $\mu =1000$ MeV and the
subtraction constant $\alpha = -2.07$. However, in Ref.~\refcite{Liang:2015twa} some  corrections to the work of Ref.~\refcite{Molina:2009ct} are done,
due to the findings of Ref.~\refcite{Liang:2014eba} concerning heavy quark
spin symmetry. It was found there that a factor $m_{D^*}/m_{K^*}$
has to be implemented in the hidden gauge coupling $g=m_\rho/2
f_\pi$ in order to account for the $D^*\to D\pi$ decay. However,
this factor should not be implemented in the Weinberg-Tomozawa terms
(coming from exchange of vector mesons) because these terms
automatically implement this factor in the vertices of vector
exchange. In Ref.~\refcite{Liang:2015twa}, $\mu=1000$ MeV and $\alpha=-2.19$ are used, by means of which a good reproduction of the masses is obtained.

We summarize here the results that we obtain for the ratios,
\begin{equation}\label{eq:ratio-1}
R_1=0.95,~~ R_2=0.96,~~ R_3=0.95,~~ R_4=0.83,~~ R_5=0.84.
\end{equation}

As we can see, all the ratios are of the order of unity. The ratios close to unity for the $\phi$ or $K^{*0}$
production are linked to the fact that the resonances are
dynamically generated from $D^* \bar D^*$ and $D^*_s \bar D^*_s$,
which are produced by the hadronization of the $c\bar c$ pair. The
ratio for the $J^P=2^+$ is even more subtle since it is linked to
the particular couplings of these resonances to $D^* \bar D^*$ and
$D^*_s \bar D^*_s$, which are a consequence of the dynamics that
generates these states. Actually, the ratios $R_1,~R_2,~R_3, ~R_4$
are based only on phase space and result from the elementary
mechanisms of Fig \ref{Fig:B2JpsiXYZ}. One gets the same ratios
as far as the resonances are $c \bar c$ based. Hence, even if these
ratios do not prove the molecular nature of the resonances, they
already provide valuable information telling us that they are  $c
\bar c$ based.

The ratio $R_5$ provides more information since it involves two
independent resonances and it is  not just a phase space ratio. If
we take into account only phase space, then $R_5\approx 4$ instead
of the value $0.84$ that we obtain.

As for the absolute rates, an analogy is established with the $\bar
B^0_s \to J/\psi f_0(980)$ decay in Ref.~\refcite{Liang:2015twa}, and branching
fractions of the order of $10^{-4}$ are obtained, which are an order of magnitude
bigger than many rates of the order of $10^{-5}$ already catalogued
in the PDG \cite{Agashe:2014kda}.

Given the fact that the ratios $R_1,~R_2,~R_3, ~R_4$ obtained are
not determining the molecular nature of the resonances, but only on
the fact that they are $c \bar c $  based, a complementary test is
proposed in the next section.

\subsection{Complementary test of the molecular nature of the resonances}

In this section we propose a test that is linked to the molecular
nature of the resonances. We study the decay $\bar B^0_s \to \phi
D^* \bar D^*$ or $\bar B^0_s \to \phi D_s^* \bar D_s^*$ close to the
$D^* \bar D^*$ and $D_s^* \bar D_s^*$ thresholds.

\begin{figure}[ht]\centering
\includegraphics[scale=0.5]{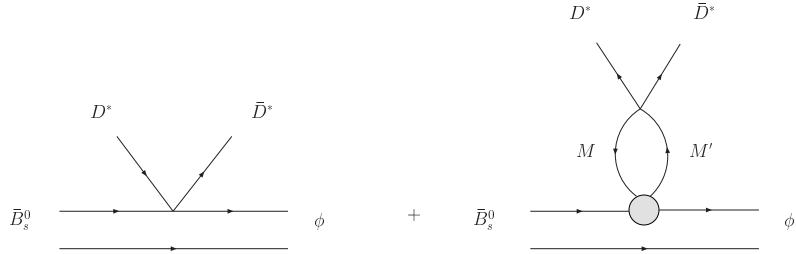}
\caption{Feynman diagrams for the $D^*D^*$ production in $B^0_s$ decays.\label{new}}
\end{figure}

\begin{figure*}[t]
\begin{center}
\includegraphics[scale=0.55]{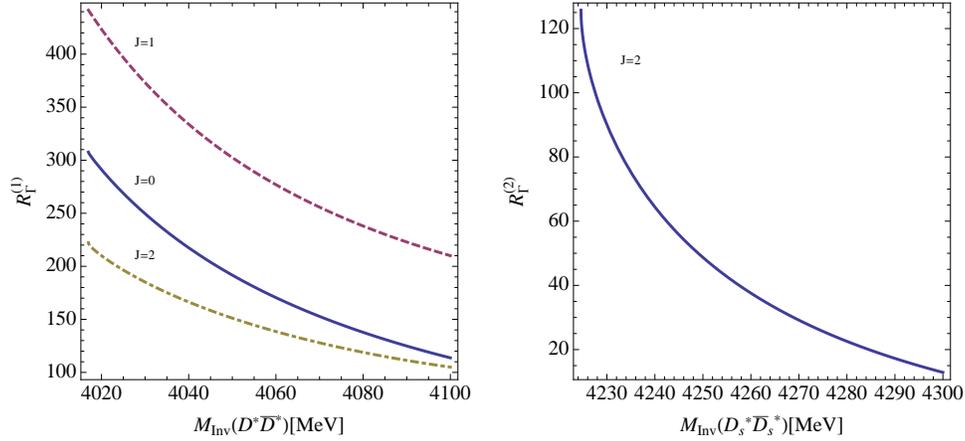}
\end{center}
\caption{Results of $R_{\Gamma}^{(1)}$ of Eq. (\ref{eq:rg1}) as a function of $M_{inv}(D^* \bar D^*)$ for the first three resonances of the Table \ref{Tab:XYZstate} (left) and $R^{(2)}_\Gamma$ as a function of $M_{inv}(D_s^* \bar D_s^*)$ (right) for the fourth resonance of the Table \ref{Tab:XYZstate}.}
\label{fig:rgam}
\end{figure*}

Let us now look to the process $\bar B^0_s \to \phi D^* \bar D^*$ depicted in Fig. \ref{new}. The production matrix for this process will be given by
\begin{equation}
t_{(\bar B^0_s \to\phi D^*\bar{D}^*)} = V_P (\sqrt{2}+\sqrt{2}\, G_{1}\,t_{(1\to 1)} + G_{2}\,t_{(2\to 1)})~,\label{eq:tmat}
\end{equation}
where $1$ and $2$ stands for the $D^*\bar{D}^*$ and
$D^*_s\bar{D}^*_s$ channels, respectively. The differential cross
section for production will be given by\cite{Liang:2014tia}:
\begin{equation}
\frac{d\Gamma}{d M_{\mathrm{inv}}}=\frac{1}{32\,\pi^3 M^2_{\bar B^0_s }}p_\phi \tilde{p}_{D^*}|t_{(\bar B^0_s \to\phi D^*D^*)}|^2 p^{2 L}_\phi,
\label{eq:invd}
\end{equation}
where $p_{\phi}$ is the $\phi$ momentum in the $\bar B^0_s$  rest
frame and $\tilde p_{D^*}$ the $D^*$ momentum in the $D^* \bar D^*$
rest frame. By comparing this equation with Eq. (\ref{eq:width}) for
the coalescence production of the resonance in $\bar B^0_s \to \phi
~R$, we find
\begin{eqnarray}
R_\Gamma =\frac{M^3_{R}}{p_\phi
\tilde{p}_{D^*}}\frac{1}{\Gamma_{R}}\frac{d \Gamma}{d
M_{\mathrm{inv}}} = \frac{M^3_{R}}{4 \pi^2}\frac{p_\phi^{2
L}(M_\mathrm{inv})}{p_\phi^{2 L+1}(M_{R})}\left| \frac{t_{(\bar
B^0_s \to\phi D^*\bar{D}^*)}}{t_{(\bar{B}^0\to R\phi)}} \right|^2,
\label{eq:rg1}
\end{eqnarray}
where we have divided the ratio of widths by the phase space factor
$p_{\phi} \tilde p_{D^*}$ and multiplied by $M_{R}^3$ to get a
constant value at threshold and a dimensionless magnitude. We apply this method for the three resonances that couple strongly to
$D^*\bar{D}^*$ (see Table \ref{Tab:XYZstate}). In the case of the
resonance $R_2$ with $J=2$, that couples mostly to the
$D^*_s\bar{D}^*_s$ channel (see Table \ref{Tab:XYZstate}), we look
instead for the production of $D^*_s\bar{D}^*_s$, for which we have:
\begin{equation}
t_{(\bar B^0_s \to\phi D^*_s\bar{D}^*_s)} = V_P (1+\sqrt{2}\,G_{1}\,t_{(1\to 2)}+ G_{2}\,t_{(2\to 2)})~,\label{eq:tmat2}
\end{equation}
and we use Eq.~\eqref{eq:rg1} but with $D^*_s\bar{D}^*_s$ instead of $D^*\bar{D}^*$ in the final state.
Equation~\eqref{eq:rg1} is then evaluated using the scattering matrices obtained in Ref.~\refcite{Molina:2009ct} modified as discussed above, together with Eqs.~\eqref{eq:tmat} and \eqref{eq:tmat2}. The results are shown in Fig. \ref{fig:rgam}.

We can see that the ratios are different for each case and have some structure. We observe that there is a fall down of the
differential cross sections as a function of energy, as it would correspond to the tail of a resonance below threshold. Note also that in the case of $D^*\bar{D}^*$, one produces the $I=0$ combination. If instead, one component like $D^{*+}D^{*-}$ is observed,
the rate should be multiplied by $1/2$. In the case of $D^*_s\bar{D}^*_s$ there is a single component and the rate predicted is fine.

\begin{figure*}[ht]
\begin{center}
\includegraphics[scale=0.55]{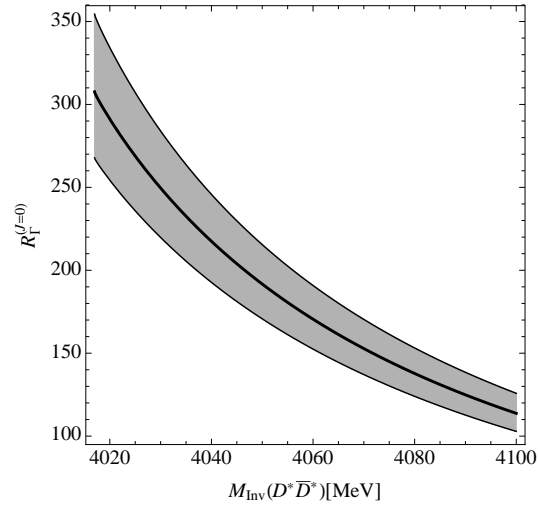}
\end{center}
\caption{Results of $R_{\Gamma}^{(1)}$
of Eq. (\ref{eq:rg1}) as a function of $M_{inv}(D^* \bar D^*)$ for spin$=0$.}
\label{fig:rgamj0}
\end{figure*}

\section[Testing the molecular nature of ${D_{s 0}^{\ast}(2317)}$ and ${D_{0}^{\ast} (2400)}$ in semileptonic ${B_s}$ and $B$ decays]{\boldmath Testing the molecular nature of ${D_{s 0}^{\ast}(2317)}$ and ${D_{0}^{\ast} (2400)}$ in semileptonic ${B_s}$ and $B$ decays}

In this section and the following one, we describe two processes for semileptonic decay, one for $B$ decay and the other for $D$ decay. The semileptonic $B$ decays will be used to test the molecular nature of the ${D_{s 0}^{\ast}(2317)}$ and ${D_{0}^{\ast} (2400)}$, while those of the $D$ mesons, to be studied in section~\ref{sec:semileptonicD_scalar_vector}, will be used to further investigate the nature of scalar and vector mesons.

\subsection[Introduction: semileptonic $B$ decays]{\boldmath Introduction: semileptonic $B$ decays}  \label{sec:semileptonic}

\begin{figure*}[htbp]
\begin{center}
\includegraphics[scale=0.2]{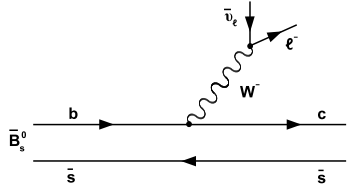}
\caption{Semileptonic decay of $\bar{B}_s^0$ into $\bar{\nu}_l l^-$
and a primary $c \bar{s}$ pair.}\label{Fig:fig1}
\end{center}
\end{figure*}

\begin{figure*}[htbp]
\begin{center}
\includegraphics[scale=0.2]{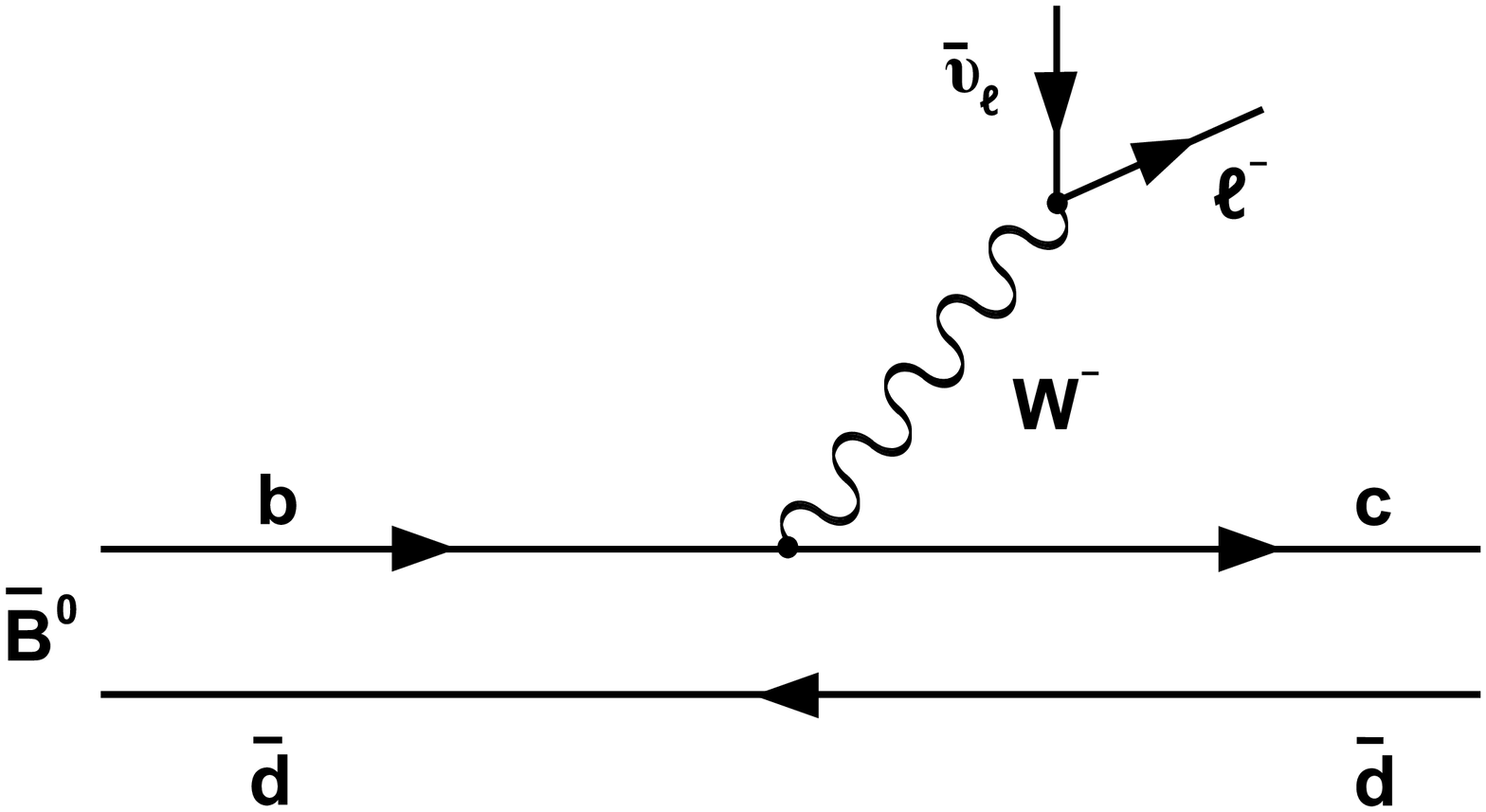}
\caption{Semileptonic decay of $\bar{B}^0$ into $\bar{\nu}_l l^-$
and a primary $c \bar{d}$ pair.}\label{Fig:fig2}
\end{center}
\end{figure*}

\begin{figure*}[htbp]
\begin{center}
\includegraphics[scale=0.2]{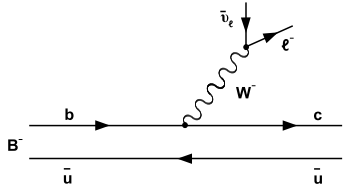}
\caption{Semileptonic decay of $B^-$ into $\bar{\nu}_l l^-$ and a
primary $c \bar{u}$ pair.} \label{Fig:fig3}
\end{center}
\end{figure*}

The formalism is very similar to the one presented in former
sections for nonleptonic $B$ decays. The basic mechanisms are depicted in Figs.~\ref{Fig:fig1},
\ref{Fig:fig2}, \ref{Fig:fig3}. In all of them, after the $W$
emission one has a $c \bar{q}$ pair. In order to have two mesons in
the final state the $c \bar{q}$ is allowed to hadronize into a pair
of pseudoscalar mesons and the relative weights of the different pairs
of mesons will be known. Once the meson pairs are produced they 
interact in the way described by the chiral unitary
model in coupled channels,  generating the $D_{s 0}^{\ast}
(2317)$ and $D_{0}^{\ast} (2400)$ resonances. 

We will consider the semileptonic $B$ decays into $D$
resonances in the following decay modes:
\begin{equation}
\begin{split}
& \bar{B}_{s}^{0} \to D_{s 0}^{\ast} (2317)^{+} \bar{\nu} _{l} l^{-} , \\
& \bar{B}^{0} \to D_{0}^{\ast} (2400)^{+} \bar{\nu} _{l} l^{-} , \\
& B^{-} \to D_{0}^{\ast} (2400)^{0} \bar{\nu} _{l} l^{-} ,
\end{split}
\end{equation}
where the lepton flavor $l$ can be $e$ and $\mu$. With respect to
the former sections we have now a different dynamics which we
discuss below, together with the hadronization process.

\subsection{Semileptonic decay widths}
\label{sec:decay_width}

The decay amplitude of $B \to \bar{\nu} l^{-} \text{hadron(s)}$, $T_{B}$, is given by:
\begin{align}
- i T_{B} =  & \overline{u}_{l} i \frac{g_{\rm W}}{\sqrt{2}} \gamma
^{\alpha} \frac{1 - \gamma _{5}}{2} v_{\nu} \times \frac{- i
g_{\alpha \beta}}{p^{2} - M_{W}^{2}} \nonumber \\ 
& \times \overline{u}_{c} i
\frac{g_{\rm W} V_{b c}}{\sqrt{2}} \gamma ^{\beta} \frac{1 - \gamma
_{5}}{2} u_{b} \times (- i V_{\rm had})~,
\end{align}
where $u_{l}$, $v_{\nu}$, $u_{c}$, and $u_{b}$ are Dirac spinors
corresponding to the lepton $l^{-}$, neutrino, charm quark, and
bottom quark, respectively, $g_{\rm W}$ is the coupling constant of
the weak interaction, $V_{b c}$ is the CKM
matrix element, and $M_{W}$ is the $W$ boson mass.  The factor
$V_{\rm had}$ describes the hadronization process and it
will be evaluated in the sections below. Ignoring the squared
three-momentum of the $W$ boson ($p^{2}$) which is much smaller than
$M_{W}^{2}$ in the $B$ decay process, the decay amplitude becomes
\begin{align}
T_{B} = & - i \frac{G_{\rm F} V_{b c}}{\sqrt{2}}
L^{\alpha} Q_{\alpha} \times V_{\rm had} ,
\label{eq:TB}
\end{align}
where the Fermi coupling constant $G_{\rm F} \equiv g_{\rm W}^{2} /
(4 \sqrt{2} M_{W}^{2})$ is introduced, and we define the lepton and
quark parts of the $W$ boson couplings as:
\begin{equation}
L^{\alpha} \equiv \overline{u}_{l} \gamma ^{\alpha} ( 1 - \gamma _{5} ) v_{\nu} ,
\quad
Q_{\alpha} \equiv \overline{u}_{c} \gamma _{\alpha} ( 1 - \gamma _{5} ) u_{b} ,
\label{eq:LQ}
\end{equation}
respectively. 

In the calculation of the decay widths, one needs the average and
sum of $|T_B|^2$ over the polarizations of the initial-state quarks
and final-state leptons and quarks. In terms of the
amplitude in Eq.~\eqref{eq:TB}, one can obtain the squared decay
amplitude as
\begin{equation}
\frac{1}{2} \sum _{\rm pol} | T_{B} |^{2}
= \frac{| G_{\rm F} V_{b c} V_{\rm had} |^{2}}{4}
\sum _{\rm pol} | L^{\alpha} Q_{\alpha} |^{2}
\end{equation}
where the factor $1/2$ comes from the average of the bottom quark
polarization. Finally with some algebra discussed in
Ref.~\refcite{Navarra:2015iea} one obtains the squared decay amplitude:
\begin{equation}
\frac{1}{2} \sum _{\rm pol} | T_{B} |^{2}
= \frac{4 | G_{\rm F} V_{b c} V_{\rm had} |^{2}}{m_{\nu} m_{l} m_{B} m_{R}}
( p_{B} \cdot p_{\nu} ) ( p_{R} \cdot p_{l} ) .
\label{amp2}
\end{equation}

Using the above squared amplitude we can calculate the decay width. We
will be interested in two types of decays: three-body decays, such
as $\bar{B}_s^0 \rightarrow D_{s0}^+ \, \bar{\nu_l} \, l^- $, and
four-body decays, such as $\bar{B}_s^0 \rightarrow D^+ \, K^{0} \,
\bar{\nu_l} \, l^- $ and also for the similar $\bar{B}^0$ and $B^-$
initiated processes. As it will be seen, both decay types can be
described by the amplitude $T_{B}$ with different assumptions for
$V_{\rm had}$.

\subsection{Hadronization} \label{sec:hadronization}

\begin{figure*}[htbp]
\begin{center}
\includegraphics[scale=0.25]{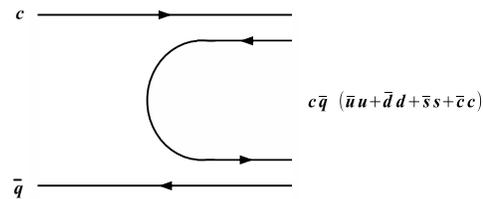}
\caption{Schematic representation of the hadronization $c\bar q \to
c \bar q \, (\bar{u} u + \bar{d} d + \bar{s} s + \bar{c}
c)$.}\label{Fig:fig4}
\end{center}
\end{figure*}

For the conversion of quarks into hadrons in the final stage of hadron
reactions we follow the same procedure as in former sections and assume that the  matrix element for this process can be represented
by  an unknown constant. Explicit evaluations, where usually one must 
parametrize some information, have been discussed in subsection \ref{sec:relotherapproaches}. 
Since the energies involved are of the order of
a few GeV or  less, this is a  non-perturbative process.
In some cases one can develop an approach
based on effective Lagrangians\cite{Braaten:2001uu,Braaten:2002yt} to study hadronization.  Here we describe  hadronization
as depicted in Fig.~\ref{Fig:fig4}. An extra $\bar{q}q$ pair with the
quantum numbers of the vacuum, $\bar{u} u  + \bar{d} d + \bar{s} s +
\bar{c} c$, is added to the already existing quark pair. The
probability of producing the pair is assumed to be given by a number
which is the same for all light flavors and which will cancel out
when taking ratios of decay widths. We can write this $c \bar q \,
(\bar{u} u  + \bar{d} d + \bar{s} s  + \bar{c}  c)$  combination in
terms of pairs of mesons. For this purpose we follow the procedure
of the former sections and find the correspondence, with $\phi$
given by Eq.~\eqref{eq:phimatrixAlba},
\begin{align}
c\bar s  \,  (\bar{u} u  + & \bar{d} d + \bar{s} s  + \bar{c}  c)  \equiv  \left( \phi \cdot \phi \right)_{43}  \nonumber\\
= & D^0 K^+ + D^+ K^0 + D_s^+ \left ( - \frac{1}{\sqrt{3}} \eta + \sqrt{\frac{2}{3}} \eta ^{\prime} \right ) +  \eta_c D_s^+~, \label{eq:4}\\
c\bar d  \,  (\bar{u} u  + & \bar{d} d + \bar{s} s  + \bar{c}  c)  \equiv  \left( \phi \cdot \phi \right)_{42} \nonumber\\
= & D^0 \pi^+ + D^+  \left ( - \frac{1}{\sqrt{2}} \pi^0 + \frac{1}{\sqrt{3}} \eta + \frac{1}{\sqrt{6}} \eta ^{\prime} \right )  +  D_s^+ \bar K^0 + \eta_c D^+~,\label{eq:5}\\
c\bar u  \,  (\bar{u} u  + & \bar{d} d + \bar{s} s  + \bar{c}  c)  \equiv  \left( \phi \cdot \phi \right)_{41} \nonumber\\
= & D^0 \left ( \frac{1}{\sqrt{2}} \pi^0 + \frac{1}{\sqrt{3}} \eta + \frac{1}{\sqrt{6}} \eta ^{\prime} \right ) +  D^+ \pi^-  +  D_s^+  K^- + \eta_c D^0~.\label{eq:6}
\end{align}
for $D_{s 0}^{\ast} (2317)^{+}$, $D_{0}^{\ast} (2400)^{+}$, and
$D_{0}^{\ast} (2400)^{0}$ production, respectively. As it was pointed out in
Ref.~\refcite{Gamermann:2006nm}, the most important channels for
the description of $D_{s 0}^{\ast} (2317)$ ($D_{0}^{\ast} (2400)$)
are $D K$ and $D_{s} \eta$ ($D \pi$ and $D_{s} \bar{K}$).
Therefore, the weights of the channels to generate the $D$
resonances can be written in terms of  ket vectors as:
\begin{align}
| ( \phi \phi )_{43} \rangle &
= \sqrt{2} | D K ( 0 , \, 0) \rangle
- \frac{1}{\sqrt{3}} | D_{s} \eta ( 0 , \, 0) \rangle~,\nonumber\\
| ( \phi \phi )_{42} \rangle &
= - \sqrt{\frac{3}{2}} | D \pi ( 1/2, \, 1/2) \rangle
+ | D_{s} \bar{K} ( 1/2, \, 1/2) \rangle~,\label{eq:phiphi_41}\\
| ( \phi \phi )_{41} \rangle &
= \sqrt{\frac{3}{2}} | D \pi ( 1/2, \, -1/2) \rangle
- | D_{s} \bar{K} ( 1/2, \, -1/2) \rangle~,\nonumber
\end{align}
where we have used two-body states in the isospin basis, which are
specified as $(I, \, I_{3})$. Due to the isospin
symmetry, both the charged and neutral $D_{0}^{\ast}(2400)$ are
produced with the weight of $| (\phi \phi )_{42} \rangle = - | (\phi
\phi )_{41} \rangle$, which means that the ratio of the decay widths
into the charged and neutral $D_{0}^{\ast}(2400)$ is almost unity.
Using these weights, we can write $V_{\rm had}$ in terms of two
pseudoscalars.

\begin{figure*}[htbp]\centering
\includegraphics[scale=0.2]{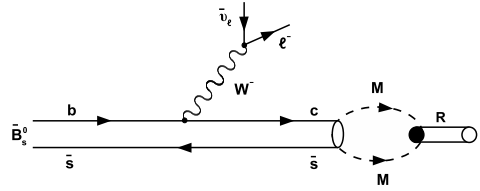}
\caption{Diagrammatic representation of $ D_{s0}^{\ast +} (2317)$
production via meson coalescence after rescattering.\label{Fig:fig5}}
\end{figure*}

\begin{figure*}[htbp]
\begin{center}
\includegraphics[scale=0.2]{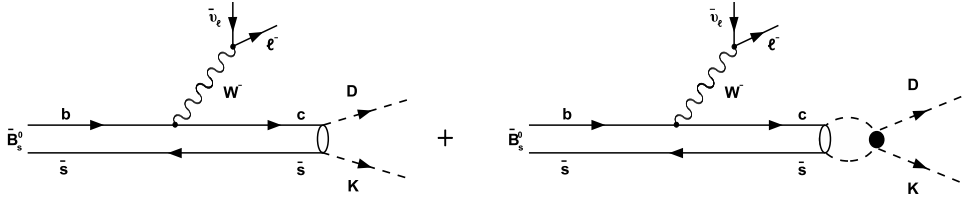}
\caption{Diagrammatic representation of $D K$ production: directly
(on the left) and via rescattering (on the right) in $\bar B^0_s$
decays.} \label{Fig:fig6}
\end{center}
\end{figure*}

After the hadronization of  the quark-antiquark pair,  two mesons are formed and  start
to interact.  The $D$ resonances can be generated as a result of
complex two-body interactions with coupled channels described by the
Bethe-Salpeter equation.  If the resonance is formed, independently  of
how it decays, the process is usually called
``coalescence''~\cite{Braaten:2004rw,Braaten:2004fk} and it is a reaction with three
particles in the final state (see Fig.~\ref{Fig:fig5}). If we look
for a specific two meson final channel we can have it by ``prompt''
or direct production (first diagram of Fig.~\ref{Fig:fig6}), and by
rescattering, generating the resonance (second diagram of
Fig.~\ref{Fig:fig6}). This process is usually called
``rescattering'' and it is a reaction with four particles in the
final state. Coalescence and rescattering will be discussed in the
next sections.

\subsection{Coalescence}  \label{sec:coalescence}

In this section we consider $D$ resonance production via meson
coalescence as depicted in Fig.~\ref{Fig:fig5}. This
process has a three-body final state with a lepton, its neutrino and
the resonance R. The hadronization factor, $V_{\rm had}$,  can be
written as
\begin{eqnarray}
V_{\text{had}} ( D_{s 0}^{\ast}(2317) ) &=& C
  \left ( \sqrt{2} G_{D K} g_{D K}
  - \frac{1}{\sqrt{3}} G_{D_{s} \eta} g_{D_{s} \eta} \right ) ,
  \label{vhads} \\
V_{\text{had}} ( D_{0}^{\ast}(2400)^{+} ) &=& C
  \left ( - \sqrt{\frac{3}{2}} G_{D \pi} g_{D \pi}
  + G_{D_{s} \bar{K}} g_{D_{s} \bar{K}} \right ) ,
\label{vhado}
\end{eqnarray}
Here $g_{i}$ is the coupling constant of the $D$ resonance to the
$i$-th two meson channel and $G_{i}$ is the loop function of two meson
propagators. As mentioned above 
$ V_{\text{had}} ( D_{0}^{\ast}(2400)^{0} ) = -  V_{\text{had}} ( D_{0}^{\ast}(2400)^{+} )$.  
We will assume that $C$ is a constant in the limited range of
invariant masses that we discuss and hence it will be  cancelled when we
take the ratio of decay widths.

The formula for the width is then given by
\begin{equation}
\Gamma _{\rm coal} = 
\frac{| G_{\rm F} V_{b c} V_{\rm had} (D^{\ast}) |^{2}}
{8 \pi ^{3} m_{B}^{3} m_{R}}
\int d M_{\rm inv}^{( \nu l)} 
p_{D}^{\rm cm} \tilde{p}_{\nu} 
[ M_{\rm inv}^{( \nu l )}]^{2} \left (
\tilde{E}_{B} \tilde{E}_{R} - \frac{\tilde{p}_{B}^{2}}{3} 
\right )~,
\label{eq:Gcoal}
\end{equation}
where the integral range of $M_{\rm inv}^{(\nu l)}$ is $[m_{l} +
m_{\nu} , \, m_{B} - m_{R} ]$. In Eq.~\eqref{eq:Gcoal}, we have:
\begin{align}
p_D^\text{cm} & = \frac{\lambda^{1/2}\left(m_B^2,\left[M_\text{inv}^{(\nu l)}\right]^2,m_R^2\right)}{2m_B}~,\\
\tilde{p}_\nu & = \frac{\lambda^{1/2}\left(\left[M_\text{inv}^{(\nu l)}\right]^2,m_\nu^2,m_l^2\right)}{2m_B}~, \label{eq:pnu_coal}\\
\tilde{E}_B & = \frac{m_B^2+\left[M_\text{inv}^{(\nu l)}\right]^2-m_R^2}{2M_\text{inv}^{(\nu l)}}~,\\
\tilde{E}_R & = \frac{m_B^2-\left[M_\text{inv}^{(\nu l)}\right]^2-m_R^2}{2M_\text{inv}^{(\nu l)}}~,
\end{align}
where $\tilde{p}_B^2 = \tilde{E}_B^2-m_B^2$. Here $m_B$ and $m_R$ are the masses of the $B$ and $D^\ast$ mesons, respectively. Further detailed can be found in Ref.~\refcite{Navarra:2015iea}.

\subsection{Rescattering}   \label{sec:rescattering}

Now we address the production of two pseudoscalars with prompt production
plus rescattering through a $D$ resonance, as depicted in the diagrams of  
 Fig.~\ref{Fig:fig6}. The hadronization
amplitude $V_{\rm had}$ in the isospin basis is given by
\begin{eqnarray}
  V_{\text{had}} ( D K ) &=& C
  \left ( \sqrt{2} + \sqrt{2} G_{D K} T_{D K \to D K}
  - \frac{1}{\sqrt{3}} G_{D_{s} \eta} T_{D_{s} \eta \to D K} \right ) ,
\label{vhadk} \\
  V_{\text{had}} ( D_{s} \eta ) &=& C
  \left ( - \frac{1}{\sqrt{3}} + \sqrt{2} G_{D K} T_{D K \to D_{s} \eta}
    - \frac{1}{\sqrt{3}} G_{D_{s} \eta} T_{D_{s} \eta \to D_{s} \eta} \right ) ,
\label{vhadeta} \\
  V_{\text{had}} ( D \pi ) &=& C
  \left ( - \sqrt{\frac{3}{2}}
    - \sqrt{\frac{3}{2}} G_{D \pi} T_{D \pi \to D \pi}
  + G_{D_{s} \bar{K}} T_{D_{s} \bar{K} \to D \pi} \right ) ,
\label{vhadpi} \\
  V_{\text{had}} ( D_{s} \bar{K} ) &=& C
  \left ( 1 - \sqrt{\frac{3}{2}} G_{D \pi} T_{D \pi \to D_{s} \bar{K}}
    + G_{D_{s} \bar{K}} T_{D_{s} \bar{K} \to D_{s} \bar{K}} \right ) .
\label{vhadsk}
\end{eqnarray}
As it can be seen,  the prefactor $C$ is the same in all the reactions.
In order to calculate decay widths in the particle basis, we need to
multiply the amplitudes by the appropriate Clebsch-Gordan coefficients.

Using Eq.~\eqref{amp2} we can compute the differential decay
width $d\Gamma _{i} / d M_{\rm inv}^{(i)}$, where $i$ represents the
two pseudoscalar states and $M_{\rm inv}^{(i)}$ is the invariant
mass of the two pseudoscalars, as
\begin{align}
\frac{d \Gamma _{i}}{d M_{\rm inv}^{(i)}}
= \frac{| G_{\rm F} V_{b c} V_{\rm had} (i) |^{2}}
{32 \pi ^{5} m_{B}^{3} M_{\rm inv}^{(i)}}
\int d M_{\rm inv}^{( \nu l)} 
P^{\rm cm} \tilde{p}_{\nu} \tilde{p}_{i}
[ M_{\rm inv}^{( \nu l )}]^{2} \left (
\tilde{E}_{B} \tilde{E}_{i} - \frac{\tilde{p}_{B}^{2}}{3} 
\right ) , 
\label{eq:dGamma_dM}
\end{align}
where $P^{\rm cm}$ is the momentum of the $\nu l$ system in the $B$
rest frame, $\tilde{p}_{\nu}$ is the momentum of the $\nu$ in the
neutrino lepton rest system [given in Eq.~\eqref{eq:pnu_coal}], and $\tilde{p}_{i}$ is the relative
momentum of the two pseudoscalars in their rest frame.

\subsection[The $D K$-$D_{s} \eta$ and $D \pi$ - $D_{s}\bar{K}$ 
scattering amplitudes]{\boldmath The $D K$-$D_{s} \eta$ and $D \pi$ - $D_{s}\bar{K}$ scattering amplitudes}
\label{sec:amplitudes}

We will now discuss the meson-meson scattering amplitudes for
the rescattering to generate the $D_{s 0}^{\ast} (2317)$ and
$D_{0}^{\ast} (2400)$ resonances in the final state of the $B$ decay.
In Ref.~\refcite{Gamermann:2006nm}  it was found that the couplings to $D
K$ and $D_{s} \eta$ are dominant for $D_{s 0}^{\ast} (2317)$ and the
couplings to $D \pi$ and $D_{s} \bar{K}$ are dominant for
$D_{0}^{\ast} (2400)$.  Therefore, in the following we concentrate on
$D K$-$D_{s} \eta$ two-channel scattering in isospin $I=0$ and $D
\pi$-$D_{s} \bar{K}$ two-channel scattering in $I=1/2$, extracting
essential portions from Ref.~\refcite{Gamermann:2006nm} and assuming
isospin symmetry.  Namely, we obtain these amplitudes by solving a
coupled-channel scattering equation in an algebraic form
\begin{equation}
T_{i j} (s) = V_{i j} (s) + \sum _{k} V_{i k} (s) G_{k} (s) T_{k j} (s) ,
\label{eq:BSEq}
\end{equation}
where $i$, $j$, and $k$ are channel indices, $s$ is the Mandelstam
variable of the scattering, $V$ is the interaction kernel, and $G$
is the two-body loop function. This generalizes what was found in
section \ref{sec:Ds02317_Alba} with just one channel.

The interaction kernel $V$ corresponds to the tree-level transition
amplitudes  obtained from phenomenological Lagrangians developed in
Ref.~\refcite{Gamermann:2006nm}. We use dimensional regularization 
in the loop function $G$. 

The  $D$ resonances can appear as poles of the
scattering amplitude $T_{i j} (s)$ with the residue $g_{i} g_{j}$:
\begin{equation}
T_{i j} ( s ) = \frac{g_{i} g_{j}}{s - s_{\rm pole}}
+ ( \text{regular at }s = s_{\rm pole} ) .
\label{eq:amp_pole}
\end{equation}
The pole is described by its position $s_{\rm pole}$ and the
constant $g_{i}$, which is the coupling constant of the $D$
resonance to the $i$ channel. Further details can be found in
Ref.~\refcite{Navarra:2015iea}.

Let us introduce the concept of compositeness, which is defined as
the contribution from the two-body part to the normalization of the
total wave function and measures the fraction of the two-body
state~\cite{Hyodo:2011qc,Aceti:2012dd,Xiao:2012vv,Hyodo:2013nka,Sekihara:2014kya}.
The expression of the compositeness is given by
\begin{equation}\label{eq:abcde}
X_{i} =
- g_{i}^{2} \left [ \frac{d G_{i}}{d s} \right ] _{s = s_{\rm pole}}~.
\end{equation}
In an analogous way we introduce  the elementariness $Z$, which measures the fraction
of missing channels and is expressed as
\begin{equation}
Z = - \sum _{i, j} g_{j} g_{i} \left [ G_{i} \frac{d V_{i j}}{d s}
G_{j} \right ] _{s = s_{\rm pole}}~.
\end{equation}
In general both the compositeness $X_{i}$ and
elementariness $Z$ are complex values for a resonance state and
hence one cannot interpret the compositeness (elementariness) as the
probability to observe a two-body (missing-channel) component inside
the resonance except for bound states. However, a striking property
is that the sum of them is exactly unity:
\begin{equation}
\sum _{i} X_{i} + Z = 1 ,
\end{equation}
which is guaranteed by a generalized Ward identity proved in
Ref.~\refcite{Sekihara:2010uz}. Therefore one can deduce the structure
by comparing the value of the compositeness with unity, on the basis
of the similarity to the stable bound state case. The values of the
compositeness and elementariness of the $D$ resonances in this
approach are listed in Table~\ref{tab:Ds}. The result indicates
that the $D_{s 0}^{\ast} (2317)$ resonance, which is obtained as a
bound state in the present model, is indeed dominated by the $D K$
component.  This has been corroborated in the recent analysis of lattice QCD
results of Ref.~\refcite{Torres:2014vna}. In contrast, we may conclude that
the $D_{0}^{\ast} (2400)$ resonance is constructed with missing
channels, although the imaginary part for each component is not
negligible.

\begin{table}
  \begin{center}
  \begin{tabular*}{8.6cm}{@{\extracolsep{\fill}}lcc|lc}
    \hline \hline
    \multicolumn{2}{c}{$D_{s 0}^{\ast} (2317)$} &  &
    \multicolumn{2}{c}{$D_{0}^{\ast} (2400)$} \\
    \hline
    $\sqrt{s_{\rm pole}}$ & $2317 \mev$ & &
    $\sqrt{s_{\rm pole}}$ & $2128 - 160 i \mev$ \\
    $g_{D K}$ & $10.58 \gev$ &  &
    $g_{D \pi}$ & $\phph 9.00 - 6.18 i \gev$ \\
    $g_{D_{s} \eta}$ & $- 6.11 \gev ~$ & &
    $g_{D_{s} \bar{K}}$ & $- 7.68 + 4.35 i \gev$ \\
    $X_{D K}$ & $0.69$ & &
    $X_{D \pi}$ & $0.34 + 0.41 i$ \\
    $X_{D_{s} \eta}$ & $0.09$ & &
    $X_{D_{s} \bar{K}}$ & $0.03 - 0.12 i$ \\
    $Z$ & $0.22$ & &
    $Z$ & $0.63 - 0.28 i$ \\
    \hline \hline
  \end{tabular*}
   \end{center}
  \caption{Pole position $\sqrt{s_{\rm pole}}$, coupling constant
    $g_{i}$, compositeness $X_{i}$, and elementariness $Z$ for the $D$
    resonances in the isospin basis.   \label{tab:Ds} }
\end{table}

\subsection{Numerical results} \label{sec:resultsDKDseta}

\begin{table}
   \begin{center}
  \begin{tabular*}{8.6cm}{@{\extracolsep{\fill}}lc}
    \hline \hline
    $R$ & $0.45$ \\
    $\Gamma_{B^{-} \to D_{0}^{\ast}(2400)^{0} \bar{\nu}_{l} l^{-}}
    / \Gamma_{\bar{B}^{0} \to D_{0}^{\ast}(2400)^{+} \bar{\nu}_{l} l^{-}}$ &
    $1.00$ \\
%
    $\mathcal{B}[\bar{B}^{0} \to D_{0}^{\ast} (2400)^{+} \bar{\nu} _{l} l^{-}]$     &    $4.5 \times 10^{-3}$ (input) \\
    $\mathcal{B}[\bar{B}^{-} \to D_{0}^{\ast} (2400)^{0} \bar{\nu} _{l} l^{-}$]     &    $4.9 \times 10^{-3}$ \\
    \hline \hline
  \end{tabular*}
  \end{center}
  \caption{Ratios of decay widths and branching fractions
    of semileptonic $B$ decays.   \label{tab:Gamma3}}
\end{table}

First we consider the coalescence case.  The numerical results are
summarized in Table~\ref{tab:Gamma3}.  The most interesting quantity
is the ratio $R = \Gamma_{\bar{B}_{s}^{0} \to D_{s0}^{\ast} (2317)^{+}
  \bar{\nu}_{l} l^{-}} / \Gamma_{\bar{B}^{0} \to D_{0}^{\ast}
  (2400)^{+} \bar{\nu}_{l} l^{-}}$ in the coalescence treatment, which
removes the unknown factor $C$ of the hadronization process. The
decay width in the coalescence is expressed by Eq.~\eqref{eq:Gcoal}.
The coupling constants of the two mesons to the $D$ resonances are
listed in Table~\ref{tab:Ds}.  Note that there are no fitting
parameters for the ratio $R$ in this scheme.  As a result, we obtain
the ratio of the decay widths as $R = 0.45$.  On the other hand, we
find that the ratio $\Gamma_{B^{-} \to D_{0}^{\ast}(2400)^{0}
\bar{\nu}_{l} l^{-}} / \Gamma_{\bar{B}^{0} \to
D_{0}^{\ast}(2400)^{+} \bar{\nu}_{l} l^{-}}$ is $1.00$, which can be
expected from the same strength of the decay amplitude to the
charged and neutral $D_{0}^{\ast} (2400)$ due to the isospin
symmetry, as discussed after Eq.~\eqref{eq:phiphi_41}.

The absolute value of the common prefactor $C$ can be determined with the help  
of experimental data on the decay width.  The branching
fraction of the semileptonic decay $\bar{B}^{0} \to D_{0}^{\ast}
(2400)^{+} \bar{\nu} _{l} l^{-}$ to the total decay is reported as
$(4.5 \pm 1.8) \times 10^{-3}$ by the Particle Data
Group~\cite{Agashe:2014kda}.  By using this mean value we find $C =
7.22$, and the fractions of decays $\bar{B}_{s}^{0} \to
D_{s0}^{\ast}(2317)^{+} \bar{\nu}_{l} l^{-}$ and $B^{-} \to
D_{0}^{\ast} (2400)^{0} \bar{\nu} _{l} l^{-}$ to the total decay
widths are obtained as $2.0 \times 10^{-3}$ and $4.9 \times 10^{-3}$,
respectively.  The values of these fractions are similar to each other.
The difference of the fractions of $\bar{B}^{0} \to D_{0}^{\ast}
(2400)^{+} \bar{\nu} _{l} l^{-}$ and $B^{-} \to D_{0}^{\ast}
(2400)^{0} \bar{\nu} _{l} l^{-}$ comes from the fact that the total
decay widths of $\bar{B}^{0}$ and $B^{-}$ are different.
\begin{table}
  \caption{Branching fraction of the process $\bar{B}_{s}^{0} \to D_{s0}^{\ast}(2317)^{+} \bar{\nu}_{l} l^{-}$ in percentage}
  \label{tab:bran2}
   \begin{center}
  \begin{tabular*}{8.6cm}{@{\extracolsep{\fill}}lc}
    \hline \hline
     Approach  &   $\mathcal{B}[\bar{B}_{s}^{0} \to D_{s0}^{\ast}(2317)^{+} \bar{\nu}_{l} l^{-}]$ \\
     \hline
     This work &     $0.20$ \\
     QCDSR + HQET \cite{Huang:2004et} &  $0.09 - 0.20$ \\
     QCDSR (SVZ)  \cite{Aliev:2006qy} &  $0.10$  \\
     LCSR         \cite{Li:2009wq}    &  $0.23 \pm 0.11$\\
     CQM          \cite{Zhao:2006at}  &  $0.49 - 0.57$ \\
     CQM          \cite{Segovia:2011dg}   &  $0.44$ \\
     CQM          \cite{Albertus:2014bfa}   &  $0.39$ \\
     \hline \hline
  \end{tabular*}
  \end{center}
\end{table}

A comparison of our  predictions for
$\mathcal{B}[\bar{B}_{s}^{0} \to D_{s0}^{\ast}(2317)^{+}
\bar{\nu}_{l} l^{-}]$ with the results obtained with other
approaches is presented in Table \ref{tab:bran2}. We emphasize that our approach is the
only one where the $D_{s0}^{\ast}(2317)^{+}$ is treated as a mesonic
molecule. Looking at Table \ref{tab:bran2} we can divide the results
in two groups: the first four numbers, which are ``small'' and the
last three, which are ``large''. In the second group, the
constituent quark models (CQM) yield larger branching fractions. In Ref.~\refcite{Navarra:2015iea} one 
can find some discussion on the origin of the differences based on the compact picture of the quark 
models versus the more extended structure of the molecular description.

Now let us discuss the rescattering process for the final-state two
mesons. We keep using the common prefactor $C = 7.22$ fixed from the
experimental value of the width of the semileptonic decay
$\bar{B}^{0} \to D_{0}^{\ast} (2400)^{+} \bar{\nu} _{l} l^{-}$.  The
meson-meson scattering amplitude was discussed in
Sec.~\ref{sec:amplitudes}, and now we include the $D_{s} \pi
^{0}$ channel as the isospin-breaking decay mode of $D_{s0}^{\ast}
(2317)$.  Namely, we calculate the scattering amplitude involving
the $D_{s} \pi ^{0}$ channel as
\begin{equation}
T_{i \to D_{s} \pi ^{0}} = \frac{g_{i} g_{D_{s} \pi ^{0}}}
{s - [ M_{D_{s0}^{\ast}} - i \Gamma _{D_{s0}^{\ast}} / 2]^{2}} ,
\end{equation}
for $i = D K$ and $D_{s} \eta$.  We take the $D_{s0}^{\ast} (2317)$
mass as $M_{D_{s0}^{\ast}} = 2317 \mev$, while we assume its decay
width as $\Gamma _{D_{s0}^{\ast}} = 3.8 \mev$, which is the upper
limit from experiments~\cite{Agashe:2014kda}.  The $D_{s0}^{\ast}
(2317)$-$i$ coupling constant $g_{i}$ ($i = D K$, $D_{s} \eta$) is
taken from Table~\ref{tab:Ds}, and the $D_{s0}^{\ast} (2317)$-$D_{s}
\pi ^{0}$ coupling constant $g_{D_{s} \pi ^{0}}$ is calculated from
the $D_{s0}^{\ast} (2317)$ decay width as
\begin{equation}
g_{D_{s} \pi ^{0}} = \sqrt{\frac{8 \pi M_{D_{s0}^{\ast}}^{2}
\Gamma _{D_{s0}^{\ast}}}{p_{\pi}}} ,
\end{equation}
with the pion center-of-mass momentum $p_{\pi}$, and we obtain
$g_{D_{s} \pi ^{0}} = 1.32$ GeV.

\begin{figure*}[htbp]
  \centering
\includegraphics[scale=0.5]{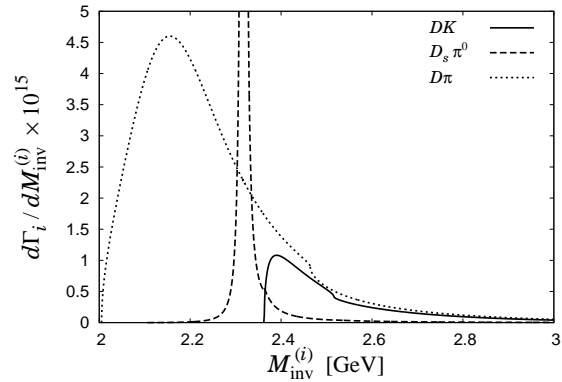}
  \caption{Differential decay width $d \Gamma _{i} / d M_{\rm
      inv}^{(i)}$ for the two pseudoscalars channel $i$ in the isospin
    basis.  Here we consider the semileptonic decays $\bar{B}_{s}^{0}
    \to (D K)^{+} \bar{\nu}_{l} l^{-}$, $(D_{s} \pi ^{0})^{+}
    \bar{\nu}_{l} l^{-}$ and $\bar{B}^{0} \to (D \pi )^{+} \bar{\nu}
    _{l} l^{-}$.  The $D K$ and $D_{s} \pi ^{0}$ channels couple to
    the $D_{s0}^{\ast}(2317)^{+}$ resonance, and $D \pi$ to the
    $D_{0}^{\ast} (2400)$ resonance.  The peak height for the $D_{s}
    \pi ^{0}$ channel is $d \Gamma _{D_{s} \pi ^{0}} / d M_{\rm
      inv}^{(D_{s} \pi ^{0})} \sim 1.5 \cdot 10^{-13}$.  }
  \label{Fig:fig7}
\end{figure*}

In Fig.~\ref{Fig:fig7} we show our predictions for the
differential decay width $d\Gamma _{i} / d M_{\rm inv}^{(i)}$ (Eq.~\eqref{eq:dGamma_dM}), 
where $i$ represents the two pseudoscalar states. In the figure we use  the isospin basis.  
When translating into the particle basis we use the following relations:  
\begin{equation}
[ D^{0} K^{+} ] = [ D^{+} K^{0} ] = \frac{1}{2} [ D K ] ,
\end{equation}
\begin{equation}
[ D_{s}^{+} \pi ^{0} ] = [ D_{s} \pi ^{0} ] ,
\end{equation}
\begin{equation}
[ D^{0} \pi ^{+} ] = 2 [ D^{+} \pi ^{0} ] = \frac{2}{3} [ D \pi ] ,
\end{equation}
where $[A B]$ is the partial decay width to the $A B$ channel.  An
interesting point is that the $D K$ mode shows a rapid increase from
its threshold $\approx 2360 \mev$ due to the existence of the bound
state, i.e., the $D_{s0}^{\ast} (2317)$ resonance.  In experiments,
such a rapid increase from the $D K$ threshold would support the
interpretation of the $D_{s0}^{\ast} (2317)$ resonance as a $D K$
bound state.  The strength of the $D K$ contribution in the $M_{\rm
  inv}^{(i)} \gtrsim 2.4 \gev$ region is similar to that of $D \pi$,
which corresponds to the ``tail'' for the $D_{0}^{\ast} (2400)$
resonance. On the other hand, the $D_{s} \pi ^{0}$ peak coming from the
$D_{s0}^{\ast} (2317)$ resonance is very sharp due to its narrow
width.

The distributions shown in Fig.~\ref{Fig:fig7} are our predictions and they can be measured at the LHCb.
They were obtained in the framework of the chiral unitary approach in coupled channels and their
experimental observation would give support to the $D_{s0}^{\ast} (2317)$ and $D_{0}^{\ast} (2400)$
as dynamically generated resonances, which is inherent to this approach.

Apart from comparing shapes and relative strength, one can make an
analysis of the $DK$ mass distribution as suggested in
Ref.~\refcite{Albaladejo:2015kea} to determine $g_{DK}$. With this value and
the use of Eq.~\eqref{eq:abcde} one can determine the amount of $DK$
component in the $D_{s0}^{\ast} (2317)$ wave function. Note that the
shape of the $DK$ mass distribution is linked to the potential, with
its associated energy dependence, and the mass of the $D_{s0}^{\ast}
(2317)$ \cite{Albaladejo:2015kea}. With the same binding of the
resonance, different models that have different amount of $DK$
component provide different shapes, leading to different values of the
$g_{DK}$ coupling, and it is possible to discriminate among models
that have a different nature for the $D_{s0}^{\ast} (2317)$ resonance.

\section[Investigating the nature of light scalar mesons with semileptonic decays of 
$D$ mesons]{\boldmath Investigating the nature of light scalar mesons with semileptonic decays of $D$ mesons}\label{sec:semileptonicD_scalar_vector}

\begin{figure}\centering
\includegraphics[height=7cm,keepaspectratio]{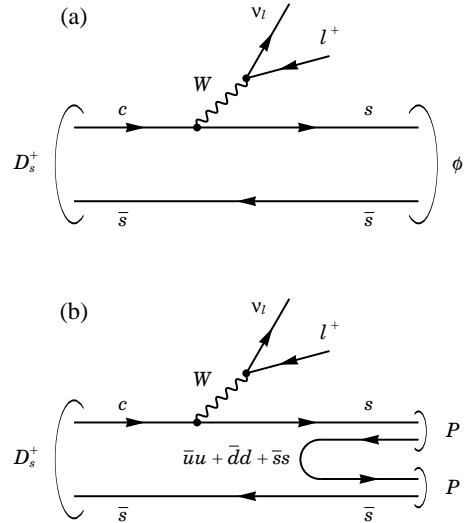}
\caption{(a) Semileptonic decay of $D^+_s$ into $l^+\nu_l$ and a primary
$s\bar{s}$ pair. (b) Semileptonic decay of $D^+_s$ into $l^+\nu_l$ and two
pseudoscalar mesons $P$ with a hadronization.\label{fig:DsTossbarlvul}}
\end{figure}

Here we consider the semileptonic decay of $D \to \text{hadron(s)} +
l^{+} \nu _{l}$, extending the work reported in the former section.
The semileptonic $D$ decays have been experimentally investigated
in, {\it e.g.}, BES~\cite{Ablikim:2006ah, Ablikim:2006hw},
FOCUS~\cite{Link:2004gp,Link:2005xe}, BaBar~\cite{Aubert:2008rs,
delAmoSanchez:2010fd}, and CLEO~\cite{Yelton:2009aa, Ecklund:2009aa,
Martin:2011rd,Yelton:2010js, CLEO:2011ab}.  In order to see
how the semileptonic decay takes place, let us consider the
$D_{s}^{+}$ meson. Since the constituent quark component of
$D_{s}^{+}$ is $c \bar{s}$, we expect a Cabibbo favored semileptonic
decay of $c \to s \, l^{+} \, \nu _{l}$ and hence the decay
$D_{s}^{+} \to ( s \bar{s} ) \, l^{+} \, \nu _{l}$ with $s \bar{s}$
being the vector meson $\phi (1020)$, which is depicted in
Fig.~\ref{fig:DsTossbarlvul}(a).  Actually this semileptonic decay mode has
been observed in experiments, and its branching fraction to the
total decay width is $\mathcal{B}[D_{s}^{+} \to \phi (1020) \, e^{+}
\, \nu _{e}] = 2.49 \pm 0.14 \%$~\cite{Agashe:2014kda} (see
Table~\ref{tab:Br}, in which we list branching fractions for the
semileptonic decays of $D_{s}^{+}$, $D^{+}$, and $D^{0}$ reported by
the Particle Data Group). In this study we consider the production
of the $f_{0} (980)$ or $f_{0} (500)$ as dynamically generated
resonances in the semileptonic $D_{s}^{+}$ decay, so we have to
introduce an extra $\bar{q} q$ pair to make a hadronization as shown
in Fig.~\ref{fig:DsTossbarlvul}(b).

\def\arraystretch{1.25}
\begin{table}
  \caption{Branching fractions for the semileptonic decays of
    $D_{s}^{+}$, $D^{+}$, and $D^{0}$ reported by the Particle Data
    Group~\cite{Agashe:2014kda}.  In this Table we only show decay
    modes relevant to this study.}
 \label{tab:Br}
    \begin{tabular}
    {@{\extracolsep{\fill}}lc}
      \multicolumn{2}{c}{$D_{s}^{+}$} \\
      \hline
      Mean life [s] & $(500 \pm 7) \times 10^{-15}$ \\
      $\mathcal{B}[\phi (1020) e^{+} \nu _{e}]$ &
      $(2.49 \pm 0.14) \times 10^{-2}$ \\
      $\mathcal{B}[\omega (782) e^{+} \nu _{e}]$ &
      $< 2.0 \times 10^{-3}$ \\
      $\mathcal{B}[K^{\ast} (892)^{0} e^{+} \nu _{e}]$ &
      $(1.8 \pm 0.7) \times 10^{-3}$ \\
      $\mathcal{B}[f_{0} (980) e^{+} \nu _{e}, \,
      f_{0} (980) \to \pi ^{+} \pi ^{-}]$ &
      $(2.00 \pm 0.32) \times 10^{-3}$ \\
      \\
      \multicolumn{2}{c}{$D^{+}$} \\
      \hline
      Mean life [s] & $(1040 \pm 7) \times 10^{-15}$ \\
      $\mathcal{B}[\bar{K}^{\ast} (892)^{0} e^{+} \nu _{e}, \,
      \bar{K}^{\ast} (892)^{0} \to K^{-} \pi ^{+}]$ &
      $(3.68 \pm 0.10) \times 10^{-2}$ \\
      $\mathcal{B}[(K^{-} \pi ^{+})_{s\text{-wave}} e^{+} \nu _{e}]$ &
      $(2.32 \pm 0.10) \times 10^{-3}$ \\
      $\mathcal{B}[\bar{K}^{\ast} (892)^{0} \mu ^{+} \nu _{\mu}, \,
      \bar{K}^{\ast} (892)^{0} \to K^{-} \pi ^{+}]$ &
      $(3.52 \pm 0.10) \times 10^{-2}$ \\
      $\mathcal{B}[\rho (770)^{0} e^{+} \nu _{e}]$ &
      $(2.18 ^{+0.17}_{-0.25}) \times 10^{-3}$ \\
      $\mathcal{B}[\rho (770)^{0} \mu ^{+} \nu _{\mu}]$ &
      $(2.4 \pm 0.4) \times 10^{-3}$ \\
      $\mathcal{B}[\omega (782) e^{+} \nu _{e}]$ &
      $(1.82 \pm 0.19) \times 10^{-3}$ \\
      $\mathcal{B}[\phi (1020) e^{+} \nu _{e}]$ &
      $< 9 \times 10^{-5}$ \\
      \\
      \multicolumn{2}{c}{$D^{0}$} \\
      \hline
      Mean life [s] & $(410.1 \pm 1.5) \times 10^{-15}$ \\
      $\mathcal{B}[K^{\ast} (892)^{-} e^{+} \nu _{e}]$ &
      $(2.16 \pm 0.16) \times 10^{-2}$ \\
      $\mathcal{B}[K^{\ast} (892)^{-} \mu ^{+} \nu _{\mu}]$ &
      $(1.90 \pm 0.24) \times 10^{-2}$ \\
      $\mathcal{B}[K^{-} \pi ^{0} e^{+} \nu _{e}]$ &
      $(1.6 ^{+1.3}_{-0.5}) \times 10^{-2}$ \\
      $\mathcal{B}[\bar{K}^{0} \pi ^{-} e^{+} \nu _{e}]$ &
      $(2.7 ^{+0.9}_{-0.7}) \times 10^{-2}$ \\
      $\mathcal{B}[\rho (770)^{-} e^{+} \nu _{e}]$ &
      $(1.9 \pm 0.4) \times 10^{-3}$ \\
    \end{tabular}
\end{table}
\def\arraystretch{1.0}

\subsection{Formulation}  \label{sec:form}

\begin{table}\centering
    \begin{tabular}{@{\extracolsep{\fill}}lc}
      \multicolumn{2}{c}{$D_{s}^{+}$}
      \\
      \hline
      $\phi (1020) \, l^{+} \, \nu _{l}$ & favored
      \\
      $K^{\ast} (892)^{0} \, l^{+} \, \nu _{l}$ & suppressed
      \\
      $\pi ^{+} \pi ^{-} \, l^{+} \, \nu _{l}$ & favored
      \\
      $K^{+} K^{-} \, l^{+} \, \nu _{l}$ & favored
      \\
      $\pi ^{-} K^{+} \, l^{+} \, \nu _{l}$ & suppressed
      \\
      \\
      \multicolumn{2}{c}{$D^{+}$}
      \\
      \hline
      $\bar{K}^{\ast} (892)^{0} \, l^{+} \, \nu _{l}$ & favored
      \\
      $\rho (770)^{0} \, l^{+} \, \nu _{l}$ & suppressed
      \\
      $\omega (782) \, l^{+} \, \nu _{l}$ & suppressed
      \\
      $\pi ^{+} \pi ^{-} \, l^{+} \, \nu _{l}$ & suppressed
      \\
      $\pi ^{0} \eta \, l^{+} \, \nu _{l}$ & suppressed
      \\
      $K^{+} K^{-} \, l^{+} \, \nu _{l}$ & suppressed
      \\
      $\pi ^{+} K^{-} \, l^{+} \, \nu _{l}$ & favored
      \\
      \\
      \multicolumn{2}{c}{$D^{0}$}
      \\
      \hline
      $K^{\ast} (892)^{-} \, l^{+} \, \nu _{l}$ & favored
      \\
      $\rho (770)^{-} \, l^{+} \, \nu _{l}$ & suppressed
      \\
      $\pi ^{-} \eta \, l^{+} \, \nu _{l}$ & suppressed
      \\
      $K^{0} K^{-} \, l^{+} \, \nu _{l}$ & suppressed
      \\
      $\pi ^{-} \bar{K}^{0} \, l^{+} \, \nu _{l}$ & favored
      \\
  \end{tabular}
  \caption{Semileptonic decay modes of $D_{s}^{+}$, $D^{+}$, and
    $D^{0}$ considered in this study.  The lepton flavor $l$ is $e$
    and $\mu$.  We also specify Cabibbo favored/suppressed process for
    each decay mode; the semileptonic decay into two pseudoscalar
    mesons is classified following the discussions given in
    Sec.~\ref{sec:had}.\label{tab:mode}}
\end{table}

In this section we formulate the semileptonic decay widths of
$D_{s}^{+}$, $D^{+}$, and $D^{0}$ into light scalar and vector mesons:
\begin{equation}
D_{s}^{+} , \, D^{+} , \, D^{0} \to
\begin{cases}
S l^{+} \nu _{l} , \quad S \to P P , \\
V l^{+} \nu _{l} ,
\end{cases}
\end{equation}
where $S$, $V$, and $P$ represent the light scalar, vector, and
pseudoscalar mesons, respectively, and the lepton flavor $l$ can be
$e$ and $\mu$.  Explicit decay modes are listed in
Table~\ref{tab:mode}.

\subsection[Amplitudes and widths of semileptonic $D$ decays]{\boldmath Amplitudes and widths of semileptonic $D$ decays} \label{sec:decay}

The calculation of the amplitudes proceeds exactly like in the
former section changing the masses and the coefficient $C$.

\subsection{Hadronizations}  \label{sec:had}

\begin{figure}[!b]
  \centering
  \PsfigII{0.185}{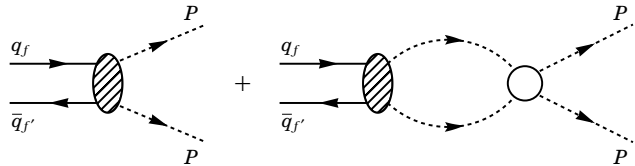}
  \caption{Diagrammatic representation of the direct plus
    rescattering processes for two pseudoscalar mesons.  The solid and
    dashed lines denote quarks and pseudoscalar mesons, respectively.
    The shaded ellipses indicate the hadronization of a
    quark--antiquark pair into two pseudoscalar mesons, while the open
    circle indicates the rescattering of two pseudoscalar mesons.}
\label{fig:3}
\end{figure}

Next we fix the mechanism for the appearance of the scalar and
vector mesons in the final state of the semileptonic decay. We note
that, for the scalar and vector mesons in the final state, the
hadronization processes should be different from each other
according to their structure.  For the scalar mesons, we employ the
chiral unitary approach, in which the scalar mesons are dynamically
generated from the interaction of two pseudoscalar mesons governed
by the chiral Lagrangians.  Therefore, in this picture the light
quark--antiquark pair after the $W$ boson emission gets hadronized
by adding an extra $\bar{q} q$ with the quantum number of the
vacuum, $\bar{u} u + \bar{d} d + \bar{s} s$, which results in two
pseudoscalar mesons in the final state [see Fig.~\ref{fig:DsTossbarlvul}(b)].
Then the scalar mesons are obtained as a consequence of the final
state interaction of the two pseudoscalar mesons as diagrammatically
shown in Fig.~\ref{fig:3}.  For the vector mesons, on the other
hand, hadronization with an extra $\bar{q} q$ is unnecessary since
they are expected to consist genuinely of a light quark--antiquark
pair [see Fig.~\ref{fig:DsTossbarlvul}(a)].

\subsubsection{Scalar mesons}

In this scheme we can calculate the weight of each pair of
pseudoscalar mesons in the hadronization.  Namely, the $s \bar{s}$
pair gets hadronized as $s \bar{s} ( \bar{u} u + \bar{d} d + \bar{s}
s ) \equiv ( \phi \cdot \phi )_{3 3}$, where
\begin{align}
( \phi \cdot \phi )_{3 3}
& =
K^{-} K^{+} + \bar{K}^{0} K^{0}
+ \frac{1}{3} \eta \eta .
\label{eq:phiphi33}
\end{align}
Here and in the following we omit the $\eta ^{\prime}$ contribution
since $\eta ^{\prime}$ is irrelevant to the description of the
scalar mesons due to its large mass. In a similar manner, the $d
\bar{s}$, $s \bar{d}$, $d \bar{d}$, $s \bar{u}$, and $d \bar{u}$
pairs get hadronized, respectively, as:
\begin{align}
( \phi \cdot \phi )_{2 3} & = \pi ^{-} K^{+} - \frac{1}{\sqrt{2}} \pi ^{0} K^{0}~,\\
( \phi \cdot \phi )_{3 2} & = K^{-} \pi ^{+} - \frac{1}{\sqrt{2}} \bar{K}^{0} \pi ^{0}~,\\
( \phi \cdot \phi )_{2 2} & = \pi ^{-} \pi ^{+} + \frac{1}{2} \pi^{0} \pi ^{0} + \frac{1}{3} \eta \eta - \sqrt{\frac{2}{3}} \pi ^{0}\eta + K^{0} \bar{K}^{0}~,\\
( \phi \cdot \phi )_{3 1} & = \frac{1}{\sqrt{2}} \pi ^{0} K^{-} + \pi ^{-} \bar{K}^{0}~,\\
( \phi \cdot \phi )_{2 1} & = \frac{2}{\sqrt{3}} \pi ^{-} \eta + K^{0} K^{-}~.
\end{align}
By using these weights, we can express the hadronization amplitude
for the scalar mesons, $V_{\rm had}^{(s)}$, in terms of two
pseudoscalar mesons.  For instance, we want to reconstruct the
$f_{0} (500)$ and $f_{0} (980)$ from the $\pi ^{+} \pi ^{-}$ system
in the $D_{s}^{+} \to \pi ^{+} \pi ^{-} \, l^{+} \, \nu _{l}$ decay.
Because of the quark configuration in the parent particle
$D_{s}^{+}$, in this decay the $\pi ^{+} \pi ^{-}$ system should be
obtained from the hadronization of the $s \bar{s}$ pair and the
rescattering process for two pseudoscalar mesons, as seen in
Fig.~\ref{fig:3}, with the weight in Eq.~\eqref{eq:phiphi33}.
Therefore, for the $D_{s}^{+} \to \pi ^{+} \pi ^{-} \, l^{+} \, \nu
_{l}$ decay mode we can express the hadronization amplitude with a
prefactor $C$ and the CKM matrix elements
$V_{c s}$ as
\begin{eqnarray}
V_{\rm had}^{(s)} [ D_{s}^{+}, \, \pi ^{+} \pi ^{-}] & = & C V_{c s}
( G_{K^{+} K^{-}} T_{K^{+} K^{-} \to \pi ^{+} \pi ^{-}}+ G_{K^{0} \bar{K}^{0}} T_{K^{0} \bar{K}^{0} \to \pi ^{+} \pi ^{-}} \nonumber \\
&& + \frac{1}{3} \cdot 2 \cdot \frac{1}{2} G_{\eta \eta} T_{\eta
\eta \to \pi ^{+} \pi ^{-}} ) .
\end{eqnarray}
In this equation, the decay mode is abbreviated as $[ D_{s}^{+}, \,
\pi ^{+} \pi ^{-}]$, and $G$ and $T$ are the loop function and
scattering amplitude of two pseudoscalar mesons, respectively. We
have introduced extra factors $2$ and $1/2$ for the identical
particles $\eta \eta$, as also discussed in former sections. The scalar mesons $f_{0} (500)$ and
$f_{0} (980)$ appear in the rescattering process and exist in the
scattering amplitude $T$ for two pseudoscalar mesons.  Note that this is a Cabibbo favored process with $V_{c s}$.
Furthermore, since the $s \bar{s}$ pair is hadronized, this is
sensitive to the component of the strange quark in the scalar
mesons. As done in former sections, we assume that $C$ is a constant, and hence
the hadronization amplitude $V_{\rm had}^{(s)}$ is a function only
of the invariant mass of two pseudoscalar mesons.  Here we emphasize
that the factor $C$ should be common to all reactions for scalar
meson production, because in the hadronization the $SU(3)$ flavor
symmetry is reasonable, i.e., the light quark--antiquark pair $q_{f}
\bar{q}_{f^{\prime}}$ hadronizes in the same way regardless of the
quark flavor $f$.  In this sense we obtain
\begin{align}
V_{\rm had}^{(s)} [ D_{s}^{+}, \, K^{+} K^{-}] & = C V_{c s} ( 1 +
G_{K^{+} K^{-}} T_{K^{+} K^{-} \to K^{+} K^{-}} +
G_{K^{0} \bar{K}^{0}} T_{K^{0} \bar{K}^{0} \to K^{+} K^{-}} \notag
\\
& + \frac{1}{3} \cdot 2 \cdot \frac{1}{2} G_{\eta \eta} T_{\eta \eta
\to K^{+} K^{-}}  ) ,
\end{align}
for the $D_{s}^{+} \to K^{+} K^{-} \, l^{+} \, \nu _{l}$ decay.  In
this case we have to take into account the direct production of the
two pseudoscalar mesons without rescattering (the first diagram in
Fig.~\ref{fig:3}), which results in the unity in the parentheses.  On
the other hand, for the $D_{s}^{+} \to \pi ^{-} K^{+} \, l^{+} \, \nu
_{l}$ decay mode the $\pi ^{-} K^{+}$ system should be obtained from
the hadronization of $d \bar{s}$ and hence this is a Cabibbo
suppressed decay mode.  The hadronization amplitude is expressed as
\begin{align}
V_{\rm had}^{(s)} [ D_{s}^{+}, \, \pi ^{-} K^{+}] = C V_{c d} \left
( 1 + G_{\pi ^{-} K^{+}} T_{\pi ^{-} K^{+} \to \pi ^{-} K^{+}}
- \frac{1}{\sqrt{2}} G_{\pi ^{0} K^{0}} T_{\pi
^{0} K^{0} \to \pi ^{-} K^{+}} \right ) . \label{eq:Ds_pipi}
\end{align}
In a similar way we can construct every hadronization amplitude for
the scalar meson.  The resulting expressions are as follows:
\begin{align}
V_{\rm had}^{(s)} [ D^{+}, \, \pi ^{+} \pi ^{-}] & = C V_{c d} \left
( 1 + G_{\pi ^{+} \pi ^{-}} T_{\pi ^{+} \pi ^{-} \to \pi ^{+} \pi
^{-}} + \frac{1}{2} \cdot 2 \cdot \frac{1}{2} G_{\pi
^{0} \pi ^{0}} T_{\pi ^{0} \pi ^{0} \to \pi ^{+} \pi ^{-}}  \right .
\notag \\
& \left . + \frac{1}{3} \cdot 2 \cdot \frac{1}{2} G_{\eta \eta}
T_{\eta \eta \to \pi ^{+} \pi ^{-}} + G_{K^{0} \bar{K}^{0}} T_{K^{0}
\bar{K}^{0} \to \pi ^{+} \pi ^{-}} \right ) ,
\end{align}
\begin{align}
V_{\rm had}^{(s)} [ D^{+}, \, \pi ^{0} \eta ] = C V_{c d} \left ( -
\sqrt{\frac{2}{3}} - \sqrt{\frac{2}{3}} G_{\pi ^{0} \eta} T_{\pi
^{0} \eta \to \pi ^{0} \eta}
  + G_{K^{0} \bar{K}^{0}} T_{K^{0} \bar{K}^{0} \to \pi ^{0} \eta}
\right ) ,
\end{align}
\begin{align}
& V_{\rm had}^{(s)} [ D^{+}, \, K^{+} K^{-}]
= C V_{c d} \left ( G_{\pi ^{+} \pi ^{-}} T_{\pi ^{+} \pi ^{-} \to K^{+} K^{-}}
\phantom{\sqrt{\frac{1}{2}}}
\right .
\notag \\
&
+ \frac{1}{2} \cdot 2 \cdot \frac{1}{2}
G_{\pi ^{0} \pi ^{0}} T_{\pi ^{0} \pi ^{0} \to K^{+} K^{-}}
+ \frac{1}{3} \cdot 2 \cdot \frac{1}{2}
G_{\eta \eta} T_{\eta \eta \to K^{+} K^{-}}
\notag \\
&
\left . \!
- \sqrt{\frac{2}{3}}
G_{\pi ^{0} \eta} T_{\pi ^{0} \eta \to K^{+} K^{-}}
+ G_{K^{0} \bar{K}^{0}} T_{K^{0} \bar{K}^{0} \to K^{+} K^{-}}
\right ) ,
\end{align}
\begin{align}
& V_{\rm had}^{(s)} [ D^{+}, \, \pi ^{+} K^{-}]
= C V_{c s} \left ( 1 + G_{\pi ^{+} K^{-}} T_{\pi ^{+} K^{-} \to \pi ^{+} K^{-}}
\phantom{\frac{1}{\sqrt{2}}}
\right .
\notag \\
&
\left .
  \quad \quad \quad \quad
- \frac{1}{\sqrt{2}}
G_{\pi ^{0} \bar{K}^{0}} T_{\pi ^{0} \bar{K}^{0} \to \pi ^{+} K^{-}}
\right ) ,
\end{align}
\begin{align}
& V_{\rm had}^{(s)} [ D^{0}, \, \pi ^{-} \eta ]
= C V_{c d} \left ( \frac{2}{\sqrt{3}}
+ \frac{2}{\sqrt{3}}
G_{\pi ^{-} \eta} T_{\pi ^{-} \eta \to \pi ^{-} \eta}
\right .
\notag \\
& \left . \phantom{\frac{2}{\sqrt{3}}}
  \quad \quad \quad \quad
  + G_{K^{0} K^{-}} T_{K^{0} K^{-} \to \pi ^{-} \eta}
\right ) ,
\end{align}
\begin{align}
& V_{\rm had}^{(s)} [ D^{0}, \, K^{0} K^{-} ]
= C V_{c d} \left ( 1
+ \frac{2}{\sqrt{3}}
G_{\pi ^{-} \eta} T_{\pi ^{-} \eta \to K^{0} K^{-}}
\right .
\notag \\
& \left . \phantom{\frac{2}{\sqrt{3}}}
  \quad \quad \quad \quad
  + G_{K^{0} K^{-}} T_{K^{0} K^{-} \to K^{0} K^{-}}
\right ) ,
\end{align}
\begin{align}
& V_{\rm had}^{(s)} [ D^{0}, \, \pi ^{-} \bar{K}^{0} ]
= C V_{c s} \left ( 1
+ \frac{1}{\sqrt{2}}
G_{\pi ^{0} K^{-}} T_{\pi ^{0} K^{-} \to \pi ^{-} \bar{K}^{0}}
\right .
\notag \\
& \left . \phantom{\frac{1}{\sqrt{2}}}
  \quad \quad \quad \quad
  + G_{\pi ^{-} \bar{K}^{0}} T_{\pi ^{-} \bar{K}^{0} \to \pi ^{-} \bar{K}^{0}}
\right ) .
\end{align}
Some of these expressions are further simplified using isospin symmetry in Ref.~\refcite{Sekihara:2015iha}.
From the above expressions one can easily specify Cabibbo favored and
suppressed processes for the semileptonic decays into two pseudoscalar
mesons, which are listed in Table~\ref{tab:mode}.

\subsubsection{Vector mesons}

Next we consider processes with the vector mesons in the final
state. As done before, they are associated to $\bar{q}q$ states.

In order to see how the production proceeds, we consider the
semileptonic decay $D_{s}^{+} \to \phi (1020) \, l^{+} \, \nu _{l}$
as an example.  The decay process is diagrammatically represented in
Fig.~\ref{fig:DsTossbarlvul}(a), and the amplitude $V_{\rm
had}^{(v)}$ can be expressed with a prefactor $C^{\prime}$ and the
CKM matrix element $V_{c s}$ as
\begin{equation}
V_{\rm had}^{(v)} [ D_{s}^{+}, \, \phi ] = C^{\prime} V_{c s} ,
\end{equation}
where the decay mode is abbreviated as $[ D_{s}^{+}, \, \phi ]$ in the
equation.  Here we emphasize that the prefactor $C^{\prime}$ should be
common to all reactions for vector meson production, as in the case
of the scalar meson production, because the $SU(3)$ flavor symmetry is
reasonable in the hadronization, i.e., the light quark--antiquark pair
$q_{f} \bar{q}_{f^{\prime}}$ hadronizes in the same way regardless of
the quark flavor $f$.  We further assume that $C^{\prime}$ is a
constant again.  This formulation is straightforwardly applied to
other vector meson productions and we obtain the hadronization
amplitude for vector mesons:
\begin{equation}
V_{\rm had}^{(v)} [ D_{s}^{+}, \, K^{\ast 0} ] = C^{\prime} V_{c d} ,
\end{equation}
\begin{equation}
V_{\rm had}^{(v)} [ D^{+}, \, \bar{K}^{\ast 0} ] = C^{\prime} V_{c s} ,
\end{equation}
\begin{equation}
V_{\rm had}^{(v)} [ D^{+}, \, \rho ^{0} ]
= - \frac{1}{\sqrt{2}} C^{\prime} V_{c d} ,
\end{equation}
\begin{equation}
V_{\rm had}^{(v)} [ D^{+}, \, \omega ]
= \frac{1}{\sqrt{2}} C^{\prime} V_{c d} ,
\end{equation}
\begin{equation}
V_{\rm had}^{(v)} [ D^{0}, \, K^{\ast -} ] = - C^{\prime} V_{c s} ,
\end{equation}
\begin{equation}
V_{\rm had}^{(v)} [ D^{0}, \, \rho ^{-} ]
= C^{\prime} V_{c d} ,
\end{equation}
where we have used $K^{\ast}$, $\rho$ and $\omega$ states in the
isospin basis. We note that these equations clearly indicate Cabibbo
favored and suppressed processes with the
CKM matrix elements $V_{c s}$ and $V_{c
  d}$, respectively.

\subsection{Numerical results} \label{sec:resultssemileptonic}

\subsubsection{Production of scalar mesons} \label{sec:scalar}

In order to calculate the branching fractions of the scalar meson
productions, we first fix the prefactor constant $C$ so as to
reproduce the experimental branching fraction which has the smallest
experimental error for the process with the $s$-wave two pseudoscalar
mesons, that is, $\mathcal{B}[D^{+} \to ( \pi ^{+} K^{-}
)_{s\text{-wave}} e^{+} \nu _{e}] = (2.32 \pm 0.10) \times 10^{-3}$.
By integrating the differential decay width, or mass distribution, $d
\Gamma _{4} / d M_{\rm inv}^{(hh)}$ in an appropriate range, in the
case of $\pi ^{+} K^{-}$ [$m_{\pi} + m_{K}$, $1 \gev$], we find that
$C = 4.6$ can reproduce the branching fraction of $( \pi ^{+} K^{-}
)_{s\text{-wave}} e^{+} \nu _{e}$.

\begin{figure}[!t]
  \centering
  \Psfig{8.6cm}{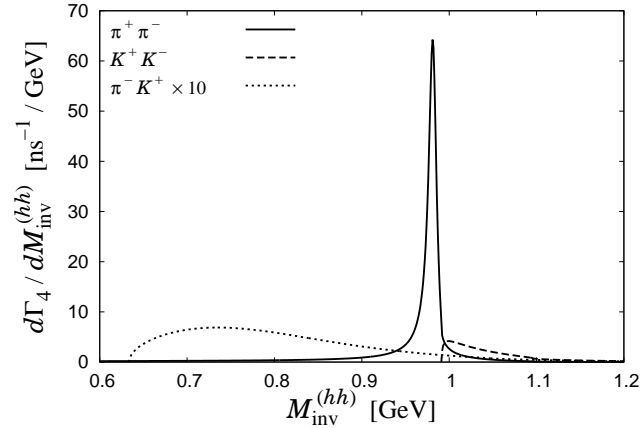}
  \caption{Meson--meson invariant mass distributions for the
    semileptonic decay $D_{s}^{+} \to P P e^{+} \nu _{e}$ with $P P =
    \pi ^{+} \pi ^{-}$, $K^{+} K^{-}$, and $\pi ^{-} K^{+}$ in $s$
    wave.  We multiply the $\pi ^{-} K^{+}$ mass distribution, which
    is a Cabibbo suppressed process, by $10$. }
\label{fig:dG4_Ds}
\end{figure}

\begin{figure}[!t]
  \centering
  \Psfig{8.6cm}{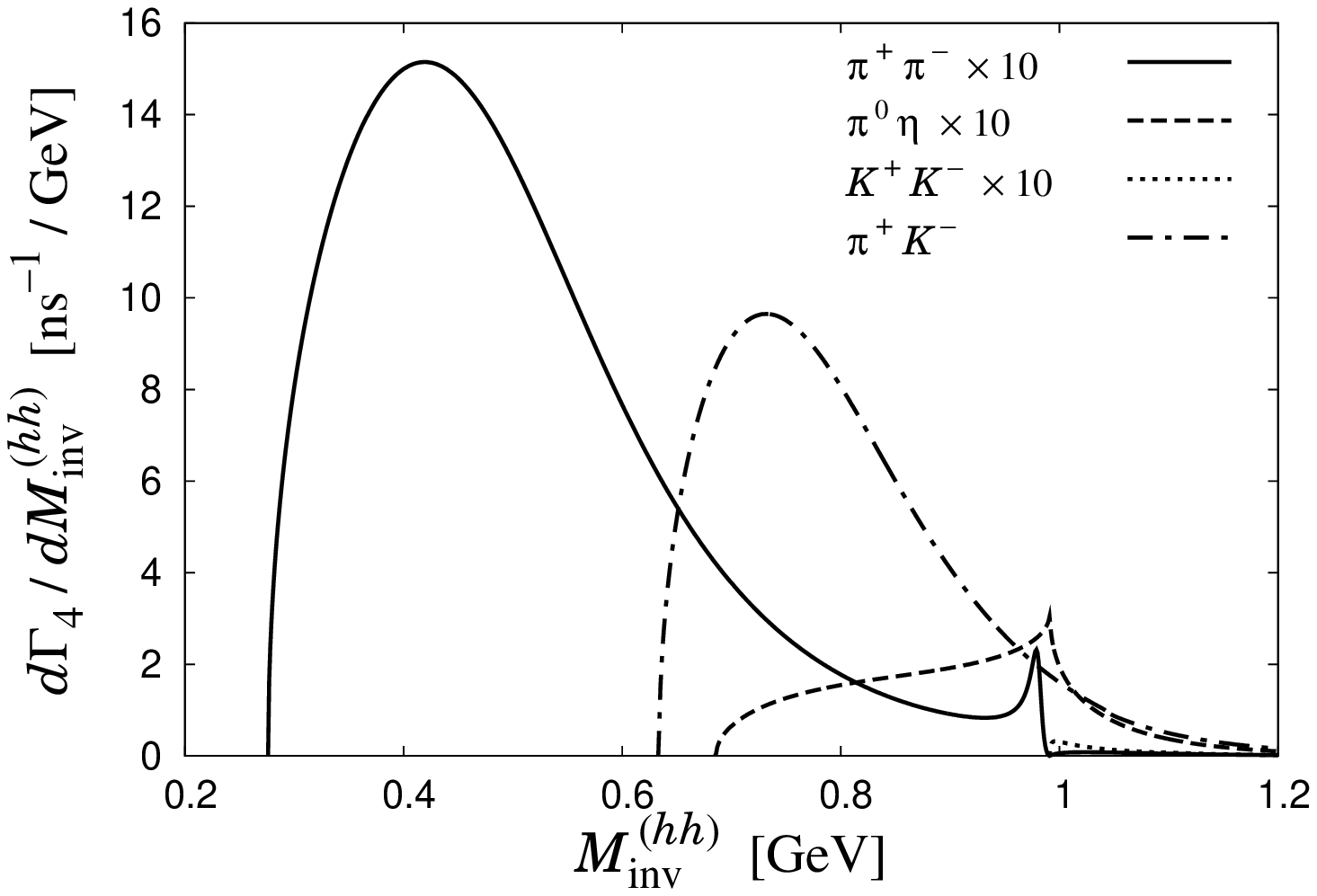}
  \caption{Meson--meson invariant mass distributions for the
    semileptonic decay $D^{+} \to P P e^{+} \nu _{e}$ with $P P = \pi
    ^{+} \pi ^{-}$, $\pi ^{0} \eta$, $K^{+} K^{-}$, and $\pi ^{+}
    K^{-}$ in $s$ wave.  We multiply the $\pi ^{+} \pi ^{-}$, $\pi
    ^{0} \eta$, and $K^{+} K^{-}$ mass distributions, which are
    Cabibbo suppressed processes, by $10$. }
\label{fig:dG4_Dp}
\end{figure}

\begin{figure}[!t]
  \centering
  \Psfig{8.6cm}{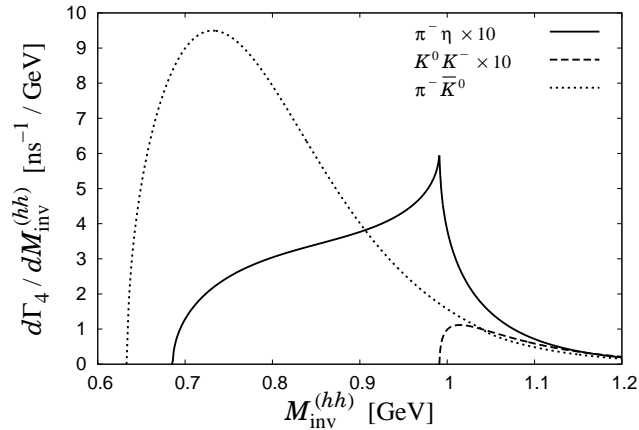}
  \caption{Meson--meson invariant mass distributions for the
    semileptonic decay $D^{0} \to P P e^{+} \nu _{e}$ with $P P = \pi
    ^{-} \eta$, $K^{0} K^{-}$, and $\pi ^{-} \bar{K}^{0}$ in $s$ wave.
    We multiply the $\pi ^{-} \eta$ and $K^{0} K^{-}$ mass
    distributions, which are Cabibbo suppressed processes, by $10$. }
\label{fig:dG4_Dz}
\end{figure}

By using the common prefactor $C = 4.6$, we can calculate the mass
distributions of two pseudoscalar mesons in $s$ wave for all scalar
meson modes, which are plotted in Figs.~\ref{fig:dG4_Ds},
\ref{fig:dG4_Dp}, and \ref{fig:dG4_Dz} for $D_{s}^{+}$, $D^{+}$, and
$D^{0}$ semileptonic decays, respectively.  We show the mass
distributions with the lepton flavor $l = e$; the contribution from
$l = \mu$ is almost the same as that from $l = e$ in each
meson--meson mode due to the small lepton masses. As one can see,
the largest value of the mass distribution $d \Gamma _{4} / d M_{\rm
inv}^{(hh)}$ is obtained in the $D_{s}^{+} \to \pi ^{+} \pi ^{-}
e^{+} \nu _{e}$ process, in which clear peak due to $f_{0} (980)$
is observed. In the $D_{s}^{+} \to \pi ^{+}
\pi ^{-} e^{+} \nu _{e}$ process we find a clear $f_{0} (980)$
signal while the $f_{0} (500)$ contribution is negligible, which
strongly indicates a substantial fraction of the strange quarks in
the $f_{0} (980)$ meson.  For the
$D_{s}^{+}$ semileptonic decay we also observe a rapid enhancement
of the $K^{+} K^{-}$ mass distribution at threshold, as a tail
of the $f_{0} (980)$ contribution, although its height is much
smaller than the $\pi ^{+} \pi ^{-}$ peak.  For the $D^{+}$ and
$D^{0}$ semileptonic decays, we can see the $\pi ^{+} K^{-}$ and
$\pi ^{-} \bar{K}^{0}$ as Cabibbo favored processes, respectively.
We note that the $\pi ^{+} K^{-}$ and $\pi ^{-} \bar{K}^{0}$ mass
distributions are almost the same due to isospin symmetry. The $\pi \eta$ mass
distributions in Figs.~\ref{fig:dG4_Dp} and \ref{fig:dG4_Dz} of the
$D^{+}$ and $D^{0}$ decays show peaks corresponding to $a_{0}(980)$, but its peak is not as high as the $f_{0} (980)$ peak
in the $\pi ^{+} \pi ^{-}$ mass distribution of the $D_{s}^{+}$
decay since they correspond to Cabibbo suppressed processes.  The
$D^{+} \to \pi ^{+} \pi ^{-} e^{+} \nu _{e}$ decay is Cabibbo
suppressed and it has a large contribution from the $f_{0} (500)$
formation and a small one of the $f_{0} (980)$, similar to what is
found in the $\bar{B}^{0} \to J / \psi \pi ^{+} \pi ^{-}$ decay in
section \ref{sec:scalarsectorBDdecays}. A different way to account for the $PP$
distribution is by means of dispersion relations, as used in
Ref.~\refcite{Kang:2013jaa} in the semileptonic decay of $B$, where the
$\pi ^{+} \pi ^{-}$ $s$-wave distribution has a shape similar to
ours.

\begin{figure}[!t]
  \centering
  \Psfig{8.6cm}{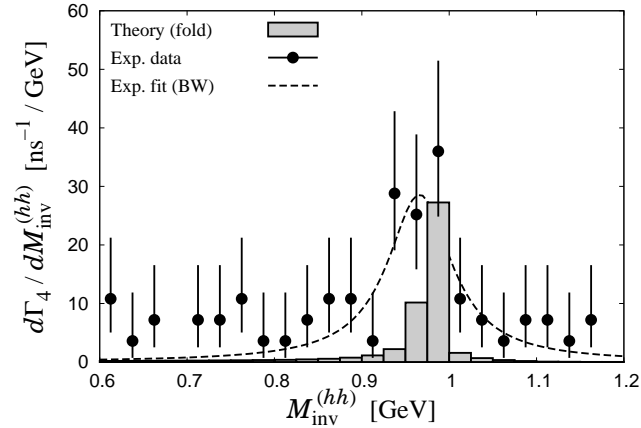}
  \caption{$\pi ^{+} \pi ^{-}$ invariant mass distribution for the
    semileptonic decay $D_{s}^{+} \to \pi ^{+} \pi ^{-} e^{+} \nu
    _{e}$.  The theoretical calculation is folded with the size of
    experimental bins, $25 \mev$.  The experimental data\cite{Ecklund:2009aa} are scaled so that the fitted
    Breit--Wigner distribution (dashed line) reproduces the branching
    fraction of $\mathcal{B}[D_{s}^{+} \to f_{0} (980) e^{+} \nu _{e},
    \, f_{0} (980) \to \pi ^{+} \pi ^{-}] = 0.2 \%$ by the Particle
    Data Group (see Table~\ref{tab:Br}). }
\label{fig:hist}
\end{figure}

The theoretical $\pi ^{+} \pi ^{-}$ mass distribution of the
semileptonic decay $D_{s} \to \pi ^{+} \pi ^{-} e^{+} \nu _{e}$ is
compared with the experimental data~\cite{Ecklund:2009aa} in
Fig.~\ref{fig:hist}.  We note that we plot the figure in unit of
$\text{ns}^{-1} / \text{GeV}$, not in arbitrary units.  The
theoretical mass distribution is folded with $25 \mev$ spans since
the experimental data are collected in bins of $25 \mev$.  The
experimental data, on the other hand, are scaled so that the fitted
Breit--Wigner distribution reproduces the branching fraction of
$\mathcal{B}[D_{s}^{+} \to f_{0} (980) e^{+} \nu _{e}, \, f_{0}
(980) \to \pi ^{+} \pi ^{-}] = 0.2 \%$~\cite{Agashe:2014kda}.  The
mass and width of the Breit--Wigner distribution are fixed as $M =
966 \mev$ and $\Gamma = 89 \mev$, respectively, taken from
Ref.~\refcite{Ecklund:2009aa}.  In Fig.~\ref{fig:hist} we can see a
qualitative correspondence between the theoretical and experimental
signals of $f_{0} (980)$.  We emphasize that, both in experimental
and theoretical results, the $\pi ^{+} \pi ^{-}$ mass distribution
shows a clear $f_{0} (980)$ signal while the $f_{0} (500)$
contribution is negligible.  This strongly indicates that the $f_{0}
(980)$ has a substantial fraction of the strange quarks while the
$f_{0} (500)$ has a negligible strange quark component.  It is
interesting to recall that the appearance of the $f_{0} (980)$ in
the case one has a hadronized $s \bar{s}$ component at the end, and
no signal of the $f_{0} (500)$, is also observed in $B_{s}^{0}$ and
$B^{0}$ decays\cite{Aaij:2011fx,Li:2011pg,Aaltonen:2011nk,Abazov:2011hv,LHCb:2012ae}. The explanation of this feature was
already discussed in section~\ref{sec:scalarsectorBDdecays}. In the experimental analysis of
Ref.~\refcite{Ecklund:2009aa} different sources of background are
considered that make up for the lower mass region of the
distribution. The width of the $f_{0} (980)$ extracted in the
analysis of Ref.~\refcite{Ecklund:2009aa} is $\Gamma = 91 ^{+30}_{-22}
\pm 3 \mev$, which is large compared to most
experiments~\cite{Agashe:2014kda}, including the LHCb experiment
of Ref.~\refcite{Aaij:2014emv}, although the admitted uncertainties are also
large.  One should also take into account that, while a
Breit--Wigner distribution for the $f_{0} (980)$ is used in the
analysis of Ref.~\refcite{Ecklund:2009aa}, the large coupling of the
resonance to $K \bar{K}$ requires a Flatte form that brings down
fast the $\pi ^{+} \pi ^{-}$ mass distribution above the $K \bar{K}$
threshold.  Our normalization in Fig.~\ref{fig:hist} is done with
the central value of the $\mathcal{B} [D^{+} \to (\pi ^{+}
K^{-})_{s\text{-wave}} e^{+} \nu _{e}]$ and no extra uncertainties
from this branching fraction are considered.  Yet, we find
instructive to do an exercise, adding to our results a
``background'' of $10 \text{ ns}^{-1} / \text{GeV}$ from different
sources that our calculation does not take into account, and then
our signal for the $f_{0} (980)$ has a good agreement with the peak
of the experimental distribution. It is instructive to see that in a
reanalysis of the data of Ref.~\refcite{Ecklund:2009aa} done in
Ref.~\refcite{Hietala:2015jqa}, taking a window of $60$ MeV around $980$
MeV and using a Flatte form, one obtains a rate about half of that
in Ref.~\refcite{Ecklund:2009aa}.

\begin{table}\centering
    \begin{tabular}{@{\extracolsep{\fill}}lccc}
      \multicolumn{4}{c}{$D_{s}^{+}$}
      \\
      Mode & Range of $M_{\rm inv}^{(hh)}$ [GeV] & $l = e$ & $l = \mu$ \\
      \hline
      $\pi ^{+} \pi ^{-}$ & [0.9, 1.0] &
      $5.10 \times 10^{-4}$  &  $4.71 \times 10^{-4}$
      \\
      $K^{+} K^{-}$ & [$2 m_{K}$, 1.2] &
      $1.42 \times 10^{-4}$ & $1.30 \times 10^{-4}$
      \\
      $\pi ^{-} K^{+}$ & [$m_{\pi} + m_{K}$, 1.0] &
      $8.11 \times 10^{-5}$ & $7.63 \times 10^{-5}$
      \\
      \\
      \multicolumn{4}{c}{$D^{+}$}
      \\
      Mode & Range of $M_{\rm inv}^{(hh)}$ [GeV] & $l = e$ & $l = \mu$ \\
      \hline
      $\pi ^{+} \pi ^{-}$ & [$2 m_{\pi}$, 1.0] &
      $5.11 \times 10^{-4}$ & $4.85 \times 10^{-4}$
      \\
      $\pi ^{0} \eta$ & [$m_{\pi} + m_{\eta}$, 1.1] &
      $6.37 \times 10^{-5}$ & $5.86 \times 10^{-5}$
      \\
      $K^{+} K^{-}$ & [$2 m_{K}$, 1.2] &
      $2.24 \times 10^{-6}$ & $2.01 \times 10^{-6}$
      \\
      $\pi ^{+} K^{-}$ & [$m_{\pi} + m_{K}$, 1.0] &
      $2.32 \times 10^{-3}$ & $2.16 \times 10^{-3}$
      \\
      \\
      \multicolumn{4}{c}{$D^{0}$}
      \\
      Mode & Range of $M_{\rm inv}^{(hh)}$ [GeV] & $l = e$ & $l = \mu$ \\
      \hline
      $\pi ^{-} \eta$ & [$m_{\pi} + m_{\eta}$, 1.1] &
      $4.93 \times 10^{-5}$ & $4.53 \times 10^{-5}$
      \\
      $K^{0} K^{-}$ & [$2 m_{K}$, 1.2] &
      $5.47 \times 10^{-6}$ & $4.88 \times 10^{-6}$
      \\
      $\pi ^{-} \bar{K}^{0}$ & [$m_{\pi} + m_{K}$, 1.0] &
      $8.99 \times 10^{-4}$ & $8.38 \times 10^{-4}$
      \\
    \end{tabular}
  \caption{Branching fractions of semileptonic $D$ decays into two
    pseudoscalar mesons in $s$ wave.  The branching fraction of
    the $D^{+} \to (\pi ^{+} K^{-})_{s\text{-wave}} e^{+} \nu _{e}$ mode
    is used as an input.\label{tab:G4}}
\end{table}

Integrating the mass distributions we calculate the branching
fractions of the semileptonic $D$ mesons into two pseudoscalar
mesons in $s$ wave, which are listed in Table~\ref{tab:G4}.  We note
that the branching fraction $\mathcal{B}[D^{+} \to (\pi ^{+}
K^{-})_{s\text{-wave}} e^{+} \nu _{e}]= 2.32 \times 10^{-3}$ is used
as an input to fix the common constant, $C = 4.6$.  Among the
listed values, we can compare the theoretical and experimental
values of the branching fraction $\mathcal{B}[D_{s}^{+} \to (K^{+}
K^{-})_{s\text{-wave}} e^{+} \nu _{e}]$.  Namely, in
Ref.~\refcite{Aubert:2008rs} this branching fraction is obtained as $(
0.22 ^{+0.12}_{-0.08} \pm 0.03) \%$ of the total $D_{s}^{+} \to
K^{+} K^{-} e^{+} \nu _{e}$, which is dominated by the $\phi (1020)$
vector meson.  Hence, together with the branching fraction
$D_{s}^{+} \to \phi (1020) e^{+} \nu _{e}$, we can estimate
$\mathcal{B}[D_{s}^{+} \to (K^{+} K^{-})_{s\text{-wave}} e^{+} \nu
_{e}] = (5.5^{+3.1}_{-2.1}) \times 10^{-5}$, and theoretically this is
$1.42 \times 10^{-4}$.  Although our value overestimates the mean
value of the experimental data, it is still in $3 \sigma$ errors of
the experimental value.

\subsubsection{Production of vector mesons} \label{sec:vector}

\begin{table}\centering
    \begin{tabular}{@{\extracolsep{\fill}}lcc}
      \multicolumn{3}{c}{$D_{s}^{+}$}
      \\
      Mode & $l = e$ & $l = \mu$ \\
      \hline
      $\phi (1020)$ &
      $2.12 \times 10^{-2}$  &  $1.94 \times 10^{-2}$
      \\
      $K^{\ast} (892)^{0}$ &
      $2.02 \times 10^{-3}$ & $1.89 \times 10^{-3}$
      \\
      \\
      \multicolumn{3}{c}{$D^{+}$}
      \\
      Mode & $l = e$ & $l = \mu$ \\
      \hline
      $\bar{K}^{\ast} (892)^{0}$ &
      $5.56 \times 10^{-2}$ & $5.12 \times 10^{-2}$
      \\
      $\rho (770)^{0}$ &
      $2.54 \times 10^{-3}$ & $2.37 \times 10^{-3}$
      \\
      $\omega (782)$ &
      $2.46 \times 10^{-3}$ & $2.29 \times 10^{-3}$
      \\
      \\
      \multicolumn{3}{c}{$D^{0}$}
      \\
      Mode & $l = e$ & $l = \mu$ \\
      \hline
      $K^{\ast} (892)^{-}$ &
      $2.15 \times 10^{-2}$ & $1.98 \times 10^{-2}$
      \\
      $\rho (770)^{-}$ &
      $1.97 \times 10^{-3}$ & $1.84 \times 10^{-3}$
      \\
    \end{tabular}
  \caption{Branching fractions of semileptonic $D$ decays into vector
    mesons.\label{tab:G3}}
\end{table}

\begin{figure}[!b]
  \centering
  \Psfig{8.6cm}{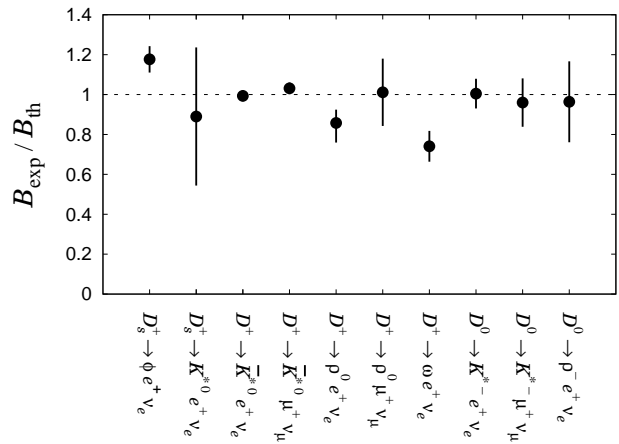}
  \caption{Ratio of the experimental to theoretical branching
    fractions for the semileptonic $D$ decays into vector mesons. }
\label{fig:vec}
\end{figure}

Let us now address the vector meson productions in the semileptonic $D$
decays.  For the vector mesons we fix the common prefactor
$C^{\prime}$ so as to reproduce the ten available experimental
branching fractions listed in Table~\ref{tab:Br}.  From the best fit
we obtain the value $C^{\prime} = 1.563 \gev$, which gives $\chi
^{2} / N_{\rm d.o.f.} = 22.8 / 9 \approx 2.53$.  The theoretical
values of the branching fractions are listed in Table~\ref{tab:G3}
and are compared with the experimental data in Fig.~\ref{fig:vec},
where we plot the ratio of the experimental to theoretical branching
fractions. We calculate the experimental branching fraction of the
$D^{+} \to \bar{K} (892)^{0} l^{+} \nu _{l}$ ($l = e$ and $\mu$)
process by dividing the value in Table~\ref{tab:Br} by the branching
fraction $\mathcal{B}[\bar{K}^{\ast} (892)^{0} \to K^{-} \pi ^{+}] =
2/3$, which is obtained with isospin symmetry.  As one can see from
Fig.~\ref{fig:vec}, the experimental values are reproduced well
solely by the model parameter $C^{\prime}$.

\begin{figure}[!t]
  \centering
  \Psfig{8.6cm}{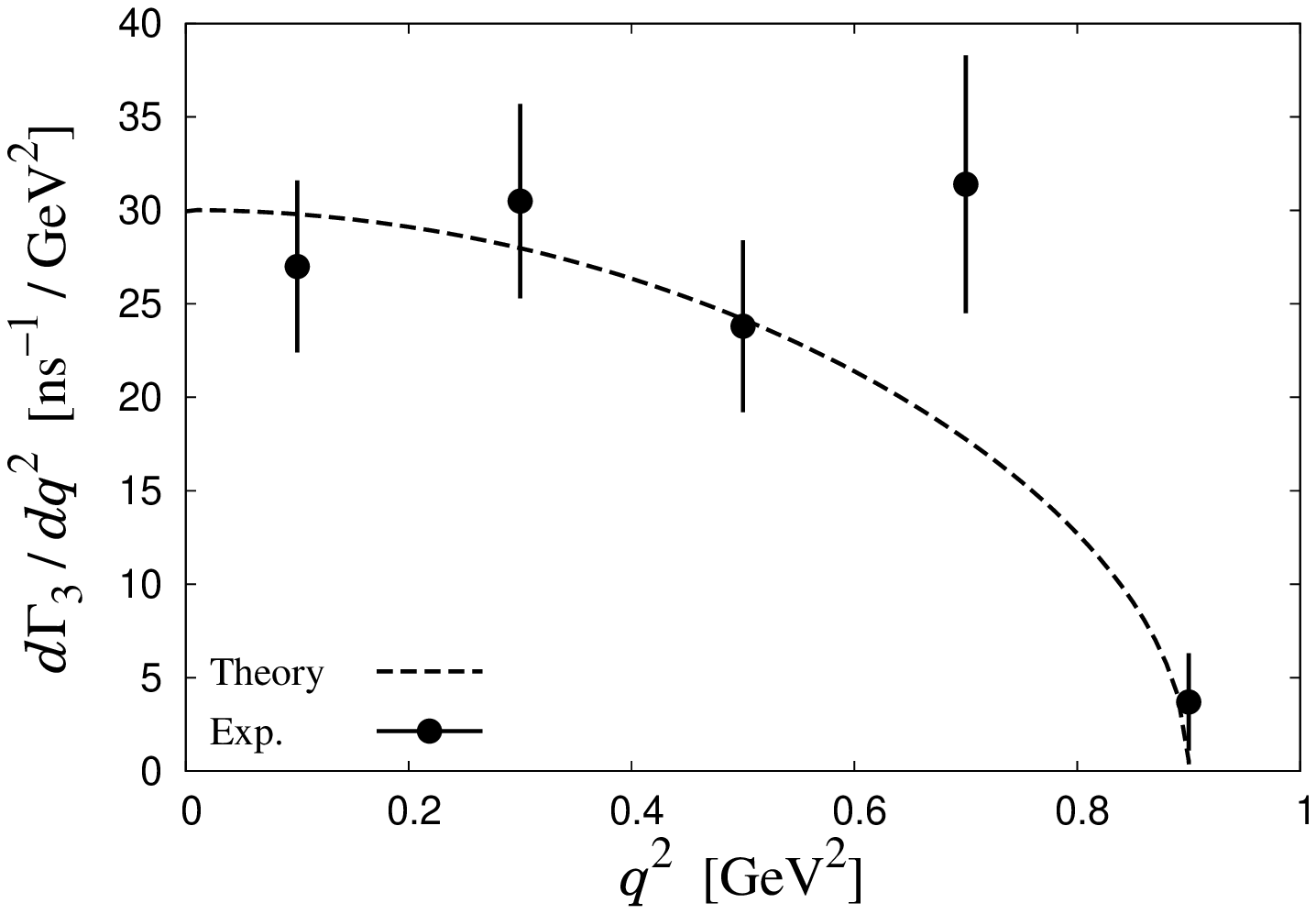}
  \caption{Differential decay width of the $D_{s}^{+} \to \phi (1020)
    e^{+} \nu _{e}$ decay mode followed by $\phi (1020) \to K^{+}
    K^{-}$, with $q^{2} = [M_{\rm inv}^{(l \nu )}]^{2}$, compared with experimental data\cite{Ecklund:2009aa}.  The
    theoretical value is multiplied by the branching fraction of the
    $\phi (1020)$ meson to $K^{+} K^{-}$, $\mathcal{B} [\phi (1020)
    \to K^{+} K^{-}] = 48.9 \%$~\cite{Agashe:2014kda}. }
\label{fig:dG3dq2}
\end{figure}

Next, for the $D_{s}^{+} \to \phi (1020) e^{+} \nu _{e}$ decay mode,
we consider the differential decay width with respect to the squared
momentum transfer $q^{2}$, which coincides with the squared
invariant mass of the lepton pair: $q^{2} = [M_{\rm inv}^{( l \nu
)}]^{2}$. This differential decay width was already measured in an
experiment~\cite{Ecklund:2009aa} for the $D_{s}^{+} \to \phi (1020)
e^{+} \nu _{e}$ decay mode. The differential decay width for the
vector meson production is expressed as\cite{Sekihara:2015iha}:
\begin{equation}
\frac{d \Gamma _{3}}{d q^{2}}
= \frac{\left | G_{\rm F} V_{\rm had}^{(v)} \right | ^{2}}
{16 \pi ^{3} m_{D}^{3} m_{V}}
P_{\rm cm} \tilde{p}_{\nu} M_{\rm inv}^{(l \nu )}
\left ( \tilde{E}_{D} \tilde{E}_{V} - \frac{1}{3} | \tilde{\bm{p}}_{D} |^{2}
\right ) .
\end{equation}
In Fig.~\ref{fig:dG3dq2} we compare our result for this reaction with
the experimental one.  As one can see, our theoretical result
reproduces the experimental value of the differential decay width
quantitatively well.  This means that our method to calculate the
semileptonic decays of $D$ mesons is good enough to describe the
decays into vector mesons.

\subsection{Comparison between scalar and vector meson contributions}
\label{sec:compare}

Finally we compare the mass distributions of the two pseudoscalar
mesons in $s$- and $p$-wave contributions.  In the present approach
the $s$-wave part comes from the rescattering of two pseudoscalar
mesons including the scalar meson contribution, while the $p$-wave appears in the decay of a vector meson.  In this study we consider three
decay modes: $D_{s}^{+} \to \pi ^{+} \pi ^{-} e^{+} \nu _{e}$,
$D_{s}^{+} \to K^{+} K^{-} e^{+} \nu _{e}$, and $D^{+} \to \pi ^{+}
K^{-} e^{+} \nu _{e}$.  The $D^{+} \to \pi ^{+} \pi ^{-} e^{+} \nu
_{e}$ decay mode would have a large $p$-wave contribution from $\rho
(770)$, but we do not consider this decay mode since it is a Cabibbo
suppressed process.

First we consider the $D_{s}^{+} \to \pi ^{+} \pi ^{-} e^{+} \nu _{e}$
decay mode.  This is a specially clean mode, since it does not have
vector meson contributions and is dominated by the $s$-wave part.
Namely, while the $\pi ^{+} \pi ^{-}$ can come from a scalar meson,
the primary quark--antiquark pair in the semileptonic $D_{s}^{+}$
decay is $s \bar{s}$, which is isospin $I = 0$ and hence the $\rho
(770)$ cannot contribute to the $\pi ^{+} \pi ^{-}$ mass distribution.
The primary $s \bar{s}$ can be $\phi (1020)$, but it decays dominantly
to $K \bar{K}$ and the $\phi (1020) \to \pi ^{+} \pi ^{-}$ decay is
negligible.  This fact enables us to observe the scalar meson peak
without a contamination from vector meson decays and discuss the quark
configuration in the $f_{0} (980)$ resonance as in
Sec.~\ref{sec:scalar}.

\begin{figure}[!t]
  \centering
  \Psfig{8.6cm}{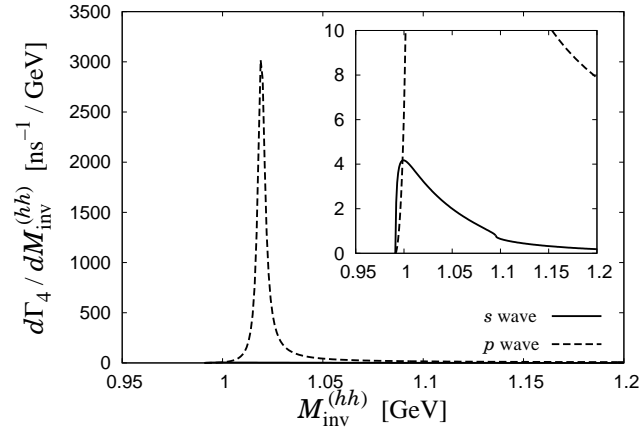}
  \caption{$K^{+} K^{-}$ invariant mass distribution for the
    semileptonic decay $D_{s}^{+} \to K^{+} K^{-} e^{+} \nu _{e}$ both
    in $s$ and $p$ waves.  }
\label{fig:KpKm}
\end{figure}

Next let us consider the $D_{s}^{+} \to K^{+} K^{-} e^{+} \nu _{e}$
decay mode.  As we have seen, the $K^{+} K^{-}$ mass distribution in
$s$ wave is a consequence of the $f_{0} (980)$ tail.  However, its
contribution should be largely contaminated by the $\phi (1020) \to
K^{+} K^{-}$ in $p$ wave, which has a larger branching fraction than
the $( K^{+} K^{-} )_{s\text{-wave}}$ in the semileptonic decay.  In
order to see this, we calculate the $p$-wave $K^{+} K^{-}$ mass
distribution for $D_{s}^{+} \to K^{+} K^{-} e^{+} \nu _{e}$, which can
be formulated as
\begin{equation}
\frac{d \Gamma_{3}}{d M_{\rm inv}^{(hh)}}
= - \frac{2 m_{V}}{\pi} \text{Im }
\frac{\Gamma_{3} \times \mathcal{B}[V \to h h]}{
[ M_{\rm inv}^{(hh)} ]^{2} - m_{V}^{2}
+ i m_{V} \Gamma _{V} ( M_{\rm inv}^{(hh)} ) } ,
\end{equation}
where $m_{V}$ is the vector meson mass and the energy dependent decay
width $\Gamma _{V} ( M_{\rm inv}^{(hh)})$ is defined as
\begin{equation}
\Gamma _{V} ( M_{\rm inv}^{(hh)}) \equiv
\bar{\Gamma}_{V}
\left ( \frac{p^{\rm off} ( M_{\rm inv}^{(hh)})}{p^{\rm on}} \right )^{3} ,
\end{equation}
\begin{equation}
p^{\rm off} ( M_{\rm inv}^{(hh)}) \equiv
\frac{\lambda ^{1/2}
( [ M_{\rm inv}^{(hh)}]^{2}, \, m_{h}^{2}, \, m_{h}^{\prime 2})}
{2 M_{\rm inv}^{(hh)}} ,
\end{equation}
\begin{equation}
p^{\rm on} \equiv
\frac{\lambda ^{1/2} ( m_{V}^{2}, \, m_{h}^{2}, \, m_{h}^{\prime 2})}
{2 m_{V}} .
\end{equation}
For the $\phi (1020)$ meson we take $\bar{\Gamma}_{\phi} = 4.27 \mev$
and $\mathcal{B}[\phi \to K^{+} K^{-}] = 0.489$~\cite{Agashe:2014kda}.
The numerical result for the $(K^{+} K^{-})_{p\text{-wave}}$ mass
distribution is shown in Fig.~\ref{fig:KpKm} together with the $(K^{+}
K^{-})_{s\text{-wave}}$.  From the figure, compared to the $(K^{+}
K^{-})_{p\text{-wave}}$ contribution we cannot find any significant
$(K^{+} K^{-})_{s\text{-wave}}$ contribution, which was already noted
in the experimental mass distribution in Ref.~\refcite{Aubert:2008rs}.
Nevertheless, we emphasize that the $(K^{+} K^{-})_{s\text{-wave}}$
fraction of the semileptonic $D_{s}^{+}$ decay is large enough to be
extracted~\cite{Aubert:2008rs}.  Actually in Ref.~\refcite{Aubert:2008rs}
they extract the $(K^{+} K^{-})_{s\text{-wave}}$ fraction by analyzing
the interference between the $s$- and $p$-wave contributions.  This
fact, and the qualitative reproduction of the branching fractions in
our model, implies that the $f_{0} (980)$ resonance couples to the $K
\bar{K}$ channel with a certain strength, which can be translated into
the $K \bar{K}$ component in $f_{0} (980)$, in a similar manner to the
$K D$ component in $D_{s 0}^{\ast} (2317)$, as discussed in sections \ref{sec:Ds02317_Alba} and \ref{sec:semileptonic}, in terms of the
compositeness~\cite{Sekihara:2014kya}. However, to be more assertive on
the structure of the $f_{0} (980)$, it is important to
reduce the experimental errors on the $(K^{+} K^{-})_{s\text{-wave}}$.

\begin{figure}[!t]
  \centering
  \Psfig{8.6cm}{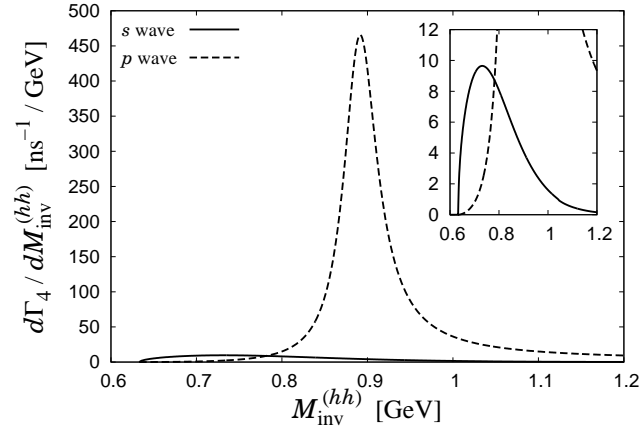}
  \caption{$\pi ^{+} K^{-}$ invariant mass distribution for the
    semileptonic decay $D^{+} \to \pi ^{+} K^{-} e^{+} \nu _{e}$ both
    in $s$ and $p$ waves.  }
\label{fig:ppKm}
\end{figure}

Finally we consider the $D^{+} \to \pi ^{+} K^{-} e^{+} \nu _{e}$
decay mode.  In this mode the $(\pi ^{+} K^{-})_{s\text{-wave}}$ from
the $K_{0}^{\ast}(800)$ and the $(\pi ^{+} K^{-})_{p\text{-wave}}$
from the $K^{\ast}(892)$ are competing with each other.  In a similar
manner to the $D_{s}^{+} \to K^{+} K^{-} e^{+} \nu _{e}$ case, we
calculate the mass distribution also for the $p$-wave $\pi ^{+} K^{-}$
contribution $d \Gamma _{3} / d M_{\rm inv}^{(hh)}$ with $\bar{\Gamma}
_{K^{\ast}} = 49.1 \mev$~\cite{Agashe:2014kda}, and the result is
shown in Fig.~\ref{fig:ppKm}.  As one can see, thanks to the width of
$\sim 50 \mev$ for the $K^{\ast} (892)$, the $s$-wave component can
dominate the mass distribution below $0.8 \gev$.  We note that one obtains essentially the result for the $D^{0} \to \pi ^{-}
\bar{K}^{0} e^{+} \nu _{e}$ decay mode due to isospin symmetry.

\section{Predictions for the $\Lambda_b \to J/\psi ~ \Lambda(1405)$ decay}\label{sec:LambdabJPsiLambda1405}

Through sections \ref{sec:LambdabJPsiLambda1405}--\ref{sec:WeakLambdaSSS} we report upon decays of $\Lambda_b$ and $\Lambda_c$
into different channels with the aim of learning about the $\Lambda(1405)$, $\Lambda(1670)$ and other resonances appearing in
the meson baryon interaction.

\subsection{Introduction}

The reason to suggest the measurement of the $\Lambda(1405)$ in
the  $\Lambda_b$ decay
is the relevance of the $\Lambda(1405)$ as the most significant example of a
dynamically generated resonance. Indeed, very early it was already suggested
that
this resonance should be a molecular state of $\bar K N$ and $\pi \Sigma$
\cite{Dalitz:1960du,Dalitz:1967fp}. This view has been also invoked in
Ref.~\refcite{Veit:1984an}. However, it was with the advent of chiral unitary theory that this idea gained
strength\cite{Kaiser:1995eg,Kaiser:1996js,Oset:1997it,Oller:2000fj,Lutz:2001yb,
Oset:2001cn,Hyodo:2002pk,Jido:2003cb,GarciaRecio:2002td,GarciaRecio:2005hy,Borasoy:2005ie,
Oller:2006jw,Borasoy:2006sr,Mai:2014xna}.

One of the surprises of these works is that two poles were found for
the $\Lambda(1405)$\footnote{In fact, one might thus speak of two $\Lambda(1405)$ particles. Indeed, in the next edition of the PDG, two distinct resonances will be officially catalogued.}. The existence of two states was hinted in Ref.~\refcite{Fink:1989uk},
using the chiral quark model, and it was found in Ref.~\refcite{Oller:2000fj} using
the chiral unitary approach. A thorough search was conducted in Ref.~\refcite{Jido:2003cb} by
looking at the breaking of SU(3) in a gradual way, confirming the existence of
these two poles and its dynamical origin. One of the consequences of this two-pole
structure is that the peak of the resonance does not always appear at the same energy,
but varies between 1420~MeV and 1480~MeV depending on the reaction used\cite{Thomas:1973uh,Hemingway:1984pz,Niiyama:2008rt,Prakhov:2004an,Moriya:2012zz,Moriya:2013eb, Zychor:2007gf,Agakishiev:2012xk}.
This is because different reactions give different weights to each of the
poles. While originally most reactions gave energies around 1400~MeV,
the origin of the nominal mass of the resonance,  the $K^- p \to \pi^0\pi^0 \Sigma^0$
was measured\cite{Prakhov:2004an} and a peak was observed around 1420~MeV,
narrower than the one observed in Refs.~\refcite{Thomas:1973uh,Hemingway:1984pz}, which
was interpreted within the chiral unitary approach in Ref.~\refcite{Magas:2005vu}. Another
illustrating experiment was the one of Ref.~\refcite{Braun:1977wd}  where  a clear
peak was observed around 1420~MeV in the $K^- d \to n \pi \Sigma$ reaction,
which was also interpreted theoretically in Ref.~\refcite{Jido:2009jf} along the same lines (see also Refs.~\refcite{Miyagawa:2012xz,Jido:2012cy}). Very recently it has
also been suggested that the  neutrino induced production
of the $\Lambda(1405)$ is a good tool to further investigate the properties and
nature of this resonance\cite{Ren:2015bsa}.

The basic feature in the dynamical generation of the $\Lambda(1405)$ in
the chiral unitary approach is the coupled channel unitary treatment of the
interaction between the coupled channels $K^-p$, $\bar{K}^0n$, $\pi^0\Lambda$,
$\pi^0\Sigma^0$, $\eta\Lambda$, $\eta\Sigma^0$, $\pi^+\Sigma^-$, $\pi^-\Sigma^+$, $K^+\Xi^-$
and $K^0\Xi^0$. The coupled channels study allows us to relate the $K^-p$ and $\pi \Sigma$
production, where the resonance is seen, and this is a unique feature of the nature of
this resonance as a dynamically generated state. It allows us to make
predictions for the $\Lambda(1405)$ production from the measured
$\Lambda_b \to J/\psi ~ K^-p$ decay.

\subsection{Formalism}\label{sec:formalismLambdab}

\begin{figure*}[t]
\begin{minipage}{0.48\linewidth}
 \includegraphics[width=\linewidth]{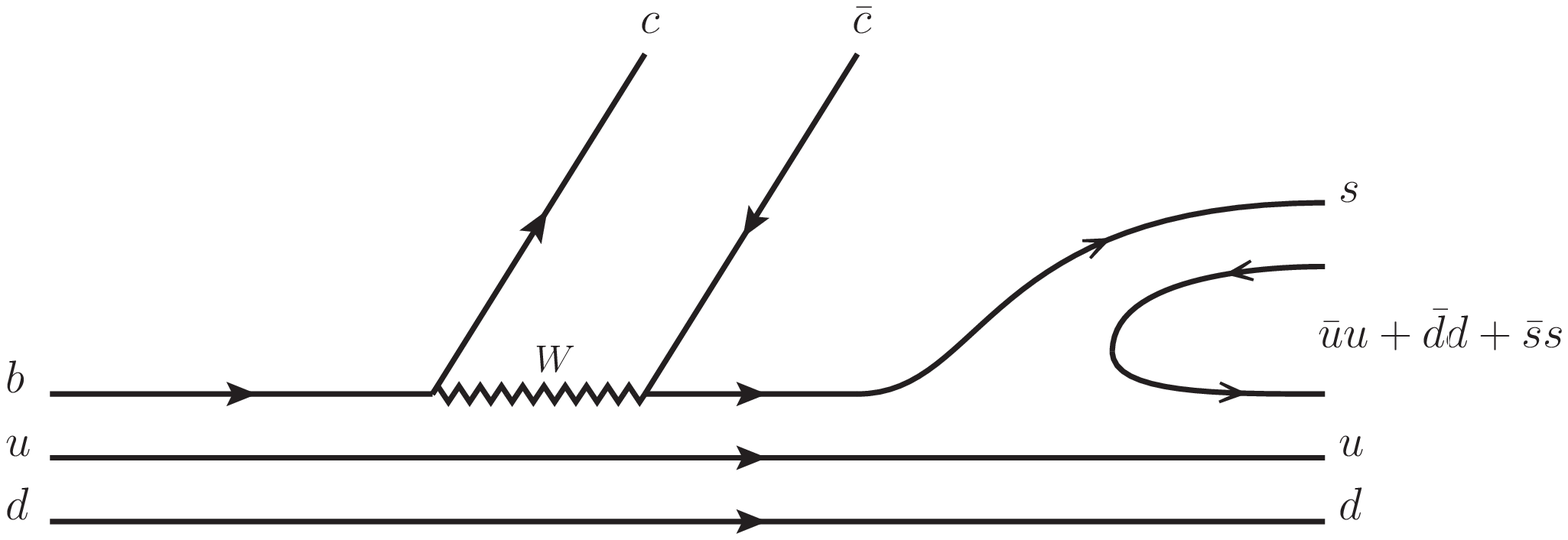}
$\hspace{0.05\linewidth}
\underbrace{\hspace{0.5\linewidth}}_{\text{Weak decay}}
\underbrace{\hspace{0.35\linewidth}}_{\text{Hadronization}}
\hspace{0.1\linewidth}$
\caption{Production of a $K^-p$ pair from the weak decay
${\Lambda_b\to\Lambda\,J/\psi}$
via a hadronization mechanism. The full and wiggly lines correspond to
quarks and the $W$-boson, respectively.}\label{fig:weak}
\end{minipage}
~~~
\begin{minipage}{0.48\linewidth}
\vspace{+1cm}
\includegraphics[width=\linewidth]{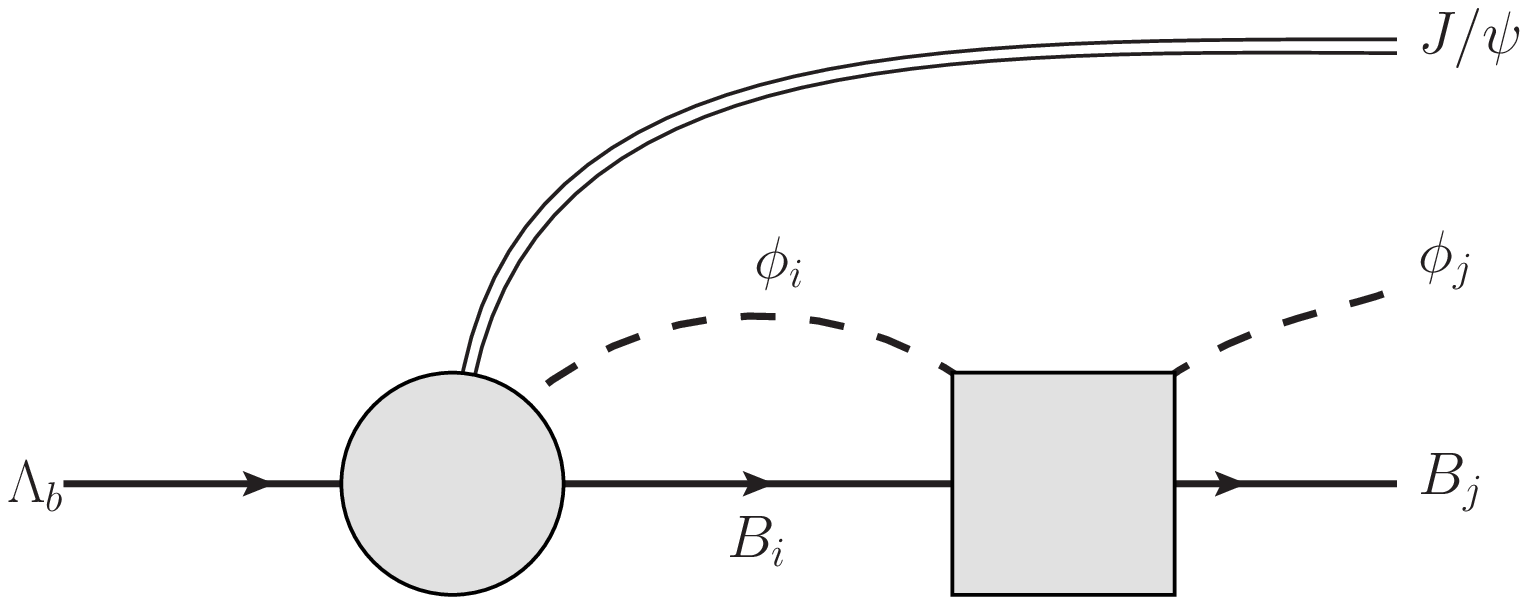}
\vspace{-0.1cm}
\caption{Final-state interaction of the meson-baryon pair, where the double,
full and dashed lines denote
the $J/\psi$, the baryons and the pseudoscalar mesons, respectively.
The circle and square denote the production mechanism of the $J/\psi B_i\phi_i$
as depicted in Fig.~\ref{fig:weak} as well as meson-baryon scattering
matrix $t_{ij}$, respectively}\label{fig:full}
\end{minipage}
\end{figure*}

In this section we describe the reaction mechanism for the process\cite{Roca:2015tea}
$\Lambda_b \to J/\psi\,\Lambda(1405)$, which is divided into three parts.
The first  two parts describe the decay mechanism $\Lambda_b\to J/\psi\,B\phi$,
with $B\phi$ the meson-baryon system of strangeness $S=-1$, in the language of
the quark model. Then, after hadronization,
the final-state interaction is described in terms of the
effective (hadronic) degrees of freedom of chiral perturbation theory (ChPT).
After a resummation of the chiral meson-baryon potential to an infinite order,
the $\Lambda(1405)$ is generated dynamically. In the following, we describe each
single step of this reaction mechanism in more detail.

\smallskip

\noindent\textbf{Weak decay:} The $b$ quark of the $\Lambda_b$ undergoes the
weak transition to a $c\bar c$ pair and an $s$-quark as depicted in the left
part of Fig.~\ref{fig:weak}. This transition is quantified by the matrix elements
of the CKM matrix $V_{cb}V_{cs}^*$ and it
is favored compared to $b\to c\bar c d$ leading to the
$\Lambda_b\to J/\psi p\pi^-$, which was observed for the first time by the
LHCb collaboration, see Ref.~\refcite{Aaij:2014zoa}.

\smallskip

\noindent\textbf{Hadronization:} The $c \bar c$ pair forms the well-known
$J/\psi$, while the virtual $uds$ three quark state  undergoes  hadronization
to form a meson-baryon pair. This happens due to the large phase space
available ($\leq 2522$~MeV for $M_{\Lambda_b}=5619$~MeV,
$M_{J/\psi}=3097$~MeV), so that a quark-antiquark pair can become real, forming
together with the three available quarks a meson-baryon pair.
In principle, different meson-baryon states can be produced in such a mechanism.
To determine their  relative significance, we assume first that the $u$ and $d$
quarks of the original $\Lambda_b$ state are moving independently in a potential well.
Further, we note that the $\Lambda_b$ ($J^p=1/2^+$)  is in the ground state of the
three-quarks $(udb)$. Therefore, all relative angular momenta
between different quarks are zero. After the weak transition, but before the
hadronization, the three-quark state $(uds)$ has to be in a p-wave since the
final $\Lambda(1405)$ is a negative-parity
state. On the other hand, since the $u$ and $d$ quarks are considered to
be spectators and they were originally in $L=0$, the only possibility is that
the $s$ quark carries the angular momentum, $L=1$. Moreover, since the final
mesons and baryons are in the ground state and in s-wave to each other,
all the angular momenta in the final state are zero. Consequently, the $\bar q q$
pair cannot be produced elsewhere, but between the $s$ quark and the $ud$ pair
as depicted in Fig.~\ref{fig:weak}. There are other possibilities to hadronize in which one of the original $u,d$ quarks goes into a meson and the $s$ quark into a baryon, followed by rescattering. However, these mechanisms are discussed in the next section and are suppressed due to large momentum transfers to the $u$ or $d$ quarks.

The flavour state of the initial $\Lambda_b$ can be written as
\begin{align*}
|\Lambda_b\rangle=\frac{1}{\sqrt{2}}|b(ud-du)\rangle\,,
\end{align*}
turning after the weak process into
\begin{align*}
\frac{1}{\sqrt{2}}|s(ud-du)\rangle\,,
\end{align*}
since the $u$ and $d$ quarks are considered to be spectators. Thus, after
hadronization, the  final quark flavor state is
\begin{align*}
|H\rangle&\equiv \frac{1}{\sqrt{2}}|s\,(\bar u u +\bar d d +\bar s
s)\,(ud-du)\rangle =\frac{1}{\sqrt{2}}
\sum_{i=1}^3{|P_{3i}q_i(ud-du)}\rangle\,,
\end{align*}
where we have defined
\begin{equation*}
q\equiv \left(\begin{array}{c}u\\d\\s\end{array}\right)~,
\end{equation*}
and $P$ denotes here the $M$ matrix defined in Eq.~\eqref{eq:qqbarmatrix}. We recall that it is in correspondence with the pseudoscalar meson matrix $\phi$ defined in Eq.~\eqref{eq:phimatrix}. The hadronized state $|H\rangle$ can now be written as
\begin{align*}
|H\rangle= \frac{1}{\sqrt{2}} \bigg( K^- u(ud&-du)+\bar K^0 d(ud-du)
+\frac{1}{\sqrt{3}}\left(-\eta+\sqrt{2}\eta'\right)s(ud-du)\bigg).
\end{align*}
We can see that these states have overlap with the mixed antisymmetric baryon octet states
\cite{Close:1979bt}:
octet baryons can be written as
\begin{align*}
|p      \rangle=&\frac{1}{\sqrt{2}}|u(ud-du)\rangle\,,\\
|n      \rangle=&\frac{1}{\sqrt{2}}|d(ud-du)\rangle\,,\\
|\Lambda\rangle=&\frac{1}{\sqrt{12}}| (usd-dsu)+(dus-uds) +2(sud-sdu)\rangle\,.
\end{align*}
Consequently, the hadronized state can be expressed in terms of the ground state
meson and baryon octets as
\begin{equation}
|H\rangle=|K^-p\rangle+|\bar K^0 n\rangle
-\frac{\sqrt{2}}{3}|\eta\Lambda\rangle+\frac{2}{3}|\eta'\Lambda\rangle\,,
\label{eq:Hflav}
\end{equation}
which provides the relative weights between the final meson-baryon channels.
Note that there is not direct production of $\pi\Sigma$ and $K\Xi$, however,
these channels are present through the intermediate loops in the final state
interaction as described below. Moreover, the final $\eta'\Lambda$ channel
will be neglected since it has a small effect due to its high mass and can be
effectively reabsorbed in the regularization parameters as will be
explained below.

\smallskip

\noindent\textbf{Formation of the {\boldmath$\Lambda(1405)$}:} After the production of
a meson-baryon pair, the final-state interaction takes place, which is
parametrized by the scattering matrix
$t_{ij}$. Thus, after absorbing the CKM matrix elements and kinematic prefactors into an
overall factor $V_p$, the amplitude $\mathcal{M}_{j}$ for the transition
$\Lambda_b\to J/\psi\,\phi_jB_j$ can be written as
\begin{align}\label{eqn:fullamplitude}
\mathcal{M}_{j}(M_{\rm inv})=V_p\left( h_j+\sum_{i}h_iG_i(M_{\rm inv})\,t_{ij}(M_{\rm inv}) \right)\,,
\end{align}
where, considering Eq.~(\ref{eq:Hflav}),
\begin{align}
&h_{\pi^0\Sigma^0}=h_{\pi^+\Sigma^-}=h_{\pi^-\Sigma^+}=0\,,~h_{\eta\Lambda}=
-\frac{\sqrt{2}}{3}~,\nonumber \\
&h_{K^-p}=h_{\bar K^0n}=1\,,~h_{K^+\Xi^-}=h_{K^0\Xi^0}=0~, \label{eq:juansaysa}
\end{align}
and $G_i$ denotes the one-meson-one-baryon loop function, chosen in
accordance with the models for the scattering matrix\footnote{More precisely,
$t_{ij}$ denotes the $s$-wave contribution to the scattering matrix.} $t_{ij}$. Further, $M_{\rm inv}$
is the invariant mass of the meson-baryon system in the final state.
Note also that the above
amplitude holds for an $s$-wave only and every intermediate particle is put on the
corresponding mass shell. Finally, the invariant mass distribution $\Lambda_b\to
J/\psi\,\phi_jB_j$ reads
\begin{align}\label{eqn:dGammadM}
\frac{d\Gamma_j}{dM_{\rm inv}}(M_{\rm inv})
=\frac{1}{(2\pi)^3}\frac{m_j}{M_{\Lambda_b}}{\rm\textbf{p}}_{J/\psi} {\rm\textbf{p}}_j\left|\mathcal{M}_{j}(M_{\rm inv})\right|^2\,,
\end{align}
where $\rm\textbf{p}_{J/\psi}$ and $\rm\textbf{p}_j$ denote the modulus of the
three-momentum
of the $J/\psi$ in the $\Lambda_b$ rest-frame and the modulus of the center-of-mass
three-momentum in the final meson-baryon system, respectively. The mass of the
final baryon is denoted by $m_j$.

As already described in the introduction, the baryonic $J^P=1/2^-$ resonance
$\Lambda(1405)$ has to be understood as a dynamically generated state from the coupled-channel effects. The
modern approach for it is referred to as chiral unitary models\cite{Kaiser:1995eg,Kaiser:1996js,Oset:1997it,Oller:2000fj,Lutz:2001yb,Oset:2001cn,Hyodo:2002pk,Jido:2003cb,GarciaRecio:2002td,GarciaRecio:2005hy,Borasoy:2005ie,Oller:2006jw,Borasoy:2006sr,Ikeda:2012au,Mai:2012dt}. In the present approach we use the scattering
amplitude from two very recent versions of such approaches, one from Ref.~\refcite{Mai:2014xna}, that we call Bonn model and the other from Refs.~\refcite{Roca:2013av,Roca:2013cca}, which we call MV model. While the basic motivation
is the same for both approaches there are some differences, such as the order of truncation the underlying chiral potential as described in Ref.~\refcite{Roca:2015tea}.

\subsection{Results}\label{sec:resultsLambdab}


After having set up the framework, we present here the predictions for the
$\pi\Sigma$ and $\bar K N$ invariant mass distributions from the $\Lambda_b$ decay.

\begin{figure*}[t]\centering
\includegraphics[width=0.85\linewidth]{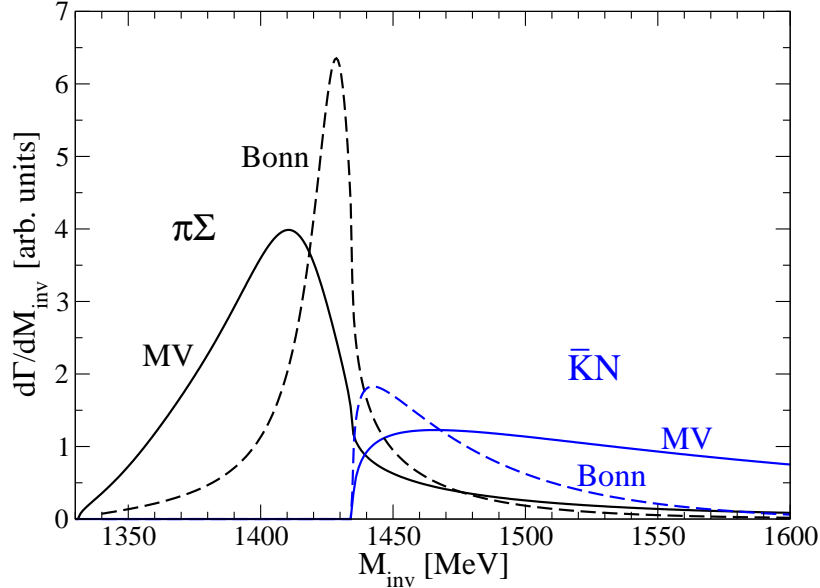}
\caption{Results for the $\pi\Sigma$ and $\bar KN$ invariant mass distributions
for the $\Lambda_b \to J/\psi\,\pi\Sigma$ and $\Lambda_b \to J/\psi\,\bar KN$ decays,
respectively, for both models considered in the present work.}\label{fig:fig1}
\end{figure*}

In Fig.~\ref{fig:fig1} we show the final results for both the Bonn and MV models.
In the $\pi\Sigma$ final state channel the peak of the $\Lambda(1405)$ is  clearly
visible.
In fact, this is mostly due to the higher mass pole of  the $\Lambda(1405)$ since the
contribution proportional to $t_{\bar K N,\pi\Sigma}$  of Eq.~(\ref{eqn:fullamplitude}) is
the dominant one.  The difference in the $\pi\Sigma$ mass distribution between both
models is
reminiscent of the fact that the Bonn model gets a narrower
($24\mev$)
highest $\Lambda(1405)$ pole than the MV model ($58\mev$).

In the $\bar K N$ final state,  the dominant contribution comes from the
part proportional to $t_{\bar K N,\bar K N}$  which again is more sensitive to the higher
mass $\Lambda(1405)$ pole. However,  in this channel only the effect of the tail of the
resonance is visible since  the threshold of the $\bar K N$ mass distribution is located
above the position  of the $\Lambda(1405)$ peak. Nevertheless, that tail is enough to
provide  a high strength close to the threshold, what makes the line shape of the  $\bar K N$
invariant mass distribution  to be very different from just  a phase-space
distribution. The
dependence on the choice of the model in this channel is due to the fact that
 the highest pole is
slightly closer to threshold in the  Bonn model compared to the MV one. Because of this
feature, the Bonn model produces  a narrower bump close to $\bar K N$ invariant mass
threshold than the MV one.  This observable is then very sensitive to the exact
position of
the resonance  pole, due to the proximity between the threshold and the pole. As
mentioned in  the introduction, different reactions can reflect different weights
for both  poles of
the $\Lambda(1405)$ resonance, depending on the particular production  dynamics. In the
present case, the highest pole is the one that shows up dominantly.

On the other hand, the agreement in Fig.~\ref{fig:fig1} of the results
between the MV and Bonn models is remarkable, given their theoretical
differences and fitting strategies used in them. Nonetheless we can
regard the difference between the models as the main source of the
theoretical uncertainty.

While the overall normalization of the invariant mass distributions is unknown,
the shape and the
ratio between the  $\pi\Sigma$ and $\bar K N$ distributions is
unchanged and it is a genuine prediction.
 Indeed, the ratio between the maximum values of
the  $\pi\Sigma$ and $\bar K N$ distribution  is 3.3 for the MV and 3.5 for
the Bonn model.
The value of that ratio as well as the shape of the distributions are then genuine
predictions of the chiral unitary approach. In conclusion, Fig.~\ref{fig:fig1} serves to predict the  invariant mass
distributions of either $\pi\Sigma$ or $\bar K N$,  once the absolute
normalization of the mass distribution of the other channel has been measured.
For instance, if the LHCb\cite{Aaij:2014zoa} and CDF\cite{Aaltonen:2014vra}
collaboration were  to measure the $K^-p$ mass distribution in the
$\Lambda_b \to J/\psi ~ K^- p$ decay, then the shape should agree with the
prediction of Fig.~\ref{fig:fig1} and once normalized,
the $\bar K N$ and $\pi\Sigma$ distributions would be given both in size and shape.

\section{The $\Lambda_b \to J/\psi ~ K ~ \Xi$ decay and the higher order chiral terms of the meson baryon interaction.}\label{sec:Lb_highchiral}

This work is complementary to the one shown in the former subsection.

\subsection{Formalism}\label{sec:formalismLambdab2}

\subsubsection{The $\Lambda_b\to J/\psi~M~B$ process}
\label{subsec:weak}

At the quark level, the
Cabibbo favored mechanism for $J/\psi$ production in $\Lambda_b$ decay is depicted by the
first part of the diagram of Fig. \ref{fig:weak}. This corresponds
to internal emission in the classification of topologies of Ref.~\refcite{Chau:1982da}, and is also the dominant mechanism in the
related $\bar B^0 \to J/\psi~ \pi~ \pi$ decay\cite{Stone:2013eaa}. As we can see in the figure, a $sud$
state is obtained after the weak decay. The next step consists in
the hadronization of this final three quark state by introducing a
$\bar q q$ pair with the quantum numbers of the vacuum, $\bar u u+
\bar d d+ \bar s s$, which will then produce an initial meson-baryon
pair. As in the former section, in this reaction the $u$ and $d$
quarks act as spectators. This means that the $ud$ pair in the final
$sud$ state after the weak decay has $I=0$ and, since the $s$ quark
also has $I=0$, the final three-quark system has total $I=0$. Hence,
even if the weak interaction allows for isospin violation, in this
case the process has filtered $I=0$ in the final state. Since
isospin is conserved in the strong hadronization process and in the
subsequent final state rescattering interaction, the final
meson-baryon component also appears in $I=0$.

As already discussed in Sec.~\ref{sec:LambdabJPsiLambda1405}, another observation concerning the hadronization is that, since the
$sud$ quark state after the weak decay has $J^P=1/2^-$ and the $ud$
quarks have the same quantum numbers as in the original $\Lambda_b$
state ($J^P=1/2^+$ each) in an independent quark model used  for the
argumentation, it is the $s$ quark the one that must carry the minus
parity, which would correspond to an $L=1$ orbit of a potential
well. Since the final meson-baryon states are all in their ground state, the $s$ quark must de-excite and hence it must participate in the hadronization. This latter process gives rise to some meson-bayon states with the weight given earlier in Eqs.~\eqref{eq:juansaysa}. As usual in these studies, we neglect the $\eta' \Lambda$ component, and we only have primary $K^-p$, $K^0n$ or $\eta\Lambda$ production. We can see that a $K \Xi$ pair is not produced in the first step.

Next, one must incorporate the final state interaction of these meson-baryon pairs, which is depicted in Fig. \ref{fig:full}. The matrix element for the production of the final state, $j$, is given by Eq.~\eqref{eqn:fullamplitude}. The factor $V_p$, which includes the common dynamics of the production of the different pairs, is unknown and we take it as constant. Finally, the invariant mass distribution $\Lambda_b\to J/\psi\,\phi_j\, B_j$ is given by Eq.~\eqref{eqn:dGammadM}.

\subsection{Results}\label{sec:resultsLambdab2}

\begin{figure}[!htb]
\centering
\includegraphics[height=9cm]{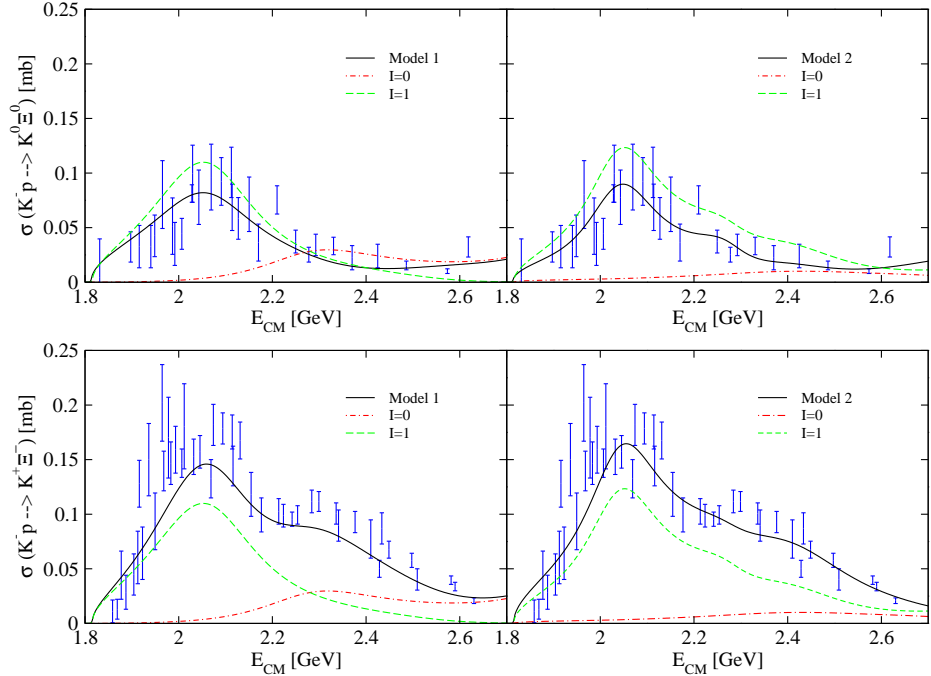}
\caption{(Color online) The total cross sections of the $K^- p\to K^0 \Xi^0$ reaction (top row) and the $K^- p\to K^- \Xi^+$ reaction (bottom row) for the two different models (Model 1 and Model 2) discussed in the text. The solid lines show the results of the full amplitude, while the dashed  and dash-dotted lines denote the $I=1$ and $I=0$ contributions, respectively. Theoretical values are compared with experimental data\cite{exp_1,exp_2,exp_3,exp_4,exp_5,exp_6,exp_7}.}
  \label{fig:iso}
\end{figure}
We start this section by presenting in Fig.~\ref{fig:iso} the cross section data of the  $K^- p\to K^0 \Xi^0$ reaction (top panels) and of the $K^- p\to K^- \Xi^+$ reaction (bottom panels), obtained employing Model 1 (left panels) or Model 2 (right panels) \cite{Feijoo:2015yja}. The figure shows the complete results (solid lines), as well as the results where only the isospin $I=1$ component (dashed lines) or the $I=0$ one (dash-dotted lines) have been retained. It is interesting to see that, in both models, the $I=1$ component is dominant and is concentrated at lower energies. The smaller $I=0$ component at higher energies adds up destructively to the cross section in the case of the $K^- p\to K^0 \Xi^0$ reaction, while it contributes to enhance the cross section in the $K^- p\to K^- \Xi^+$ process. We note that the tree-level chiral contributions to these reactions come entirely from the NLO Lagrangian and, upon inspecting the size of the coefficients, their strength in the $I=0$ channel would be nine times larger than that in the $I=1$ channel. The reversed trend observed in Fig.~\ref{fig:iso} is a consequence of the unitarization in coupled channels with coupling coefficients determined by the fit and, consequently, by the data.

As we can see in Fig.~\ref{fig:iso}, the contribution of $I=0$ in
the $K^-p \to K\Xi$ cross section has a maximum around 2300~MeV for
Model 1 or around 2400~MeV and less pronounced for Model 2, far from
the peak of the data and of the complete amplitude, around 2050~MeV.
The $K^- p\to K \Xi$ reactions contain a mixture of both isospin
components, while the decay process $\Lambda_b \to J/\psi ~ K ~
\Xi$, studied in this paper, filters $I=0$ and therefore provides
additional information to the one obtained from the scattering data.

\begin{figure}[!htb]
\centering
  \includegraphics[scale=0.35]{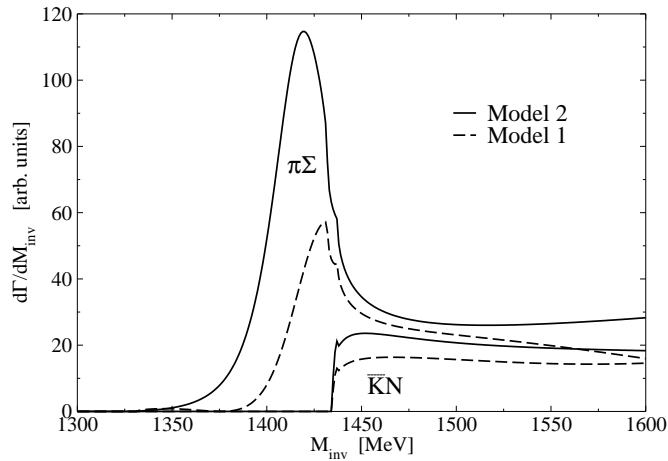}
\caption{Invariant mass distributions of $\pi\Sigma$ and $\bar{K} N$ states in the decay modes $\Lambda_b \to J/\psi ~ \pi ~\Sigma$ and  $\Lambda_b \to J/\psi ~ \bar{K} ~N$, for the two models discussed in the text: Model 1 (dashed lines) and Model 2 (solid lines). The units in the $y$ axis are obtained taking $V_p=1$. }
  \label{fig:KbarN}
\end{figure}
Since the models of Ref.~\refcite{Feijoo:2015yja} make a fitting to all $K^-p \to X$ data in a range from threshold to $K\Xi$ production, we start presenting, in Fig.~\ref{fig:KbarN},  what are the predictions of Model 1 and Model 2 for the decay reactions  $\Lambda_b \to J/\psi ~\bar{K} ~ N$ and $\Lambda_b \to J/\psi ~\pi ~ \Sigma$, already studied in the former section.  These are averaged distributions over the possible different charged states.
We can see that the results of both models are similar to those found in Sec.~\ref{sec:LambdabJPsiLambda1405}, with the shape of the $\pi \Sigma$ and $\bar K N$ distributions lying somewhat in between those of the Bonn and Murcia-Valencia models studied there (a different normalization is used in that work). We note that our $\pi \Sigma$ distributions shown in Fig.~\ref{fig:KbarN} stay over the ${\bar K}N$ ones, in contrast to what one observes in Fig.~\ref{fig:fig1}, where the $\pi\Sigma$ distributions cross below the respective ${\bar K}N$ ones just above the threshold for ${\bar K}N$ states.  This is a peculiarity of the NLO contributions, since one also obtains a crossing behavior when the interaction models are restricted to only the lowest order terms.  It is also interesting to see that the numerical results in Fig.~\ref{fig:KbarN} depend on the model, indicating their sensitivity on different parametrizations that fit equally well the  $K^-p \to X$ data. We obtain ratios of 4.9 and 3.5 for Models 1 and 2, respectively. These values are of the order of those found in the former section.

\begin{figure}[!htb]
\centering
\includegraphics[scale=0.35]{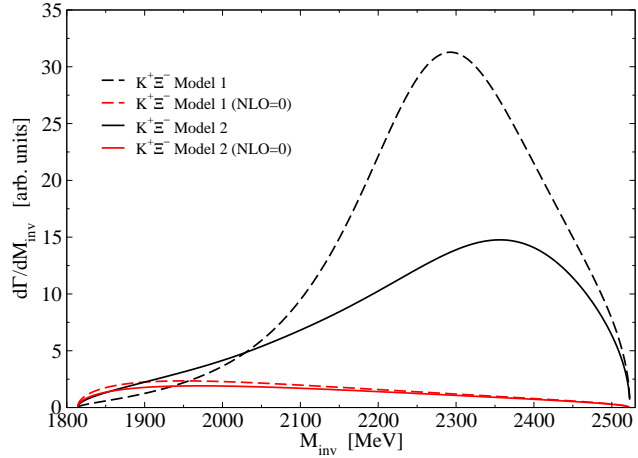}
\caption{(Color online) Invariant mass distributions of $K^+\Xi^-$
states produced in the decay $\Lambda_b \to J/\psi ~ K^+~\Xi^-$,
obtained for the two models discussed in the text: Model 1 (dashed
lines) and Model 2 (solid lines). The thin lower lines correspond to
omitting the NLO terms of the potential. The normalization is the
same as in Fig.~\ref{fig:KbarN}.}
  \label{fig:KXi}
\end{figure}

In Fig.~\ref{fig:KXi} we present the invariant mass distributions of the $K^+\Xi^-$ states from the decay process $\Lambda_b \to J/\psi ~ K^+ ~\Xi^-$. We do not show the distribution for the decay process $\Lambda_b \to J/\psi ~ K^0~ \Xi^0$, because, except for minor differences associated to the slightly different physical masses of the particles, it is identical to that of the charged channel, since these processes involve only the $I=0$ part of the strong meson-baryon amplitude. The fact that this decay filters the $I=0$ components makes the differences between Model 1 (thick dashed line) and Model 2 (thick solid line) to be more evident, not only in the strength but also in the shape of the invariant mass distribution. If, in order to eliminate the dependence on undetermined loop functions and on the unknown weak parameter $V_p$, we represented each $\Lambda_b \to J/\psi ~ K^+ ~\Xi^-$ distribution relative to its corresponding $\Lambda_b \to J/\psi ~ {\bar K} ~ N$ one shown in Fig.~\ref{fig:KbarN}, the difference would even be somewhat enhanced.
 Therefore, measuring the decay of the $\Lambda_b$ into $J/\psi ~ K^+~\Xi^-$ and into  J$/\psi ~ {\bar K} ~ N$ could help in discriminating between models that give a similar account of the scattering $K^- p \to K^0 \Xi^0, K^+ \Xi^-$ processes.  The figure also shows that the $I=0$ structure observed around 2300~MeV results from the terms of the NLO Lagrangian. When they are set to zero, the invariant mass distributions of the two models, shown by the thin dashed and thin solid lines in Fig.~\ref{fig:KXi}, become small and structureless.

\begin{figure}[!htb]
\centering
  \includegraphics[scale=0.35]{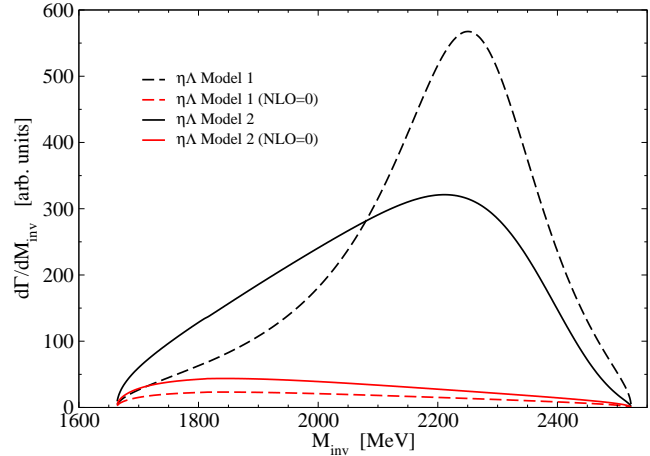}
\caption{(Color online) Invariant mass distributions of $\eta \Lambda$ states produced in the decay $\Lambda_b \to J/\psi ~ \eta ~\Lambda$, obtained for the two models discussed in the text: Model 1 (dashed lines) and Model 2 (solid lines). The thin lower lines correspond to omitting the NLO terms of the potential. The normalization is the same as in Fig.~\ref{fig:KbarN}.}
  \label{fig:etaL}
\end{figure}

We have observed a similar behaviour in the mass distributions of the reaction  $\Lambda_b \to J/\psi ~\eta ~\Lambda$ which are shown in Fig.~\ref{fig:etaL}. In this case, as the coefficient $h_{\eta\Lambda}$ does not vanish, we see from Eq.~(\ref{eqn:fullamplitude}) that the tree level term also contributes here, unlike the case of $K\Xi$ production. This makes the magnitude of the $\Lambda_b \to J/\psi ~\eta ~\Lambda$ mass distribution about twenty times bigger than that of the $\Lambda_b \to J/\psi ~ K ~ \Xi$ one.

\begin{figure}[!htb]
\centering
  \includegraphics[scale=0.6]{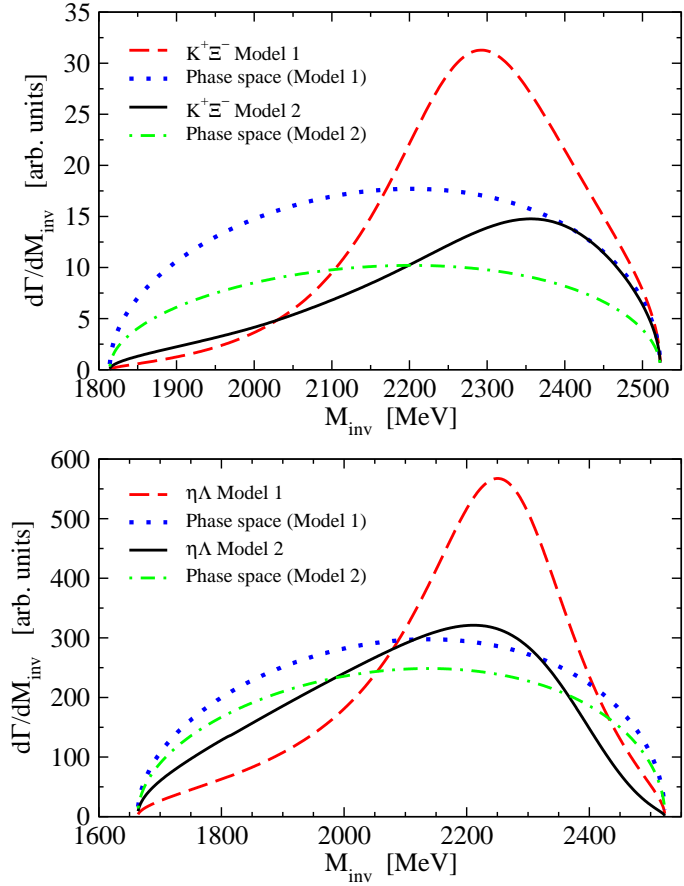}
\caption{(Color online) Comparison of the invariant mass
distributions of $K^+\Xi^-$ states (upper panel) and $\eta \Lambda$
states (lower panel) states obtained with Model 1 (dashed lines) and
Model 2 (solid lines) with a pure phase-space distribution (dotted
lines)}
  \label{fig:PS}
\end{figure}

The invariant mass distributions from the $\Lambda_b \to J/\psi ~ K^+~\Xi^-$ and $\Lambda_b \to J/\psi ~ \eta~\Lambda$ decays obtained in Models 1 and 2 are compared with phase space in Fig.~\ref{fig:PS}. The phase-space distributions (dotted lines for  Model 1 and dash-dotted lines for Model 2) are obtained by taking the amplitude $\mathcal{M}_j$ as constant in Eq.~(\ref{eqn:dGammadM}) and normalizing to the area of the invariant mass distribution of the corresponding model. The comparison allows one to see that there are dynamical features in the meson-baryon amplitudes leading to a distinct shape of the mass distributions. In the case of Model 1, we observe a peak between  2250~MeV and 2300~MeV for both $\Lambda_b \to J/\psi ~ K^+~\Xi^-$ and $\Lambda_b \to J/\psi ~ \eta~\Lambda$ distributions. The peak resembles a resonance, but we should take into account that the limitation of the phase space at about 2500~MeV produces a narrower structure than that of the cross sections of the $K^-p \to K \Xi$ reactions, as we can see from the $I=0$ contribution in Fig.~\ref{fig:iso} (left panels), which is much broader.  Actually, the $I=0$ contribution of Model 2 to the cross sections of Fig.~\ref{fig:iso} (right panels)  does not indicate any particular structure, and the very different shapes that this model predicts for  $\Lambda_b \to J/\psi ~ K^+~\Xi^-$ and $\Lambda_b \to J/\psi ~ \eta~ \Lambda$ (see the thick solid lines in Fig.~\ref{fig:PS}), peaking at about 2400~MeV and 2200~MeV respectively, do not indicate the presence of a resonance since it would necessarily appear in both final states at the same energy. In our models, it is the energy dependence in the parametrization of the next-to-leading order contribution and the interference of terms what creates this shape. In any case, what is clear is that the experimental implementation of this reaction will provide valuable information concerning the meson-baryon interaction at higher energies, beyond what present data of scattering has offered us.

Although we have given the invariant mass distributions in arbitrary units, one should bear in mind that all the figures, from Fig.~\ref{fig:KbarN} to Fig.~\ref{fig:etaL} have the same normalization. Since measurements for the $\Lambda_b \to J/\psi ~ K^- ~ p$ reaction are already available from the CDF\cite{Aaltonen:2010pj} and LHCb\cite{Aaij:2013oha, Aaij:2014zoa,Aaij:2015tga} collaborations, the measurements of the reactions proposed here could be referred to those of the $\Lambda_b \to J/\psi ~ K^-~ p$ reaction and this would allow a direct comparison with our predictions. In this spirit, we note that the recent resonance analysis of Ref.~\refcite{Aaij:2015tga} shows a $\Lambda(1405)$ contribution which lies in between the distribution found by the Bonn model in Ref.~\refcite{Roca:2015tea} and that of the Murcia-Valencia model in Ref.~\refcite{Roca:2015tea} or the Barcelona models presented here. Further details and discussions on the reaction and the results can be seen in Ref.~\refcite{Feijoo:2015cca}.

\section{Weak decay of $\Lambda_{c}^+$ for the study of $\Lambda(1405)$ and $\Lambda(1670)$}\label{sec:WeakLambdaSSS}
\subsection{Formulation}  \label{sec:formulation}  

\begin{figure}[tb]\centering
\includegraphics[scale=0.4]{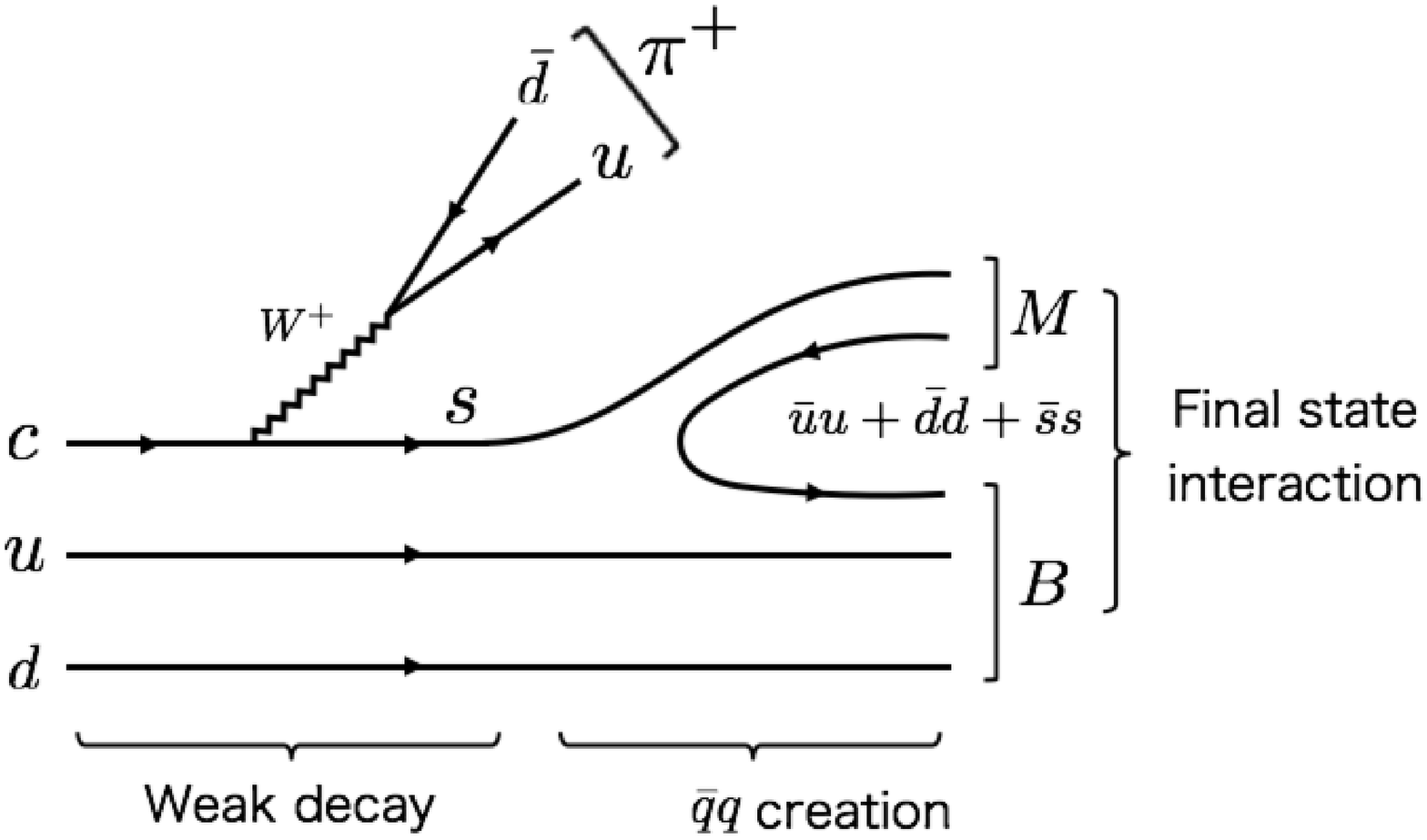}
\caption{The dominant diagram for the $\Lambda_c^+\to \pi^+MB$ decay. The solid lines and the wiggly line show the quarks and the $W$-boson, respectively.\label{fig:Lambdac_decay}}
\end{figure}

We consider the decay process $\Lambda_c^+\to\pi^+\Lambda^*\to\pi^+MB$, where $MB$ stands for the final meson-baryon states such as $\pi\Sigma$ and $\bar{K}N$. We show that, when the $MB$ invariant mass is restricted in the $\Lambda(1405)$ region, the dominant contribution of this decay process is given by the diagram shown in Fig.~\ref{fig:Lambdac_decay}.
\begin{figure}[tb]\centering
\includegraphics[scale=0.33]{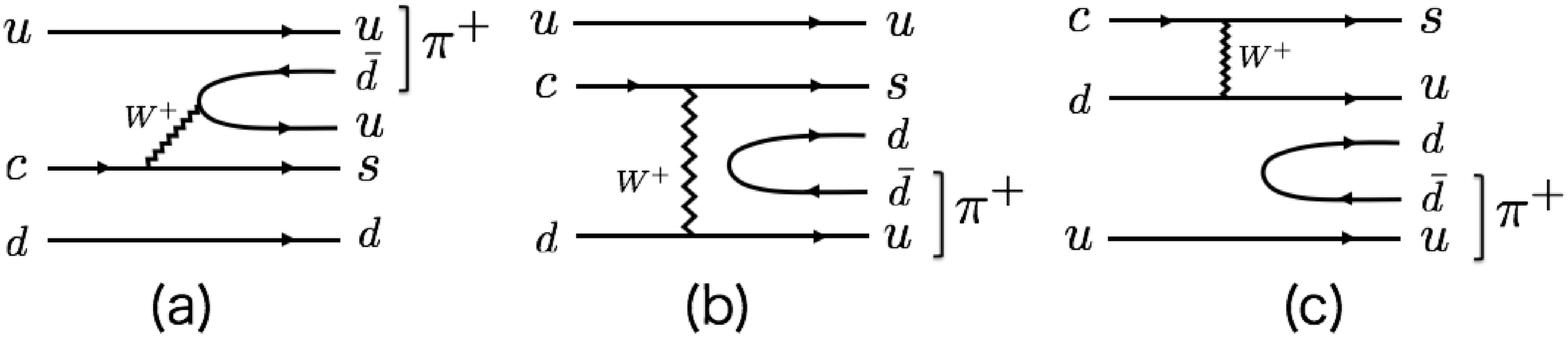}
\caption{The dominant diagram for the $\Lambda_c^+\to \pi^+MB$ decay. The solid lines and the wiggly line show the quarks and the $W$-boson, respectively.\label{fig:other_decay}}
\end{figure}
First, the charm quark in $\Lambda_c^+$ turns into the strange quark with the $\pi^+$ emission by the weak decay. Second, the $\bar{q}q$ creation occurs to form $M$ ($B$) from the $s$ quark ($ud$ diquark). Finally, considering the final state interactions of the hadrons, we obtain the invariant mass distribution for the $\Lambda_c^+\to\pi^+MB$ process. In the following, we will discuss these three steps separately.

\subsubsection{Weak decay}


We first discuss the decay of $\Lambda_{c}$ to produce $\pi^{+}$ and the $sud$ cluster in the final state. The Cabibbo favored weak processes are given by
\begin{align}
   &c\to s + u+\bar{d} \quad \text{: $c$ decay}\label{eq:cdecay} ,\\
   &c + d \to s + u \quad \text{: $cd$ scattering}\label{eq:cdscatt} .
\end{align}
The diagram shown in Fig.~\ref{fig:Lambdac_decay} is obtained by the $c$  decay process. Another contribution with the $c$ decay is to form $\pi^{+}$ using the $u$ quark in $\Lambda_{c}$ [Fig.~\ref{fig:other_decay}(a)]. In this process, however, the color of the $u\bar{d}$ pair from the $W^{+}$ decay is constrained to form the color singlet $\pi^{+}$. This process is therefore suppressed by the color factor in comparison with Fig.~\ref{fig:Lambdac_decay}. In addition, because the $ud$ diquark in $\Lambda_{c}$ is the most attractive ``good'' diquark~\cite{Jaffe:2004ph}, the process to destruct the strong diquark correlation [Fig.~\ref{fig:other_decay}(a)] is not favored. The contribution from the $cd$ scattering Eq.~\eqref{eq:cdscatt} [Fig.~\ref{fig:other_decay}(b) and (c)] is suppressed by the additional creation of a $\bar{q}q$ pair not attached to the $W$-boson as well as the $1/N_{c}$ suppression, compared with Fig.~\ref{fig:Lambdac_decay}. Diagrams ~\ref{fig:other_decay}(b) and ~\ref{fig:other_decay}(c) are called ``absorption diagrams" in the classification of Ref.~\refcite{Chau:1982da}, and they are two body processes, involving two quarks of the original $\Lambda_c$,
which are suppressed compared to the one body process (Fig.~\ref{fig:Lambdac_decay}) involving only the $c$ quark, the $u$, $d$ quarks acting as
spectators. A discussion of this suppression is done in section~\ref{sec:lowscalarD0toK0S}.

As discussed in Ref.~\refcite{Hyodo:2011js}, the kinematical condition also favors the diagram shown in Fig.~\ref{fig:Lambdac_decay}. When the $\Lambda_{c}$ decays into $\pi^{+}$ and $MB$ system with the invariant mass of 1400 MeV, the three momentum of the final state is $\sim 700$ MeV in the rest frame of $\Lambda_{c}$. Thus, the $\pi^{+}$ should be emitted with a large momentum. It is kinematically favored to create the fast pion from the quarks involved by the weak process, because of the large mass of the $c$ quark. The diagrams Fig~\ref{fig:other_decay}(a) and (c) are suppressed because one of the spectator quarks is required to participate in the $\pi^{+}$ formation.

In this way, the diagram in Fig.~\ref{fig:Lambdac_decay} is favored from the viewpoint of the CKM matrix, color suppression, the diquark correlation, and the kinematical condition. We note that this diagram has a bigger strength than the dominant one of the $\Lambda_{b}\to J/\psi \Lambda(1405)$ decay discussed in the two former sections, in which the weak process contains the necessary Cabibbo suppressed $b\to c$ transition and proceeds via internal emission\cite{Chau:1982da} where the color of every quark in the weak process is fixed.

In this process, because the $ud$ diquark in $\Lambda_{c}$ is the spectator, the $sud$ cluster in the final state is combined as
\begin{align}
\frac{1}{\sqrt{2}}|s(ud-du)\rangle .  \notag
\end{align}
This combination is a pure $I=0$ state. Because the $\bar{q}q$ creation does not change the isospin, we conclude that the dominant contribution for the $\Lambda_{c}\to \pi^{+}MB$ process produces the $MB$ pair in $I=0$. We note that the unfavored diagrams that we neglect can produce the $sud$ state with $I=1$. We will revisit the $I=1$ contribution at the end of Sec.~\ref{sec:result}.

\subsubsection{$\bar{q}q$ creation}

To create the $MB$ final state, we must proceed to hadronize the $sud$ state, creating an extra $\bar{q}q$ pair, as we have done in the former sections. Since the total spin-parity of the $MB$ pair is $J^P=1/2^-$, the $s$ quark should be in $L=1$ after the $c$ quark decay, together with the spectator $ud$ diquark. To achieve the final $MB$ state where all quarks are in the ground state, the $\bar{q}q$ creation must involve the $s$ quark to deexcite into $L=0$. Then the hadronization proceeds as depicted in Fig.~\ref{fig:Lambdac_decay}, where the $s$ quark goes into the meson $M$ and the $ud$ diquark is used to form the baryon $B$. Another possibility, the formation of the baryon from the $s$ quark, is not favored because of the correlation of the good $ud$ diquark and the suppression discussed above by forcing a spectator quark from the $\Lambda_c$ to form the emerging meson. Other possibilities of hadronization are also discussed in Ref.~\refcite{Miyahara:2015cja}, concluding that they are suppressed.

To evaluate the relative fractions of the $MB$ state, we follow the same procedure with Ref.~\refcite{Roca:2015tea}.
Using these hadronic representations, we obtain the meson-baryon states after the $\bar{q}q$ pair production as
\begin{align}
|MB\rangle &= |K^-p\rangle + |\bar{K}^0n\rangle -\frac{\sqrt{2}}{3}|\eta\Lambda\rangle.  \label{eq:hadronstate}
\end{align}
Here we neglect the irrelevant $\eta^{\prime}\Lambda$ channel because its threshold is above 2 GeV. We can see that we obtain the isospin $I=0$ $\bar{K}N$ combination in the phase convention that we use where $|K^-\rangle = -|I=1/2,I_z=-1/2\rangle$.

\subsubsection{Final state interaction}

\begin{figure}[tb]\centering
\includegraphics[scale=0.3]{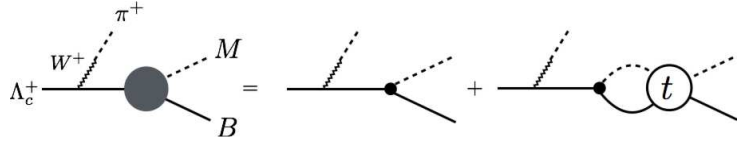}
\caption{The diagram for the meson-baryon final state interaction (filled circle) as the sum of the tree part (dot) and the rescattering part with the meson-baryon scattering amplitude (unfilled circle).}
\label{fig:FSI}
\end{figure}

Here we derive the decay amplitude $\mathcal{M}$, taking the final state
interaction of the $MB$ pair into account. As shown in
Fig.~\ref{fig:FSI}, the final state interaction consists of the tree
part and the rescattering part.
The rescattering of the $MB$ pair is described by the chiral unitary approach~\cite{Kaiser:1995eg,Oset:1997it,Oller:2000fj,Lutz:2001yb,Hyodo:2011ur}, which is based on the chiral Lagrangians and is constructed to treat the non-perturbative phenomena. Though only the $K^-p,\ \bar{K^0}n,\ \eta\Lambda$ states appear in Eq.~\eqref{eq:hadronstate} in the tree-level production, the coupled-channel scattering leads to the other $MB$ states, $\pi^0\Sigma^0$, $\pi^-\Sigma^+$, $\pi^+\Sigma^-$, $\pi^{0}\Lambda$, $K^-p$, $\bar{K^0}n$, $\eta\Lambda$, $\eta\Sigma^0$, $K^+\Xi^-$, $K^0\Xi^0$.\footnote{The $\pi^{0}\Lambda$ and $\eta\Sigma^0$ channels are accessible only through the isospin breaking processes.} The decay amplitude for the $\Lambda_{c}\to \pi^{+}(MB)_{j}$ with the meson-baryon channel $j$ can then be written as Eq.~\eqref{eqn:fullamplitude}, with the same weights for $h_i$. The weak decay and the $\bar{q}q$ pair creation are represented by the factor $V_P$ in Eq.~\eqref{eqn:fullamplitude}, which is assumed to be independent of the invariant mass $M_{\rm inv}$ in the limited range of invariant masses that we consider. Explicit forms for the $t$-matrices of Eq.~\eqref{eqn:fullamplitude} can be found in different works\cite{Kaiser:1995eg,Oset:1997it,Oller:2000fj,Lutz:2001yb,Hyodo:2011ur}. It is also instructive for practical calculations to show the amplitude in the isospin basis. If we assume the isospin symmetry, the amplitude of the decay to the $\pi\Sigma$ and $\bar{K}N$ channels are written as
\begin{align}
\mathcal{M}_{\pi^0\Sigma^0}&=\mathcal{M}_{\pi^-\Sigma^+}=\mathcal{M}_{\pi^+\Sigma^-}
= V_P\left(-\sqrt{\frac{2}{3}} G_{\bar{K}N}t^{I=0}_{\bar{K}N,\pi\Sigma}+\frac{\sqrt{2}}{3\sqrt{3}}G_{\eta\Lambda}t^{I=0}_{\eta\Lambda,\pi\Sigma} \right),  \label{eq:amplitude_isobasis_piSig} \\
\mathcal{M}_{K^-p}&=\mathcal{M}_{\bar{K^0}n} =
V_P\left(1+G_{\bar{K}N}t^{I=0}_{\bar{K}N,\bar{K}N}-\frac{1}{3}G_{\eta\Lambda}t^{I=0}_{\eta\Lambda,\bar{K}N}
\right).  \label{eq:amplitude_isobasis_KbarN}
\end{align}

The partial decay width of the $\Lambda_{c}$ into the $\pi^{+}(MB)_{j}$ channel is given by
\begin{align}
\Gamma_j &=\int d\Pi_{3}|\mathcal{M}_j|^2 ,
\end{align}
where $d\Pi_{3}$ is the three-body phase space. The invariant mass
distribution is obtained as the derivative of the partial width with
respect to $M_{\rm inv}$. In the present case, because the amplitude
$\mathcal{M}_j$ depends only on $M_{\rm inv}$, the mass distribution
${\rm d}\Gamma_{j}/{\rm d}M_{\rm inv}$ is obtained by integrating
the phase space as
\begin{align}
\frac{{\rm d}\Gamma_j}{{\rm d}M_{\rm inv}}
&=\frac{1}{(2\pi)^3}\frac{p_{\pi^+}\tilde{p}_jM_{\Lambda_c^+}M_j}{M_{\Lambda_c^+}^2}
|\mathcal{M}_j|^2,  \label{eq:massdis}
\end{align}
where $M_j$ is the baryon mass, and $p_{\pi^+}$ and $\tilde{p}_j$
represent the magnitude of the three momentum of the emitted $\pi^+$
by the weak decay in the $\Lambda_c$ rest frame and of the meson of
the final meson-baryon state in the meson-baryon rest frame, respectively.

Since the $\Lambda(1405)$ is mainly coupled to the $\pi\Sigma$ and
$\bar{K}N$ channels, we calculate the invariant mass distribution of
the decay to the  $\pi\Sigma$ and $\bar{K}N$ channels. For the study
of the $\Lambda(1670)$, we also calculate the decay to the $\eta
\Lambda$ channel.

\subsection{Results}  \label{sec:result}  

We present the numerical results of the $MB$ invariant mass spectra
in the $\Lambda_{c}\to \pi^{+}MB$ decay. We first show the results
in the energy region near the $\bar{K}N$ threshold where the
$\Lambda(1405)$ resonance plays an important role. We then discuss
the spectra in the higher energy region with the emphasis of the
$\Lambda(1670)$ resonance. The decay branching fractions of
$\Lambda_{c}$ into different final states are discussed at the end
of this section.

\subsubsection{Spectrum near the $\bar{K}N$ threshold}

To calculate the region near the $\bar{K}N$ threshold quantitatively, the final state interaction of the $MB$ system should be constrained by the new experimental data from the SIDDHARTA collaboration~\cite{Bazzi:2011zj,Bazzi:2012eq}, because the precise measurement of the energy-level shift of kaonic hydrogen significantly reduces the uncertainty of the scattering amplitude below the $\bar{K}N$ threshold. Here we employ the meson-baryon amplitude in Refs.~\refcite{Ikeda:2012au,Ikeda:2011pi}, which implements the next-to-leading order terms in chiral perturbation theory to reproduce the low-energy $\bar{K}N$ scattering data, including the SIDDHARTA constraint. The isospin symmetry breaking is introduced by the physical values for the hadron masses. In this model, the two resonance poles of the $\Lambda(1405)$ are obtained at $1424-26i$ MeV and $1381-81i$ MeV.

\begin{figure}[tb]
\begin{center}
\includegraphics[scale=0.3]{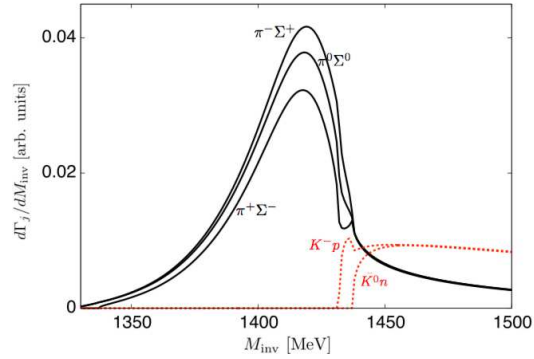}
\caption{(Color online) Invariant mass distribution of the decay $\Lambda_c^+\to \pi^+ MB$ near the $\bar{K}N$ threshold.  The solid line represents the spectrum for $\pi\Sigma$ channels and the dashed line for $\bar{K}N$ channels. The meson-baryon amplitude is taken from the work of Ikeda {\it et al.}\cite{Ikeda:2012au}.}
\label{fig:massdis_IHW}  
\end{center}
\end{figure}

We show the spectra of three $\pi\Sigma$ channels in Fig.~\ref{fig:massdis_IHW}. From this figure, we find the $\Lambda(1405)$ peak structure around 1420 MeV. It is considered that the peak mainly reflects the pole at $1424-26i$ MeV. Because the initial state contains the $\bar{K}N$ channel with vanishing $\pi\Sigma$ component as shown in Eq.~\eqref{eq:hadronstate}, the present reaction puts more weight on the higher energy pole which has the strong coupling to the $\bar{K}N$ channel.

To proceed further, let us recall the isospin decomposition of the $\pi\Sigma$ channels~\cite{Nacher:1998mi}. The particle basis and the isospin basis are related as follows,
\begin{align}
|\pi^0\Sigma^0\rangle &= -\frac{1}{\sqrt{3}}|\pi\Sigma\rangle^{I=0}-\sqrt{\frac{2}{3}}|\pi\Sigma\rangle^{I=2}, \notag \\
|\pi^-\Sigma^+\rangle &= -\frac{1}{\sqrt{3}}|\pi\Sigma\rangle^{I=0}-\frac{1}{\sqrt{2}}|\pi\Sigma\rangle^{I=1}-\frac{1}{\sqrt{6}}|\pi\Sigma\rangle^{I=2}, \label{eq:piSig_rel} \\
|\pi^+\Sigma^-\rangle &= -\frac{1}{\sqrt{3}}|\pi\Sigma\rangle^{I=0}+\frac{1}{\sqrt{2}}|\pi\Sigma\rangle^{I=1}-\frac{1}{\sqrt{6}}|\pi\Sigma\rangle^{I=2}. \notag
\end{align}
In general reactions, the initial state of the $MB$ amplitude is a mixture of the $I=0$ and $I=1$ components.\footnote{In most cases, the small effect of $I=2$ can be neglected.}  The charged $\pi\Sigma$ spectra thus contain the $I=1$ contribution as well as the interference effects of different isospin components.

It is therefore remarkable that all the $\pi\Sigma$ channels have the same peak position in Fig.~\ref{fig:massdis_IHW}. This occurs because the present reaction picks up the $I=0$ initial state selectively, as explained in Sec.~\ref{sec:formulation}. In this case, the $I=1$ contamination is suppressed down to the isospin breaking correction, and hence all the charged states exhibit almost the same spectrum.\footnote{The small deviation is caused by the isospin violation effect in the meson-baryon loop functions.} The differences of the spectra, because of the $I=0$ filter in the present reaction, are much smaller than in photoproduction \cite{Moriya:2013eb,Moriya:2013hwg}, where the explicit contribution of the $I=0$ and $I=1$ channels makes the differences between the different $\pi\Sigma$ channels much larger, even changing the position of the peaks. In this respect, the $\Lambda_{c}\to\pi^{+}\pi\Sigma$ reaction is a useful process to extract information on the $\Lambda(1405)$, because even in the charged states (the $\pi^0 \Sigma^0$ automatically projects over $I=0$) one filters the $I=0$ contribution and the charged states are easier to detect in actual experiments.

The spectra for the $\bar{K}N$ channels are also shown in Fig.~\ref{fig:massdis_IHW}. In the $\bar{K}N$ channels, the peak of the $\Lambda(1405)$ cannot be seen, because the $\bar{K}N$ threshold is above the $\Lambda(1405)$ energy. However, the  enhancement near the threshold that we observe in Fig.~\ref{fig:massdis_IHW} is related to the tail of the $\Lambda(1405)$ peak. The shape of the $\bar{K}N$ spectrum, as well as its ratio to the $\pi\Sigma$ one, is the prediction of the meson-baryon interaction model. The detailed analysis of the near-threshold behavior of the $\bar{K}N$ spectra, together with the $\pi\Sigma$ spectra, will be important to understand the nature of the $\Lambda(1405)$.

\subsubsection{Spectrum above the $\bar{K}N$ threshold}

The spectrum above the $\bar{K}N$ threshold is also interesting. The LHCb collaboration has found that a substantial amount of $\Lambda^{*}$s is produced in the $K^{-}p$ spectrum in the $\Lambda_{b}\to J/\psi K^{-}p$ decay~\cite{Aaij:2015tga}. Hence, the $K^{-}p$ spectrum in the weak decay process serves as a new opportunity to study the excited $\Lambda$ states.

For this purpose, here we adopt the model in Ref.~\refcite{Oset:2001cn} for the meson-baryon final state interaction, which reproduces the $\Lambda(1670)$ as well as the $\Lambda(1405)$ in the $I(J^P)=0(1/2^-)$ channel. The pole position of the $\Lambda(1670)$ is obtained at $1678-20i$ MeV.\footnote{The present pole position is different from the one of the original paper\cite{Oset:2001cn}. This is because the original pole position is calculated with physical basis though the present position is with isospin basis.}
Since the width of the $\Lambda(1670)$ is narrow, the pole of the $\Lambda(1670)$ also affects the invariant mass distribution of the $\Lambda_c^+$ decay.

\begin{figure}[tb]
\begin{center}
\includegraphics[scale=0.4]{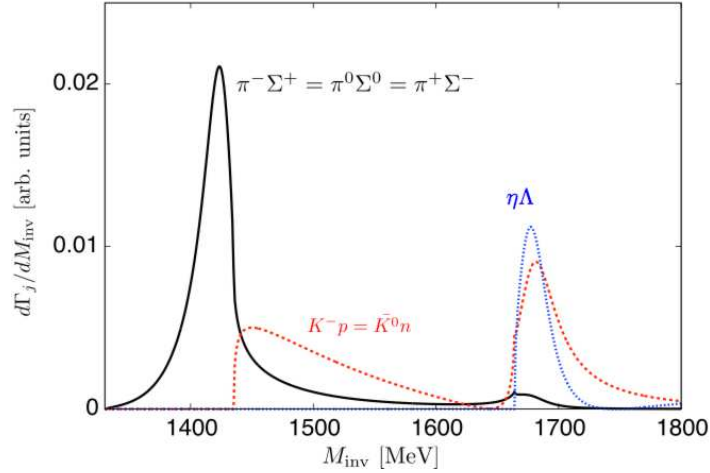}
\caption{(Color online) Invariant mass distribution of the decay $\Lambda_c^+\to \pi^+ MB$. The solid, dotted, and dash-dotted lines represent the $\bar{K}N=\{K^-p,\ \bar{K^0}n\}$, $\pi\Sigma=\{\pi^0\Sigma^0,\ \pi^-\Sigma^+,\ \pi^+\Sigma^-\}$, and $\eta\Lambda$ channels, respectively. The meson-baryon amplitude is taken from Oset {\it et al.}\cite{Oset:2001cn} where the $\Lambda(1520)$ contribution in $d$ wave is not included.}
\label{fig:massdis_ORB}  
\end{center}
\end{figure}

In Fig.~\ref{fig:massdis_ORB}, we show the invariant mass distribution of the $\Lambda_c^+$ decay into the $\pi\Sigma$, $\bar{K}N$ and $\eta\Lambda$ channels. Because the meson-baryon amplitude in Ref.~\refcite{Oset:2001cn} does not include the isospin breaking effect, all the isospin multiplets $\{K^{-}p,\bar{K}^{0}n\}$, $\{\pi^{0}\Sigma^{0},\pi^{+}\Sigma^{-},\pi^{-}\Sigma^{+}\}$ provide an identical spectrum. Because the $\Lambda(1520)$ resonance in $d$ wave is not included in the amplitude, such contribution should be subtracted to compare with the actual spectrum.

As in the previous subsection, we find the $\Lambda(1405)$ peak structure in the $\pi\Sigma$ channel and the threshold enhancement in the $\bar{K}N$ channel. Furthermore, in the higher energy region, we find the additional peak structure of the $\Lambda(1670)$ around 1700 MeV in all channels. Especially, the peak is clearly seen in the $\bar{K}N$ and $\eta\Lambda$ channels, as a consequence of the stronger coupling of the $\Lambda(1670)$ to these channels than to the $\pi\Sigma$ channel\cite{Oset:2001cn}. The $\eta\Lambda$ channel is selective to $I=0$, and the $\Lambda(1520)$ production is kinematically forbidden.

We expect that the structure of the $\Lambda(1670)$ can be analyzed from the measurements of the $\Lambda_c^+$ decay to the $\bar{K}N$ and $\eta\Lambda$ channels.
\subsubsection{Branching fractions}

Experimentally, the decay branching fractions of $\Lambda_{c}\to \pi^{+}MB$ are determined as\cite{Agashe:2014kda}:
\begin{align}
\Gamma(\Lambda_{c}\to pK^{-}\pi^{+},\text{nonresonant}) &= 2.8\pm 0.8 \%
\label{eq:pKpi}\\
\Gamma(\Lambda_{c}\to \Sigma^{+}\pi^{+}\pi^{-}) &= 3.6\pm 1.0 \% \\
\Gamma(\Lambda_{c}\to \Sigma^{-}\pi^{+}\pi^{+}) &= 1.7\pm 0.5 \% \\
\Gamma(\Lambda_{c}\to \Sigma^{0}\pi^{+}\pi^{0}) &= 1.8\pm 0.8 \%
\end{align}
where the nonresonant component is obtained by subtracting the contributions from the $K^{*}(892)^{0}$, $\Delta(1232)^{++}$, and $\Lambda(1520)$ in the $K^{-}\pi^{+}$, $p\pi^{+}$, and $pK^{-}$ spectra, respectively. Taking the ratios of the central values, we obtain
\begin{align}
\frac{\Gamma(\Lambda_{c}\to \Sigma^{+}\pi^{+}\pi^{-})}
{\Gamma(\Lambda_{c}\to pK^{-}\pi^{+},\text{nonresonant}) } &= 1.29 \\
\frac{\Gamma(\Lambda_{c}\to \Sigma^{-}\pi^{+}\pi^{+})}
{\Gamma(\Lambda_{c}\to pK^{-}\pi^{+},\text{nonresonant}) } &= 0.61 \\
\frac{\Gamma(\Lambda_{c}\to \Sigma^{0}\pi^{+}\pi^{0})}
{\Gamma(\Lambda_{c}\to pK^{-}\pi^{+},\text{nonresonant}) } &= 0.64~.
\end{align}

In principle, these ratios can be calculated in the present model by integrating Eq.~\eqref{eq:massdis} over $M_{\rm inv}$. However, in the present calculation, we consider the process which is dominant in the small $M_{\rm inv}$ region, as explained in Sec.~\ref{sec:formulation}. At large $M_{\rm inv}$ region, processes other than those considered can contribute. Also, higher excited $\Lambda^{*}$ states and resonances in the $\pi^{+} M$ and $\pi^{+} B$ channels may play an important role, as shown in the former section.\footnote{The largest contributions from $K^{*}$, $\Delta$ and $\Lambda(1520)$ are subtracted in the data of Eq.~\eqref{eq:pKpi}.} In this way, the validity of the present framework is not necessarily guaranteed for the large $M_{\rm inv}$ region.

Nevertheless, it is worth estimating the branching ratios by simply extrapolating the present model. The theoretical estimate of the ratio of the decay fraction is obtained as
\begin{align}
\frac{\Gamma_{\pi^-\Sigma^+}}{\Gamma_{K^-p}} =
\begin{cases}
1.05 & (\text{Ref.~\refcite{Ikeda:2012au}}) \\
0.95 & (\text{Ref.~\refcite{Oset:2001cn}})
\end{cases} .
\label{eq:decayratio}
\end{align}
Given the uncertainties in the experimental values and the caveats
in the extrapolation discussed above, it is fair to say that the
gross feature of the decay is captured by the present model. We note
that the difference of the charged $\pi\Sigma$ states in our model
is of the order of the isospin breaking correction. The large
deviation in the experimental data, albeit nonnegligible
uncertainties, may indicate the existence of mechanisms which are not included in the present framework. It is worth noting that in
the theoretical model of Ref.~\refcite{Ikeda:2012au} the $\pi^-
\Sigma^+ \pi^+$ channel has the largest strength as in the
experiment.

Let us also mention the measured value of the branching fraction $\mathcal{B}(\Lambda_{c}\to \Lambda\pi^{+}\pi^{0}) = 3.6\pm 1.3
\%$~\cite{Agashe:2014kda}. Because $\pi^{0}\Lambda$ is purely in
$I=1$, the present model does not provide this decay mode. The
finite fraction of this mode indicates the existence of other
mechanisms than the present process. In other words, the validity of
the present mechanism for the $I=0$ filter can be tested by
measuring the $\pi^{0}\Lambda$ spectrum in the small $M_{\rm inv}$
region. We predict that the amount of the $\pi^{0}\Lambda$ mode
should be smaller than the $\pi\Sigma$ mode, as long as the small
$M_{\rm inv}$ region is concerned.

\section{Repercussions for the pentaquark state of LHCb}

 Although baryons with open charm and open beauty have already been found, the recent experiment of Ref.~\refcite{Aaij:2015tga} that finds a neat peak in the $J/\psi\,p$ invariant mass distribution from the $\Lambda_b \to J/\psi K^- p$ decay, is the first one to report on a hidden charm baryon state. Although two states are reported from the $J/\psi\, p$ invariant mass distribution, the first one, at lower energies, is quite broad and one does not see any peak in that distribution. However, broad peaks are seen when cuts are done in the $K^- p$ invariant mass. On the other hand, the hidden charm state around 4450~MeV,
called pentaquark $P_c(4450)^+$  in the experimental work Ref.~\refcite{Aaij:2015tga},
shows up as a clear peak in this distribution, with a width of about 39$\pm5\pm19$ MeV, and this is the state we would like to discuss in this section.  We shall take the work of Ref.~\refcite{Xiao:2013yca} as reference. We find there, in the $I=1/2$ sector, one state  of $J^P=3/2^-$ mostly made of $\bar{D}^*\Sigma_c$ at 4417 MeV, with a width of about 8 MeV, which has a coupling to $J/\psi\, N$, $g=0.53$, and another one, mostly made of $\bar{D}^*\Sigma_c^*$ at 4481 MeV and with a width of about 35 MeV, which has a coupling to $J/\psi\, N$, $g=1.05$. The $3/2^-$ signature is one of the possible spin-parity assignments of the observed state and its mass is in between these two predictions, although one must take into account that a mixture of states with $\bar{D}^* \Sigma_c$ and  $\bar{D}^*\Sigma^*_c$ is possible according to Ref.~\refcite{Garcia-Recio:2013gaa,Uchino:2015uha}.

 On the other hand, in section \ref{sec:LambdabJPsiLambda1405} we have discussed the  $\Lambda_b \to J/\psi K^- p$ reaction and more concretely, $\Lambda_b \to J/\psi \Lambda(1405)$. Interestingly, the work of Ref.~\refcite{Aaij:2015tga} also sees a bump in the $K^- p$ invariant mass distribution just above the  $K^- p$ threshold which is interpreted as due to $\Lambda(1405)$ production.

In this section we combine the information obtained from the experiment on the
$K^- p$ invariant mass distribution close to threshold and the strength of the peak in the $J/\psi\, p$ spectrum and compare them to the theoretical results that one obtains combining the  results of these two former works. We find a  $K^- p$ invariant mass distribution above the  $K^- p$ threshold mainly due to the $\Lambda(1405)$ which is in agreement with experiment, and the strength of this distribution together with the coupling that we find for the theoretical hidden charm state, produces a peak in the $J/\psi\,p$ spectrum which agrees with the one reported in the experiment. These facts together provide support to the idea that the state found could be a hidden charm molecular state of  $\bar{D}^*\Sigma_c-\bar{D}^*\Sigma_c^*$ nature, predicted before by several theoretical groups.


\begin{figure}[tbp]
     \centering
          \includegraphics[width=.6\linewidth]{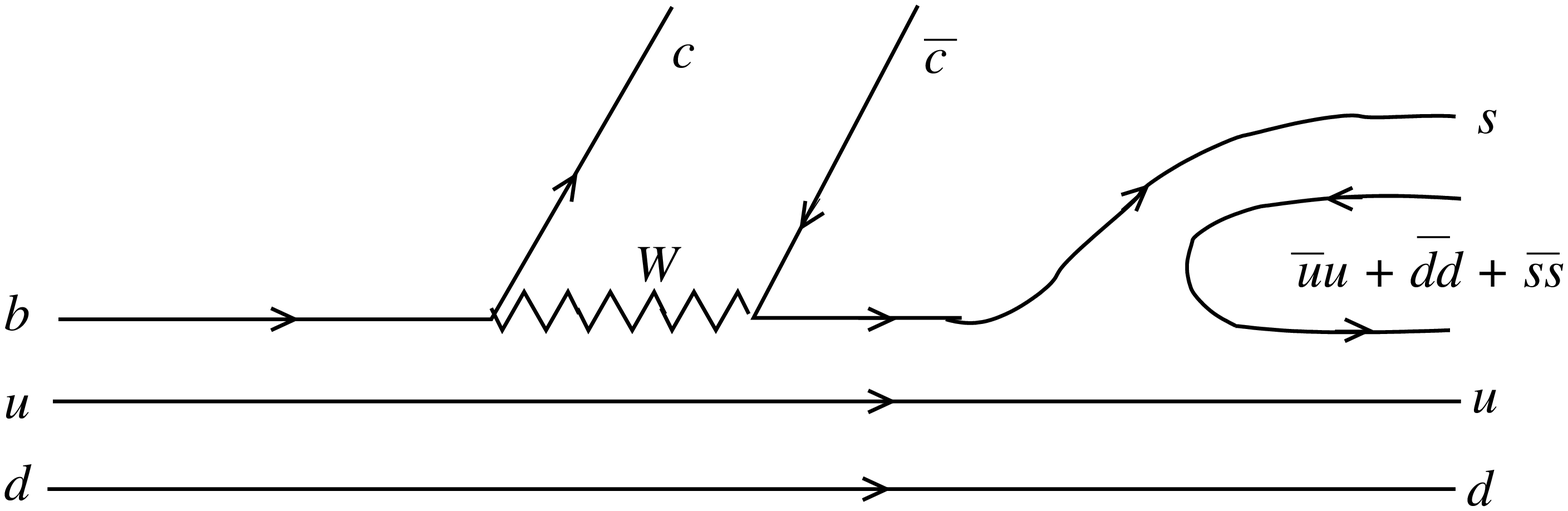}\\
          \includegraphics[width=.3\linewidth]{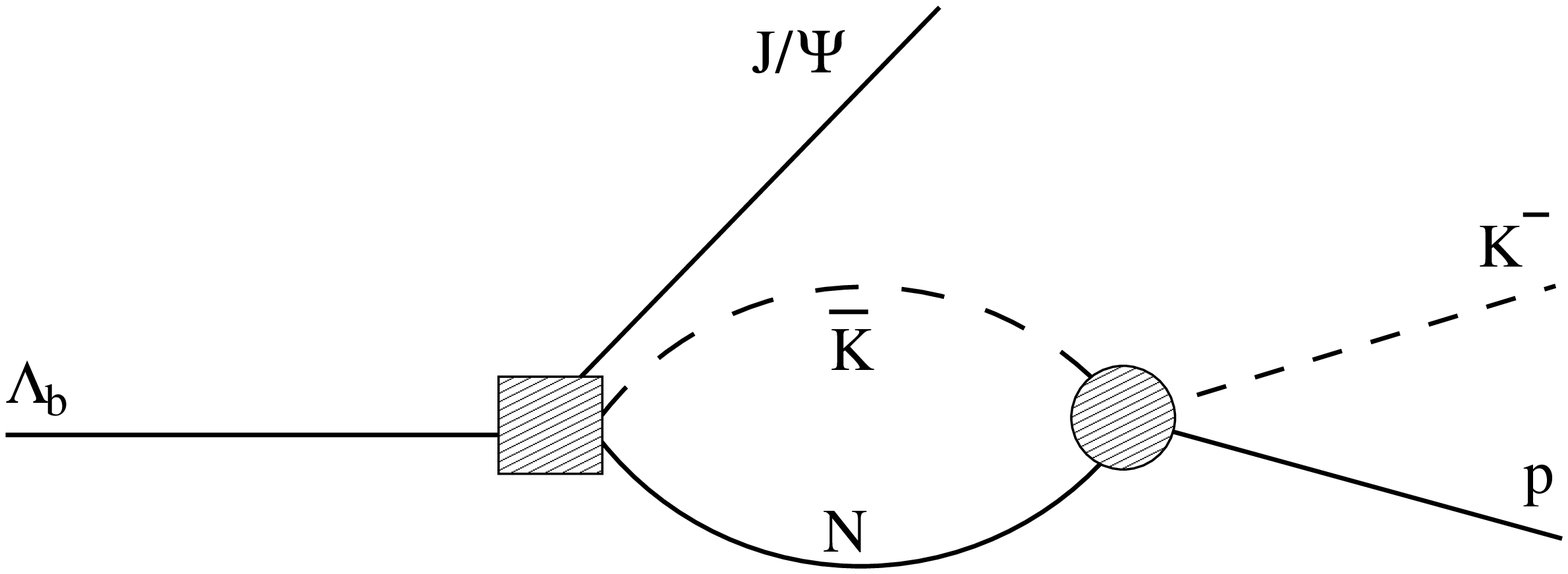}
          \includegraphics[width=.3\linewidth]{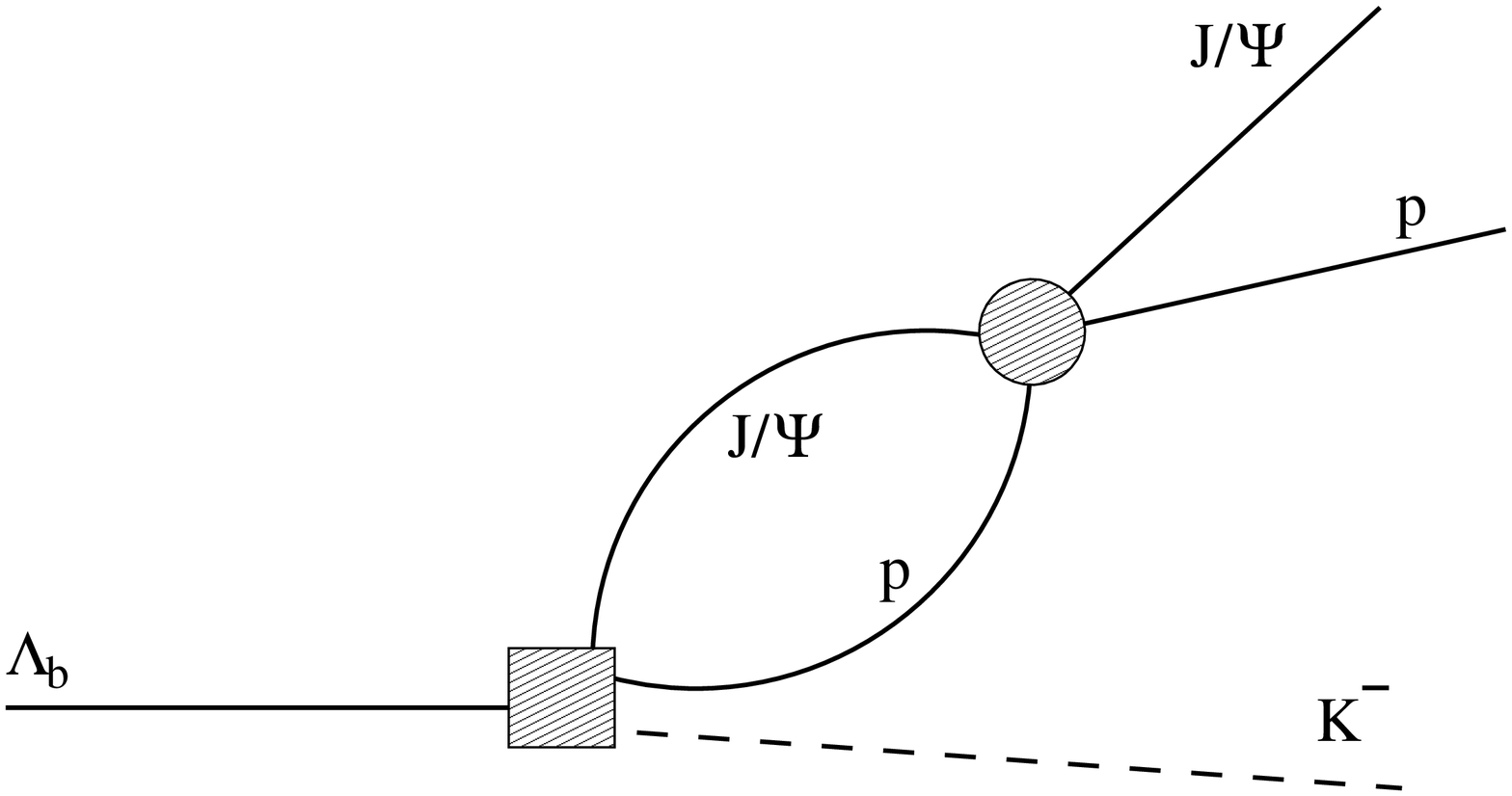}
    \caption{Mechanisms for the $\Lambda_b\to J/\psi K^-p$ reaction implementing the final state interaction\label{fig:diags_LambdabJpsiKp}}
\end{figure}

In Ref.~\refcite{Roca:2015tea}, described in section \ref{sec:LambdabJPsiLambda1405}, it was shown that the relevant mechanisms
for the $\Lambda(1405)$ production in the
decay
are those depicted in
Fig.~\ref{fig:diags_LambdabJpsiKp}. The upper figure shows the basic process to produce a $K^-p$ pair from the weak decay
of the
${\Lambda_b}$. The $u$ and $d$ quarks of the $\Lambda_b$ remain as spectators in the process and carry isospin $I=0$, as in the initial state, producing, together with the $s$ quark, an $I=0$ baryon after the weak process, and hence a meson-baryon system in $I=0$ after the hadronization of the $sud$ state. The  final meson-baryon state then undergoes final state interaction in coupled channels, as shown in the lower left part of Fig.~\ref{fig:diags_LambdabJpsiKp}, from where the $\Lambda(1405)$ is
dynamically produced.
Therefore the contribution to the $\Lambda_b\to J/\psi\,K^- p$ amplitude from the $\Lambda(1405)$ resonance is given by (see Sec.~\ref{sec:LambdabJPsiLambda1405}):

\ba
\label{eqn:MKN} T^{(K^-p)}(M_{K^-p}) = V_p\bigg( h_{K^-p}+ \sum_{i}h_iG_i(M_{K^-p})\,t_{i\,K^-p}(M_{K^-p}) \bigg)\,,
\ea
where $M_{K^-p}$ is the $K^-p$ invariant mass,
$h_i$ are numerical $SU(3)$ factors relating the production of the different meson-baryon channels $i$ in the hadronization [see Eq.~\eqref{eq:juansaysa}],
and $V_p$ accounts for CKM matrix elements and kinematic  prefactors.
Since we do not need the absolute normalization of the invariant mass distributions, the value of $V_p$ can be taken as appropriate, as explained below when discussing the results.
In Eq.~\eqref{eqn:MKN},
$G_i$ represents the meson-baryon loop function and
$t_{ij}$ stands for the s-wave meson-baryon unitarized  scattering amplitudes from Ref.~\refcite{Roca:2013cca}.
 Note that the $\Lambda(1405)$ is not included as an explicit degree of freedom but it appears dynamically in the highly non-linear dynamics involved in the unitarization procedure leading to the $t_{ij}$ amplitudes. Actually two poles are obtained for the $\Lambda(1405)$ resonance
at $1352-48i$~MeV
and $1419-29i$~MeV\cite{Roca:2013cca}. The highest mass $\Lambda(1405)$,
coupling mostly to $\bar K N$, is the one of relevance in the present work since it is closer to the energy region of concern.

On the other hand, in Refs.~\refcite{Xiao:2013yca,Wu:2010vk},
 it was shown that the $J/\psi\,N$ final state interaction in coupled channels, considering also the $\bar D^* \Lambda_c$, $\bar D^* \Sigma_c$, $\bar D \Sigma^*_c$ and $\bar D^* \Sigma^*_c$, produces poles in the $J^P=3/2^-$, $I=1/2$, sector at
$4334+19i\mev$, $4417+4i\mev$ and $4481+17i\mev$ which couple sizeably to $\JP\,p$ (see table II in Ref.~\refcite{Xiao:2013yca}).
 Therefore we can expect to see a resonance shape in the $J/\psi\, p$ invariant mass distribution in the $\Lambda_b\to J/\psi\,K^- p$ decay, maybe a mixture of the different poles. The mechanism for the final
 $\JP\,N$ state interaction is depicted in the lower right part of Fig.~\ref{fig:diags_LambdabJpsiKp}.
The filled circle in that figure represents the final $\JP\,p \to\JP\, p$ unitarized scattering amplitude.
 Since the shape of this amplitude in the real axis is very close to a Breit-Wigner\cite{Xiao:2013yca}, for the numerical evaluation in the present work we can effectively account for it by using


\noindent
\be
t_{\JP\,p \to \JP\,p}= \frac{g^2_{\JP\,p}}{M_{\JP\,p}-M_R+i \frac{\Gamma_R}{2}}
\label{eq:tJpsi}
\ee
where $M_{\JP\,p}$ is the $\JP\,p$ invariant mass and $M_R$ ($\Gamma_R$) the mass (width) of the $P_c(4450)^+$. The amplitudes in Refs.~
\refcite{Xiao:2013yca,Wu:2010vk} provide poles from where $M_R$ and $\Gamma_R$ can be directly obtained, but we fine tune these values to the  experimental results of Ref.~\refcite{Aaij:2015tga}, $M_R=4449.8\mev$ and $\Gamma_R=40\mev$ which lay indeed in between the two heaviest poles obtained in
Ref.~\refcite{Xiao:2013yca}, as quoted above.
In Eq.~\eqref{eq:tJpsi}, $g_{\JP\,p}$ stands for the coupling
 of the dynamically generated resonance to $\JP\,p$, for which
 a range of values from about 0.5 to 1 are obtained in Refs.~\refcite{Xiao:2013yca,Wu:2010vk}, which are genuine and non-trivial predictions of the theory.

The contribution of the $\JP\,p$ final state interaction to the amplitude is then

\ba T^{(\JP\,p)}(M_{\JP\,p}) = V_p\, h_{K^- p}G_{\JP\,p}(M_{\JP\,p})
t_{\JP\,p \to \JP\,p}(M_{\JP\,p})\,, \label{eqn:MJpsip} \ea with
$G_{\JP\,p}$ the $\JP\,p$ loop function regularized by dimensional
regularization as in Ref.~\refcite{Xiao:2013yca}.

Since the main building blocks of the $P_c(4450)^+$ state in Ref.~\refcite{Xiao:2013yca} are
 $\bar D^* \Sigma_c$ and  $\bar D^* \Sigma^*_c$, in principle the main sequence to produce this baryon should be
of the type $\Lambda_b\to K^- \bar D^* \Sigma^*_c \to K^- p J/\psi$ (the argument that follows hold equally for $\Sigma_c$), where one produces  $K^- \bar D^* \Sigma^*_c$ in the first step and the  $\bar D^* \Sigma^*_c \to  p J/\psi$ transition would provide the resonant amplitude accounting for the $P_c(4450)^+$ state in the $\JP\,p$ spectrum. However, as discussed in the former sections and in Ref.~\refcite{Roca:2015dva}, these alternative mechanisms are rather suppressed, and one is thus left to produce the $P_c(4450)^+$ resonance from rescattering of $J/\psi\,p$ after the primary production of $\Lambda_b \to J/\psi K^- p$ through the mechanism depicted in Fig. \ref{fig:diags_LambdabJpsiKp} discussed above. This feature of the reaction is important and is what allows us to relate the $P_c(4450)^+$ production with the $K^- p$ production, i.e, the factor $V_p\, h_{K^- p}$ enters the production of both the $\Lambda(1405)$, via Eq. (\ref{eqn:MKN}), and of the  $P_c(4450)^+$, via Eq. (\ref{eqn:MJpsip}).

\begin{figure}[tbp]
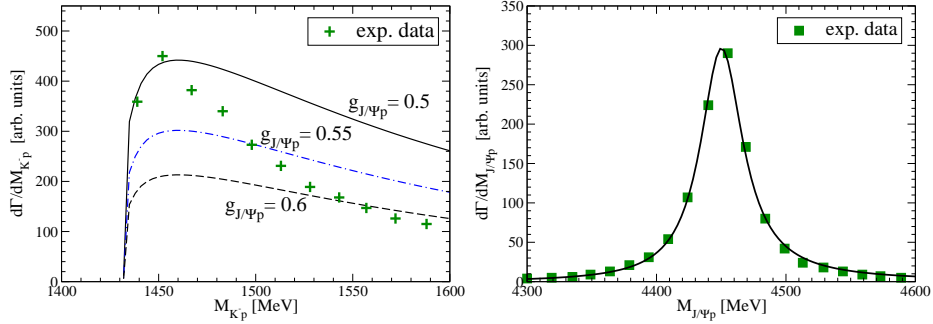

     \centering
          \label{fig:res1}
          \includegraphics[width=.48\linewidth]{figure2a.eps}
              \label{fig:res2}
          \includegraphics[width=.48\linewidth]{figure2b.eps}
    \caption{Results for the $K^- p$ and $\JP\,p$ invariant mass distributions compared to the data\cite{Aaij:2015tga}.}
    \label{fig:res}
\end{figure}

In Fig.~\ref{fig:res} we show the results for the $K^- p$  and
$\JP\,p$  invariant mass distributions compared to the experimental
data of Ref.~\refcite{Aaij:2015tga}. The absolute normalization is arbitrary but
the same for both panels. In the data shown for the $K^- p$ mass
distribution  only the $\Lambda(1405)$ contribution is included,
i.e., it shows the result of the $\Lambda(1405)$ component of the
experimental analysis carried out in Ref.~\refcite{Aaij:2015tga}. Therefore, in order
to compare to this data set, only the amplitude of
Eq.~\eqref{eqn:MKN} is considered. Similarly, the experimental
$\JP\,p$ mass distribution shown in Fig.~\ref{fig:res} (right panel) only
considers the contribution from the $P_c(4450)^+$ and, thus, the
theoretical calculation for Fig.~\ref{fig:res} (right panel)  only includes
the amplitude of Eq.~(\ref{eqn:MJpsip}).

The different curves are evaluated considering different values for
the coupling of the $P_c(4450)^+$  to $J/\psi\,p$,
($g_{\JP\,p}=0.5$, 0.55 and 0.6). For each value of $g_{\JP\,p}$,
$V_P$ has been normalized such that the peak of the $\JP\,p$
distribution agrees with experiment, and this is why there is only
one curve for the $\JP\,p$ mass distribution. Since
the higher  $\Lambda(1405)$ resonance lays below the $K^- p$
threshold, the accumulation of strength close to threshold
is due to the tail of that resonance.

The results are very sensitive to the value of the $\JP\,p$ coupling
since the $J/\psi\,p$ partial decay width is proportional to
$g_{\JP\,p}^4$. We can see in the figure that a value for the
coupling of about 0.5 can account fairly for the relative strength
between the $ \JP\,p$ and $K^- p$ mass distributions. This value of
the coupling is of the order obtained in the extended local hidden
gauge unitary approach of Refs.~\refcite{Xiao:2013yca,Wu:2010vk} which
is a non-trivial output of the theoretical model since the
value of this coupling is a reflection of the highly non-linear
dynamics involved in the unitarization of the scattering amplitudes.

It is also worth noting that the values of $g_{\JP\,p}$ used, lead
to a partial decay width of $P_c(4450)^+$ into $\JP\,p$ ($\Gamma=M_N
g_{\JP\,p}^2~p_{J/\psi}/(2\pi M_R)$) of 6.9 MeV, 8.3 MeV, 9.9 MeV,
which are of the order of the experimental width, but smaller as it
should be, indicating that this channel is one of the relevant ones
in the decay of the $P_c(4450)^+$ state.

The fact that we can fairly reproduce the relative strength of the mass distributions with values of the coupling in the range predicted by the coupled channels unitary approach, provides  support to the  interpretation of the $P_c(4450)^+$ state as dynamically generated from the coupled channels considered and to the $3/2^-$ signature of the state.

 The findings of Ref.~\refcite{Aaij:2015tga} prompted the work of Ref.~\refcite{Chen:2015loa} where, using a boson exchange model\cite{Yang:2011wz}, molecular structures of  $\bar D^* \Sigma_c$ and $\bar D^* \Sigma^*_c$ are also obtained with similarities to our earlier work of Ref.~\refcite{Xiao:2013yca}. However, the interrelation between the $J/\psi\,p$ and $K^- p$ invariant mass distributions is not addressed in Ref.~\refcite{Chen:2015loa}.
 
 The experimental observation of Ref.~\refcite{Aaij:2015tga} has prompted quite a few works aiming at interpreting those results with different models. It is not our purpose to discuss them here. A compilation of all these different works can be seen in Ref.~\refcite{Wang:2015wsa}.

\section{Further developments}
   The developments in this area in the last year have been spectacular, as shown by the different problems discussed in this review. The agreement of the results with experiment when data are available, using the approach discussed all along, has been reasonably good, and many predictions have been made for other observables that are at reach in the different facilities where the experiments have been performed. The fast experimental developments in the present  facilities and the prospects for new facilities that are now under construction, make it a fertile land to apply these theoretical tools and there is much to learn. 

  In this last section we would like to make a very short review of other problems that we have not reviewed here and which are under study or just recently finished at the time the review was written. 

   In Ref.~\refcite{miguemarina}  the $B^+$ decay into $D_s^- K^+ \pi ^+ $ is been studied in order to learn about the $D_0^*(2400)$ resonance.  

   In Ref.~\refcite{miguejido} the $B^+ \to \bar D^0 D^0 K^+ $, $B^0 \to \bar D^- D^0 K^+ $ and  
$B_s^+ \to \bar D^0  K^- \pi ^+$ are studied. In this case the aim is to see how the $D_{s0}^*(2317)$ resonance is formed and learn about the $KD$ molecular structure which has been determined in lattice calculations\cite{Torres:2014vna}. 

    Further developments are done in Ref.~\refcite{liangrio} where the  $B^0$ and  $B_s^0$ decays to $J/\psi K \bar K$ are investigated to compare with measurements done and under analysis at LHCb.

     The advent of the LHCb pentaquark experiment has also prompted the investigation of another reaction\cite{Chen:2015sxa}, $\Xi_b^- \to J/\psi K^- \Lambda$, where using the results of Ref.~\refcite{Wu:2010vk}, where a hidden charm with strangeness is predicted, invariant mass distributions of $K^- \Lambda$ and $J/\psi \Lambda$ are evaluated and a neat peak in the $J/\psi \Lambda$ invariant mass distribution is observed.

  The semileptonic $\Lambda_c \to \nu_l l^+ \Lambda(1405)$ is  addressed in Ref.~\refcite{ikeno}.

  The $B^0 \to D^0 \bar D^0 K^0$ reaction is studied in Ref.~\refcite{daivalen} in order to find evidence for a bound state of $D \bar D$ predicted in Ref.~\refcite{Gamermann:2006nm}.

    The $D_s^+ \to \pi \pi \pi$ and $D_s^+ \to  \pi K \bar K$ reactions are investigated in Ref.~\refcite{jorgivan} to compare with existing and coming data of LHCb. 

   A study is also conducted for the $B_s^0 \to J/\psi f_1(1285)$ reaction in Ref.~\refcite{raqueli} suggesting a model independent method to find the molecular component of resonances.

  The $\Lambda_b  \to \bar D_s \Lambda_c(2595)$ is also investigated in Ref.~\refcite{melaliang}.

   Finally, an incursion is also done in $B_c$ states\cite{pedrosun}, studying the $B_c \to J/\psi D_s^{*-}$ reaction in order to learn about the $D_{s0}^*(2317)$ state. 

\section{Conclusions}

  We do not want to draw conclusions on each of the many subjects dealt along this review. We can recall the basic lessons learned from this general overview. The decays studied have shown that  
weak decays, even when they do not conserve parity and isospin, are many times better filters for isospin or other quantities than strong or electromagnetic interactions. Selection rules as OZI, Cabibbo allowed or suppressed processes, details on the hadronization, etc\ldots, have as a consequence that one can isolate certain quantum numbers at the end, allowing a better study of some resonances or aspects of the hadron interaction. The selection rules and the hadronization of the quarks formed in the primary step lead to pairs of hadrons with very specific weights which allow to understand the basic features of some reactions. Particular relevance have some processes where one looks for a pair of mesons which are not produced in a primary step. In this case it is only the rescattering of the primary mesons produced what gives rise to this hadron pair in the final state. Hence, the amplitude for the process is directly proportional to the scattering amplitudes of these hadrons and one gets rid of unwanted background which could blur the interpretation of the process. If resonances are produced, this gives us a way to learn about their couplings to these primary channels. 

    We have seen that one can learn about properties of resonances, and particularly, when one is dealing with resonances which are deemed as dynamically generated from the interaction of other hadrons, one can even find from the data the amount of molecular component. 

   When dealing with charmed particles, the study of these processes allows to learn about the interaction of these hadrons. In the absence of D-meson beams, unlike for pions or kaons, the study of this final state interaction is our only source of information on the interaction of the charmed hadrons. 

    As to light mesons, the study done here presents further evidence to that gathered from other processes, that the light scalars are generated from the interaction of pseudoscalar mesons, while the vector mesons respond very well to the standard picture of $q \bar q$ states. Other mesons, scalar and tensor, that are theoretically produced from the interaction of vector mesons or a vector and a pseudoscalar, were also investigated, and support for this picture was also obtained.

\section*{Acknowledgments}
We would like to thank C. Hanhart and S. Stone for valuable comments
on the manuscript.  
One of us, E. O., wishes to acknowledge support
from the Chinese Academy of Science in the Program of
Visiting Professorship for Senior International Scientists
(Grant No. 2013T2J0012). This work is partly supported by the Spanish Ministerio
de Economia y Competitividad and European FEDER funds under the
contract numbers FIS2011-28853-C02-01, FIS2011- 28853-C02-02,
FIS2014-57026-REDT, FIS2014-51948-C2- 1-P, and FIS2014-51948-C2-2-P,
and the Generalitat Valenciana in the program Prometeo II-2014/068.
This work is also partly supported by the National Natural Science
Foundation of China under Grant Nos. 11165005, 11565007, 11475227, 11375080 and 11575076. We acknowledge the support of the European
Community-Research Infrastructure Integrating Activity Study of
Strongly Interacting Matter (acronym HadronPhysics3, Grant Agreement
n. 283286) under the Seventh Framework Programme of EU. It is also
supported by the Open Project Program of State Key Laboratory of
Theoretical Physics, Institute of Theoretical Physics, Chinese
Academy of Sciences, China (No.Y5KF151CJ1). M.~D. gratefully acknowledges support from the NSF/PIF grant No. PHY 1415459 and the NSF/Career grant No. 1452055.

\end{document}